\documentclass{article}
\usepackage[T1]{fontenc}
\usepackage{authblk}
\usepackage{geometry}
\usepackage{amsfonts}
\usepackage{graphicx,amssymb,mathrsfs,amsmath,color,fancyhdr}
\usepackage{subcaption}
\usepackage{amsthm}
\usepackage{manfnt}
\usepackage{cases}
\usepackage{mathbbol}
\usepackage{bm}
\usepackage{setspace}
\usepackage[justification=centering]{caption}
\usepackage{hyperref}
\hypersetup{colorlinks = True,linkcolor = blue,anchorcolor =red,citecolor = blue,filecolor = red,urlcolor = red, pdfauthor=author}
\newtheorem{theorem}{Theorem}
\newtheorem{definition}{Definition}

\newtheorem{proposition}{Proposition}
\newtheorem{remark}{Remark}

\title{Trading Large Orders in the Presence of Multiple High-Frequency Anticipatory Traders}
\author[1]{Ziyi Xu}
\author[1]{Xue Cheng \thanks{chengxue@pku.edu.cn}}
\affil[1]{\footnotesize LMEQF, Department of Financial Mathematics, \authorcr School of Mathematical Sciences, \authorcr
Peking University, Beijing 100871, China.}
\date{}
\begin{document}
\maketitle
\begin{abstract}
We investigate a market with a normal-speed informed trader (IT) who may employ mixed strategy and multiple anticipatory high-frequency traders (HFTs) who are under different inventory pressures, in a three-period Kyle's model. The pure- and mixed-strategy equilibria are considered and the results provide recommendations for IT's randomization strategy with different numbers of HFTs. Some surprising results about investors' profits arise: the improvement of anticipatory traders' speed or a more precise prediction may harm themselves but help IT.
\end{abstract}

\noindent\textbf{Keywords:} High-frequency anticipatory trading; Multiple high-frequency traders; Mixed strategy; Randomization; Small informed trader; Round-Tripper

\noindent\textbf{JEL Classification: }G14, G17

\tableofcontents

\newpage
\section{Introduction}
As mentioned in SEC's technical report \cite{sec2010}, highly automated exchange systems have changed the current market structure, of which high-frequency trading is a dominant component by any measure. So high-frequency traders (HFTs) have become counterparties of many investors and affected nearly all aspects of market performance.
One important way for HFTs to make profits and interact with other traders is through anticipatory trading. As empirically found in Kirilenko et al. (2017) \cite{2017The} and Hirschey (2020) \cite{hirschey2021high}, HFTs predict non-HFTs' order flow and trade ahead of them. How other investors are affected by HFTs' anticipatory trading and how they should respond to it are widely concerned issues in both research and industry.

Yang and Zhu (2020) \cite{yang2020back} is the cornerstone of literature on randomizing large orders to counteract anticipatory trading, in which the back-runners learn information from past informed orders and utilize it in subsequent trading. While, informed traders may add endogenous noise into their orders to confuse back-runners, if the latters' signal about order flow is sufficiently accurate. \cite{yang2020back} theoretically proves that randomization is quite likely to happen and a series of empirical works have verified this: Sa{\u{g}}lam (2020) \cite{sauglam2020order} illustrates that large traders attempt to engage in encryption strategies; Gu et al. (2021) \cite{gu2021strategic} shows that insiders will disguise their orders by splitting; Broggard et al. \cite{brogaard2022preventing} (2022) proves that institutions use multiple brokers to mitigate information leakage; Chakrabarty et al. (2022) \cite{chakrabarty2022order} finds evidence that investors employ hidden orders to conceal information.

As proved by Ro{\c{s}}u (2019) \cite{rocsu2019fast} and Xu and Cheng (2023) \cite{xu2023high}, depending on HFTs' different inventory aversion, they may adopt two different kinds of anticipatory strategies, which are respectively defined as \textit{Small-IT} and \textit{Round-Tripper} in \cite{xu2023high}. Both Small-IT and Round-Tripper trade in the same direction as the predicted order in advance. But when the predicted order arrives, Round-Tripper trades in the opposite direction to provide liquidity back, since she is more restricted by inventory, while Small-IT continues to take liquidity away. 

In this paper, we are concerned about the equilibrium among an informed large trader (IT), multiple anticipatory HFTs and some competitive and risk-neutral dealers, where IT may disguise her order by randomization and HFTs have to manage their inventories. Equilibria under various conditions are investigated and theoretical guidance for large trader's response to HFTs is provided.

The interactions between IT and HFTs are modeled through a three-period Kyle's model \cite{kyle1985continuous}. Specifically, IT trades twice in total, and after IT's first trading, HFTs predict her second order and preempt this order by trading ahead. During the process, HFTs may play the role of Small-IT, Round-Tripper, or both and IT may randomize her orders. So the mixed-strategy or pure-strategy equilibrium exists, depending on model parameters.

Compared to back-runners in \cite{yang2020back} who only trade once with IT's second order, anticipatory traders in this paper are faster. Round-Trippers' behavior is totally different from back-runners in the sense that they provide liquidity for IT's second trade. In contrast, Small-ITs take liquidity as back-runners, but their higher trading speed makes IT more inclined to adopt a mixed strategy. To be specific, the parameter range for the existence of mixed-strategy equilibrium is expanded and IT should add more noise.
Surprisingly, although anticipatory traders become faster, their profits may instead decrease while the profit of IT may actually increase, in the mixed-strategy region.
In addition, we find that more accurate predictions may reduce HFTs' but raise IT's profit. A similar phenomenon appears in \cite{yang2020back},
the difference is that it becomes more common here: it happens for fewer anticipatory traders. 
Furthermore, it is theoretically proved that as the size of high-speed noise trading goes to zero, Small-ITs give up their speed advantage and the equilibrium converges to the one with back-runners.

When HFTs all play the role of Round-Tripper, the above phenomenon still exists.
When there is little high-speed noise trading, the market does not provide enough shelter for IT, increasing HFTs' signal accuracy forces her to trade passively, which causes less price impact for HFTs to make profits.
When there is abundant high-speed noise trading, the market provides enough shelter for IT, in this case, a more accurate signal makes HFTs more willing to provide liquidity and thus may improve IT's profit. 

We also investigate the influences of the number of HFTs on the results. If there are no more than $3$ HFTs and all of them play the role of Round-Tripper, IT will always take a pure strategy, regardless of HFT's signal precision, when the market is active enough.
If all HFTs act as Small-IT and they have precise signals, their number does not influence IT's strategy: IT always takes mixed strategy.
If there are both types of HFTs in the market, IT randomizes less with a higher proportion of Round-Tripper. But randomization is unnecessary only when there are no more than $3$ HFTs and they are all Round-Trippers. Therefore, IT should take a mixed strategy as long as the number of HFTs is greater than $3$.

Another interesting phenomenon found in the market with both types of HFTs is that, when the size of high-speed noise trading decreases to a certain extent, HFTs with little inventory constraint will become ``inverse Round-Trippers'': they supply liquidity in preemptive trading and consume liquidity later. 

As will be discussed in Section \ref{secrl}, compared with parallel literature, the contribution of this paper mainly lies in a systematical study of the equilibria between a large trader who can adopt a mixed strategy and multiple HFTs who bear inventory pressures, under a dynamic Kyle's model. The influences of HFTs' types, numbers, compositions, and trading speed are studied and suggestions for large trader's randomization in the above specific situations are provided.

The paper is organized as follows. Section \ref{secrl} illustrates some related literature. In Section \ref{secmodel}, we set up the model. In Section \ref{secmain}, we discuss the equilibria with Small-ITs, Round-Trippers, or both. Section \ref{secconclusion} concludes
and all the proofs are displayed in the \nameref{secappendix}.

\section{Related Literature} 
\label{secrl}
This paper considers the implementation of informed trader's order randomization, as in Huddart, Hughes and Levine (2001) \cite{huddart2001public}, Buffa (2013) \cite{buffa2013insider} and Yang and Zhu (2020) \cite{yang2020back}. But in this paper, the mixed strategy is used to disturb anticipatory HFTs, which is different from \cite{huddart2001public} and \cite{buffa2013insider}, where IT randomizes to respond to disclosure requirements. Compared to \cite{yang2020back}, where anticipatory traders (back-runners) have the same trading speed as IT, they (HFTs) race to trade ahead of IT in this paper and can manage their inventories.

This paper also relates to the theoretical studies on anticipatory trading, which include but are not limited to Brunnermeier and Pedersen (2005) \cite{brunnermeier2005predatory}, Li (2018) \cite{li2018high}, Ro{\c{s}}u (2019) \cite{rocsu2019fast}, Baldauf and Mollner (2020) \cite{baldauf2020high} and Xu and Cheng (2023) \cite{xu2023high}. The most relevant one is \cite{xu2023high}, which only considers pure-strategy equilibrium and a single HFT. This paper considers pure- and mixed-strategy equilibria with 
multiple HFTs.

The comparison between Small-IT in this paper and back-runner in \cite{yang2020back} illustrates that trading speed matters in anticipatory trading. Literature belongs to this topic includes Ro{\c{s}}u (2019) \cite{rocsu2019fast} and Baldauf and Mollner (2020) \cite{baldauf2020high}.  \cite{rocsu2019fast} studies investors who differ in the speed of processing information. 
Only fast traders are able to use immediate information, hence,
the latter's order flow is predicted by the former's.
In \cite{baldauf2020high}, snipers and dealers are fast enough to implement aggressive and passive order anticipation. In this paper, Small-IT and Round-Tripper's fast trading belongs to aggressive order anticipation, while the Round-Tripper's liquidity provision can be seen as passive order anticipation.

This paper also relates to the modeling of high-frequency trading. For this topic, see Hoffmann (2014) \cite{hoffmann2014dynamic}, Budish, Cramton and Shim (2015) \cite{budish2015high}, Foucault, Hombert and Ro{\c{s}}u (2016) \cite{foucault2016news}, Baldauf and Mollner (2020) \cite{baldauf2020high}, and the comprehensive review in Menkveld (2016) \cite{menkveld2016economics}.

\section{The Model and Equilibrium}
\label{secmodel}
In this paper, we consider a four-time-points model, where $t=0,1,2$ are ordinary trading times, while $t=1_{+}\in(1,2)$ is the time at which HFTs preempt normal-speed traders.

A risky asset is traded by four types of participants: competitive and risk-neutral dealers, a normal-speed large informed trader (IT), $J$ HFTs with inventory aversion  $\{\gamma_j\}_{j=1}^J$ and noise traders.

The true value of the asset, $v$, is assumed to be normally distributed as
\begin{equation*}
    v\sim N(p_0,\sigma_v^2),
\end{equation*}
which will be revealed after $t=2.$




For IT, she receives the signal $v$ at $t=0$. In order to reduce market impact, she splits the transaction into two child orders $i_1$ and $i_2.$ $i_1$ is submitted after $t=0$ when she knows $v$ and is executed at $t=1$. $i_2$ is submitted after the execution of $i_1$ and is completed at $t=2.$

For HFTs, after the transaction of $i_1$, they
detect IT's trading and get some information about it. 
The $j$-th HFT is supposed to receive a signal:
\begin{equation*}
\Tilde{i}_{1j}=i_1+\varepsilon_j,
\end{equation*}
where $\varepsilon_j\sim N(0,\sigma_{\varepsilon j}^2)$ is independent of each other and other random variables. We assume $\sigma_{\varepsilon j}=\sigma_{\varepsilon},j=1,...,J,$ which means that different HFTs have little difference in their ability to obtain information. In practice, the information leakage is hardly new.
For example, in Sa{\u{g}}lam (2020) \cite{sauglam2020order}, signals based on the pattern of child orders are employed to predict the transaction of large orders. 

Due to the advantage of trading speed, HFTs can trade twice during $(1,2]$, which are completed at $t=1_+$ and $t=2$ respectively. The corresponding orders of the $j$-th HFT are denoted by $x_{1j}$ and $x_{2j}.$

\begin{remark}
IT is slower than HFTs, in the sense that there may be delays in her executions. This can be from (1) the submission delay: IT sends the order during $(1,1_+)$ but it does not arrive at the exchange until $t=2$, as mentioned in Baldauf and Mollner (2020) \cite{baldauf2020high}; (2) the information delay: IT receives information $y_1$ which is used to decide $i_2$ after $t=1_+$, as mentioned in Foucault, Hombert and Ro{\c{s}}u (2016) \cite{foucault2016news}. Consequently, $i_2$ can be submitted before or after $t=1_+,$ but it cannot be completed before $t=2.$
\end{remark}

IT's and HFTs' orders are accompanied by noise orders, $u_1\sim N(0,\sigma_1^2),u_{1_+}\sim N(0,\sigma_{1_+}^2),u_2\sim N(0,\sigma_2^2)$, are independent of each other and other random variables. We naturally assume that $\sigma_1,\sigma_{1_+},\sigma_2>0.$ The order flow at $t=1,1_+,2$ are:
\begin{equation*}
\begin{aligned}
&y_1=i_1+u_1,\\
&y_{1_+}=\sum_{j=1}^J x_{1j}+u_{1_+},\\
&y_2=i_2+\sum_{j=1}^J x_{2j}+u_2.
\end{aligned}
\end{equation*}
\begin{figure}[!htbp]
    \centering
    \includegraphics[width = 0.6\textwidth]{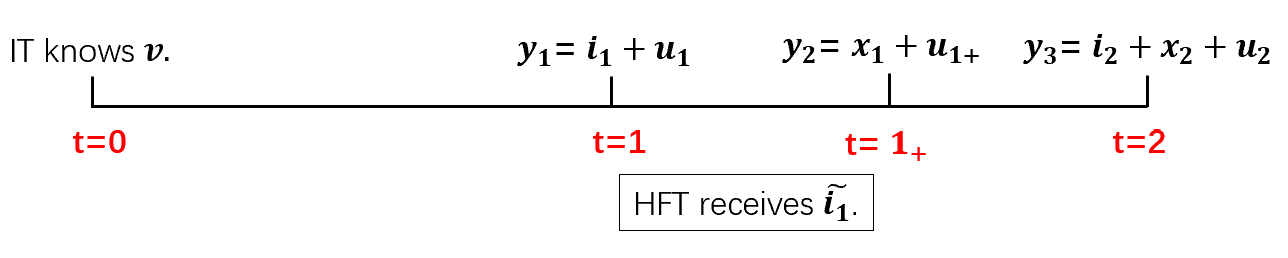}
    \caption{Timeline.}
\end{figure}

We now define the equilibrium.
\begin{definition}
The equilibrium is defined as a collection of strategies of dealers, IT and HFTs:
\begin{equation*}
    \{p_1, p_{1_+},p_2,i_1,i_2,\{x_{1j}\}_{j=1}^J,\{x_{2j}\}_{j=1}^J\},
\end{equation*}
such that the following conditions are satisfied:
\begin{enumerate}
    \item Given IT's strategies $i_1,i_2$ and HFTs' strategies $\{x_{1j}\}_{j=1}^J,\{x_{2j}\}_{j=1}^J$, dealers set price according to the weak-efficiency rule:
    \begin{equation*}
\begin{aligned}
  &p_1=\mathbb{E}(v|y_1),\\
  &p_{1_+}=\mathbb{E}(v|y_1,y_{1_+}),\\
  &p_2=\mathbb{E}(v|y_1,y_{1_+},y_2).\\
\end{aligned}
    \end{equation*}
\item Given dealers' pricing rule $p_1,p_{1_+},p_2$ and HFTs' strategies $\{x_{1j}\}_{j=1}^J,\{x_{2j}\}_{j=1}^J$,
\begin{enumerate}
    \item during $(1,2),$ IT maximizes the second-stage expected profit over all measurable strategies $i_2=i_2(v,i_1,y_1):$
    \begin{equation*}
i_2^*=\arg\max\mathbb{E}(\pi^{\text{IT}}_2|v,i_1,y_1),
    \end{equation*}
        where $\pi^{\text{IT}}_2=(v-p_2)i_2;$
\item during $(0,1),$ in the \textbf{mixed-strategy equilibrium}, IT should be indifferent among all realizations of pure strategies, i.e., $\forall i_1=i_1(v),$
\begin{equation*}
\mathbb{E}(\pi^{\text{IT}}_1+\pi^{\text{IT}}_2|v)
    \end{equation*}
are same, where $\pi^{\text{IT}}_1=i_1(v-p_1).$ In the \textbf{pure-strategy equilibrium}, IT maximizes the total expected profit over all measurable strategies $i_1=i_1(v):$
\begin{equation*}
i_1^*=\arg\max\mathbb{E}(\pi^{\text{IT}}_1+\pi^{\text{IT}}_2|v).
    \end{equation*}
    \end{enumerate}
\item Given dealers' pricing rule $p_1,p_{1_+},p_2$ and IT's strategies $i_1,i_2,$
\begin{enumerate}
    \item during $(1_+,2),$ the $j$-th HFT maximizes the second-stage expected profit with inventory penalty over all measurable strategies $x_{2j}=x_{2j}(\Tilde{i}_{1j},y_1,y_{1_+})$:
        \begin{equation*}
x_{2j}^*=\arg\max\mathbb{E}(\pi_{2j}^{\text{HFT}}-\gamma_j(x_{1j}+x_{2j})^2|\Tilde{i}_{1j},y_1,y_{1_+}),
    \end{equation*}
        where $\pi_{2j}^{\text{HFT}}=(v-p_2)x_{2j};$
\item during $(1,1_+),$ the $j$-th HFT maximizes the total expected profit with inventory penalty over all measurable strategies $x_{1j}=x_{1j}(\Tilde{i}_{1j},y_1)$:
        \begin{equation*}
x_{1j}^*=\arg\max\mathbb{E}(\pi_{1j}^{\text{HFT}}+\pi_{2j}^{\text{HFT}}-\gamma_j(x_{1j}+x_{2j})^2|\Tilde{i}_{1j},y_1),
    \end{equation*}
where $\pi_{1j}^{\text{HFT}}=(v-p_{1_+})x_{1j}.$
\end{enumerate}
\end{enumerate}
\end{definition}


To solve the equilibrium, we apply the conjecture and verify method. In the mixed-strategy equilibrium, we conjecture that market participants take the following strategies:
\begin{equation*}
\begin{aligned}
&p_1=p_0+\lambda_1y_1,\\
&p_{1_+}=p_1+\lambda_{1_+}y_{1_+},\\
&p_2=p_1+\lambda_{21}y_{1_+}+\lambda_{22}y_2;\\
&i_1=\alpha_1(v-p_0)+z,\\
&i_2=\alpha_{21}(v-\mathbb{E}(v|y_1))+\alpha_{22}(i_1-\mathbb{E}(i_1|y_1));\\
&x_{1j}=\beta_{1j}(\Tilde{i}_{1j}-\mathbb{E}(\Tilde{i}_{1j}|y_1)),\\
&x_{2j}=\beta_{21j}(\Tilde{i}_{1j}-\mathbb{E}(\Tilde{i}_{1j}|y_1))+\beta_{22j}(\sum_{k\neq j}x_{1k}+u_{1_+})+\beta_{23j}x_{1j},
\end{aligned}
\end{equation*}
where $z\sim N(0,\sigma_z^2)$ is the endogenous noise added by IT, and $\sigma_z=0$ in the pure-strategy equilibrium.

For $x_{2j},$ the first term is HFT's reliance on the signal $\Tilde{i}_{1j}$. The second term is the adjustment to the information disclosed by other investors' trading at $t=1_+$: $y_{1_+}-x_{1j}$. The last term is the adjustment to (1) the inventory penalty, since $x_{1j}$ has established some positions; (2) the information disclosed by HFT's own trading. 
$y_{1_+}$ and $y_2$ are both independent of $y_1,$ but $y_2$ is not independent of $y_{1_+}$, it is because the preemptive transaction of HFT has contained part of information in $y_2$.

Based on the conjecture, the $j$-th HFT's strategies can also be written as 
\begin{equation*}
\begin{aligned}
&x_{1j}=\beta_{1j}(i_1-\mathbb{E}(i_1|y_1))+\beta_{1j}\varepsilon_j,\\
&x_{2j}=(\beta_{21j}+\beta_{22j}\sum_{k\neq j}\beta_{1k}+\beta_{23j}\beta_{1j})(i_1-\mathbb{E}(i_1|y_1))+(\beta_{21j}+\beta_{23j}\beta_{1j})\varepsilon_j+\beta_{22j}\sum_{k\neq j}\beta_{1k}\varepsilon_k+\beta_{22j}u_{1_+},
\end{aligned}
\end{equation*}
where $\beta_{1j}$ and $\beta_{21j}+\beta_{22j}\sum_{k\neq j}\beta_{1k}+\beta_{23j}\beta_{1j}$ represent the trading directions of $x_{1j}$ and $x_{2j}$, respectively. If the coefficient is positive, we say that HFT tends to trade in the same direction as IT. 


The parameters below will be useful in following sections:
\begin{equation}
\label{params}
\begin{aligned}
&\theta_{1_+}=\sigma_{1_+}^2/\sigma_1^2,\ \text{the relative size of time-}1_+\text{ market;}\\
&\theta_2=\sigma_2^2/\sigma_1^2,\ \text{the relative size of time-2 market;}\\
&\theta_\varepsilon=\sigma_{\varepsilon}^2/\sigma_1^2,\ \text{signal accuracy;}\\
&\Gamma_j=\frac{\gamma_j}{\sigma_v/\sigma_1},\ \text{the dimensionless form of inventory aversion;}\\
&J_j,\ \text{the number of HFTs with inventory aversion } \Gamma_j.
\end{aligned}
\end{equation}
Noise orders come from both normal- and high-speed investors, while noise orders at $1_+$ are mainly from fast traders. Therefore, $\theta_{1_+}$ also measures the size of high-speed noise trading.

\section{Main Results}
\label{secmain}
\textbf{Dealer's quotes.} Competitive and risk-neutral dealers set transaction prices as expectations of $v$, conditioned on the order flow information. 

At $t=1,$ when IT's order $i_1$ is being executed, the transaction price is 
\begin{equation*}
p_1=p_0+\lambda_1y_1,
\end{equation*}
where, by the linear conjecture and projection theorem,
\begin{equation}
\label{lam1}
\lambda_1=\frac{\sigma_{v}}{\sigma_{y_1}}\rho_{(v,y_1)}=\frac{\alpha_1\sigma_v^2}{\alpha_1^2\sigma_v^2+\sigma_z^2+\sigma_1^2}.
\end{equation}

At $t=1_+,$ when HFTs' orders $\{x_{1j}\}_{j=1}^J$ are being executed, the transaction price is
\begin{equation*}
p_{1_+}=p_1+\lambda_{1_+}y_{1_+},
\end{equation*}
where 
\begin{equation}
\label{lam22}
\lambda_{1_+}=\frac{\sigma_{v}}{\sigma_{y_{1_+}}}\rho_{(v,y_{1_+})}=\frac{\alpha_1\sigma_v^2\sigma_1^2\sum_{j=1}^J\beta_{1j}}{(\alpha_1^2\sigma_v^2+\sigma_z^2)[(\sum_{j=1}^J\beta_{1j})^2\sigma_1^2+\sum_{j=1}^J\beta_{1j}^2\sigma_\varepsilon^2+\sigma_{1_+}^2]+\sigma_1^2(\sum_{j=1}^J\beta_{1j}^2\sigma_\varepsilon^2+\sigma_{1_+}^2)}.
\end{equation}

At $t=2,$ when IT's order $i_2$ and HFTs' orders  $\{x_{2j}\}_{j=1}^J$ are being executed, the transaction price is
\begin{equation*}
p_2=p_1+\lambda_{21}y_{1_+}+\lambda_{22}y_2,
\end{equation*}
where 
\begin{equation}
\label{lam32}
\lambda_{21}=\frac{\sigma_{v}}{\sigma_{y_{1_+}}}  \frac{\rho_{(v,y_{1_+})}-\rho_{(y_{1_+},y_2)} \rho_{(v,y_2)}}{1-\rho^{2}_{(y_{1_+},y_2)}},
\end{equation}
\begin{equation}
\label{lam33}
\lambda_{22}=\frac{\sigma_{v}}{\sigma_{y_2}}  \frac{\rho_{(v,y_2)}- \rho_{(v,y_{1_+})}\rho_{(y_{1_+},y_2)}}{1-\rho^{2}_{(y_{1_+},y_2)}},
\end{equation}
\begin{equation*}
\begin{aligned}
&\kappa_1\triangleq\alpha_{21}+\frac{\alpha_1[-\alpha_{21}\alpha_1\sigma_v^2+\sigma_1^2(\alpha_{22}+\sum_{j=1}^J\beta_{21j}+\beta_{23j}\beta_{1j}+\beta_{22j}\sum_{k\neq j}\beta_{1k})]}{\alpha_1^2\sigma_v^2+\sigma_z^2+\sigma_1^2},\\
&\kappa_2\triangleq\frac{\kappa_1-\alpha_{21}}{\alpha_1},\\
&\kappa_3\triangleq\frac{\alpha_{21}\alpha_1\sigma_v^2+(\alpha_1^2\sigma_v^2+\sigma_z^2)(\alpha_{22}+\sum_{j=1}^J\beta_{21j}+\beta_{23j}\beta_{1j}+\beta_{22j}\sum_{k\neq j}\beta_{1k})}{\alpha_1^2\sigma_v^2+\sigma_z^2+\sigma_1^2},\\
&\sigma_{y_{1_+}}^2=\frac{(\alpha_1^2\sigma_v^2+\sigma_z^2)\sigma_1^2}{\alpha_1^2\sigma_v^2+\sigma_z^2+\sigma_1^2}(\sum_{j=1}^J\beta_{1j})^2+\sigma_\varepsilon^2\sum_{j=1}^J\beta_{1j}^2+\sigma_{1_+}^2,\\
&\sigma_{y_2}^2=\sigma_v^2\kappa_1^2+\sigma_z^2\kappa_2^2+\sigma_1^2\kappa_3^2+\sigma_\varepsilon^2\sum_{j=1}^J(\beta_{21j}+\beta_{23j}\beta_{1j}+\beta_{1j}\sum_{k\neq j}\beta_{22k})^2+\sigma_{1_+}^2(\sum_{j=1}^J\beta_{22j})^2+\sigma_2^2,\\
&\text{Cov}(y_{1_+},y_2)=\frac{\sigma_v^2\sigma_1^2\alpha_1\kappa_1\sum_{j=1}^J\beta_{1j}+\sigma_z^2\sigma_1^2\sum_{j=1}^J\beta_{1j}+(\alpha_1^2\sigma_v^2+\sigma_z^2)\sigma_1^2\kappa_3\sum_{j=1}^J\beta_{1j}}{\alpha_1^2\sigma_v^2+\sigma_z^2+\sigma_1^2}\\
&\quad\quad\quad\quad+\sigma_{1_+}^2\sum_{j=1}^J\beta_{22j}+\sigma_\varepsilon^2\sum_{j=1}^J\beta_{1j}(\beta_{21j}+\beta_{23j}\beta_{1j}+\beta_{1j}\sum_{k\neq j}\beta_{22k}),\\
&\text{Cov}(v,y_{1_+})=\frac{\sigma_v^2\sigma_1^2\alpha_1\sum_{j=1}^J\beta_{1j}}{\alpha_1^2\sigma_v^2+\sigma_z^2+\sigma_1^2},\\
&\text{Cov}(v,y_2)=\sigma_v^2\kappa_1,\\
&\rho_{(y_{1_+},y_2)}=\frac{\text{Cov}(y_{1_+},y_2)}{\sigma_{y_{1_+}}\sigma_{y_2}},\rho_{(v,y_{1_+})}=\frac{\text{Cov}(v,y_{1_+})}{\sigma_{v}\sigma_{y_{1_+}}},\rho_{(v,y_2)}=\frac{\text{Cov}(v,y_2)}{\sigma_{v}\sigma_{y_2}}.
\end{aligned}
\end{equation*}

\textbf{HFT's strategies.} Given dealers' quotes $p_1,p_{1_+},p_2$ and IT's strategies $i_1,i_2$, the $j$-th HFT's objective function in period 2 is
\begin{equation*}
\begin{aligned}
&\mathbb{E}\left(\left.\pi_{2j}^{\text{HFT}}-\gamma_j(x_{1j}+x_{2j})^2\right|\Tilde{i}_{1j},y_1,y_{1_+}\right)\\
=&-(\lambda_{22}+\gamma_j)x_{2j}^2+x_{2j}\mathbb{E}(v-p_1-\lambda_{21}(x_{1j}+\sum_{k\neq j}x_{1k}+u_{1_+})-\lambda_{22}(i_2+\sum_{k\neq j}x_{2k})|\Tilde{i}_{1j}-\mathbb{E}(\Tilde{i}_{1j}|y_1),\sum_{k\neq j}x_{1k}+u_{1_+}).
\end{aligned}
\end{equation*}
It is maximized at $x_{2j}=\beta_{21j}(\Tilde{i}_{1j}-\mathbb{E}(\Tilde{i}_{1j}|y_1))+\beta_{22j}(\sum_{k\neq j}x_{1k}+u_{1_+})+\beta_{23j}x_{1j},$ if the SOC
\begin{equation}
\label{SOC1}
\lambda_{22}+\gamma_j>0
\end{equation}
holds, where 
\begin{equation}
\label{beta21j}
\begin{aligned}
 \beta_{21j}&=\frac{1}{2(\lambda_{22}+\gamma_j)}\{(1-\lambda_{22}\alpha_{21})\eta_{21j}\\
 &-\lambda_{22}[\alpha_{22}\mu_{21j}+\sum_{k\neq j}(\beta_{21k}+\beta_{23k}\beta_{1k})\theta_{21jk}+\sum_{k\neq j}\beta_{22k}\sum_{l\neq j,k}\beta_{1l}\theta_{21jl}+\sum_{k\neq j}\beta_{22k}\delta_{21j}]\},   
\end{aligned}
\end{equation}

\begin{equation}
\label{beta22j}
\begin{aligned}
 \beta_{22j}&=\frac{1}{2(\lambda_{22}+\gamma_j)}\{(1-\lambda_{22}\alpha_{21})\eta_{22j}-\lambda_{21}\\
 &-\lambda_{22}[\alpha_{22}\mu_{22j}+\sum_{k\neq j}(\beta_{21k}+\beta_{23k}\beta_{1k})\theta_{22jk}+\sum_{k\neq j}\beta_{22k}\sum_{l\neq j,k}\beta_{1l}\theta_{22jl}+\sum_{k\neq j}\beta_{22k}\delta_{22j}]\},
\end{aligned}
\end{equation}

\begin{equation}
\label{beta23j}
\beta_{23j}=-\frac{\lambda_{21}+2\gamma_j+\lambda_{22}\sum_{k\neq j}\beta_{22k}}{2(\lambda_{22}+\gamma_j)},
\end{equation}
\begin{equation}
\label{etamu}
\begin{aligned}
&\sigma_{1j}^2=\frac{(\alpha_1^2\sigma_v^2+\sigma_z^2)\sigma_1^2}{\alpha_1^2\sigma_v^2+\sigma_z^2+\sigma_1^2}+\sigma_\varepsilon^2,\\
&\sigma_{2j}^2=\frac{(\alpha_1^2\sigma_v^2+\sigma_z^2)\sigma_1^2}{\alpha_1^2\sigma_v^2+\sigma_z^2+\sigma_1^2}(\sum_{k\neq j}\beta_{1k})^2+\sum_{k\neq j}\beta_{1k}\sigma_\varepsilon^2+\sigma_{1_+}^2,\\
&\sigma_{12j}=\frac{(\alpha_1^2\sigma_v^2+\sigma_z^2)\sigma_1^2}{\alpha_1^2\sigma_v^2+\sigma_z^2+\sigma_1^2}\sum_{k\neq j}\beta_{1k},\\
&\Sigma_j\triangleq\begin{pmatrix}
\sigma_{1j}^2 & \sigma_{12j} \\
\sigma_{12j} & \sigma_{2j}^2
\end{pmatrix},\\
&
\begin{pmatrix}
 \eta_{21j}&\eta_{22j}   
\end{pmatrix}=\frac{\alpha_1\sigma_v^2\sigma_1^2}{\alpha_1^2\sigma_v^2+\sigma_z^2+\sigma_1^2}
\begin{pmatrix}
 1&\sum_{k\neq j}
\beta_{1k}   
\end{pmatrix}
\Sigma_j^{-1},\\
&
\begin{pmatrix}
 \mu_{21j}&\mu_{22j}   
\end{pmatrix}=\frac{(\alpha_1^2\sigma_v^2+\sigma_z^2)\sigma_1^2}{\alpha_1^2\sigma_v^2+\sigma_z^2+\sigma_1^2}
\begin{pmatrix}
    1&\sum_{k\neq j}\beta_{1k}
\end{pmatrix}
\Sigma_j^{-1},\\
&
\begin{pmatrix}
 \theta_{21jk}&\theta_{22jk}   
\end{pmatrix}
=\frac{\sigma_1^2}{\alpha_1^2\sigma_v^2+\sigma_z^2+\sigma_1^2}
\begin{pmatrix}
\alpha_1^2\sigma_v^2+\sigma_z^2&\sigma_\varepsilon^2\beta_{1k}+(\alpha_1^2\sigma_v^2+\sigma_z^2)\sum_{k\neq j}\beta_{1k}  
\end{pmatrix}
\Sigma_j^{-1},\\
&\begin{pmatrix}
 \delta_{21j}&\delta_{22j}   
\end{pmatrix}
=\begin{pmatrix}
   0&\sigma_{1_+}^2 
\end{pmatrix}
\Sigma_j^{-1}.\\
\end{aligned}
\end{equation}
HFT's objective function at $t=1_+$ is 
\begin{equation*}
\begin{aligned}
&\mathbb{E}\left(\left.\pi_{1j}^{\text{HFT}}+\pi_{2j}^{\text{HFT}}-\gamma_j(x_{1j}+x_{2j})^2\right|\Tilde{i}_{1j},y_1\right)\\
=&-(\lambda_{1_+}+\gamma_j)x_{1j}^2+(\lambda_{22}+\gamma
_j)\mathbb{E}(x_{2j}^2|\Tilde{i}_{1j},y_1).
\end{aligned}
\end{equation*}
It is maximized at $x_{1j}=\beta_{1j}(\Tilde{i}_{1j}-\mathbb{E}(\Tilde{i}_{1j}|y_1))$ if the SOC
\begin{equation}
\label{SOC2}
\lambda_{1_+}+\gamma_j-(\lambda_{22}+\gamma_j)\beta_{23j}^2>0
\end{equation}
holds, where 
\begin{equation}
\label{beta1j}
\beta_{1j}=\frac{\eta-\lambda_{1_+}\mu\sum_{k\neq j}\beta_{1k}+2(\lambda_{22}+\gamma_j)\beta_{23j}(\beta_{21j}+\beta_{22j}\sum_{k\neq j}\beta_{1k}\mu)}{2[\lambda_{1_+}+\gamma_j-(\lambda_{22}+\gamma_j)\beta_{23j}^2)]},
\end{equation}
\begin{equation*}
\begin{aligned}
&\eta=\frac{\alpha_1\sigma_v^2}{\alpha_1^2\sigma_v^2+\sigma_z^2+\sigma_1^2}\frac{\sigma_1^2}{\sigma_{1j}^2},\\
&\mu=\frac{\alpha_1^2\sigma_v^2+\sigma_z^2}{\alpha_1^2\sigma_v^2+\sigma_z^2+\sigma_1^2}\frac{\sigma_1^2}{\sigma_{1j}^2}.\\
\end{aligned}
\end{equation*}

\textbf{IT's strategies.} Given dealer's quotes $p_1,p_{1_+},p_2$ and HFT's strategies $\{x_{1j}\}_{j=1}^J,\{x_{2j}\}_{j=1}^J,$ IT's profit in period 2 is
\begin{equation*}
\mathbb{E}\big(\left.\pi_2^{\text{IT}}\right|v,i_1,y_1\big)=-\lambda_{22}i_2^2+i_2\mathbb{E}\big(v-p_1-\lambda_{21}\sum_{j=1}^J x_{1j}-\lambda_{22}\sum_{j=1}^J x_{2j}|v,i_1,y_1\big).
\end{equation*}
It is maximized at $i_2=\alpha_{21}(v-\mathbb{E}(v|y_1))+\alpha_{22}(i_1-\mathbb{E}(i_1|y_1))$ if the SOC
\begin{equation}
\label{SOC3}
\lambda_{22}>0
\end{equation}
holds, where
\begin{equation}
\label{alpha21}
\alpha_{21}=\frac{1}{2\lambda_{22}},
\end{equation}
\begin{equation}
\label{alpha22}
\alpha_{22}=-\frac{\lambda_{21}\sum_{j=1}^J\beta_{1j}+\lambda_{22}(\sum_{j=1}^J\beta_{21j}+\beta_{23j}\beta_{1j}+\beta_{22j}\sum_{k\neq j}\beta_{1k})}{2\lambda_{22}}.
\end{equation}
IT's objective function in period 1 is
\begin{equation*}
\begin{aligned}
\mathbb{E}\left(\left.\pi_1^{\text{IT}}+\pi_2^{\text{IT}}\right|v\right)=&-[\lambda_1-\lambda_{22}\left(\frac{\alpha_{22}\sigma_1^2-\alpha_{21}\alpha_1\sigma_v^2}{\alpha_1^2\sigma_v^2+\sigma_z^2+\sigma_1^2}\right)^2]i_1^2\\
&+[1+2\lambda_{22}\alpha_{21}\frac{\alpha_{22}\sigma_1^2-\alpha_{21}\alpha_1\sigma_v^2}{\alpha_1^2\sigma_v^2+\sigma_z^2+\sigma_1^2}]i_1(v-p_0)\\
&+\frac{\sigma_1^2}{(\alpha_1^2\sigma_v^2+\sigma_z^2+\sigma_1^2)^2}[\alpha_{21}\alpha
_1\sigma_v^2+\alpha_{22}(\alpha_1^2\sigma_v^2+\sigma_z^2)]. 
\end{aligned}
\end{equation*}
In the mixed-strategy equilibrium, $\sigma_z>0$, we have
\begin{equation}
\label{z1}
\lambda_1-\lambda_{22}\left(\frac{\alpha_{22}\sigma_1^2-\alpha_{21}\alpha_1\sigma_v^2}{\alpha_1^2\sigma_v^2+\sigma_z^2+\sigma_1^2}\right)^2=0,
\end{equation}
\begin{equation}
\label{z2}
1+2\lambda_{22}\alpha_{21}\frac{\alpha_{22}\sigma_1^2-\alpha_{21}\alpha_1\sigma_v^2}{\alpha_1^2\sigma_v^2+\sigma_z^2+\sigma_1^2}=0.
\end{equation}
In the pure-strategy equilibrium, $\sigma_z=0,$ it is maximized at $i_1=\alpha_1(v-p_0)$ if the SOC
\begin{equation}
\label{SOC4}
\lambda_1-\lambda_{22}\left(\frac{\alpha_{22}\sigma_1^2-\alpha_{21}\alpha_1\sigma_v^2}{\alpha_1^2\sigma_v^2+\sigma_z^2+\sigma_1^2}\right)^2>0
\end{equation}
holds, where
\begin{equation}
\label{alpha1}
\alpha_1=\frac{1+2\lambda_{22}\alpha_{21}\frac{\alpha_{22}\sigma_1^2-\alpha_{21}\alpha_1\sigma_v^2}{\alpha_1^2\sigma_v^2+\sigma_z^2+\sigma_1^2}}{2[\lambda_1-\lambda_{22}\left(\frac{\alpha_{22}\sigma_1^2-\alpha_{21}\alpha_1\sigma_v^2}{\alpha_1^2\sigma_v^2+\sigma_z^2+\sigma_1^2}\right)^2]}.
\end{equation}

\textbf{Equilibrium.} The mixed-strategy equilibrium is determined by an equality-inequality system, which consists of \eqref{lam1}, \eqref{lam22}, \eqref{lam32}, \eqref{lam33}, \eqref{beta21j}, \eqref{beta22j}, \eqref{beta23j}, \eqref{beta1j}, \eqref{alpha21}, \eqref{alpha22}, \eqref{z1}, \eqref{z2} and SOCs \eqref{SOC1}, \eqref{SOC2}, \eqref{SOC3}. The pure-strategy equilibrium is determined by the system which consists of \eqref{lam1}, \eqref{lam22}, \eqref{lam32}, \eqref{lam33}, \eqref{beta21j}, \eqref{beta22j}, \eqref{beta23j}, \eqref{beta1j}, \eqref{alpha21}, \eqref{alpha22}, \eqref{alpha1} and SOCs \eqref{SOC1}, \eqref{SOC2}, \eqref{SOC3} \eqref{SOC4}. 

Given asset's volatility $\sigma_v$ and time-1 size of noise trading $\sigma_1$, the equilibrium only depends on the following dimensionless variables:
\begin{equation}
\label{ratio}
\begin{aligned}
&\Lambda_1=\frac{\lambda_1}{\sigma_v/\sigma_1},\Lambda_{1_+}=\frac{\lambda_{1_+}}{\sigma_v/\sigma_1},\Lambda_{21}=\frac{\lambda_{21}}{\sigma_v/\sigma_1},\Lambda_{22}=\frac{\lambda_{22}}{\sigma_v/\sigma_1},\\
&A_1=\frac{\alpha_1}{\sigma_1/\sigma_v},\theta_z=\frac{\sigma_z^2}{\sigma_1^2},A_{21}=\frac{\alpha_{21}}{\sigma_1/\sigma_v},\alpha_{22},\\
&\{\beta_{1j},\beta_{21j},\beta_{22j},\beta_{23j}\}_{j=1}^J.
\end{aligned}
\end{equation}

\subsection{Small-IT and Round-Tripper}
In this section, we assume that there is only one HFT and investigate the anticipatory strategies she takes. We are going to show that in equilibrium, HFT may perform the role of Small-IT or Round-Tripper, which will be defined later.
The equilibrium conditions are simplified in Theorem \ref{thmj=1}:
\begin{theorem}[Simplification of the equilibrium with an HFT]
 \label{thmj=1}    
The mixed-strategy equilibrium is characterized by system \eqref{systemmixedj=1-1}, \eqref{systemmixedj=1-2}, \eqref{systemmixedj=1-3} and \eqref{systemmixedj=1-4}; the pure-strategy equilibrium is characterized by system \eqref{systempurej=1}, \eqref{systempurej=1-2} and \eqref{systempurej=1-3}.
\end{theorem}

If HFT tends to trade in the same direction as IT in both transactions, then she appears to invest in assets and accumulate positions. So we call her \textbf{Small-IT}. In contrast, if she tends to first trade in the same direction and then in the opposite direction as IT, then she appears to exploit price impact and control inventories. We call her \textbf{Round-Tripper}. 

These two notions are also mentioned in Xu and Cheng (2023) \cite{xu2023high}. Actually, the current case, where HFT predicts IT's future order through the past order flow and IT may take mixed strategies, is an extension of \cite{xu2023high}. 
\begin{theorem}
\label{thmj=1strategy}
For any $\theta_{1_+},\theta_2>0,\theta_\varepsilon\geq0,$ there exists a critical $\Bar{\Gamma},$ if $\Gamma\in[0,\Bar{\Gamma}),$ HFT will play the role of Small-IT; if $\Gamma\in[\Bar{\Gamma},\infty],$ HFT will play the role of Round-Tripper. 
\end{theorem}

As long as $\Gamma\in[0,\Bar{\Gamma})$ or $\Gamma\in[\Bar{\Gamma},\infty]$, the role played by HFT remains unchanged. Given such continuity of $\Gamma's$ impact, without loss of generality, we assume $\Gamma$ to be $0$ for HFTs who accumulate positions and to be $\infty$
for HFTs who control inventories. In the market considered below, there are $J_1\geq0$ HFTs with $\Gamma=0$ and $J_2\geq0$ HFTs with $\Gamma=\infty.$

We focus on the equilibrium with multiple anticipatory HFTs who may deal with inventory in two different ways, rather than the continuous change of $\Gamma$. In fact, this setting matches the empirical finding of Kirilenko et al. (2017) \cite{2017The}: in the real market, fast traders may act as ``opportunistic traders'' or ``high-frequency traders'', both of them trade at a high speed, while the former adjusts holdings less frequently and with larger fluctuations.

What's more, it is verified that $\Gamma=\infty$ is equivalent to require HFT to clear positions in period 2:
\begin{proposition}
\label{propj=1gam=inf}
If $\Gamma=\infty,$
\begin{equation*}
\beta_{21}=\beta_{22}=0,\ \beta_{23}=-1.
\end{equation*}
The equilibrium is the same as that where HFT maximizes:
\begin{equation}
\label{j=1HFTobjgam=inf}
    x_{1}^*=\arg\max\mathbb{E}(x_1(p_2-p_{1_+})|\Tilde{i}_{1},y_1).
\end{equation} 
\end{proposition}

In Section \ref{subsecSIT} - \ref{subsecgam10gam2inf}, we will investigate three kinds of markets, where HFTs are all Small-ITs, or all Round-Trippers, or both. In each case, a simplification theorem for equilibrium is provided and the equilibrium is solved through numerical methods.

\subsection{A market with Small-ITs}
\label{subsecSIT}
In this section, we assume that $J_1=J\geq1$ and $J_2=0$. All HFTs play the role of Small-IT and in equilibrium, the $j$-th HFT takes the following strategy:
\begin{equation*}
\begin{aligned}
&x_{1j}=\beta_{11}(\Tilde{i}_{1j}-\mathbb{E}(\Tilde{i}_{1j}|y_1)),\\
&x_{2j}=\beta_{21}(\Tilde{i}_{1j}-\mathbb{E}(\Tilde{i}_{1j}|y_1))+\beta_{22}(\sum_{k\neq j}x_{1k}+u_{1_+})+\beta_{23}x_{1j}.\\
\end{aligned}
\end{equation*}

\begin{theorem}[Simplification of the equilibrium with Small-ITs]
\label{thmgam0}
The mixed-strategy equilibrium is characterized by system \eqref{systemmixedgam0} and \eqref{systemmixedgam0-2}; the pure-strategy equilibrium is characterized by system \eqref{systempuregam0} and \eqref{systempuregam0-2}.
\end{theorem}

Next, we investigate the equilibrium through numerical methods. It is reasonable to assume that $\theta_{1_+}\in(0,1], \theta_2=1,$ since as shown in \eqref{ratio}, $\theta_{1_+}$ represents the size of high-speed noise trading relative to aggregate noise trading, while $\theta_2$ represents the relative size of aggregate noise trading at different times.
What's more, according to van Kervel and Menkveld (2019) \cite{van2019high}, we set $J\leq10.$

The trading behavior of Smalll-ITs is quite similar to the back-runners in Yang and Zhu (2020) \cite{yang2020back}, they both steal IT's private information to construct their own positions. The only difference is that Small-ITs trade at a higher frequency: they race to trade in front of IT's future order $i_2$.

\begin{figure}[!htbp]
    \centering
\subcaptionbox{$\theta_{1_+}=10^{-4}$}{
    \includegraphics[width = 0.27\textwidth]{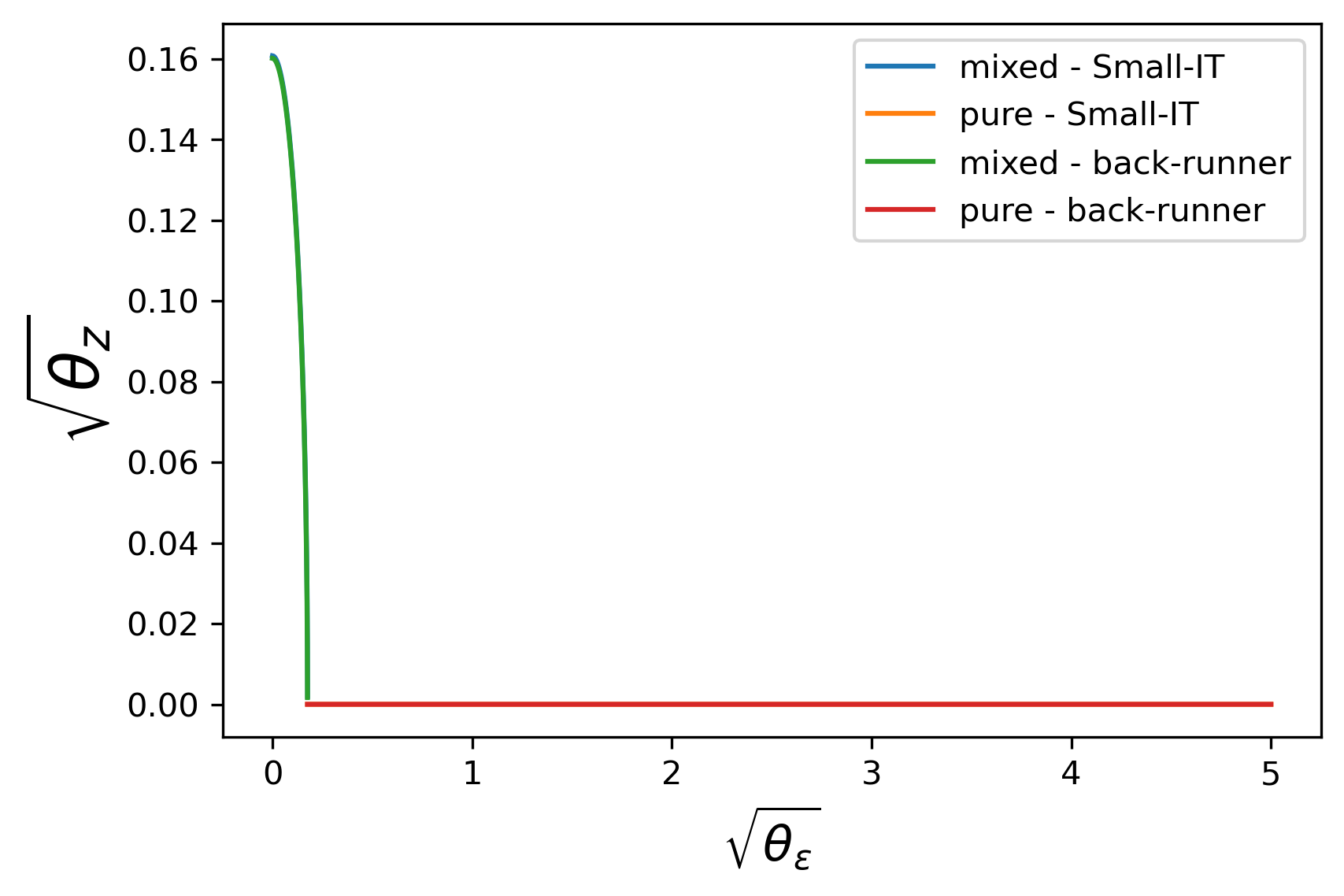}
    }
\subcaptionbox{$\theta_{1_+}=0.1$}{
    \includegraphics[width = 0.27\textwidth]{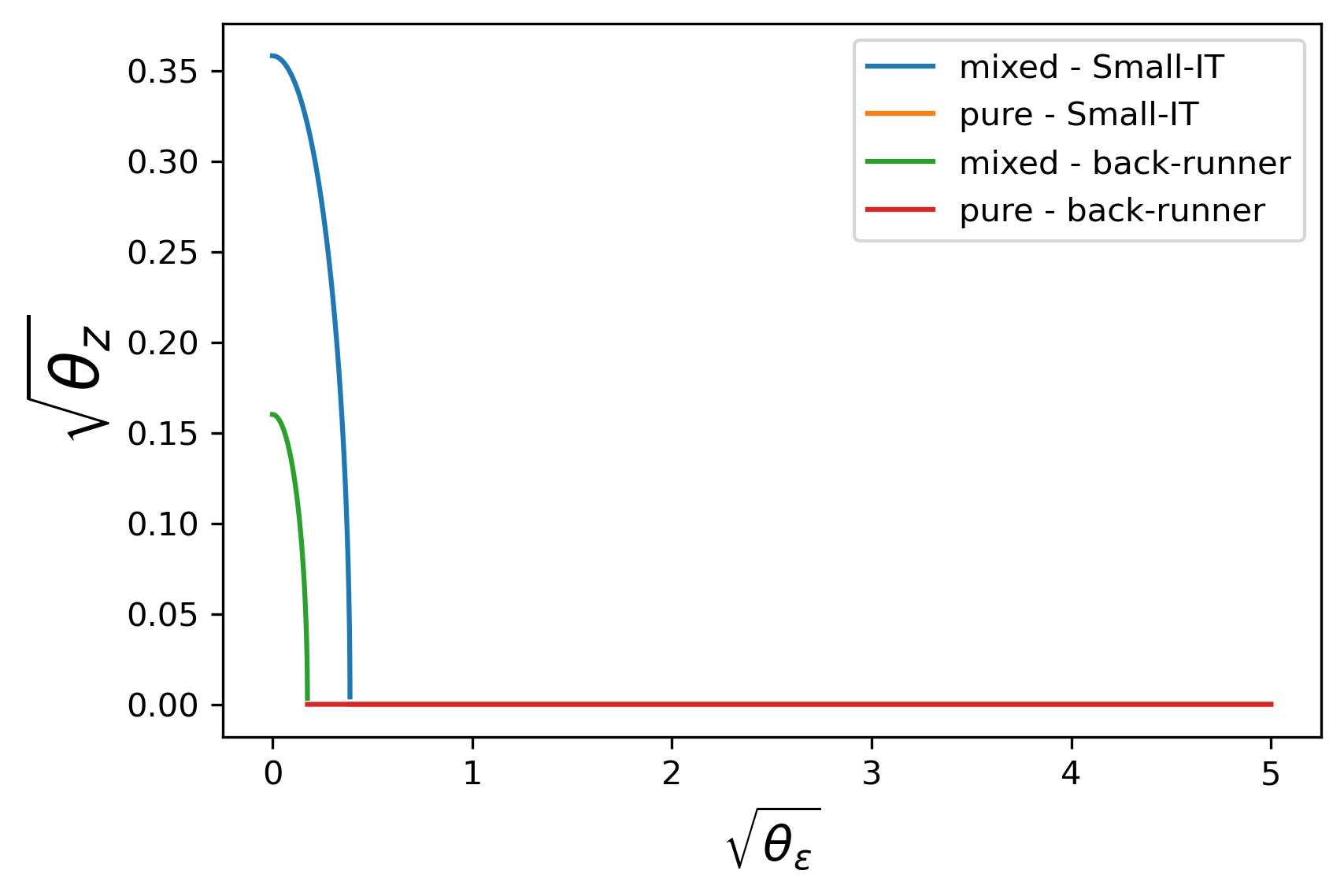}
    }
\subcaptionbox{$\theta_{1_+}=1$}{
    \includegraphics[width = 0.27\textwidth]{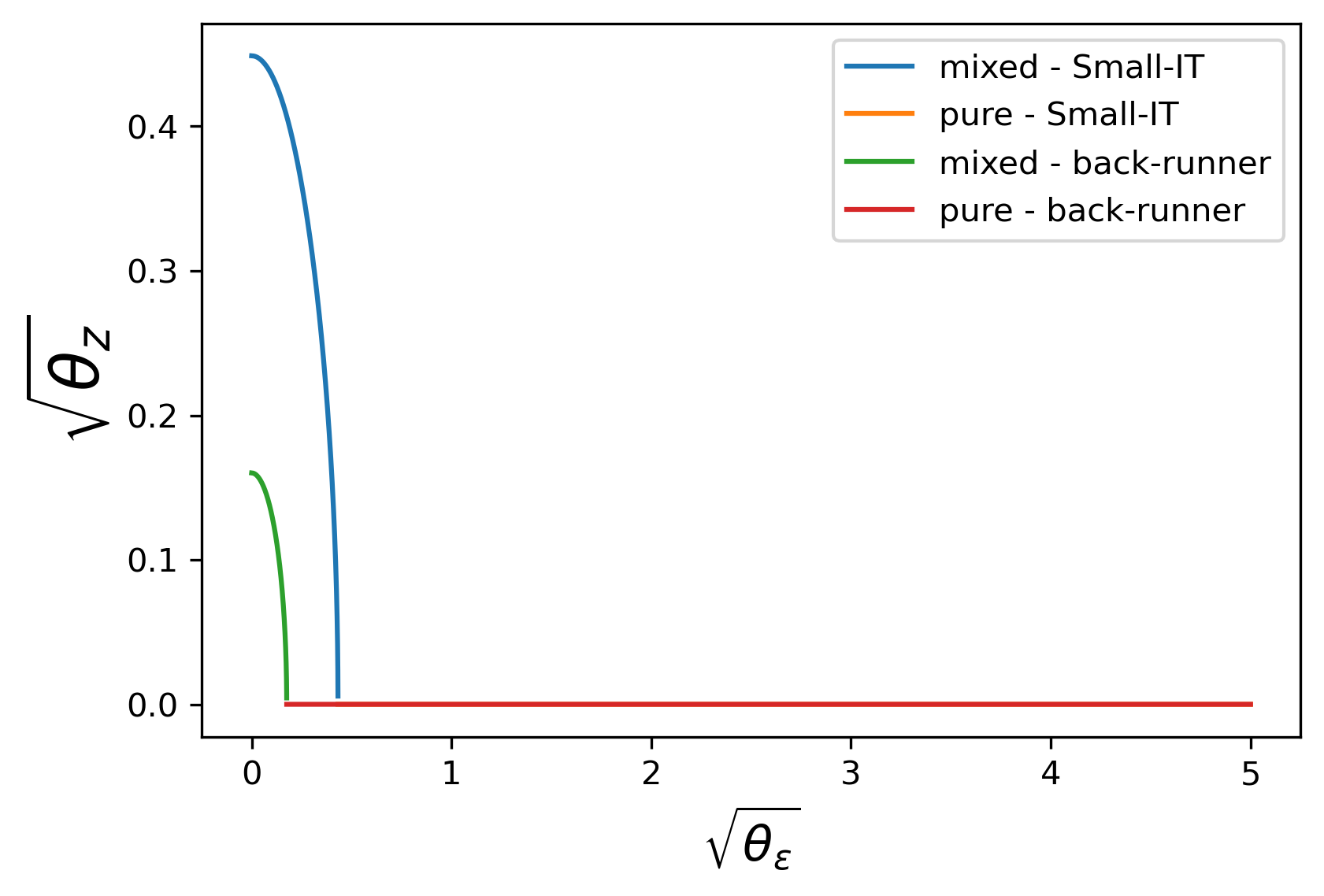}
    }
    \caption{$J=1,$ IT's mixed strategy.}
    \label{figj=1z}
\end{figure}

\begin{figure}[!htbp]
    \centering
\subcaptionbox{$\theta_{1_+}=10^{-4}$}{
    \includegraphics[width = 0.27\textwidth]{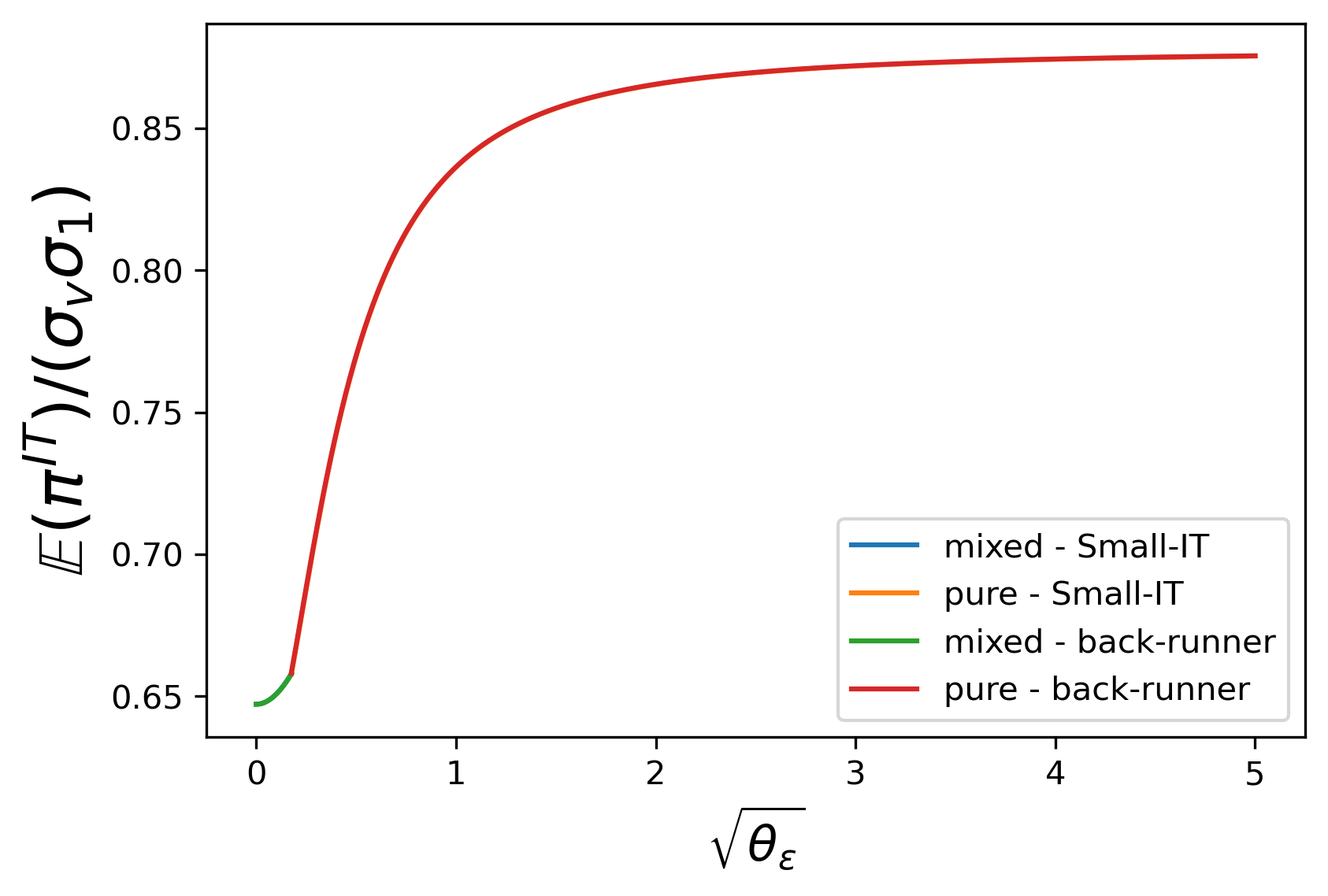}
    }
\subcaptionbox{$\theta_{1_+}=0.1$}{
    \includegraphics[width = 0.27\textwidth]{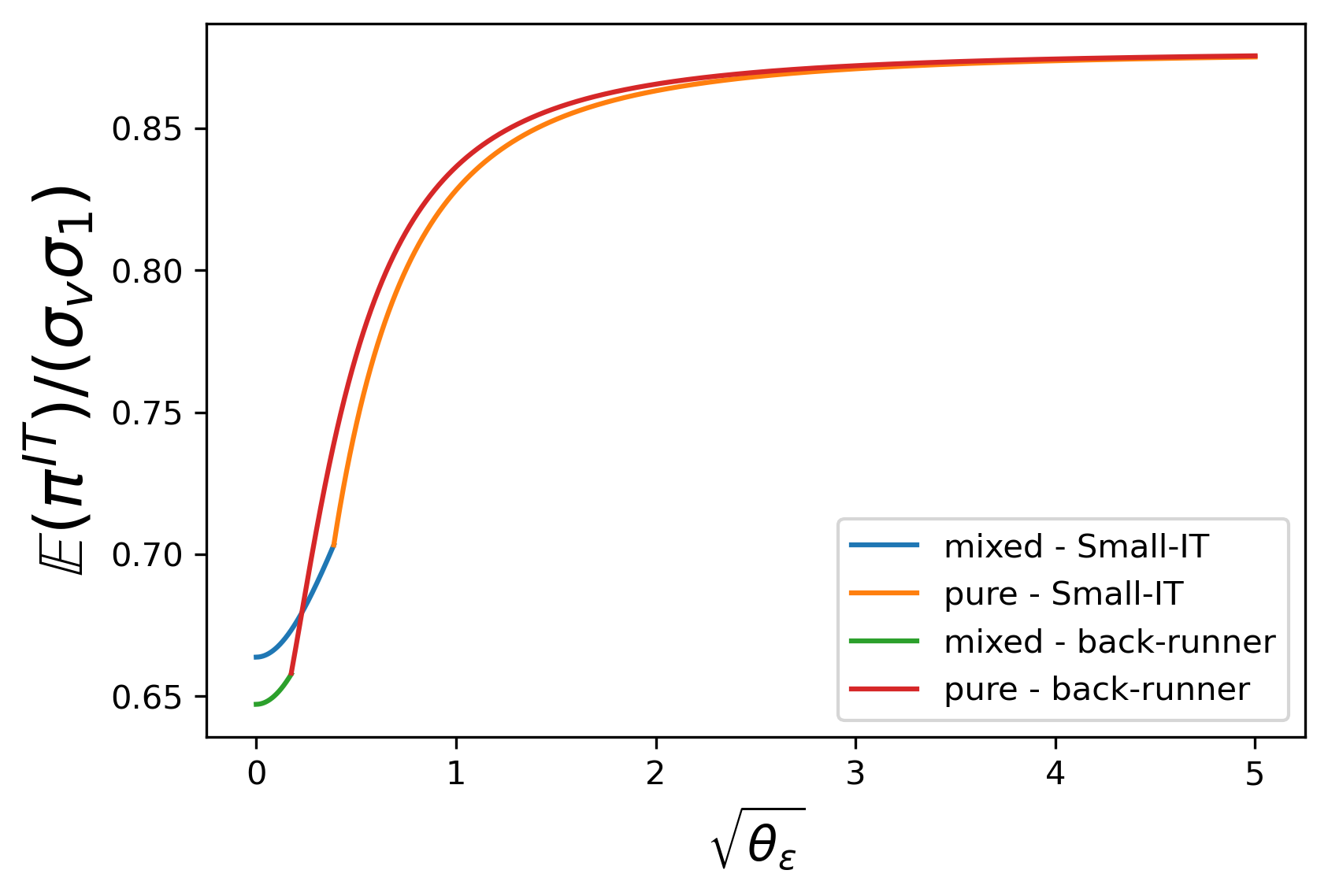}
    }
\subcaptionbox{$\theta_{1_+}=1$}{
    \includegraphics[width = 0.27\textwidth]{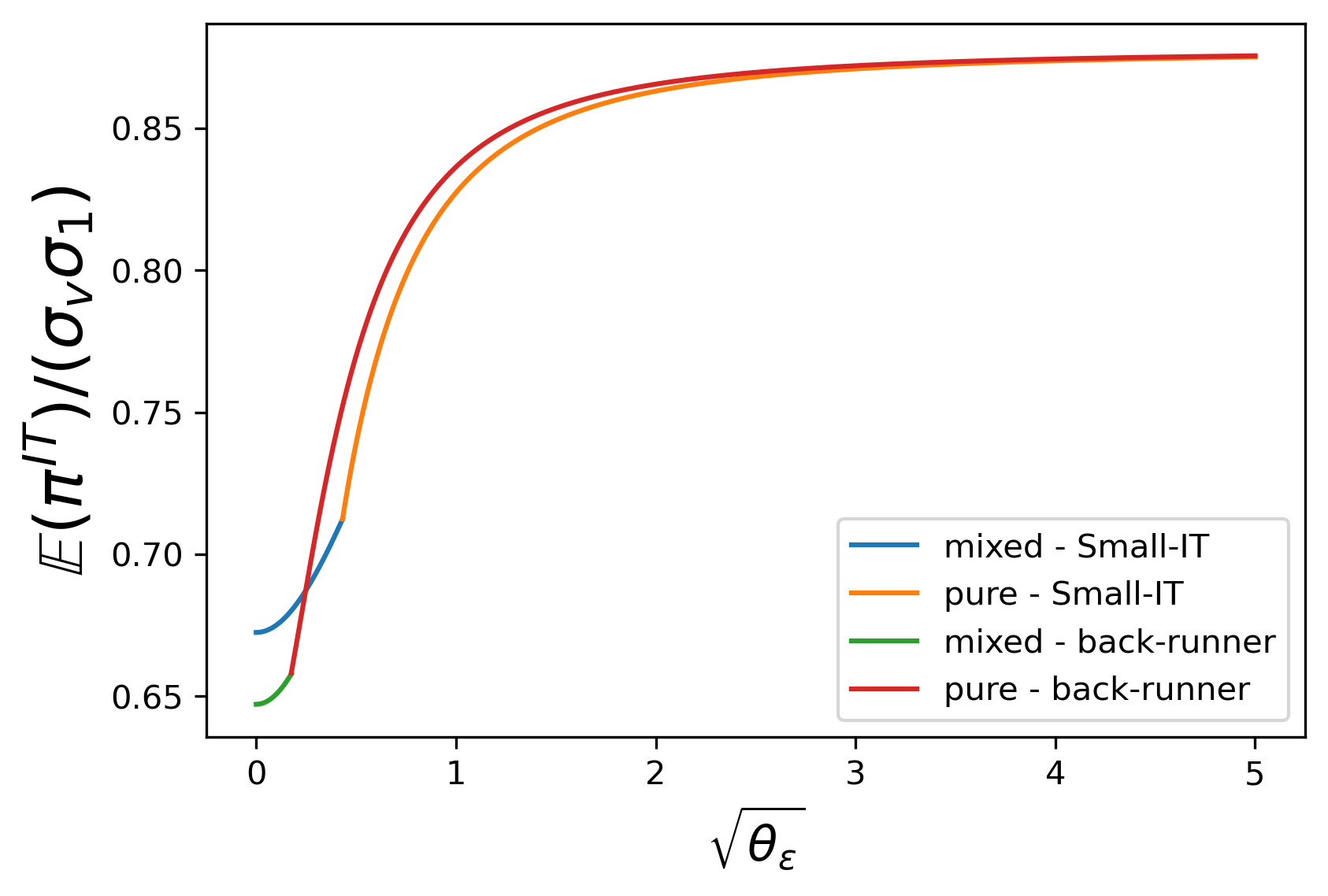}
    }
    \caption{$J=1,$ IT's profit.}
    \label{figj=1piIT}
\end{figure}

\begin{figure}[!htbp]
    \centering
\subcaptionbox{$\theta_{1_+}=10^{-4}$}{
    \includegraphics[width = 0.27\textwidth]{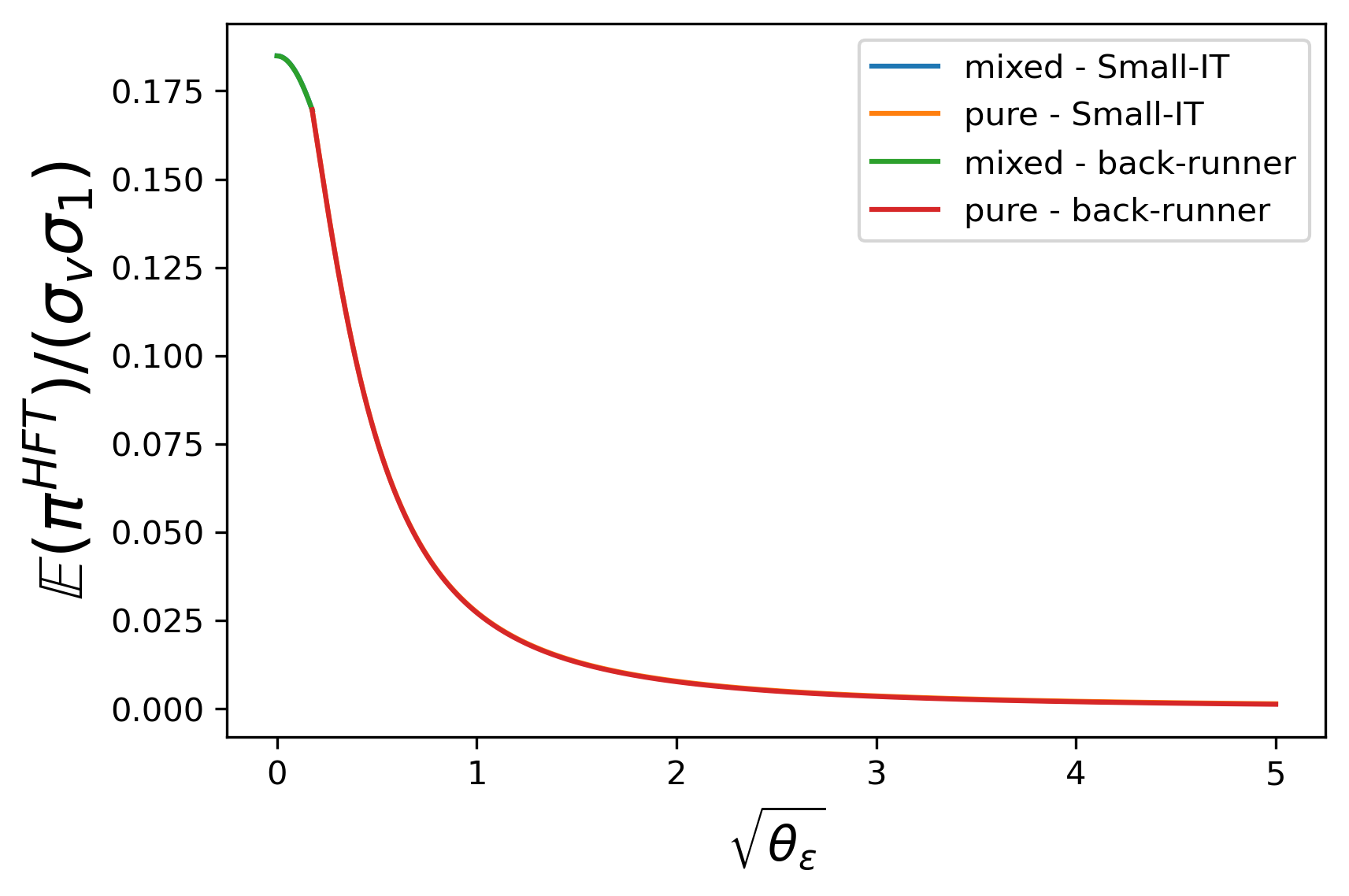}
    }
\subcaptionbox{$\theta_{1_+}=0.1$}{
    \includegraphics[width = 0.27\textwidth]{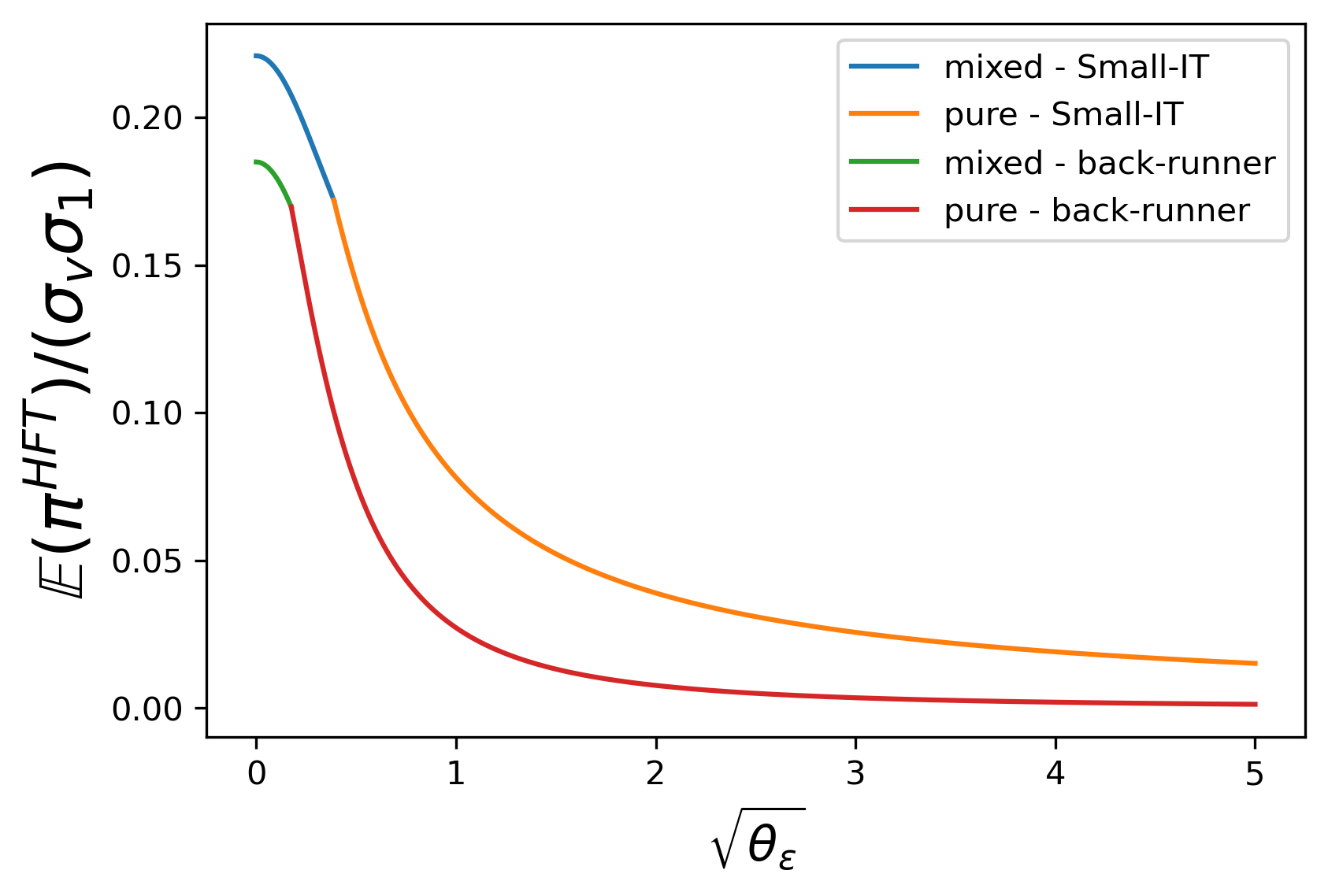}
    }
\subcaptionbox{$\theta_{1_+}=1$}{
    \includegraphics[width = 0.27\textwidth]{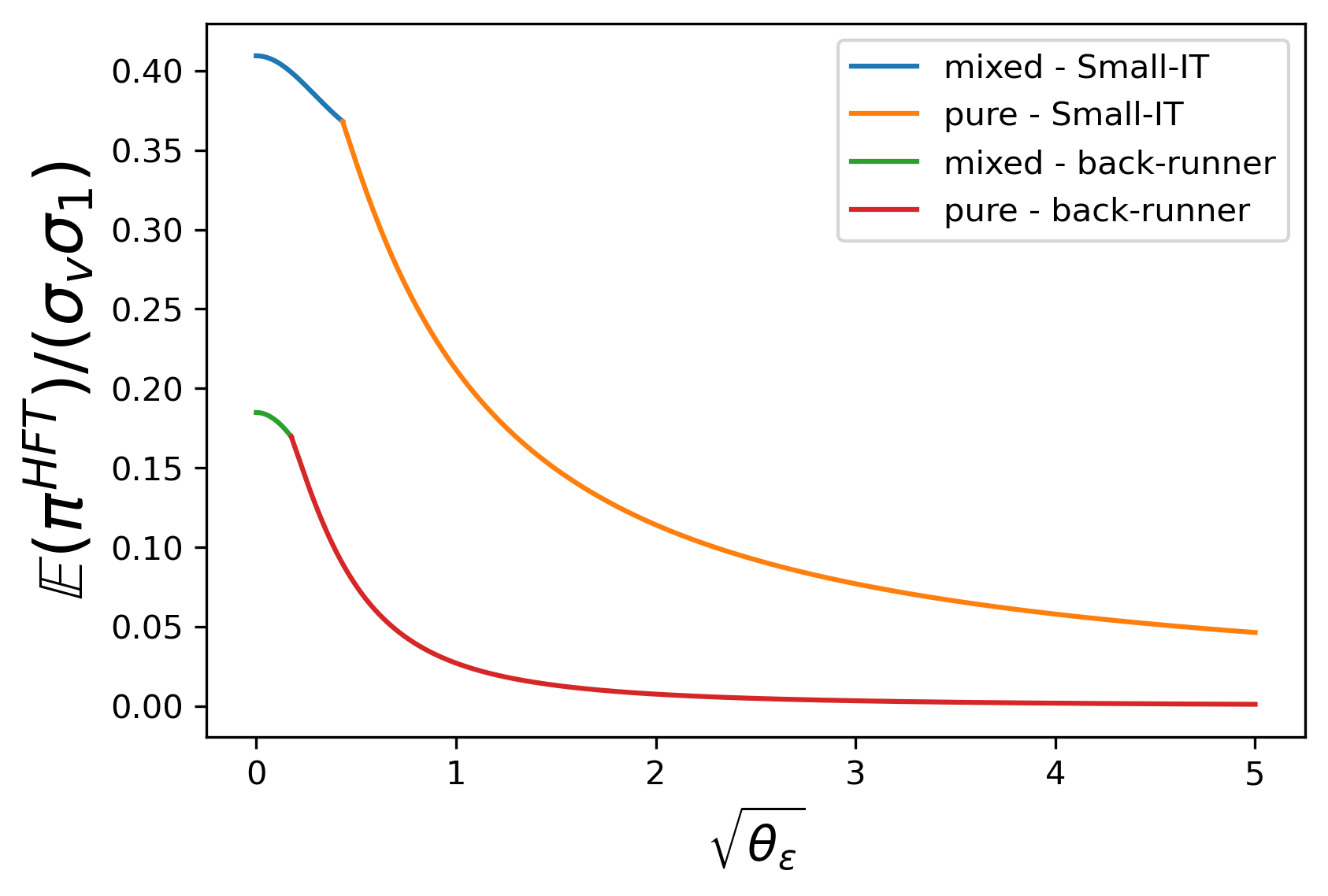}
    }
    \caption{$J=1,$ anticipatory trader's profit.}
    \label{figj=1piHFT}
\end{figure}

\begin{figure}[!htbp]
    \centering
\subcaptionbox{$\theta_{1_+}=0.01$}{
    \includegraphics[width = 0.27\textwidth]{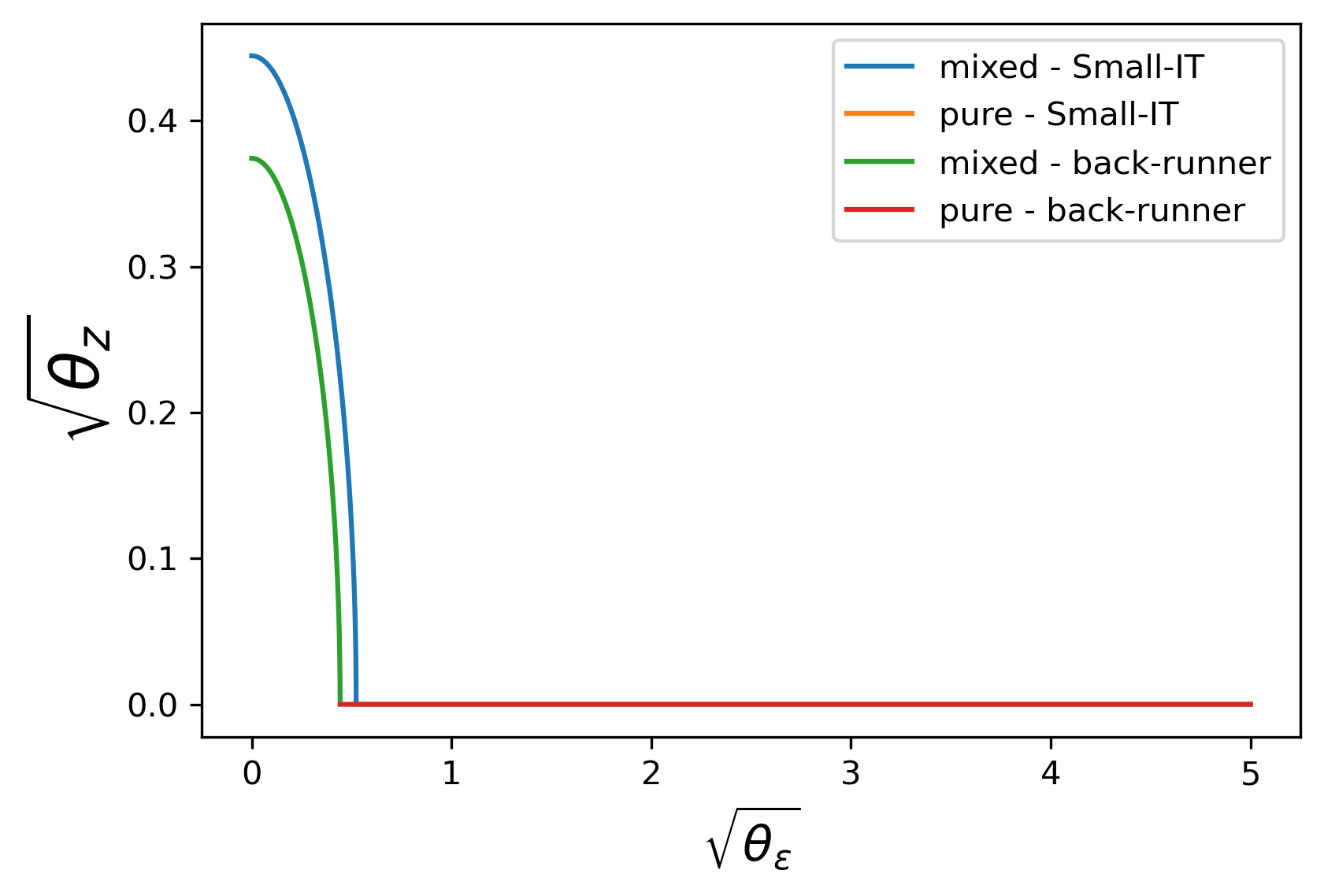}
    }
\subcaptionbox{$\theta_{1_+}=0.1$}{
    \includegraphics[width = 0.27\textwidth]{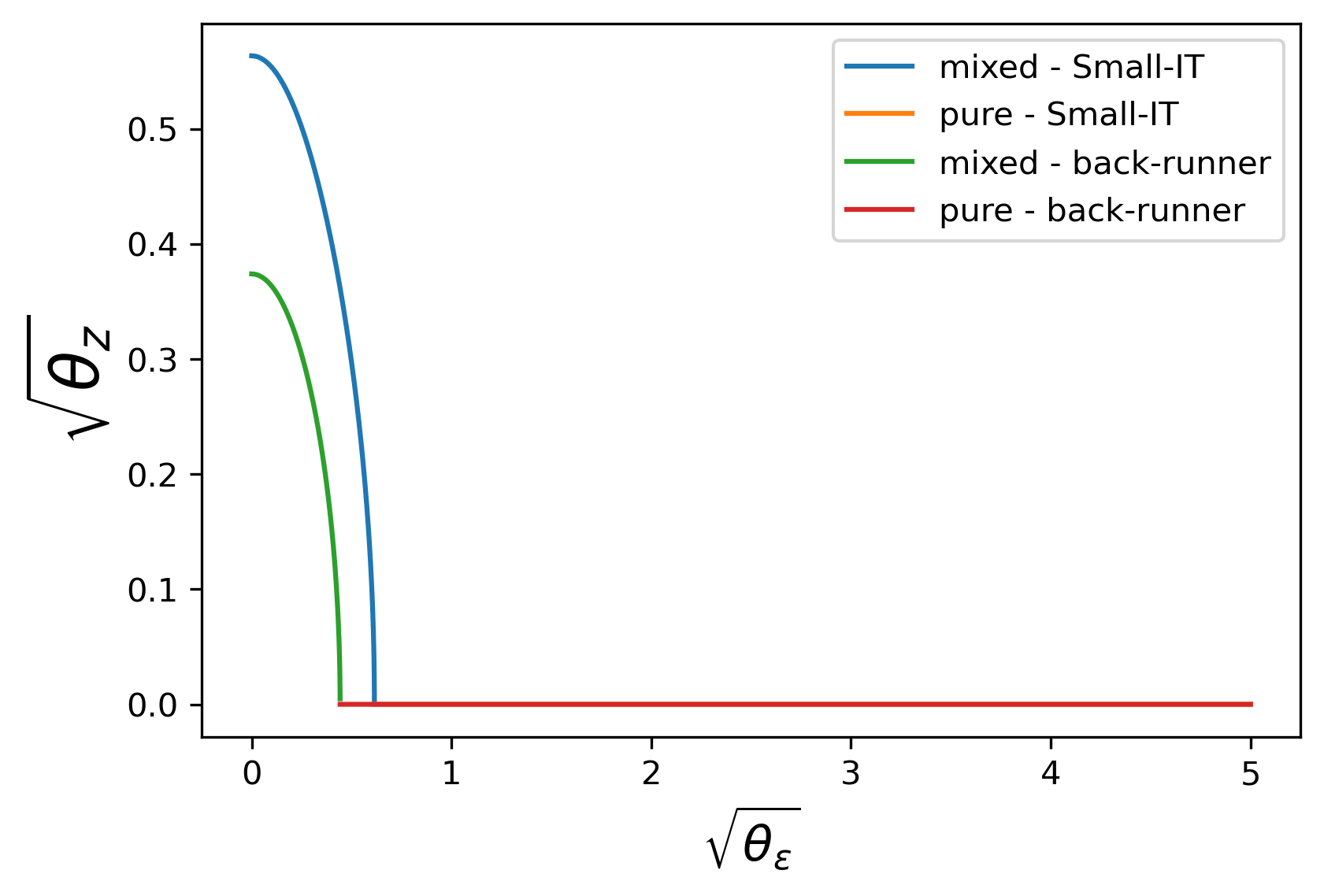}
    }
\subcaptionbox{$\theta_{1_+}=1$}{
    \includegraphics[width = 0.27\textwidth]{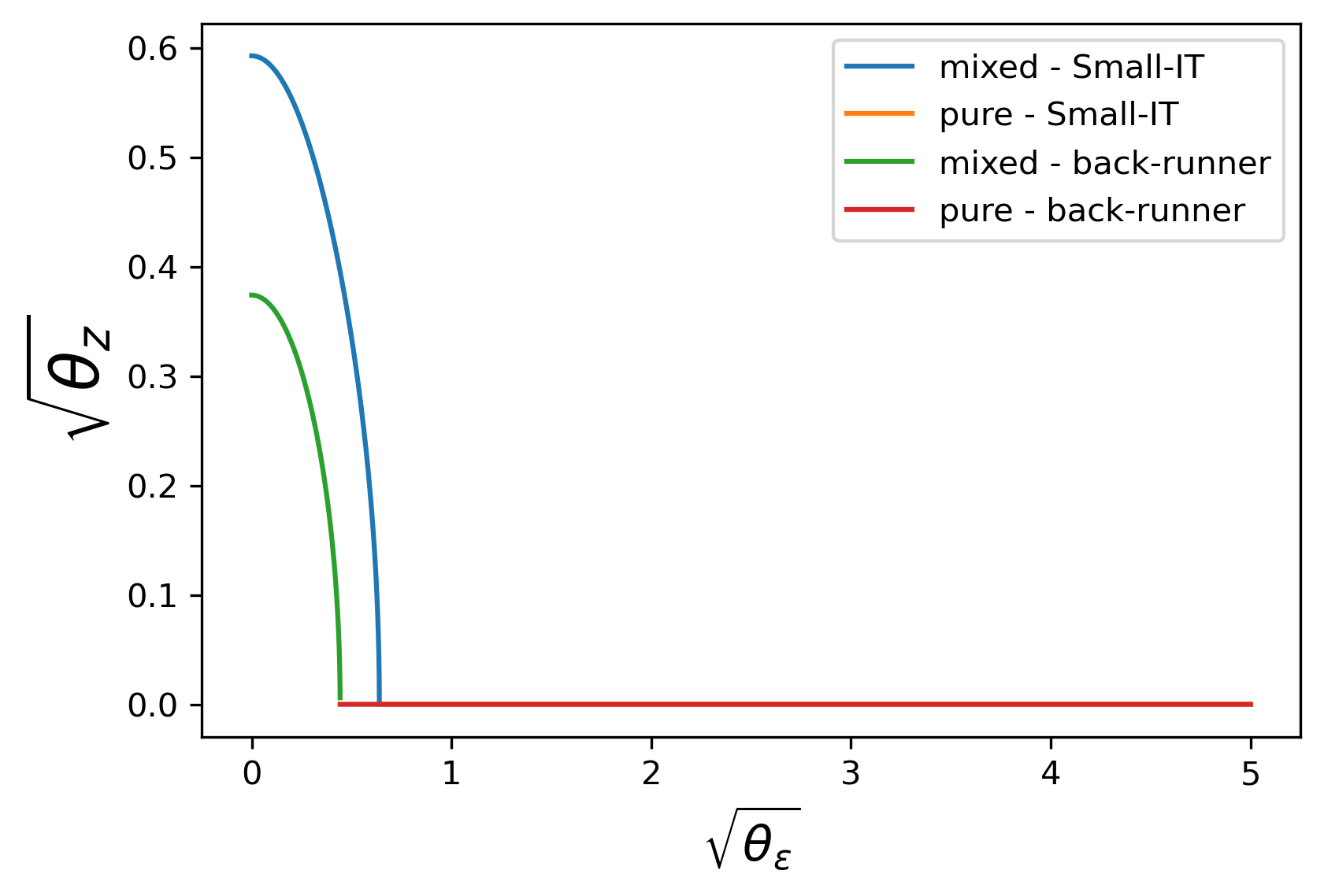}
    }
    \caption{$J=2,$ IT's mixed strategy.}
    \label{figj=2z}
\end{figure}

\begin{figure}[!htbp]
    \centering
\subcaptionbox{$\theta_{1_+}=0.01$}{
    \includegraphics[width = 0.27\textwidth]{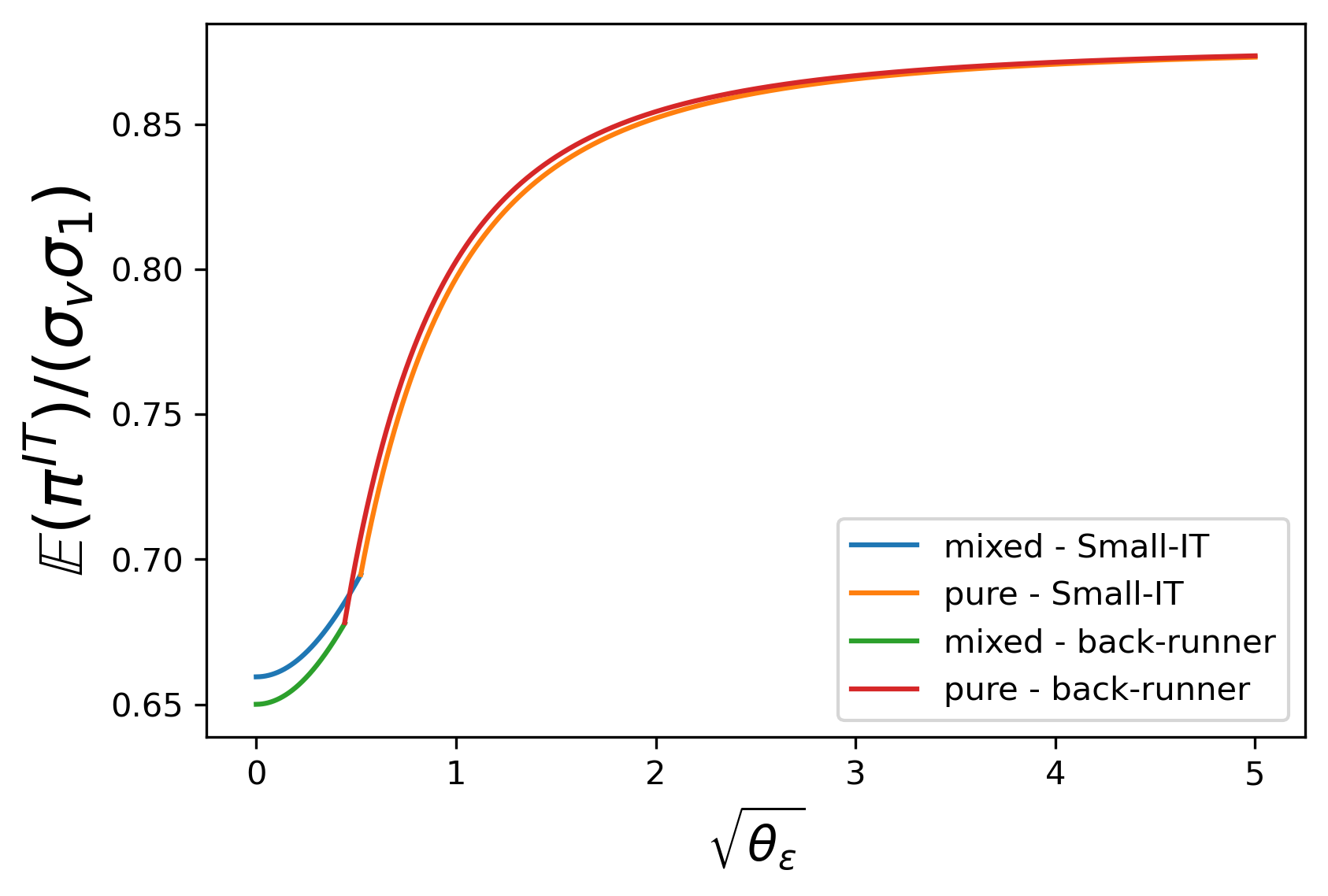}
    }
\subcaptionbox{$\theta_{1_+}=0.1$}{
    \includegraphics[width = 0.27\textwidth]{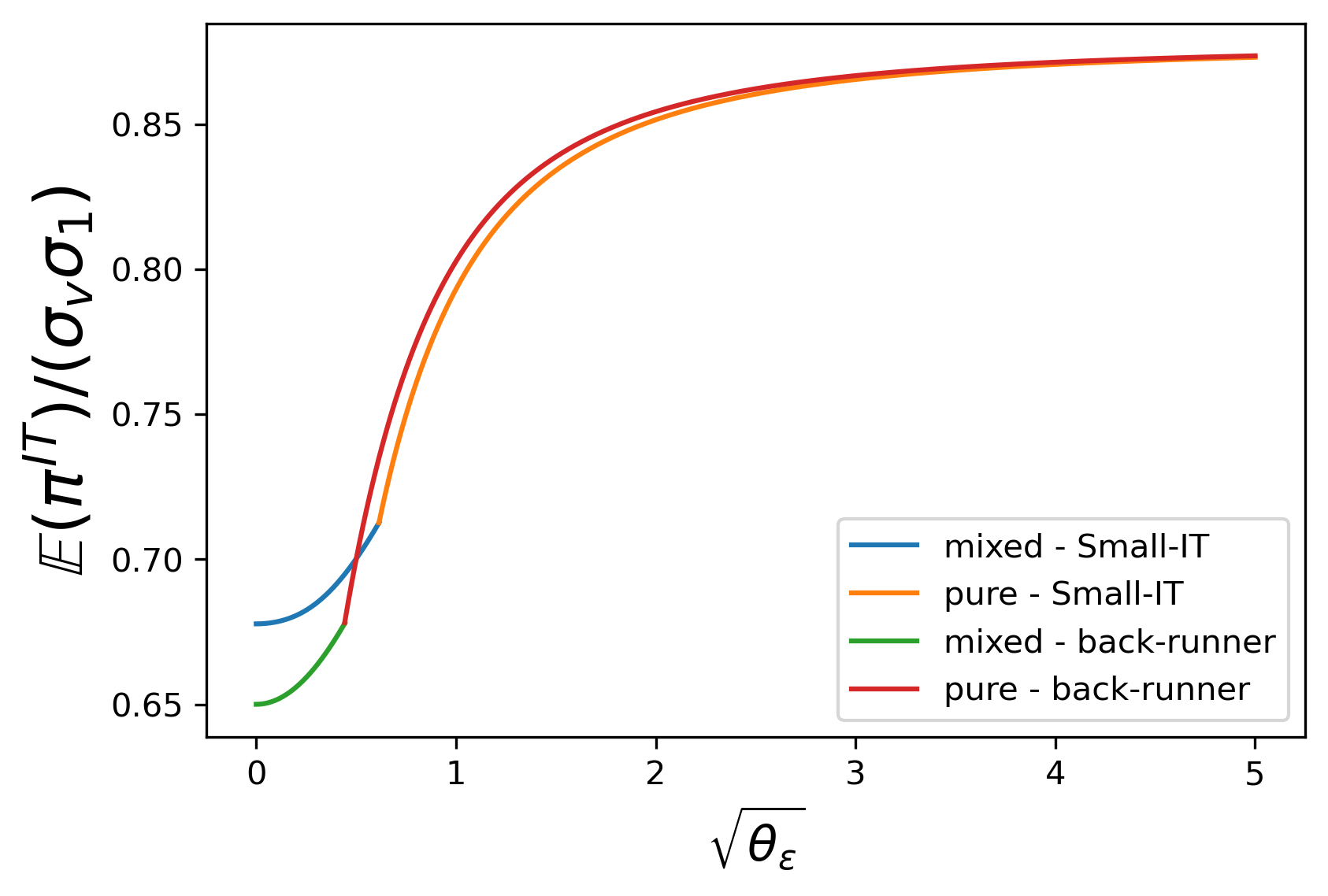}
    }
\subcaptionbox{$\theta_{1_+}=1$}{
    \includegraphics[width = 0.27\textwidth]{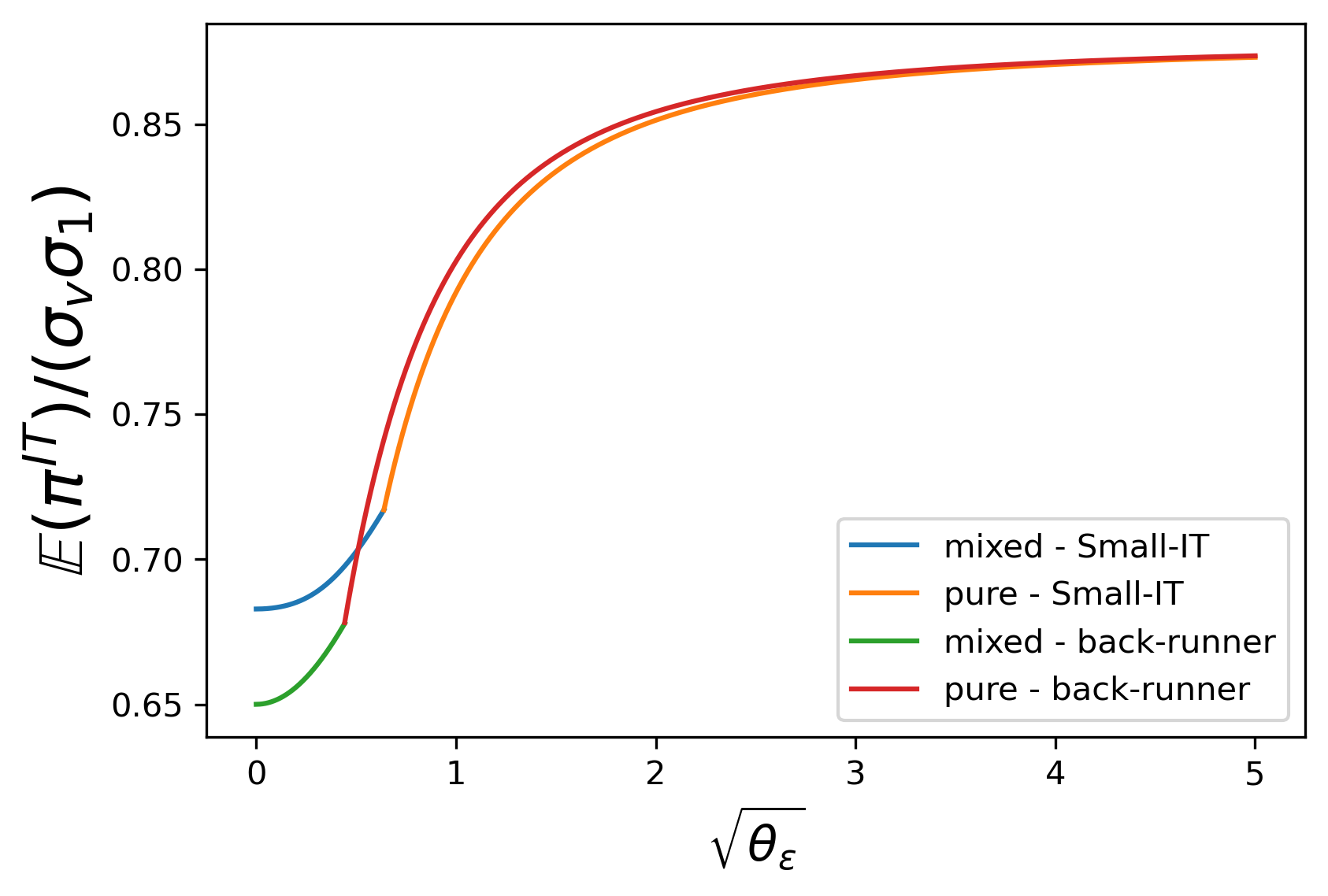}
    }
    \caption{$J=2,$ IT's profit.}
    \label{figj=2piIT}
\end{figure}

\begin{figure}[!htbp]
    \centering
\subcaptionbox{$\theta_{1_+}=0.01$}{
    \includegraphics[width = 0.27\textwidth]{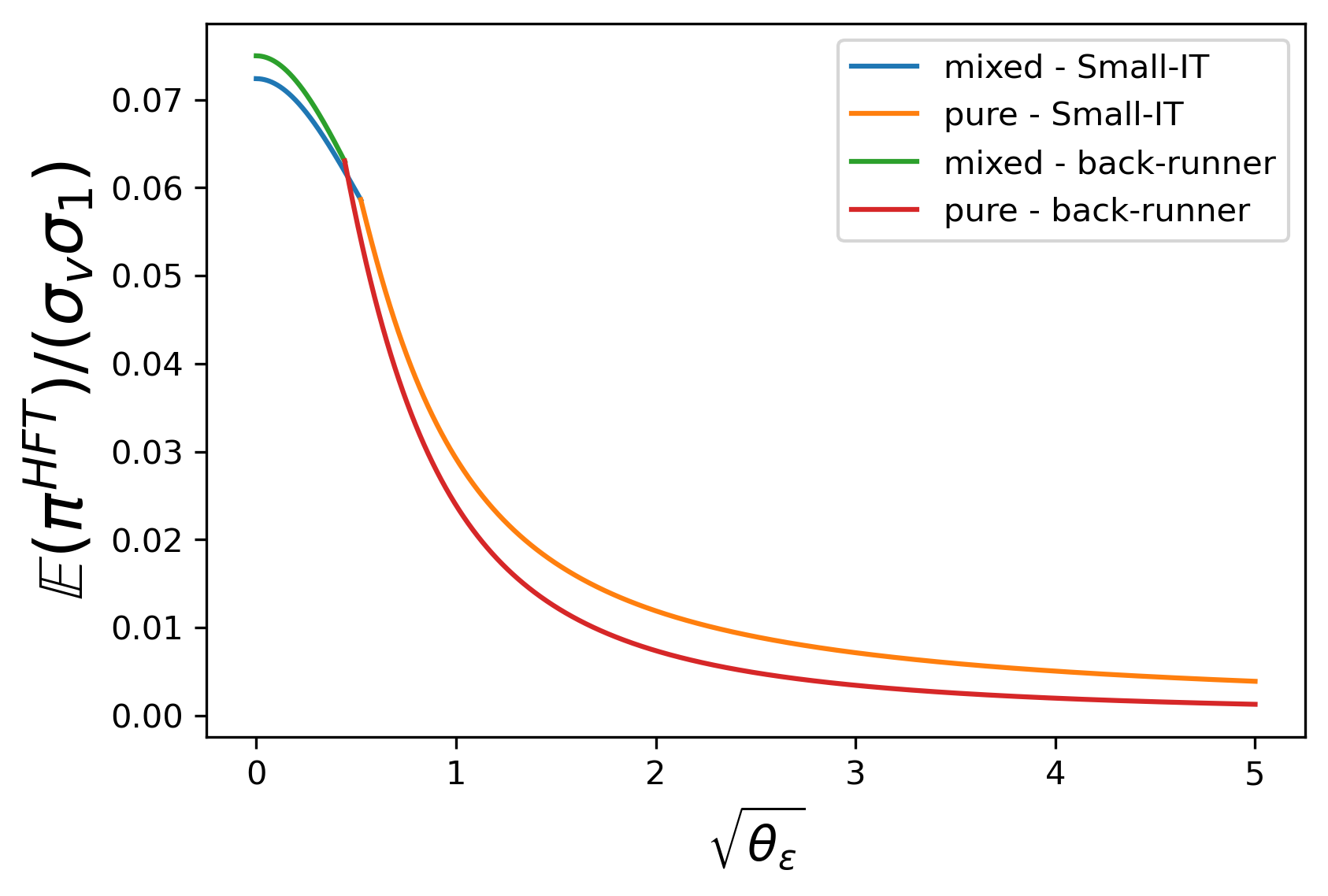}
    }
\subcaptionbox{$\theta_{1_+}=0.1$}{
    \includegraphics[width = 0.27\textwidth]{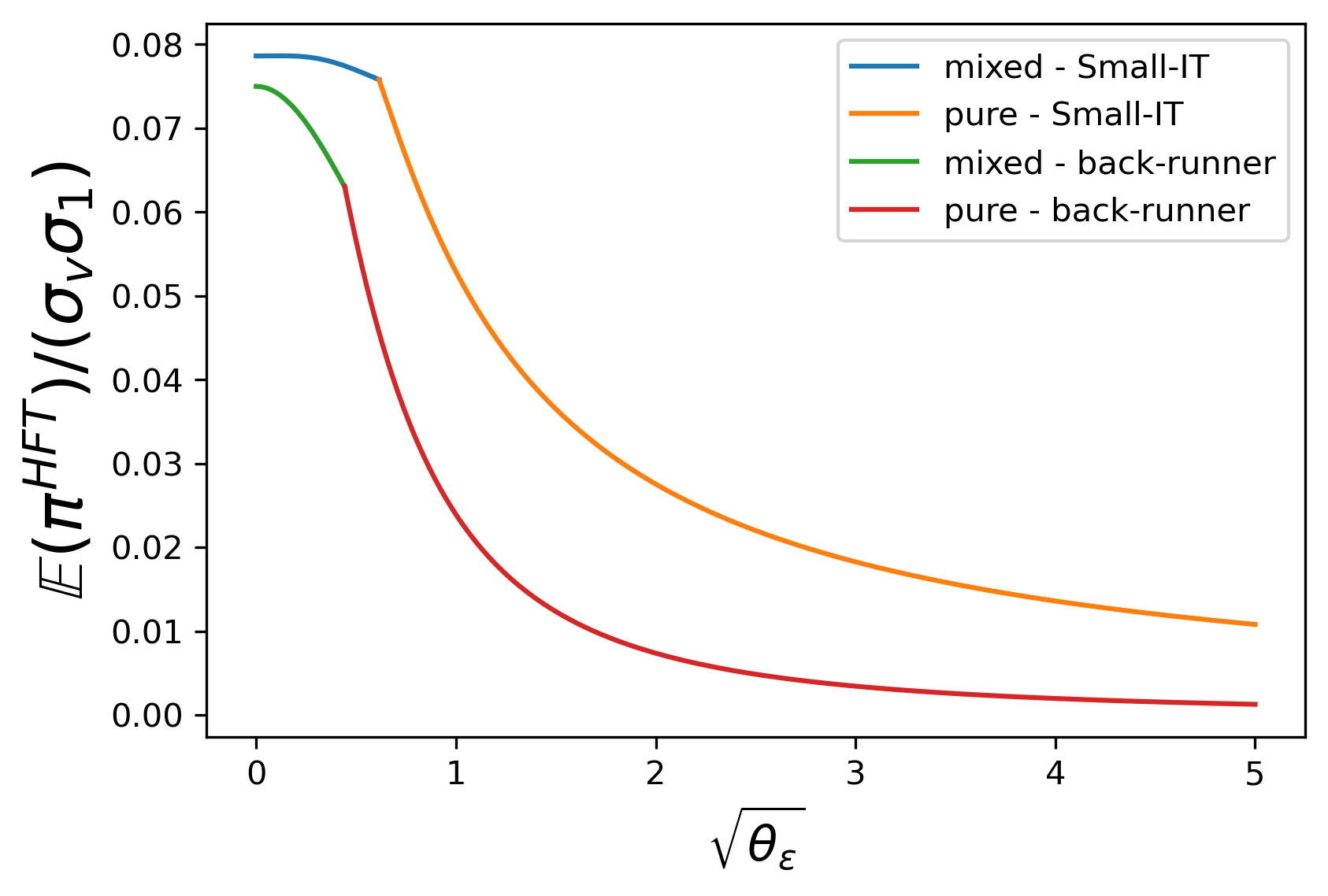}
    }
\subcaptionbox{$\theta_{1_+}=1$}{
    \includegraphics[width = 0.27\textwidth]{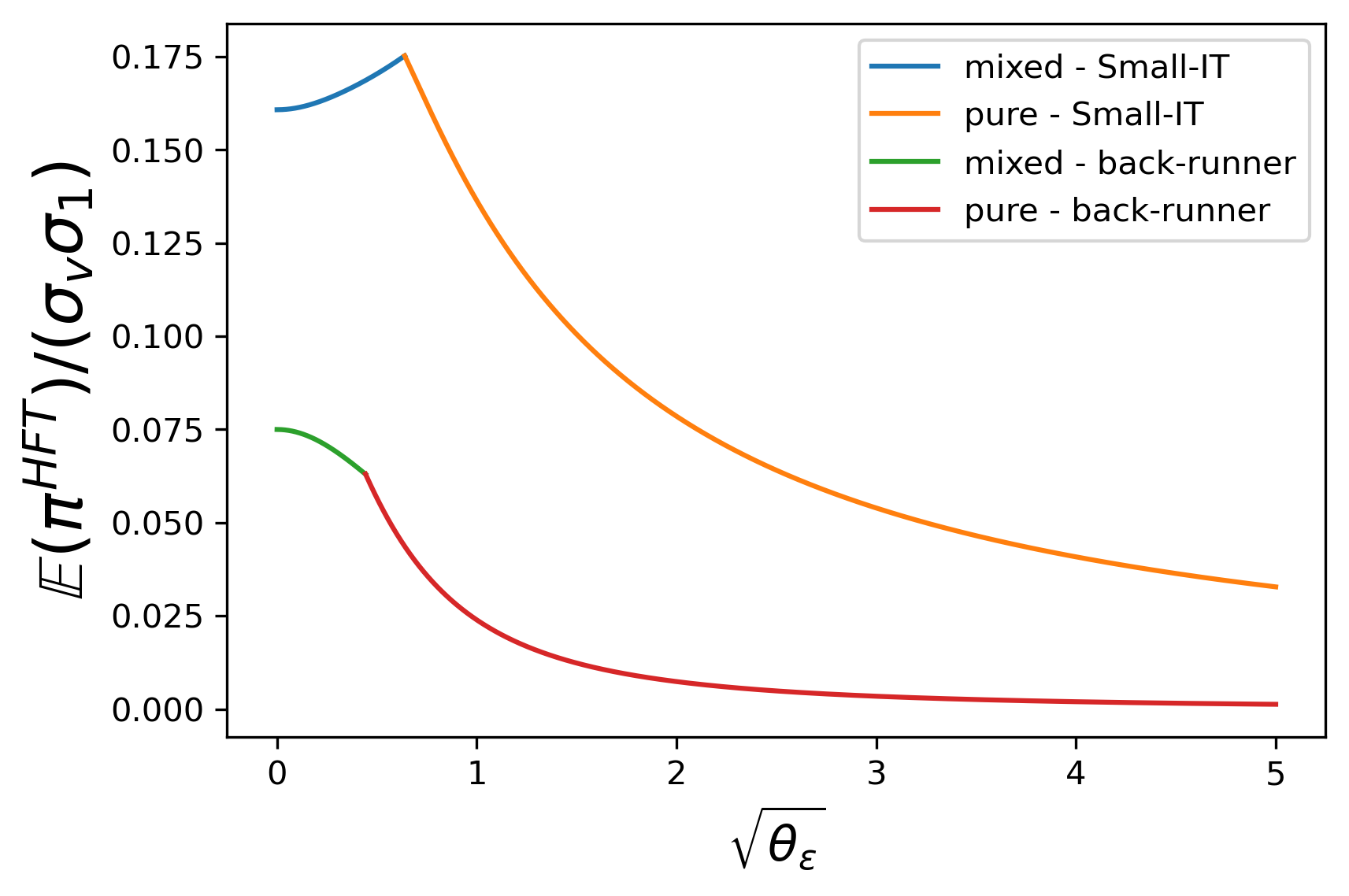}
    }
    \caption{$J=2,$ anticipatory traders' profit.}
    \label{figj=2piHFT}
\end{figure}

\begin{figure}[!htbp]
    \centering
\subcaptionbox{$\theta_{1_+}=0.01$}{
    \includegraphics[width = 0.27\textwidth]{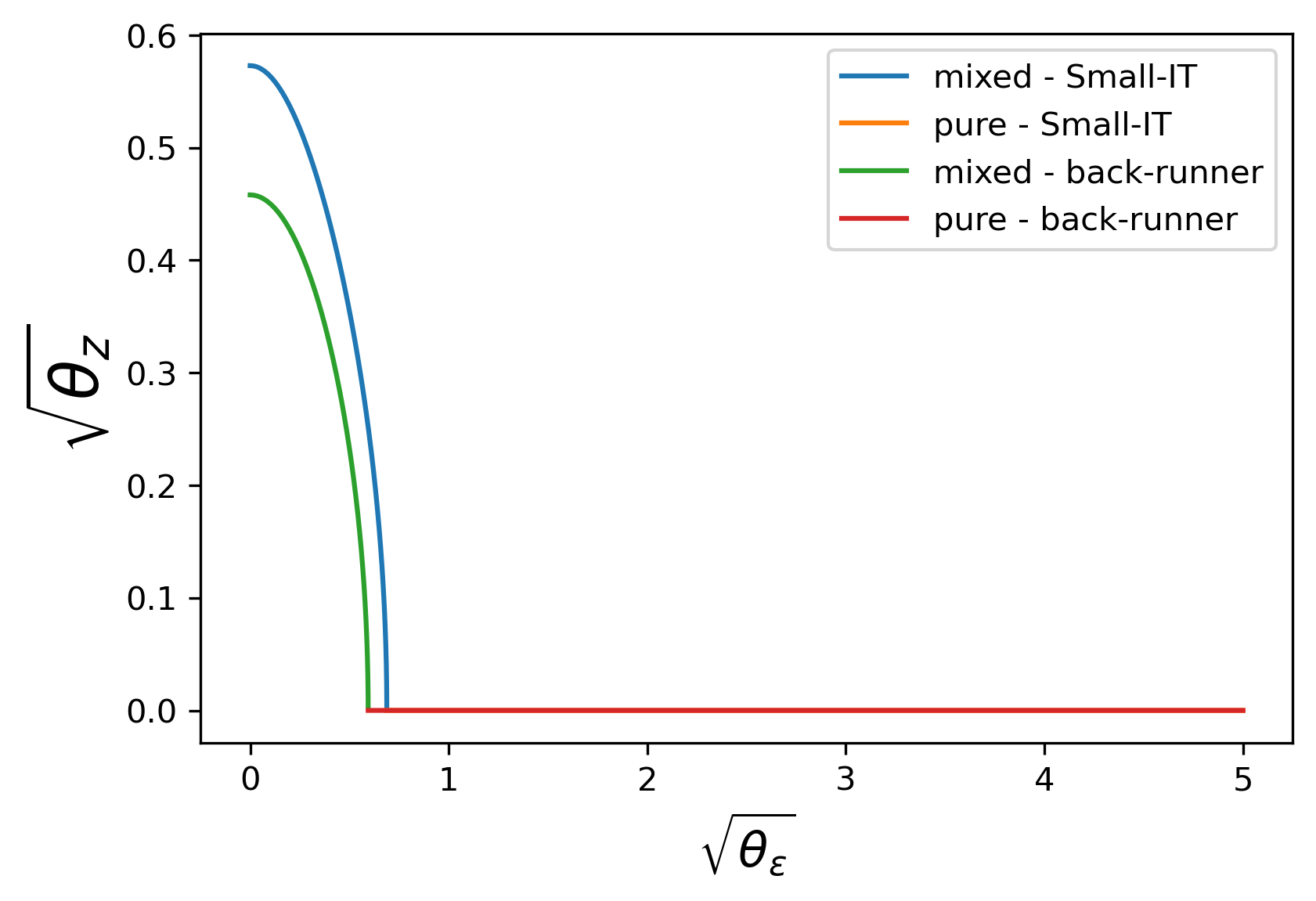}
    }
\subcaptionbox{$\theta_{1_+}=0.1$}{
    \includegraphics[width = 0.27\textwidth]{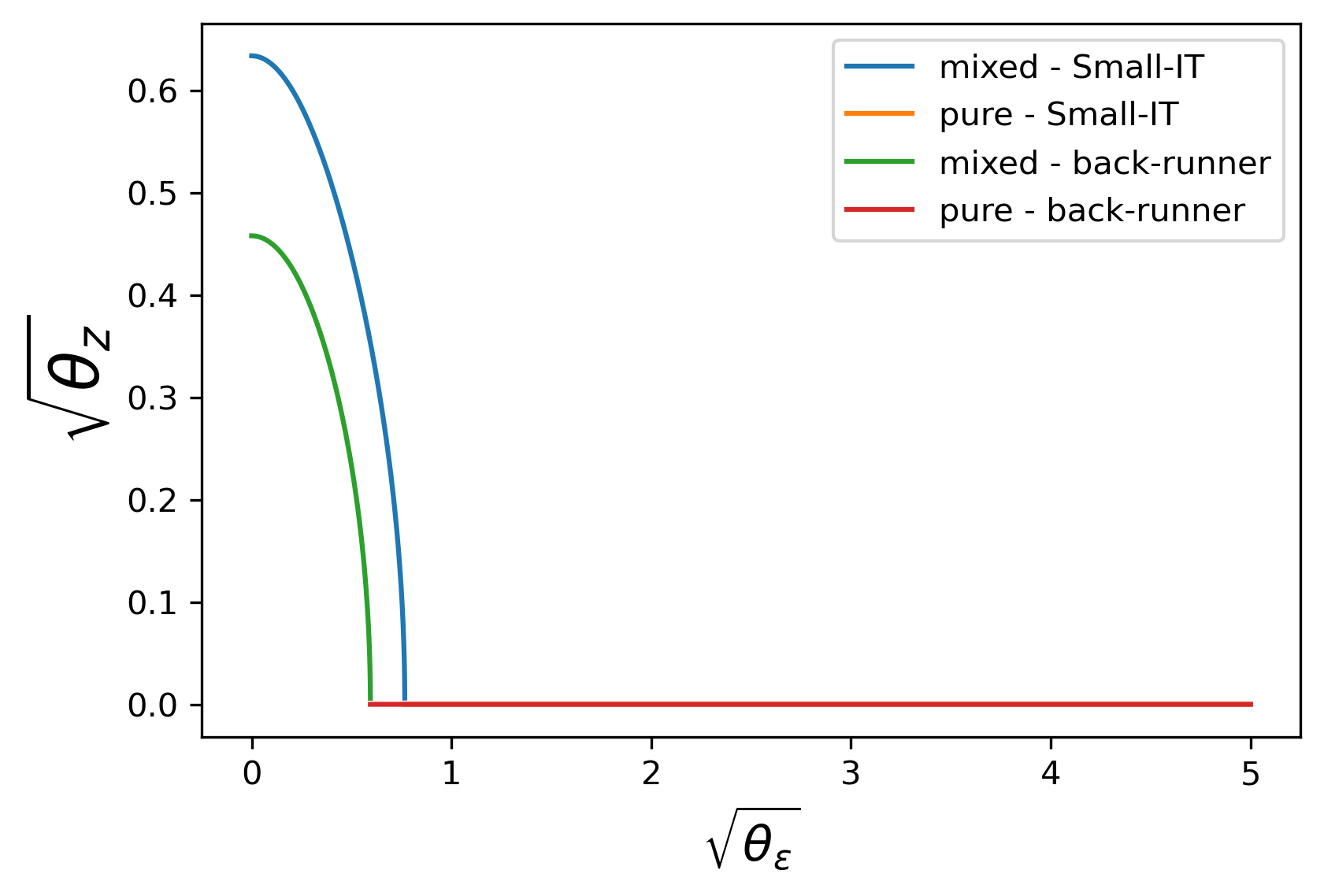}
    }
\subcaptionbox{$\theta_{1_+}=1$}{
    \includegraphics[width = 0.27\textwidth]{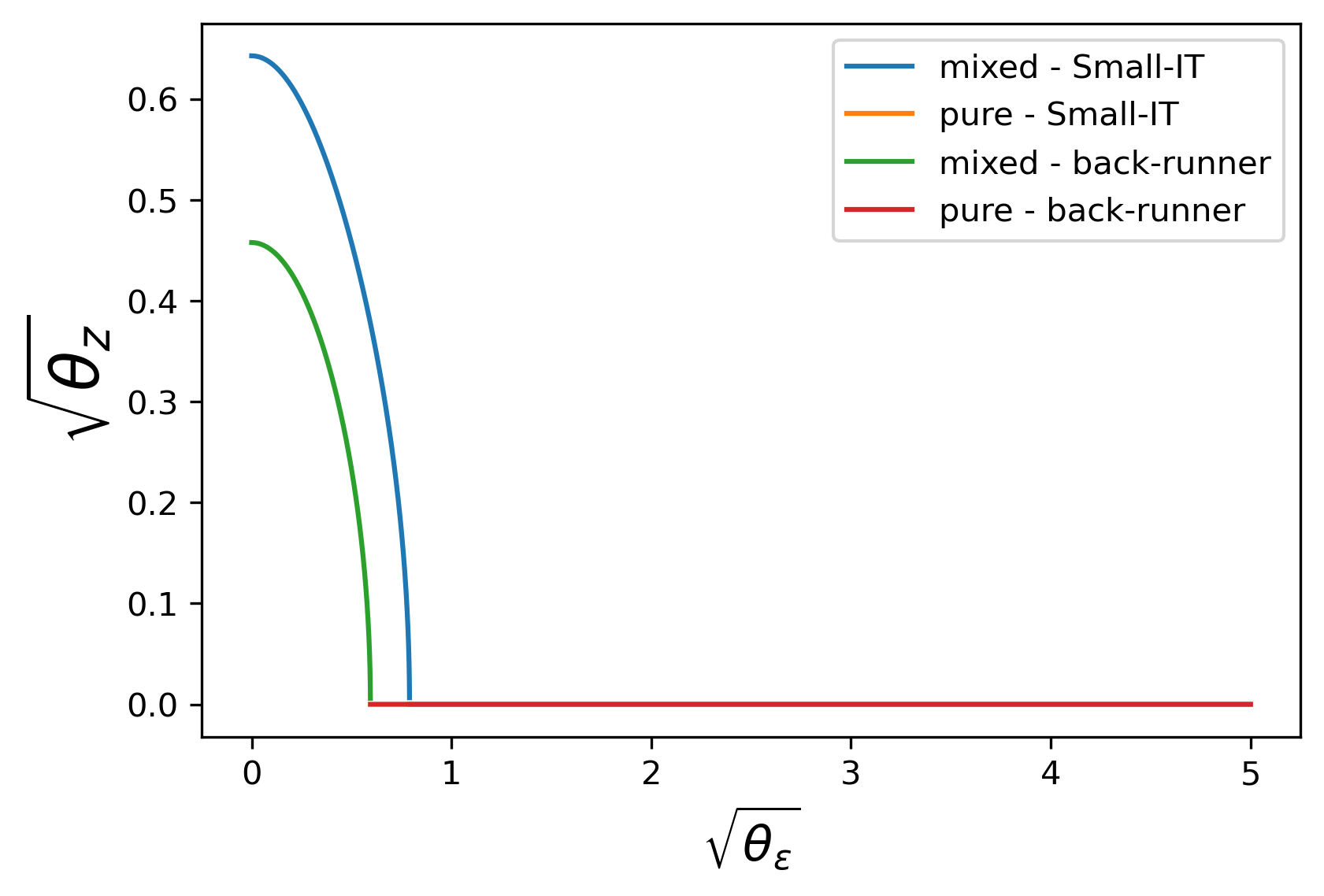}
    }
    \caption{$J=3,$ IT's mixed strategy.}
    \label{figj=3z}
\end{figure}

\begin{figure}[!htbp]
    \centering
\subcaptionbox{$\theta_{1_+}=0.01$}{
    \includegraphics[width = 0.27\textwidth]{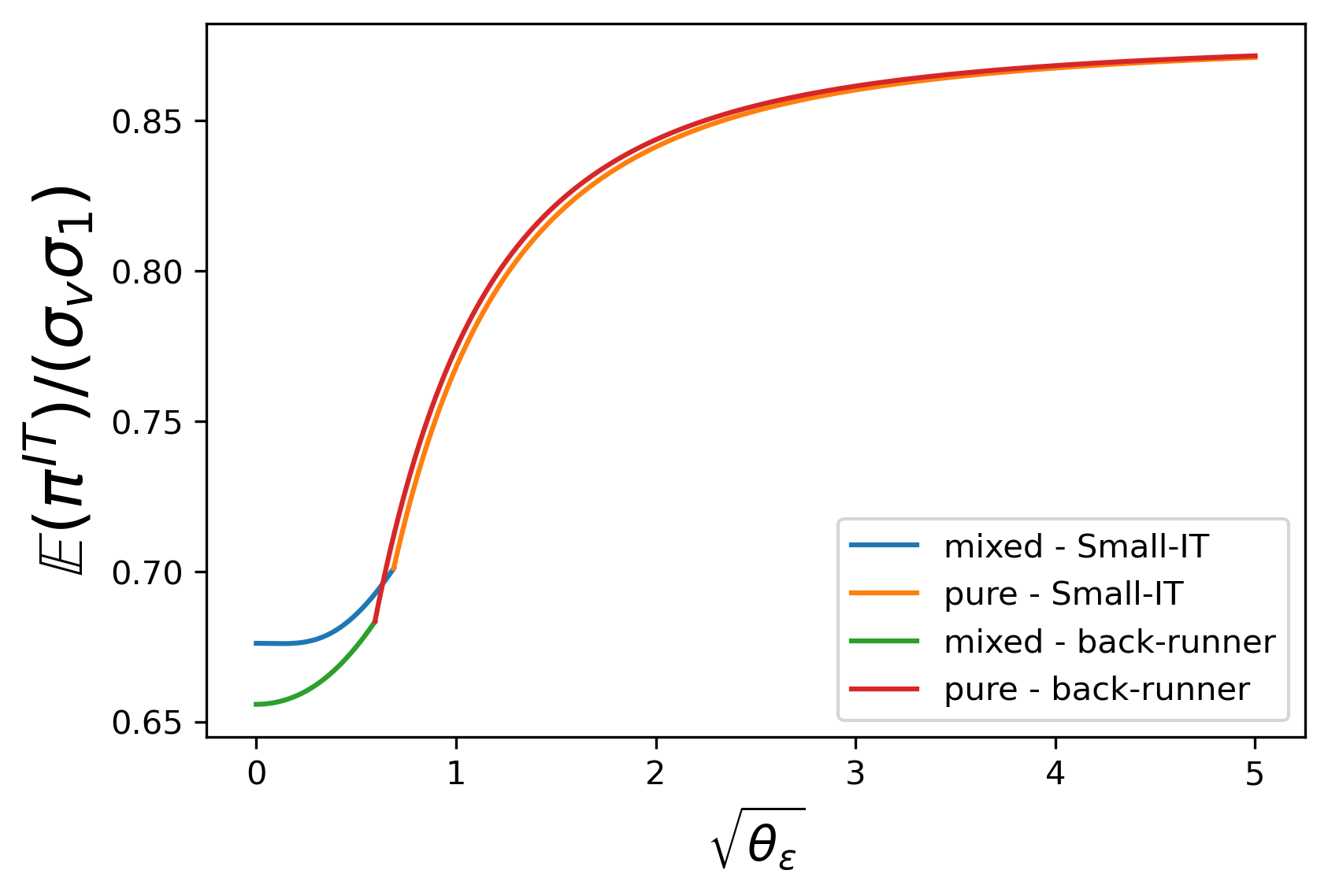}
    }
\subcaptionbox{$\theta_{1_+}=0.1$}{
    \includegraphics[width = 0.27\textwidth]{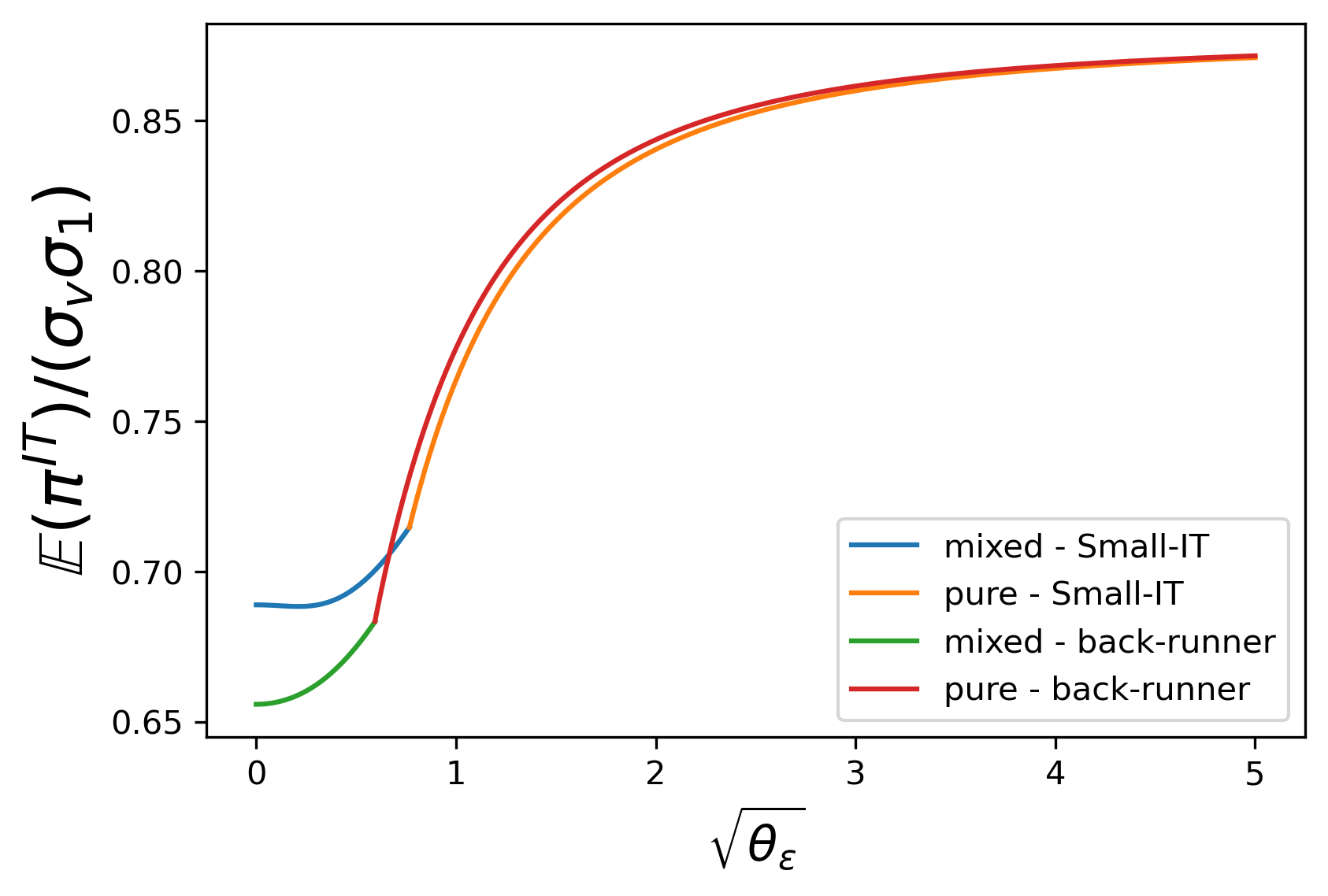}
    }
\subcaptionbox{$\theta_{1_+}=1$}{
    \includegraphics[width = 0.27\textwidth]{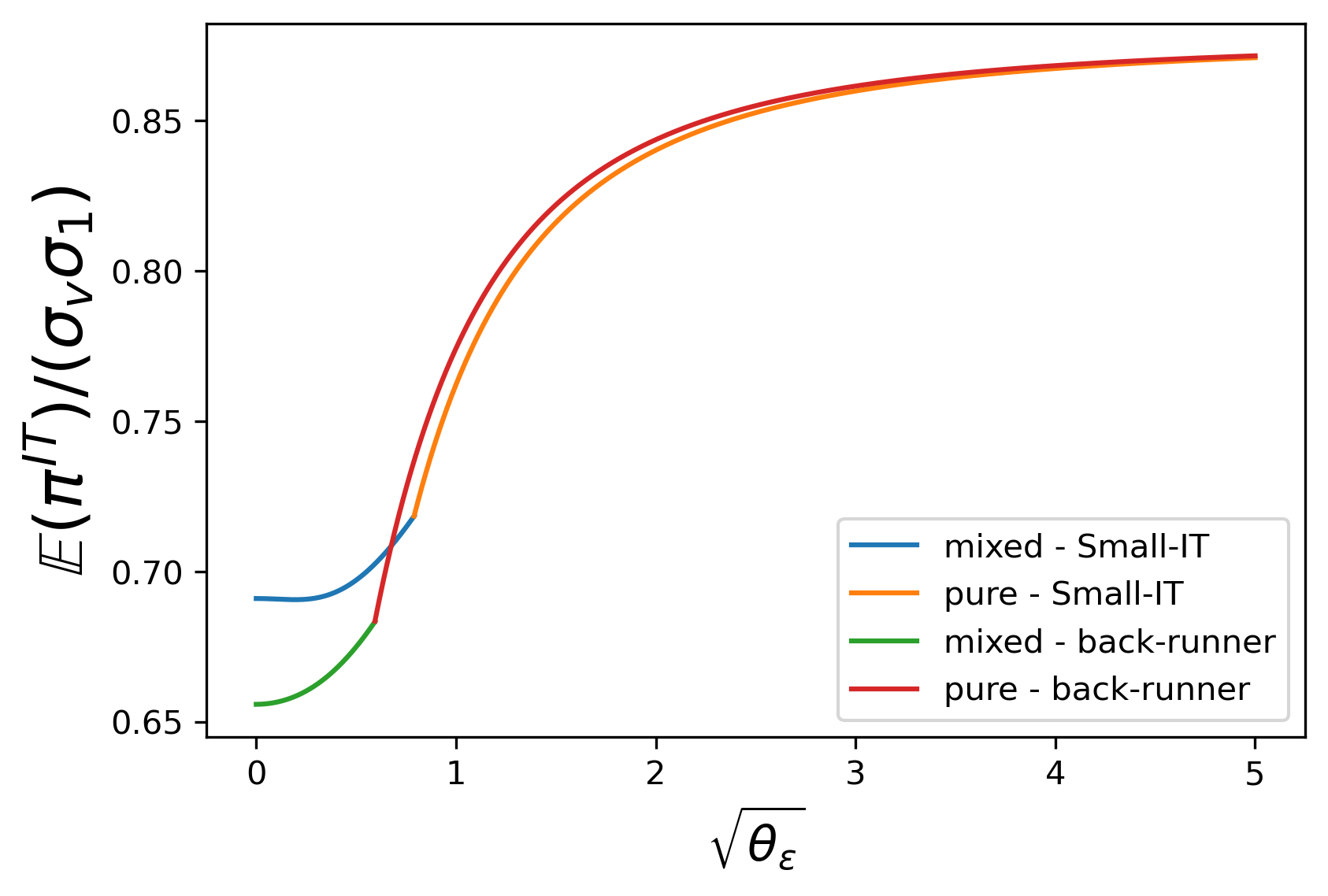}
    }
    \caption{$J=3,$ IT's profit.}
    \label{figj=3piIT}
\end{figure}

\begin{figure}[!htbp]
    \centering
\subcaptionbox{$\theta_{1_+}=0.01$}{
    \includegraphics[width = 0.27\textwidth]{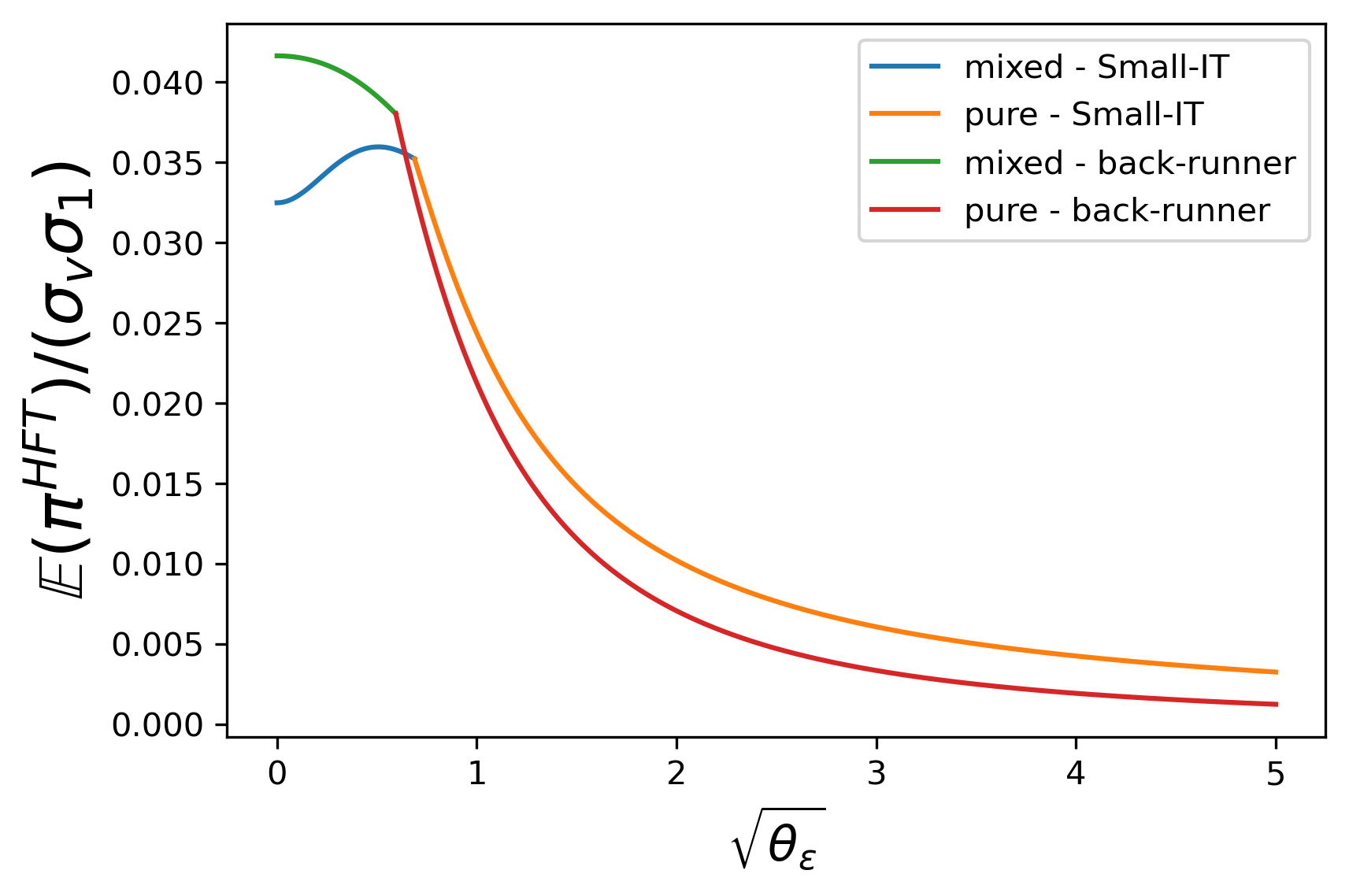}
    }
\subcaptionbox{$\theta_{1_+}=0.1$}{
    \includegraphics[width = 0.27\textwidth]{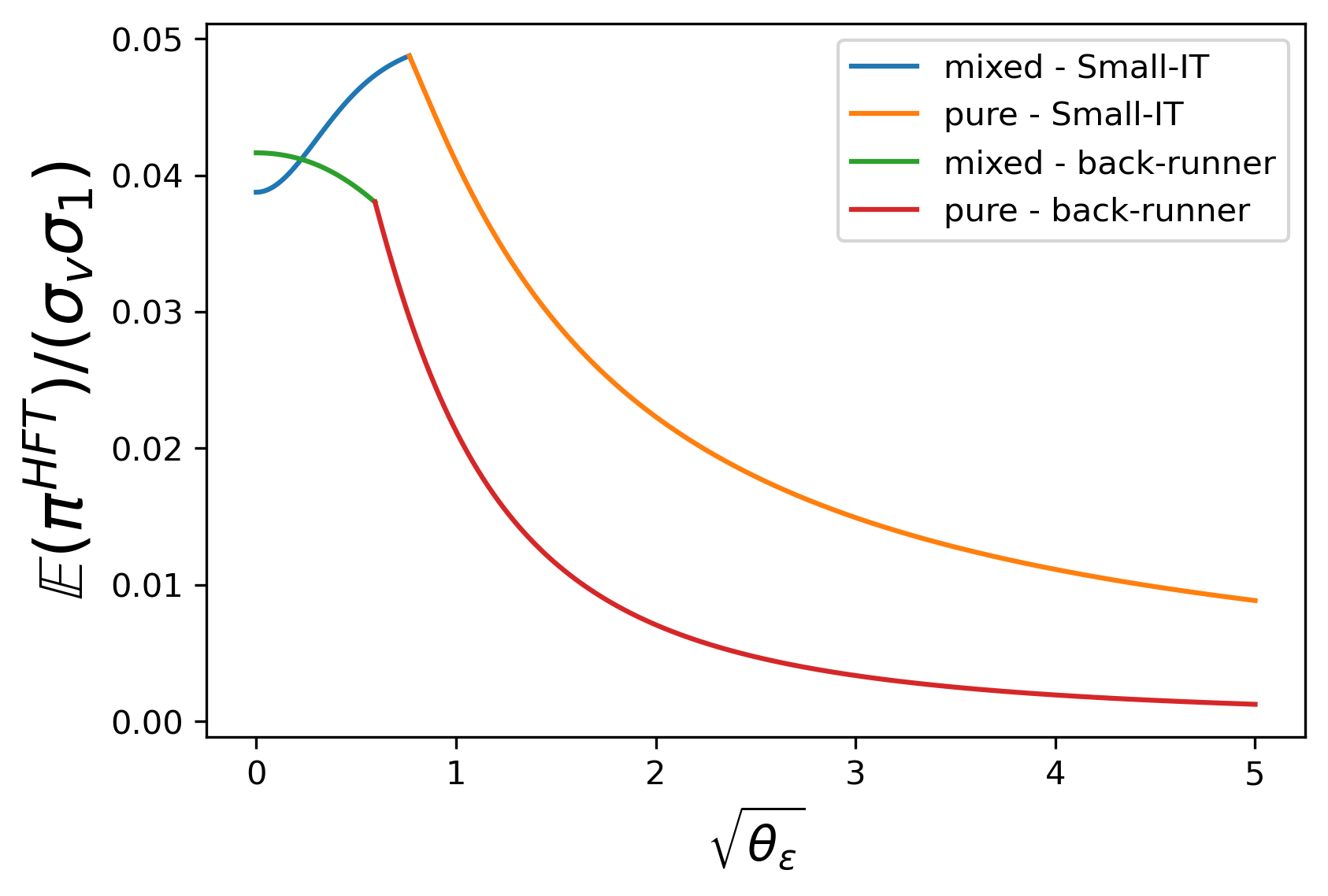}
    }
\subcaptionbox{$\theta_{1_+}=1$}{
    \includegraphics[width = 0.27\textwidth]{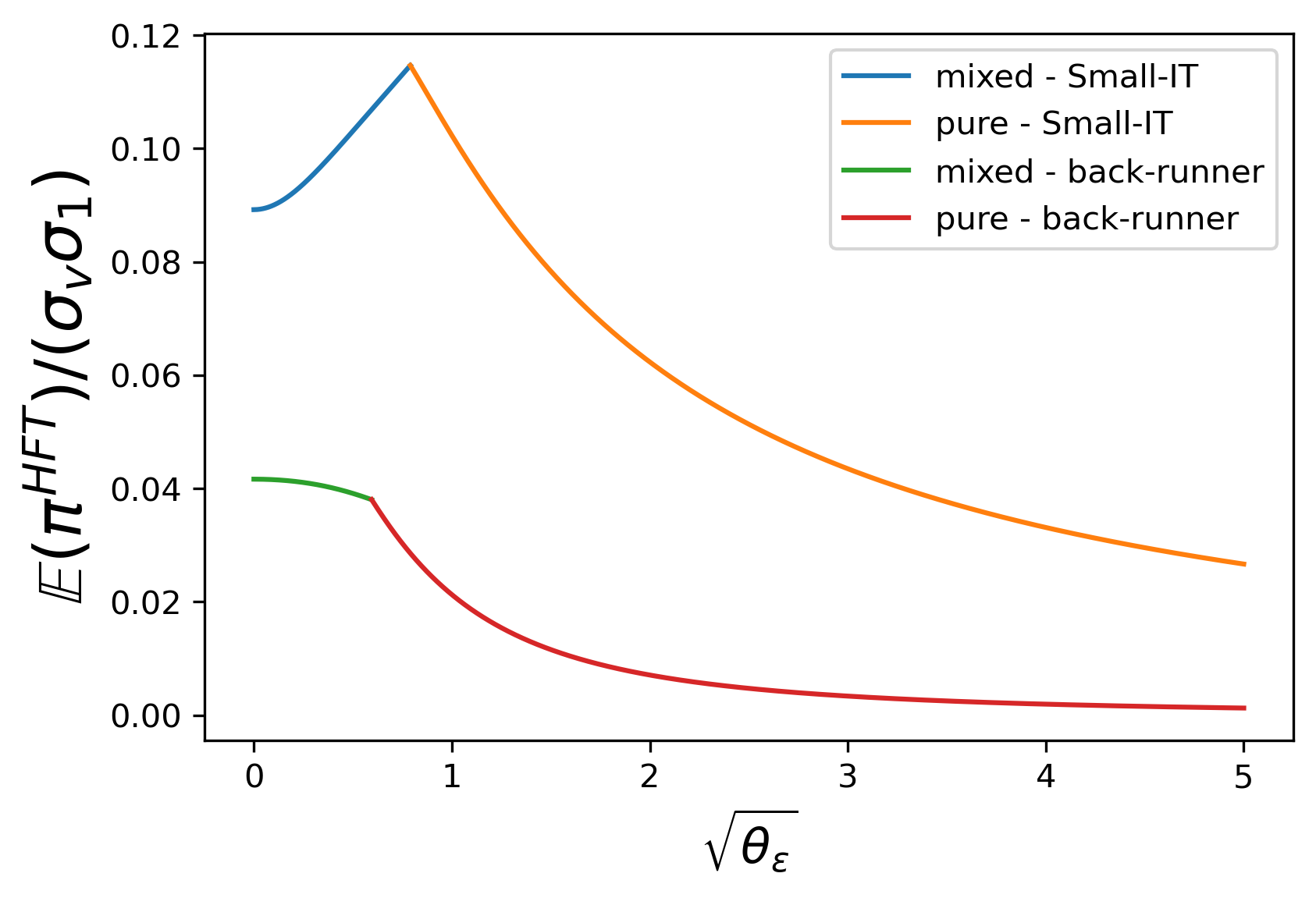}
    }
    \caption{$J=3,$ anticipatory traders' profit.}
    \label{figj=3piHFT}
\end{figure}

\begin{figure}[!htbp]
    \centering
\subcaptionbox{$\theta_{1_+}=0.01$}{
    \includegraphics[width = 0.27\textwidth]{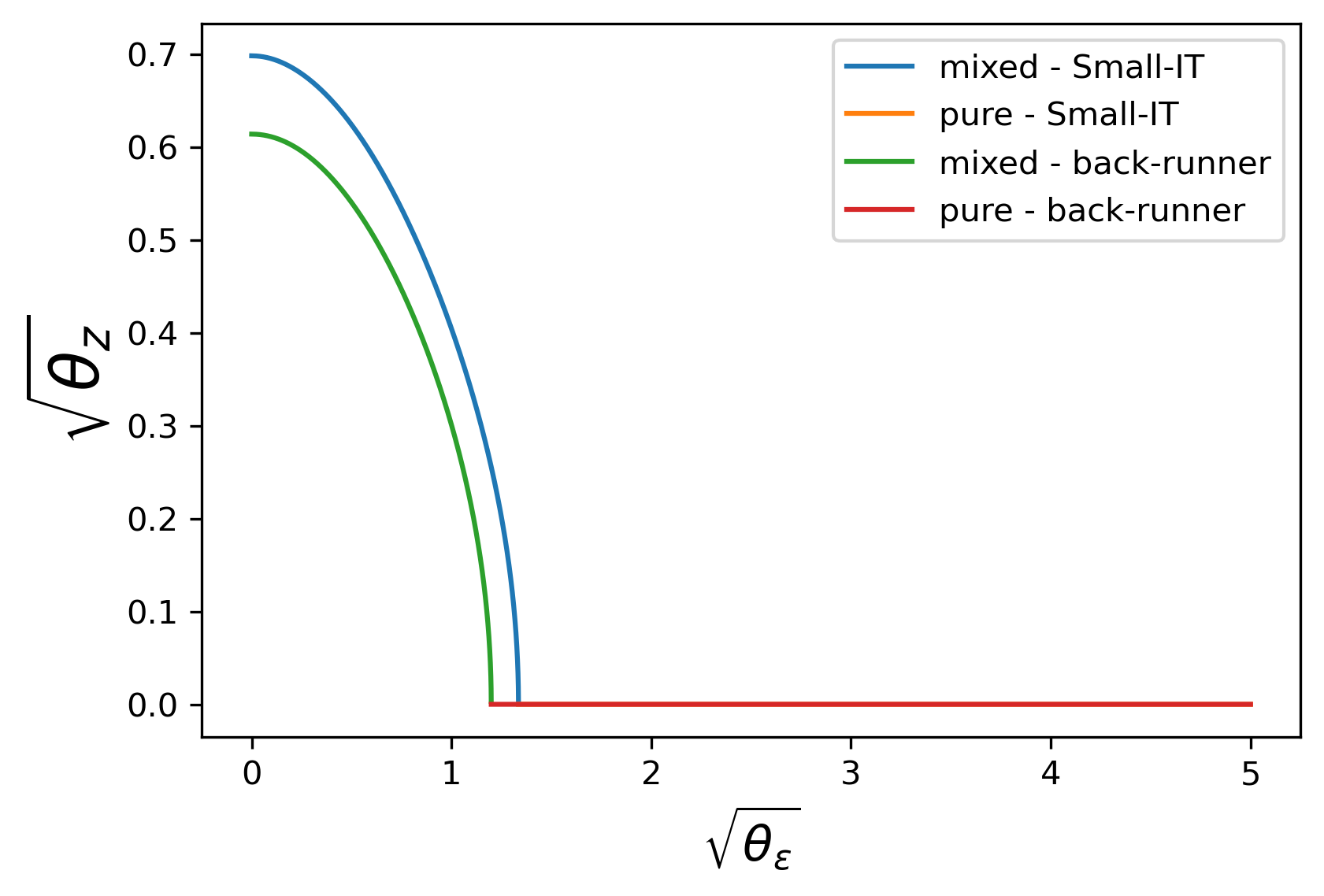}
    }
\subcaptionbox{$\theta_{1_+}=0.1$}{
    \includegraphics[width = 0.27\textwidth]{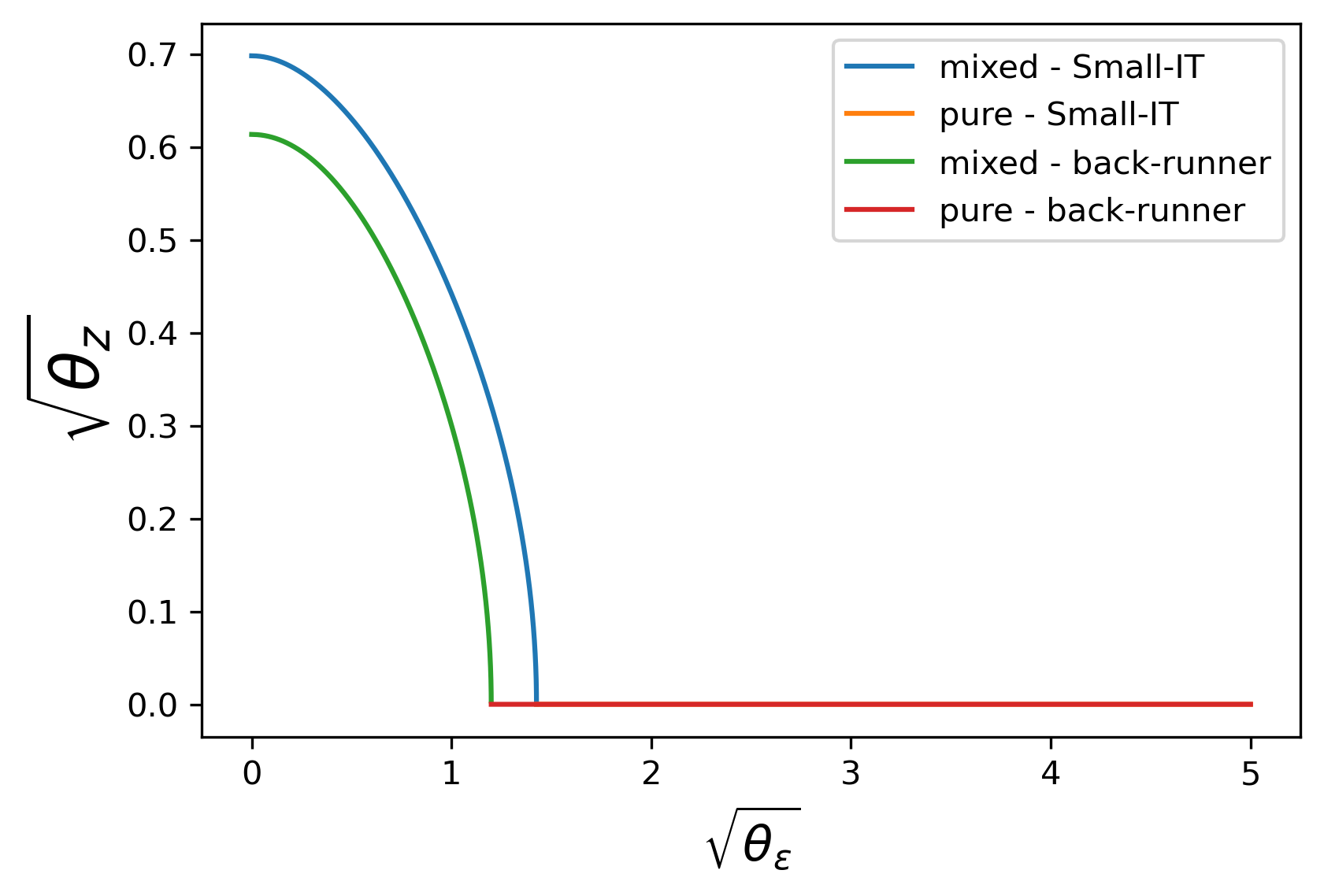}
    }
\subcaptionbox{$\theta_{1_+}=1$}{
    \includegraphics[width = 0.27\textwidth]{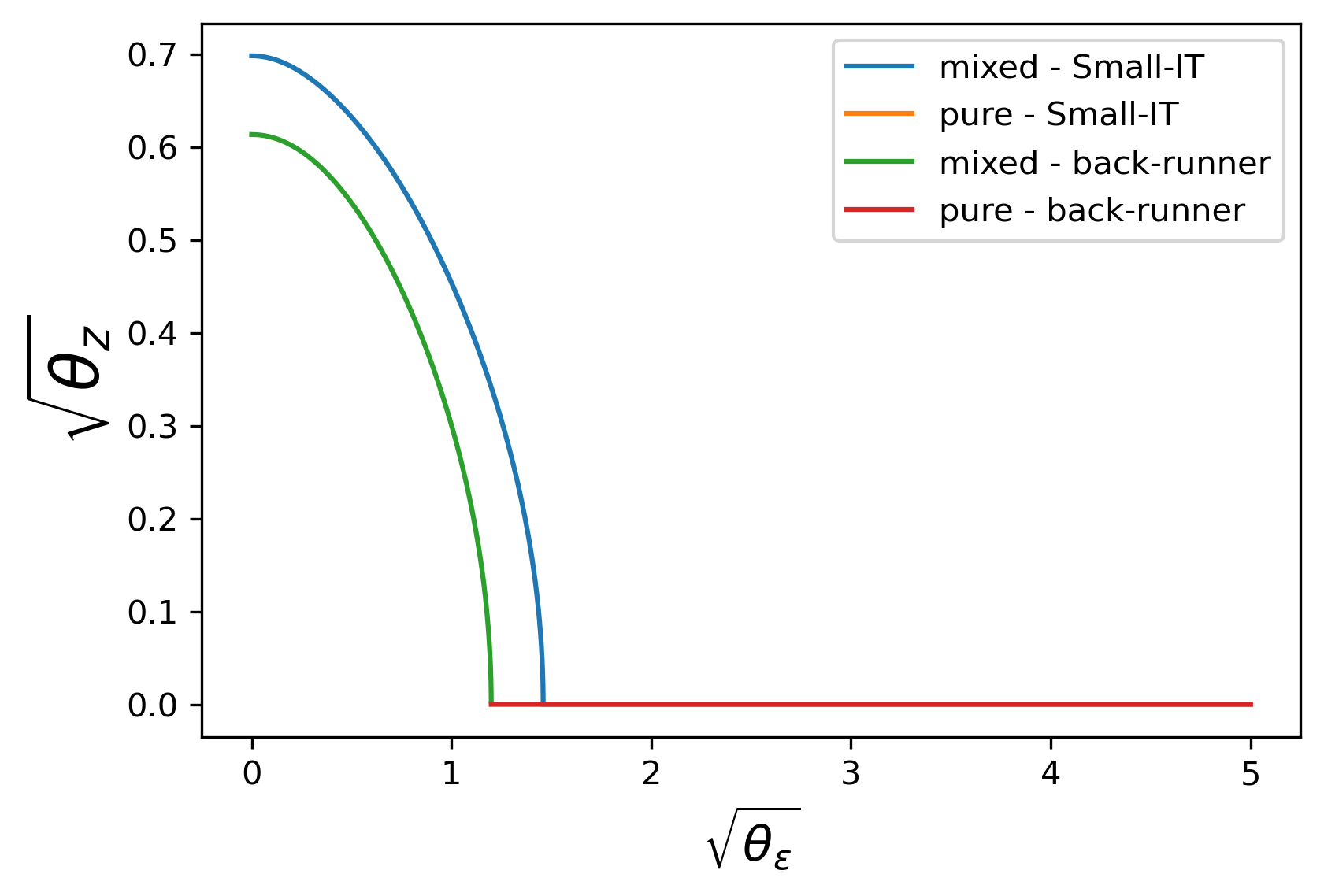}
    }
    \caption{$J=10,$ IT's mixed strategy.}
    \label{figj=10z}
\end{figure}

\begin{figure}[!htbp]
    \centering
\subcaptionbox{$\theta_{1_+}=0.01$}{
    \includegraphics[width = 0.27\textwidth]{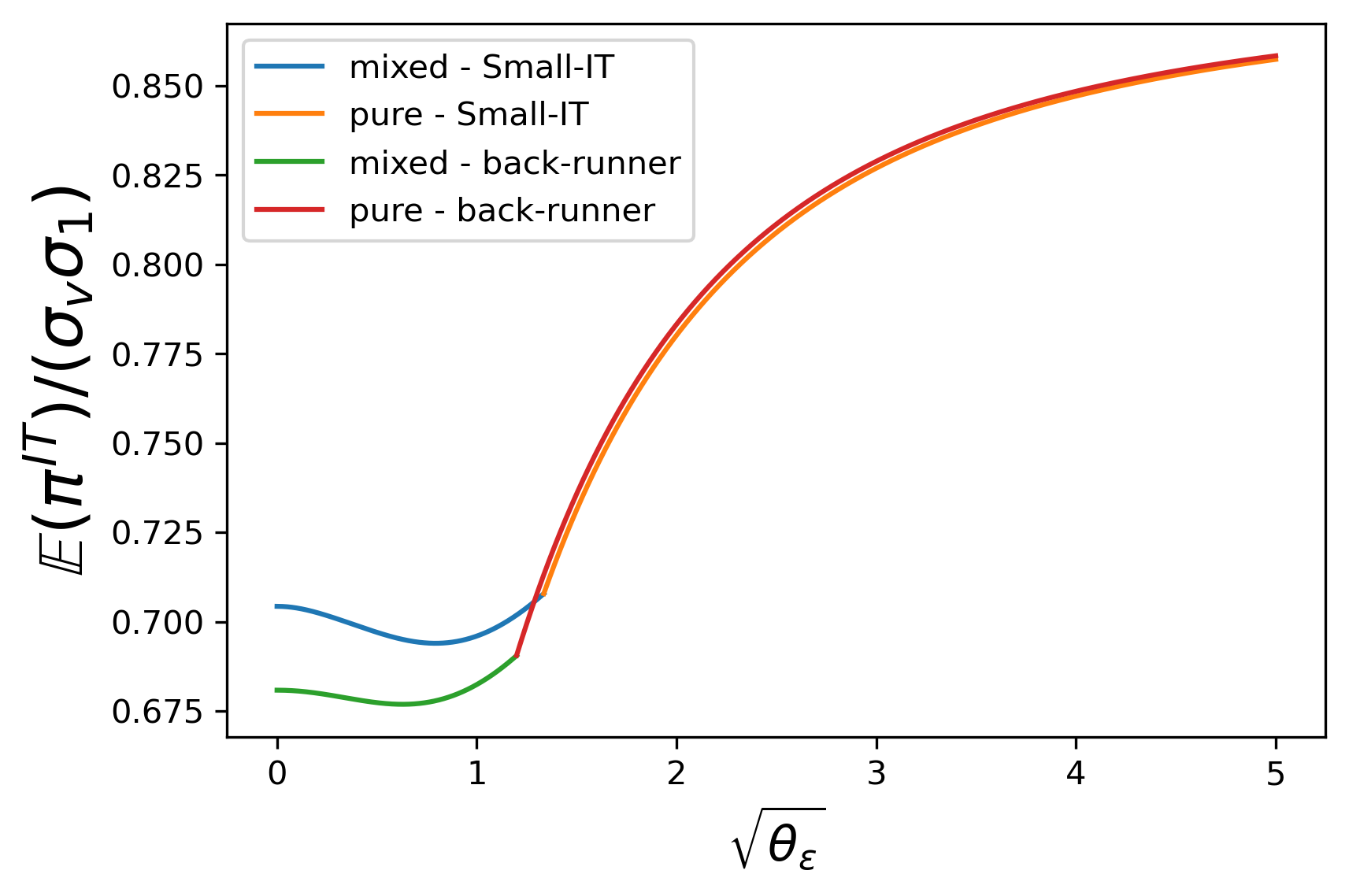}
    }
\subcaptionbox{$\theta_{1_+}=0.1$}{
    \includegraphics[width = 0.27\textwidth]{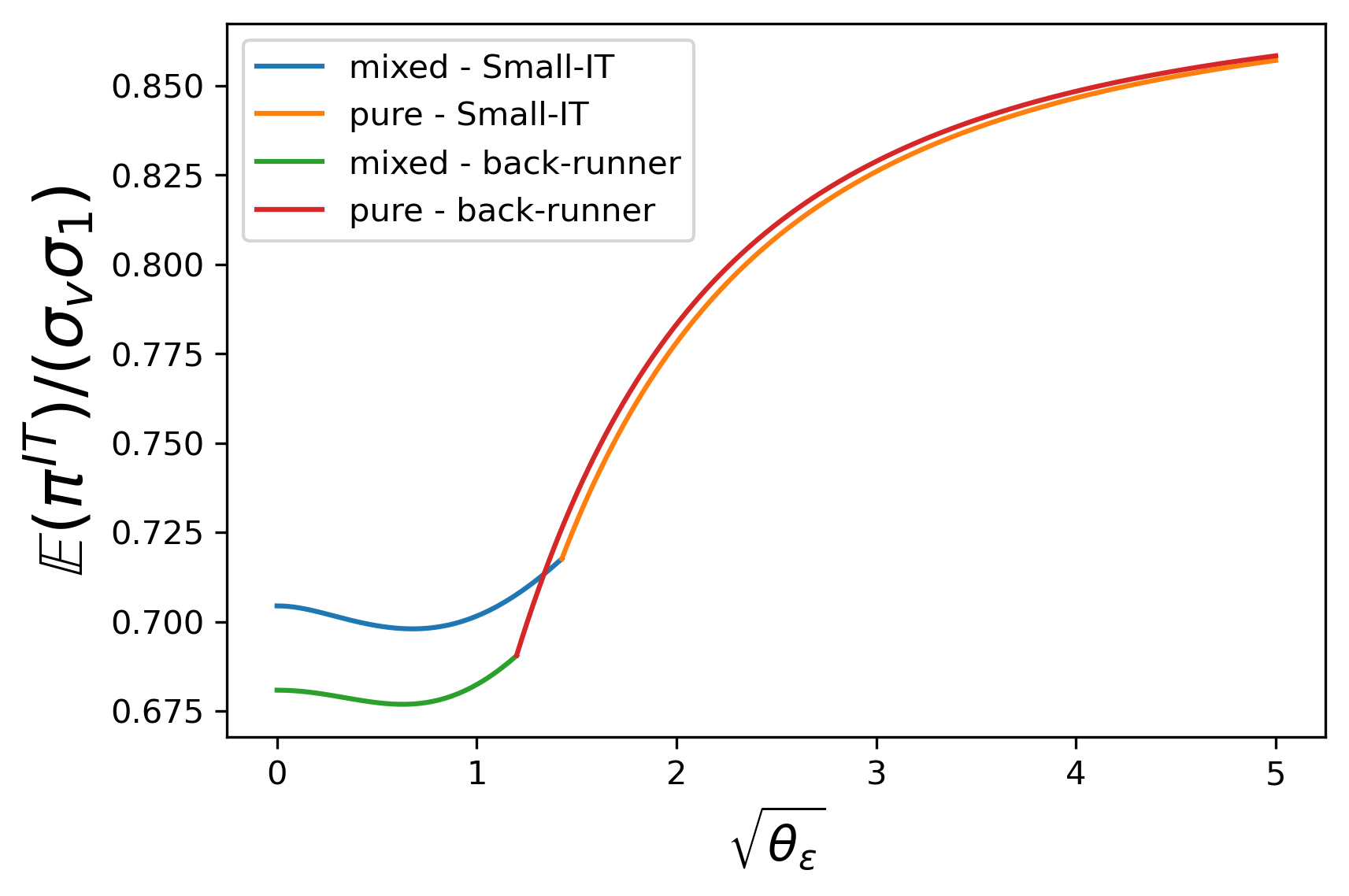}
    }
\subcaptionbox{$\theta_{1_+}=1$}{
    \includegraphics[width = 0.27\textwidth]{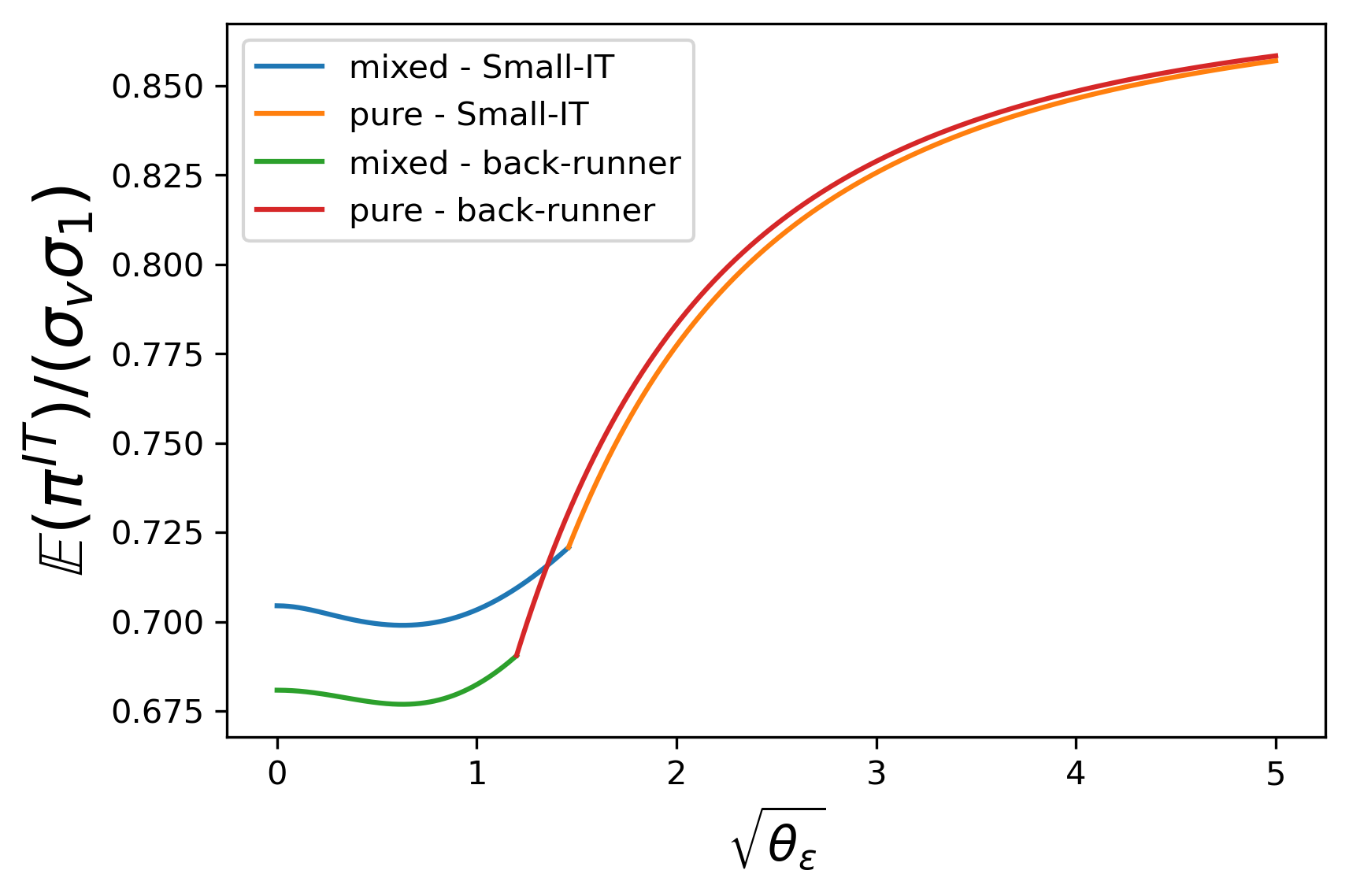}
    }
    \caption{$J=10,$ IT's profit.}
    \label{figj=10piIT}
\end{figure}

\begin{figure}[!htbp]
    \centering
\subcaptionbox{$\theta_{1_+}=0.01$}{
    \includegraphics[width = 0.27\textwidth]{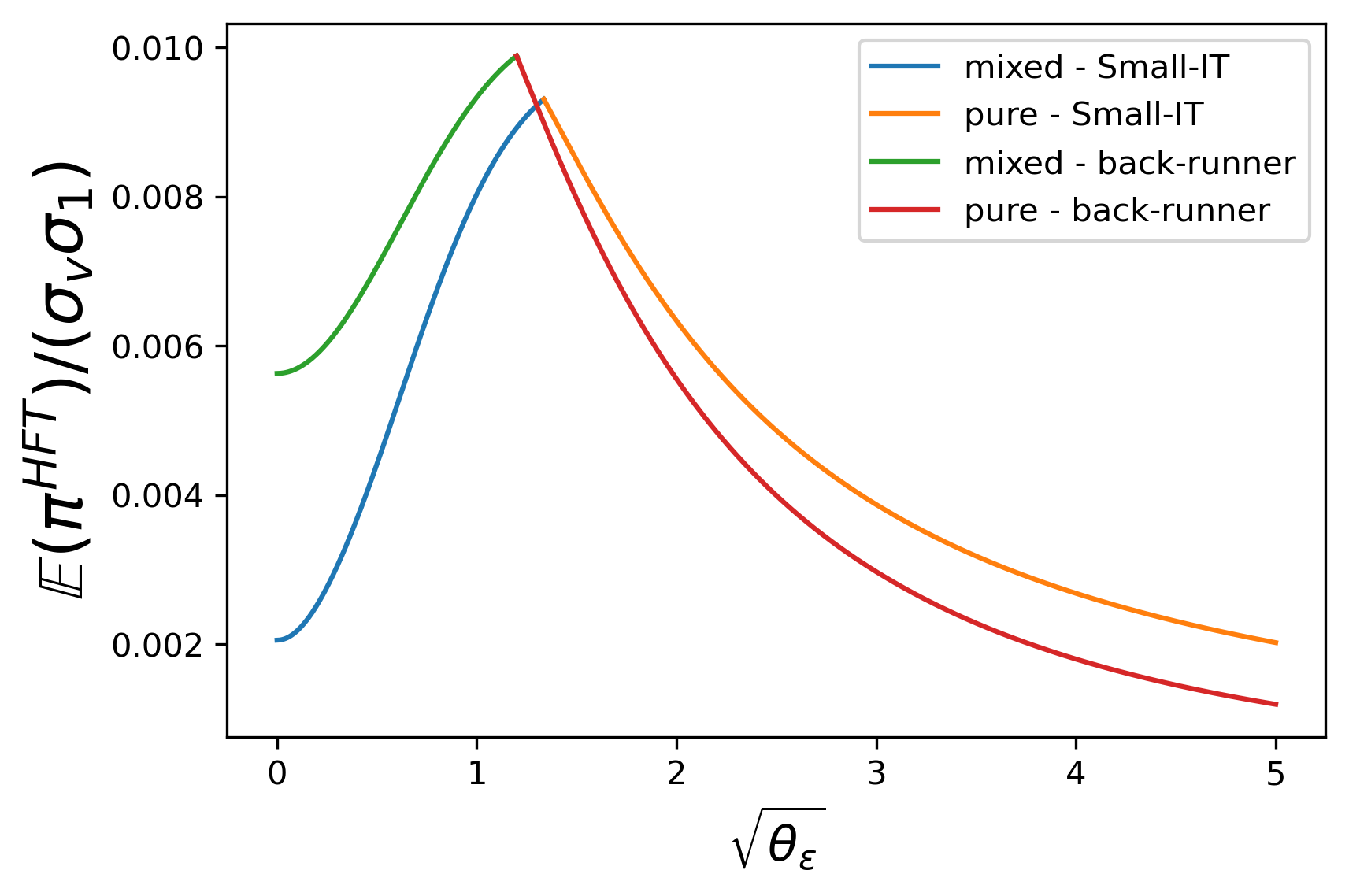}
    }
\subcaptionbox{$\theta_{1_+}=0.1$}{
    \includegraphics[width = 0.27\textwidth]{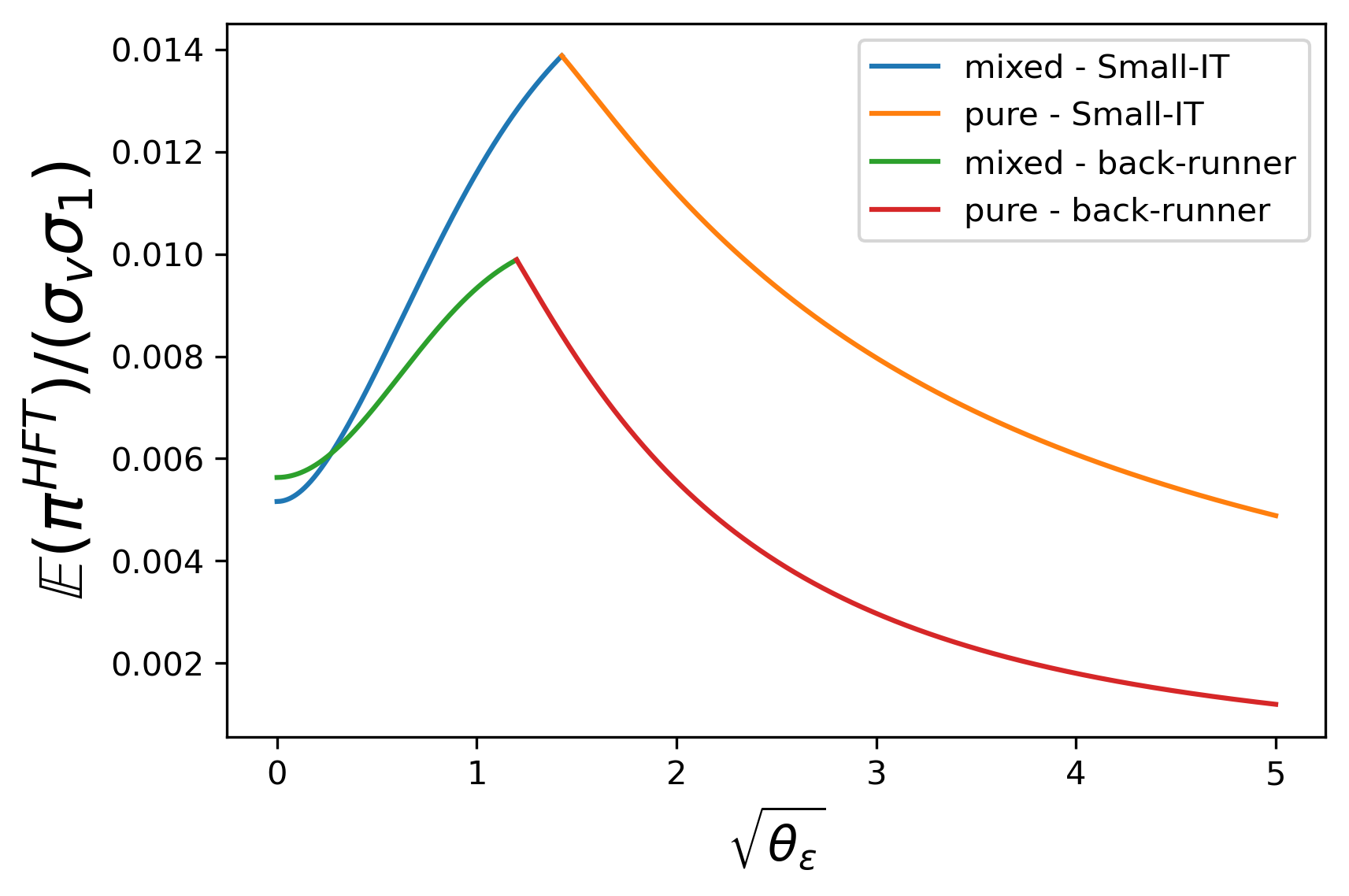}
    }
\subcaptionbox{$\theta_{1_+}=1$}{
    \includegraphics[width = 0.27\textwidth]{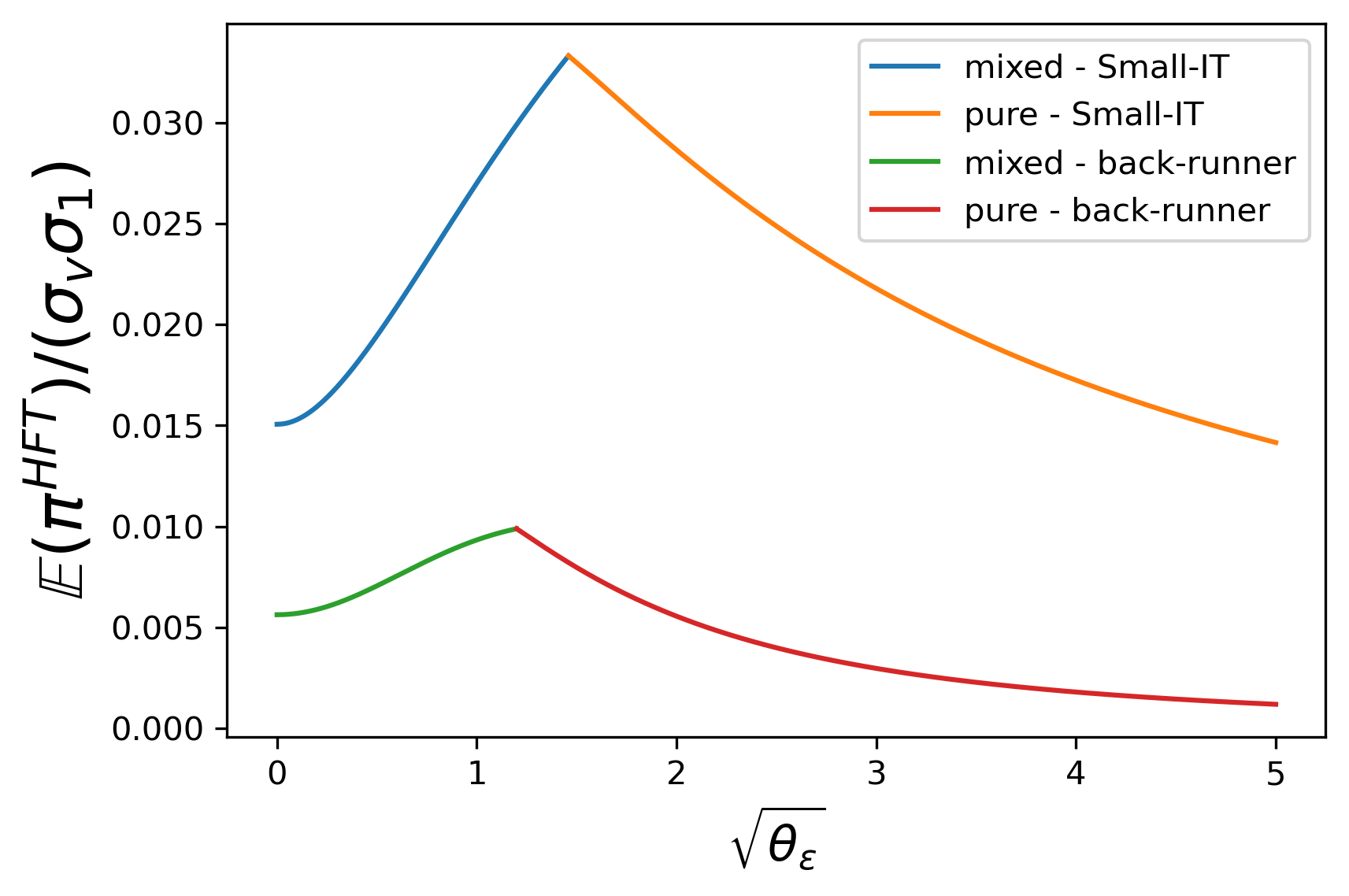}
    }
    \caption{$J=10,$ anticipatory traders' profit.}
    \label{figj=10piHFT}
\end{figure}

We do comparative static analyses on $\theta_{\varepsilon}$ in different markets distinguished by $\theta_{1_+}$ and $J$. In Figure \ref{figj=1z} - \ref{figj=10piHFT}, the blue and orange (green and red) lines represent different situations in mixed-strategy and pure-strategy equilibrium when IT meets Small-IT (back-runner).

$\theta_{1_+}=10^{-4}$ stands for a market with little high-speed noise trading, where Small-IT's fast trading $x_1$ would bring her large costs. This almost forces her to give up the speed competition. Consequently, in (a) of Figure \ref{figj=1z} - \ref{figj=1piHFT}, investors' performances are nearly the same as those in \cite{yang2020back}. It actually holds for all $J\leq10$. From (b) and (c) of Figure \ref{figj=1z}, IT increases the intensity $\theta_z$ of randomization and expands the mixed-strategy region. This indicates that when the opponents' trading speed improves, it is optimal for IT to apply significantly more sophisticated randomization to disturb them.

As displayed in Figure \ref{figj=1piIT}, we find surprisingly that when equilibria with Small-IT and back-runner are both in the mixed-strategy region, the improvement of anticipatory trader's speed promotes IT's profit instead.
On the one hand, Small-IT's additional same-direction trading $x_1$ definitely increases IT's cost, and the growth of $\theta_z$ also makes IT unable to implement her investment accurately. However, on the other hand, a larger $\theta_z$ makes it more difficult for Small-IT to detect $v$ from the signal $\Tilde{i}_1.$ As a result, she trades more conservatively in period 2. When the second effect goes beyond the former one, IT profits more with a Small-IT. When equilibria with a Small-IT and a back-runner are both within the pure-strategy region, expectedly, the improvement of anticipatory trader's speed lowers IT's profit. 

As for anticipatory trader's profit shown in Figure \ref{figj=1piHFT}, she is benefited by the higher speed. 

Comparing Figure \ref{figj=1piIT}
and \ref{figj=1piHFT}, as $\theta_\varepsilon$ grows, IT's profit increases since with higher exogenous noise $\theta_\varepsilon$, she doesn't need to add too much endogenous noise, which helps her exploit private information more precisely; in contrast, the anticipatory trader receives a less accurate signal and her profit decreases.

When $J=2,$ with the promotion of anticipatory trader's speed, IT's intensity of randomization and the mixed-strategy region still grows, as shown in Figure \ref{figj=2z}; her profit with Small-ITs is still higher when signals are precise, as shown in Figure \ref{figj=2piIT}.

Variations appear in Figure \ref{figj=2piHFT} for anticipatory traders' profit. From (a) in Figure 
\ref{figj=2piHFT}, we find that if both $\theta_{1_+}$ and $\theta_\varepsilon$ are small, i.e., high-speed noise trading is not active and anticipatory traders' signal are of high accuracy, the improvement of trading speed will reduce anticipatory traders' profit. The former condition makes Small-ITs' preemptive trading bring significant transaction costs to each other. And the latter condition makes IT add greater noise. Both of the above circumstances are bad for anticipatory traders.

An interesting result is presented 
by the blue lines in (b) and (c) of Figure \ref{figj=2piHFT}: Small-IT's profit increases with $\theta_{\varepsilon}$ when $\theta_\varepsilon$ is not large. 
It is because, with a larger $\theta_{\varepsilon}$, Small-IT is less disturbed by the endogenous noise added by IT, which outweighs the disadvantage of more exogenous noise. A similar pattern appears for back-runners in \cite{yang2020back} when $J\geq4$, compared to $J=2$ here. The improvement of anticipatory traders' speed drives IT to decrease $\theta_z$ more sharply when $\theta_\varepsilon$ increases, as displayed by (b) and (c) in Figure \ref{figj=2z}, and thus the pattern appears with a smaller number of anticipatory traders.

In Figure \ref{figj=2piHFT}, when $\theta_{1_+}=0.01,$ this pattern does not appear. In fact, it holds for $J\geq3,$ as shown in (a) of Figure \ref{figj=3piHFT}. The reason is that a smaller $\theta_{1_+}$ suppresses the activity of Small-IT, $\theta_z$'s sensitivity to $\theta_\varepsilon$ needs to be amplified by a larger number of Small-ITs. 

When $J=3$, as displayed in Figure \ref{figj=3piIT}, in the mixed-strategy equilibrium with Small-ITs, IT's profit first decreases with $\theta_{\varepsilon}$ then increases with it. In \cite{yang2020back}, it appears when $J\geq6$. It is also because the growth of anticipatory traders' speed makes the intensity of randomization more sensitive to changes in signal noise.

The case $J=10$ is displayed to verify the reached conclusions. When IT's past trading is leaked and her future trading is anticipated by faster traders: (1) IT mixes more and may profit more in the mixed-strategy region, but profits less in the pure-strategy region; (2) anticipatory trader profits more in the pure-strategy region, but in the mixed-strategy region, if both $\theta_{1_+}$ and $\theta_\varepsilon$ are relatively small, anticipatory trader may profit less; (3) the phenomena that IT's profit first decreases with $\theta_\varepsilon$ and anticipatory trader's profit first increases with $\theta_\varepsilon$ appear for smaller $J$ compared to \cite{yang2020back}.

\newpage
\subsection{A market with Round-Trippers}
\label{subsecRT}
In this section, we assume that $J_2=J\geq1$ and $J_1=0.$ It is proved in Theorem \ref{thmgaminf} that HFTs are all Round-Trippers who take liquidity first and later provide equal liquidity back. The $j$-th HFT takes the following strategy:
\begin{equation*}
\begin{aligned}
&x_{1j}=\beta_{12}(\Tilde{i}_{1j}-\mathbb{E}(\Tilde{i}_{1j}|y_1)),\\
&x_{2j}=-x_{1j}.
\end{aligned}
\end{equation*}

\begin{theorem}[Simplification of the equilibrium with Round-Trippers]
\label{thmgaminf}
The mixed-strategy equilibrium is characterized by system \eqref{systemmixedgaminf} and \eqref{systemmixedgaminf-2}; the pure-strategy equilibrium is characterized by system \eqref{systempuregaminf} and \eqref{systempuregaminf-2}. What's more, $\beta_{12}>0.$
\end{theorem}

Next, the equilibrium is solved and studied by numerical experiments. Given the number of Round-Trippers $J$, the equilibrium is only decided by market noise $\theta_{1_+}$ and signal noise $\theta_\varepsilon$. Therefore, we analyze how these two parameters affect investors' strategy and profits and show them in 3D graphs.

\noindent\textbf{IT's strategy.} 
\begin{figure}[!htbp]
    \centering
\subcaptionbox{$J=1$}{
    \includegraphics[width = 0.27\textwidth]{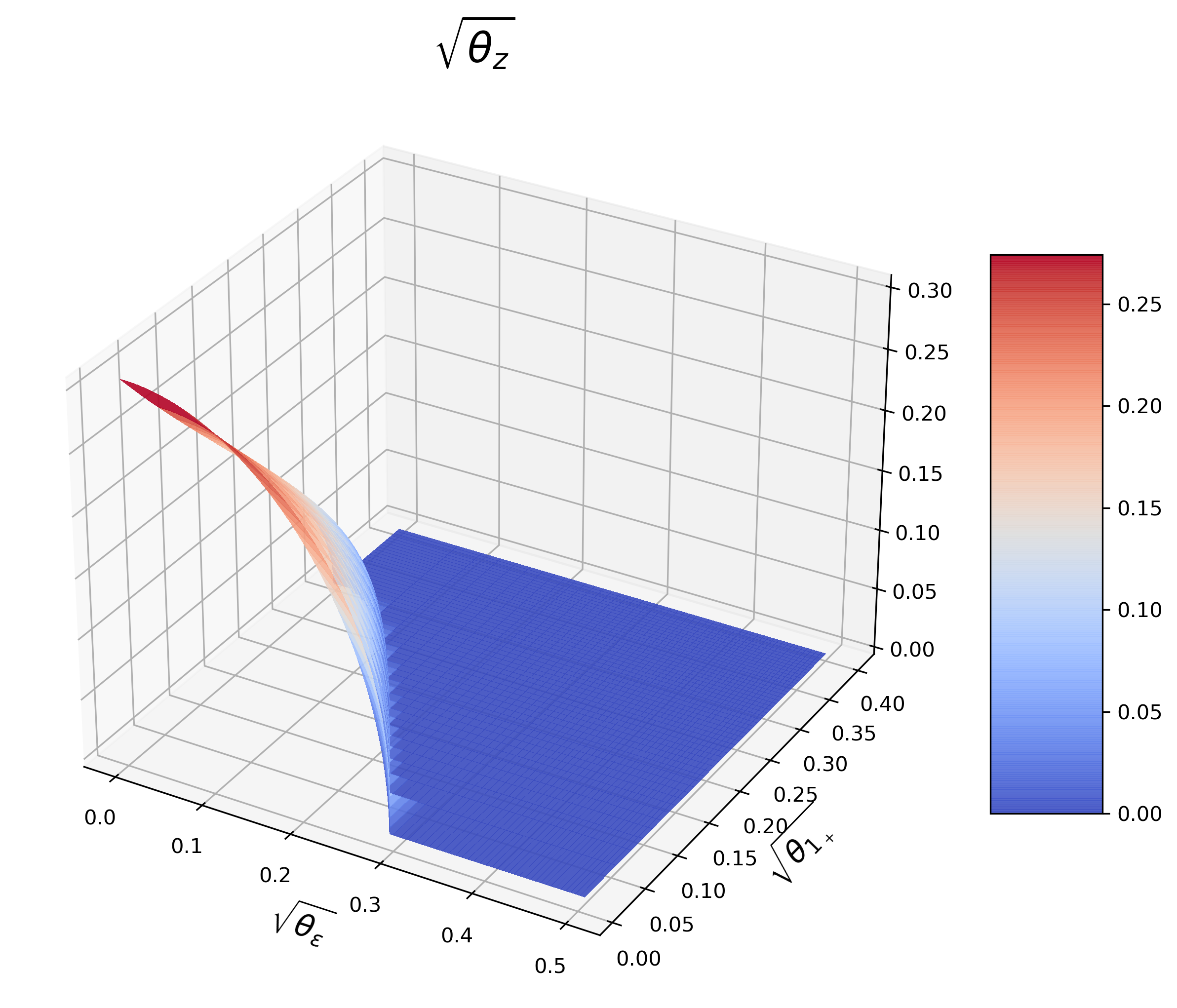}
    }
\subcaptionbox{$J=5$}{
    \includegraphics[width = 0.27\textwidth]{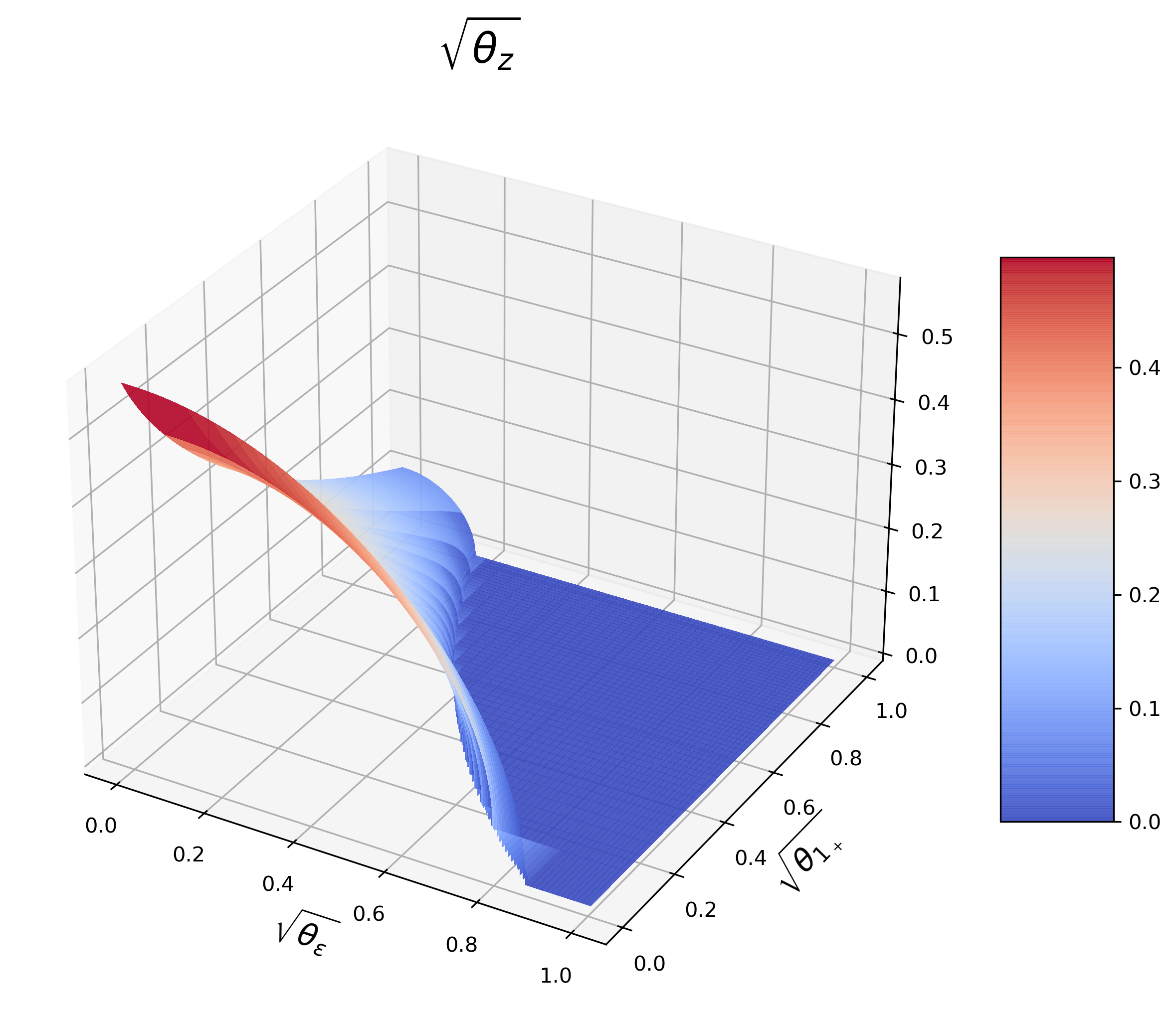}
    }
\subcaptionbox{$J=10$}{
    \includegraphics[width = 0.27\textwidth]{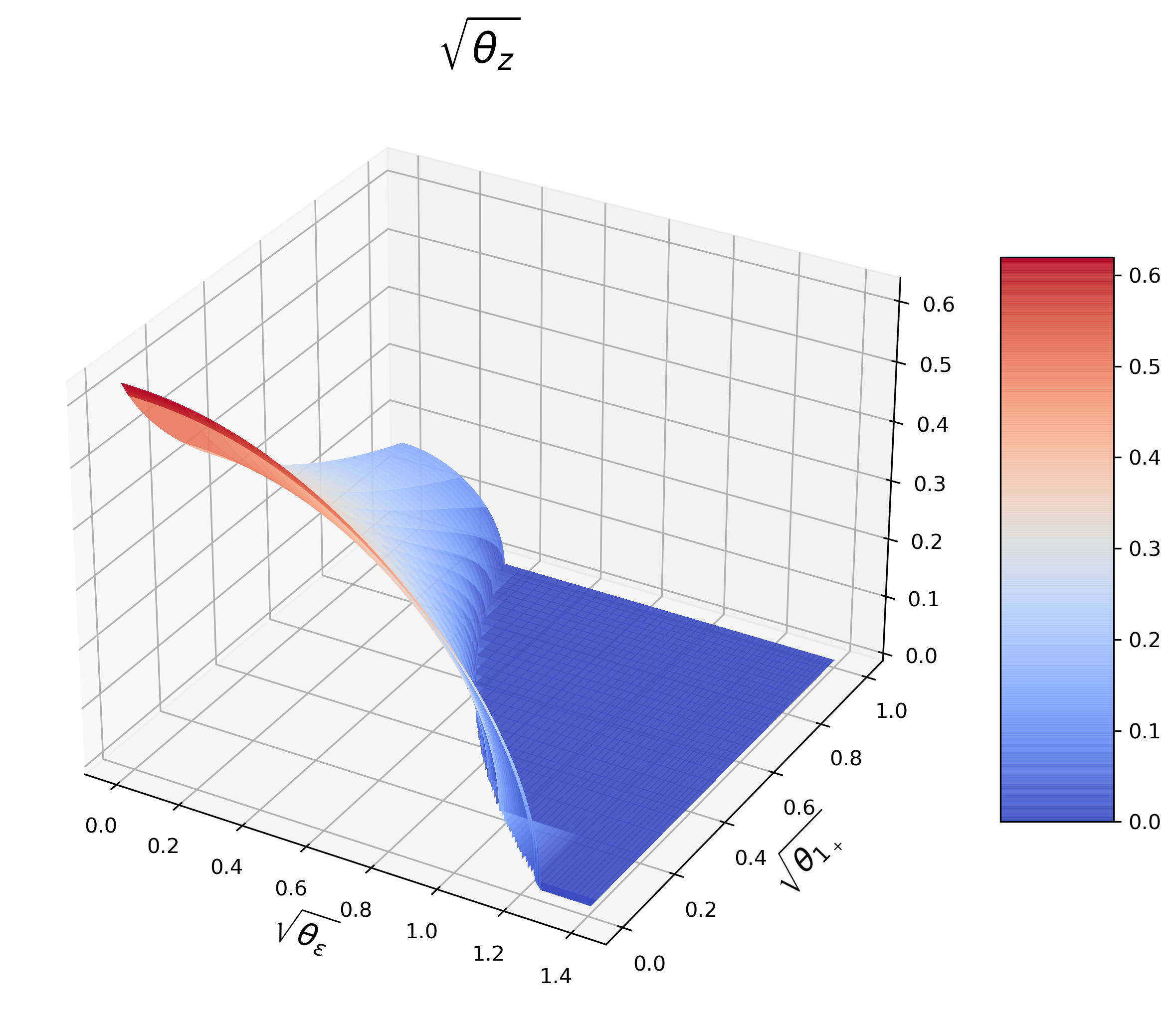}
    }
    \caption{IT's mixed strategy.}
    \label{3Dz}
\end{figure}

\begin{itemize}
    \item Given $J\geq1$ and $\theta_{1_+}>0,$ there exists a $\underline{\theta}_\varepsilon\geq0,$ if $\theta_\varepsilon>\underline{\theta}_\varepsilon,$ $\theta_z=0.$
    \item Given $J\geq1$ and $\theta_\varepsilon\geq0,$ there exists a $\underline{\theta}_{1_+}>0,$ if $\theta_{1_+}>\underline{\theta}_{1_+},$ $\theta_z=0.$
    \item Given $J\geq1,$ there exists a $\Bar{\theta}_{1_+},$ if $\theta_{1_+}>\Bar{\theta}_{1_+}$, $\theta_z=0$ for any $\theta_\varepsilon\geq0.$ Given $J\leq3,$ $\Bar{\theta}_{1_+}\in(0,1).$
\end{itemize}
\begin{figure}[!htbp]
    \centering
    \includegraphics[width = 0.27\textwidth]{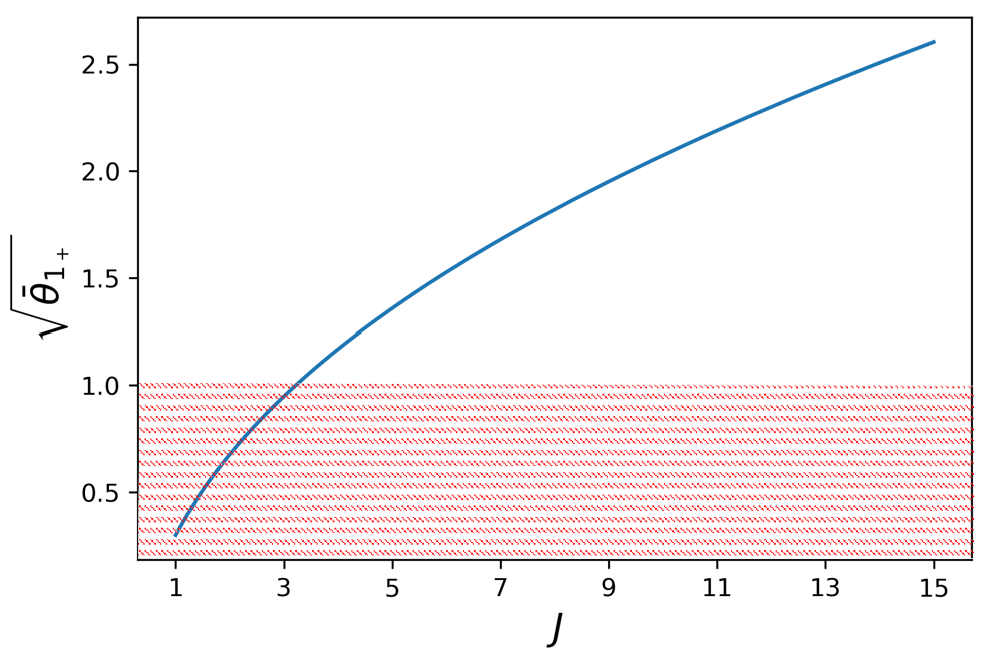}
    \caption{The critical $\Bar{\theta}_{1_+}$ such that IT always takes pure strategy.}
    \label{figgaminfty-bartheta2}
\end{figure}

The third conclusion is further demonstrated in Figure \ref{figgaminfty-bartheta2}, where the dotted area represents cases with $\Bar{\theta}_{1_+}\leq1,$ which may occur in the actual markets.
The critical $\Bar{\theta}_{1_+}$ falls into this region when $J\leq3.$
That is to say, when the number of Round-Trippers is small and the market is quite active, even if IT's intention is detected by Round-Trippers perfectly, she does not need to randomize orders. But when there is a large number of Round-Trippers, it is better for IT to adopt the mixed strategy to protect herself.

\noindent\textbf{IT's profit.} 
\begin{figure}[!htbp]
    \centering
\subcaptionbox{$J=1$}{
    \includegraphics[width = 0.27\textwidth]{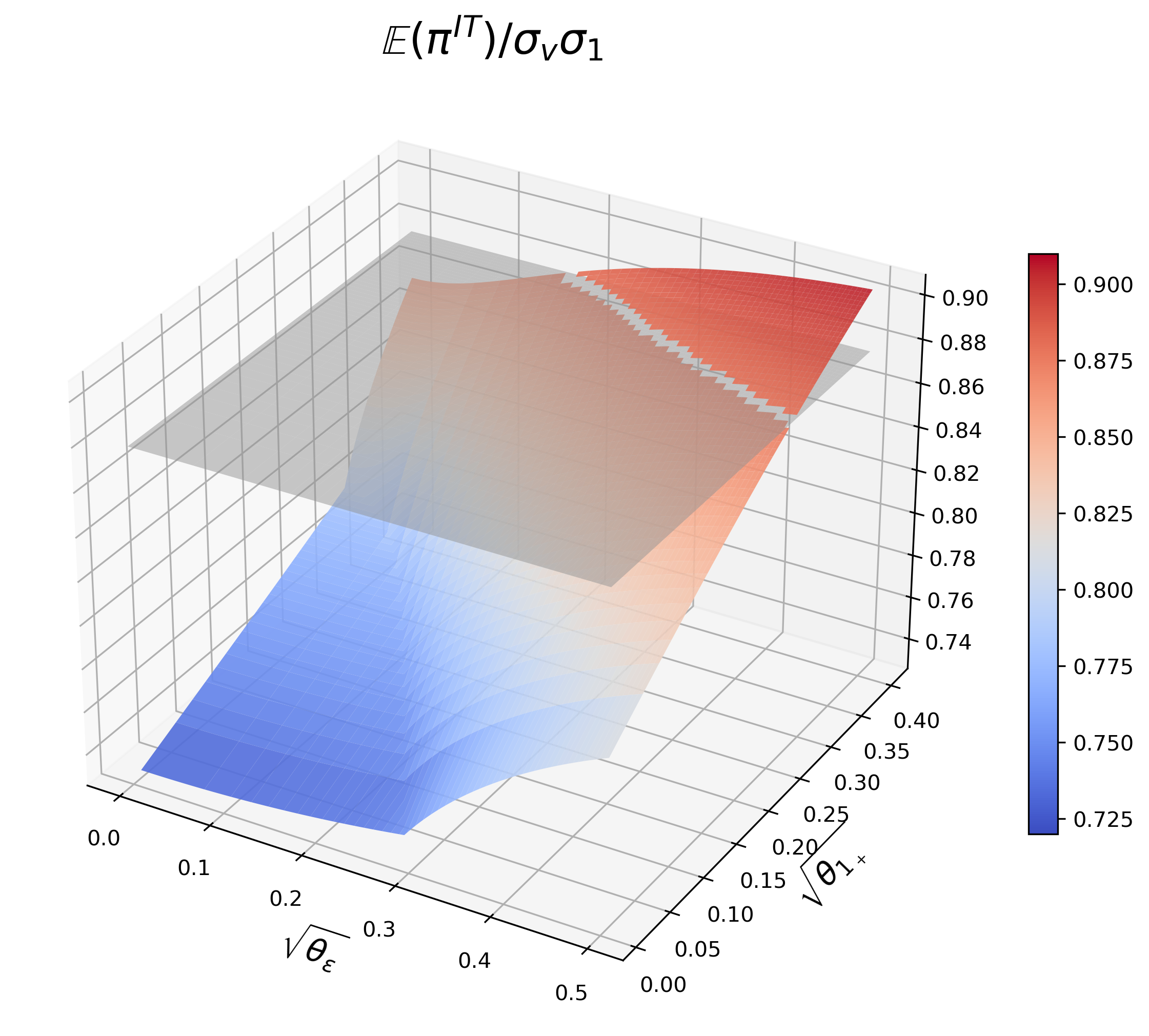}
    }
\subcaptionbox{$J=5$}{
    \includegraphics[width = 0.27\textwidth]{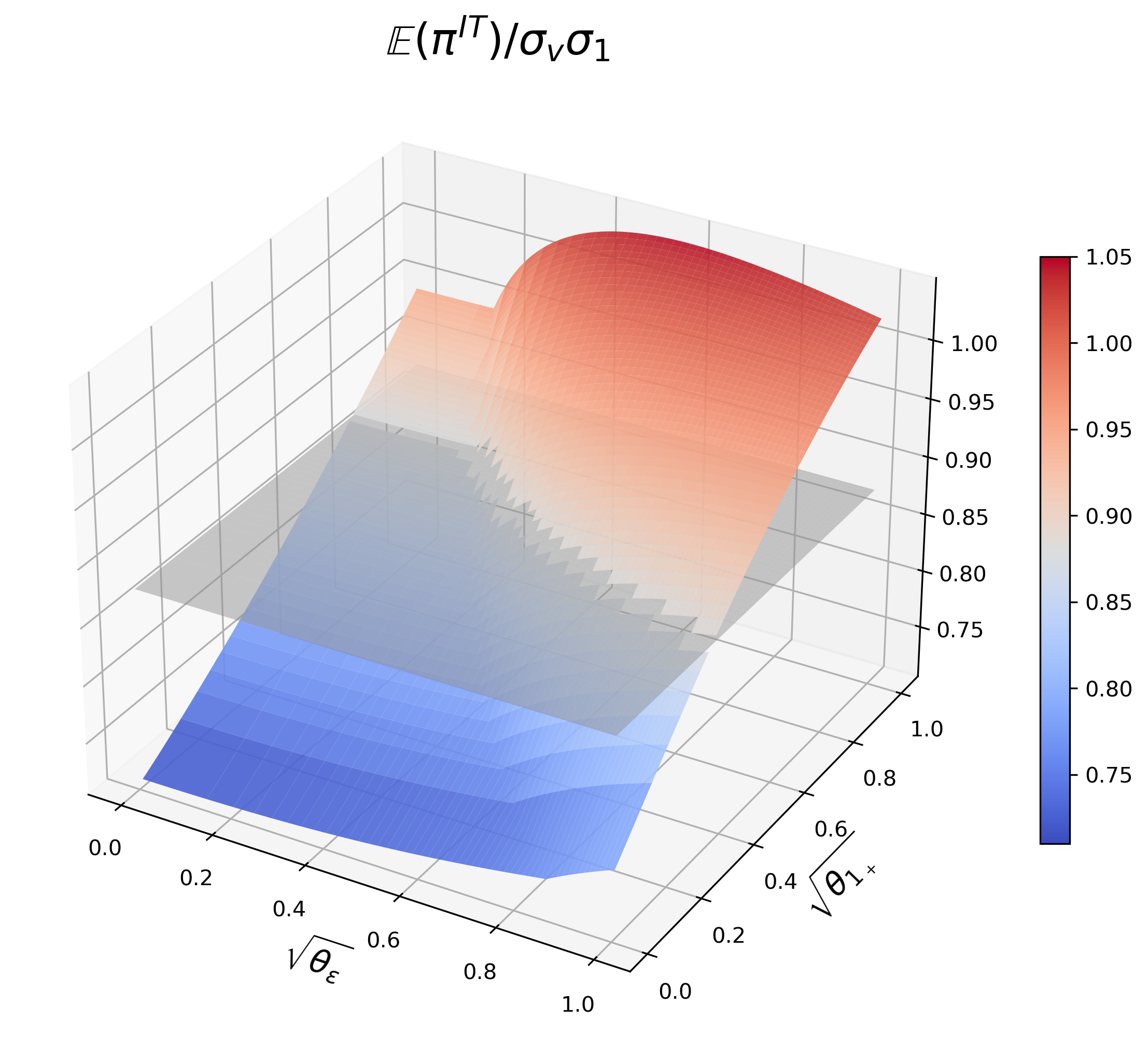}
    }
\subcaptionbox{$J=10$}{
    \includegraphics[width = 0.27\textwidth]{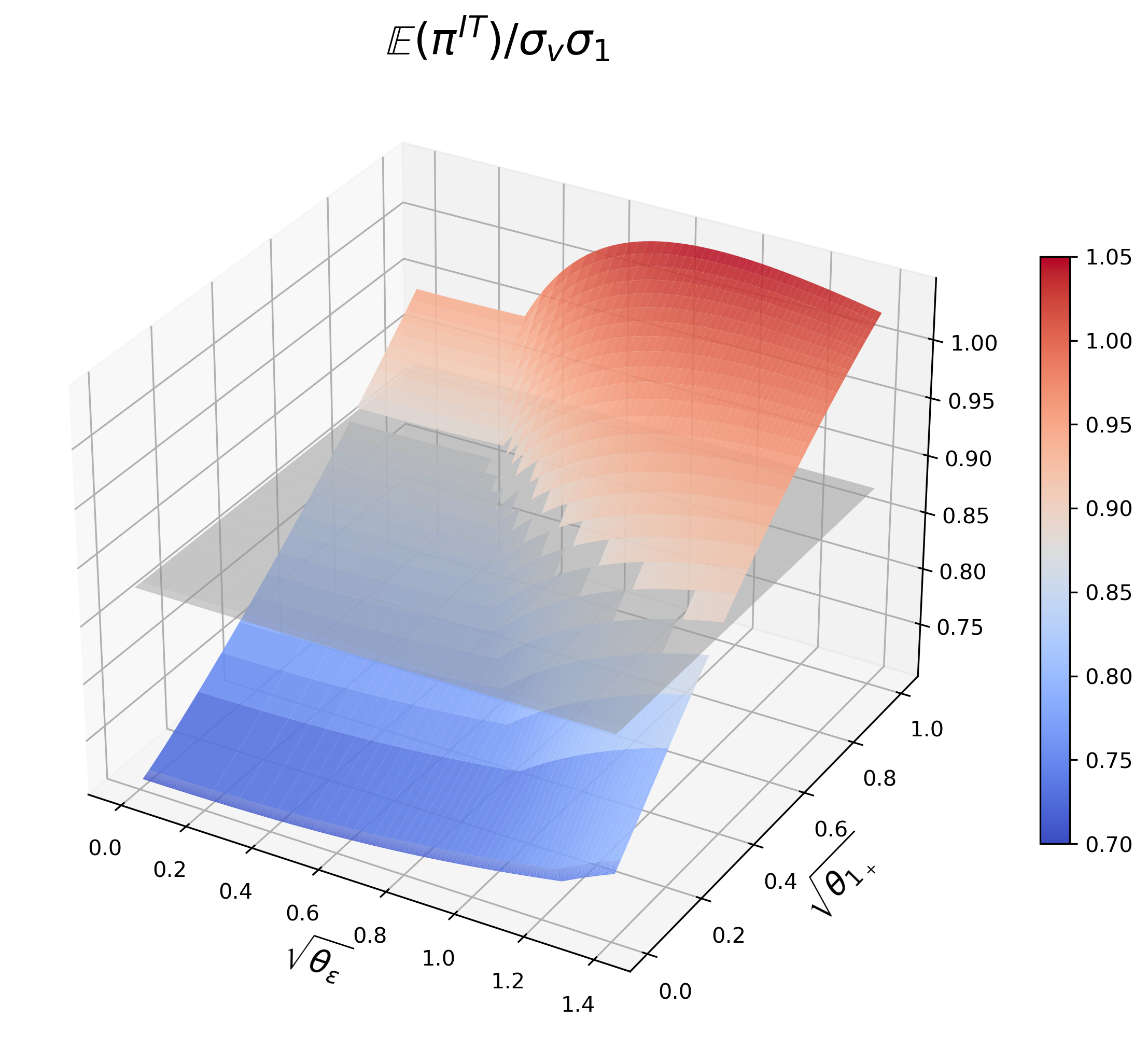}
    }
    \caption{IT's profit. The grey plane represents IT's profit without Round-Trippers.}
    \label{3DpiIT}
\end{figure}

Given $J\geq1$ and  $\theta_{1_+}>0,$ \begin{itemize}
    \item there exists a $\Tilde{\theta}_\varepsilon\geq0,$ if $\theta_\varepsilon>\Tilde{\theta}_\varepsilon,$ IT's profit is higher than that without Round-Trippers;
    \item there exists a $\Hat{\theta}_\varepsilon\geq\Tilde{\theta}_\varepsilon,$ if $\theta_\varepsilon>\Hat{\theta}_\varepsilon,$ IT's profit decreases with $\theta_\varepsilon.$
\end{itemize}

Given $J\geq1$ and $\theta_\varepsilon\geq0,$\begin{itemize}
    \item there exists a $\Tilde{\theta}_{1_+}>0,$ if $\theta_{1_+}>\Tilde{\theta}_{1_+},$ IT's profit is higher than that without Round-Trippers;
    \item IT's profit always increases with $\theta_{1_+}.$
\end{itemize}

\begin{figure}[!htbp]
    \centering
\subcaptionbox{$J=1$}{
    \includegraphics[width = 0.27\textwidth]{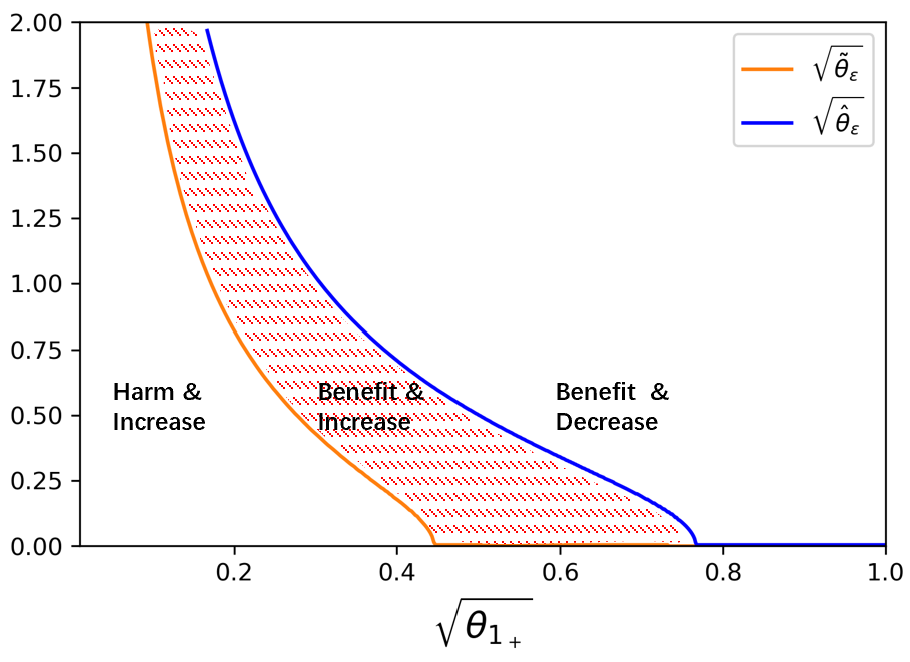}
    }
\subcaptionbox{$J=5$}{
    \includegraphics[width = 0.27\textwidth]{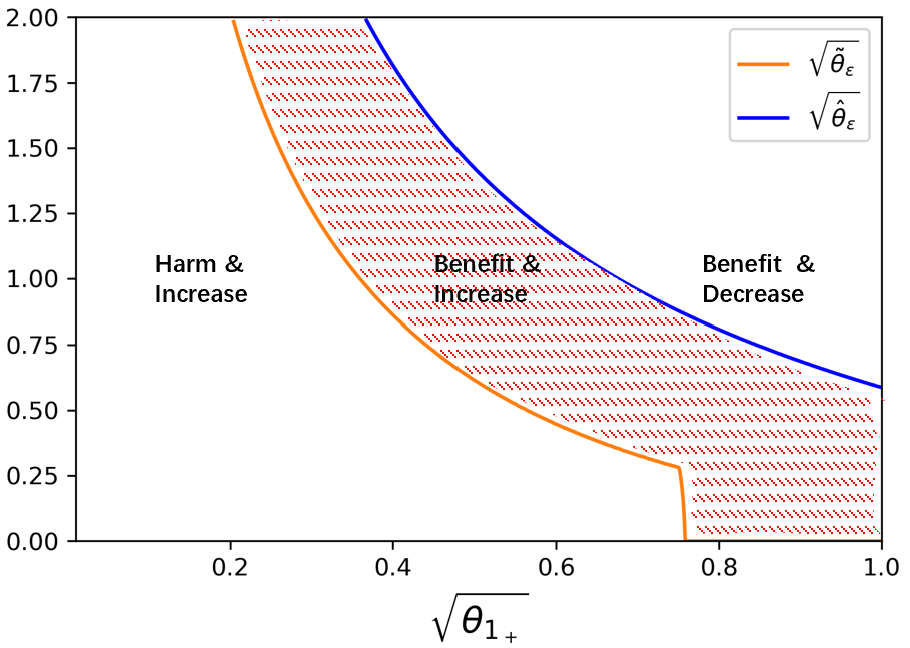}
    }
\subcaptionbox{$J=10$}{
    \includegraphics[width = 0.27\textwidth]{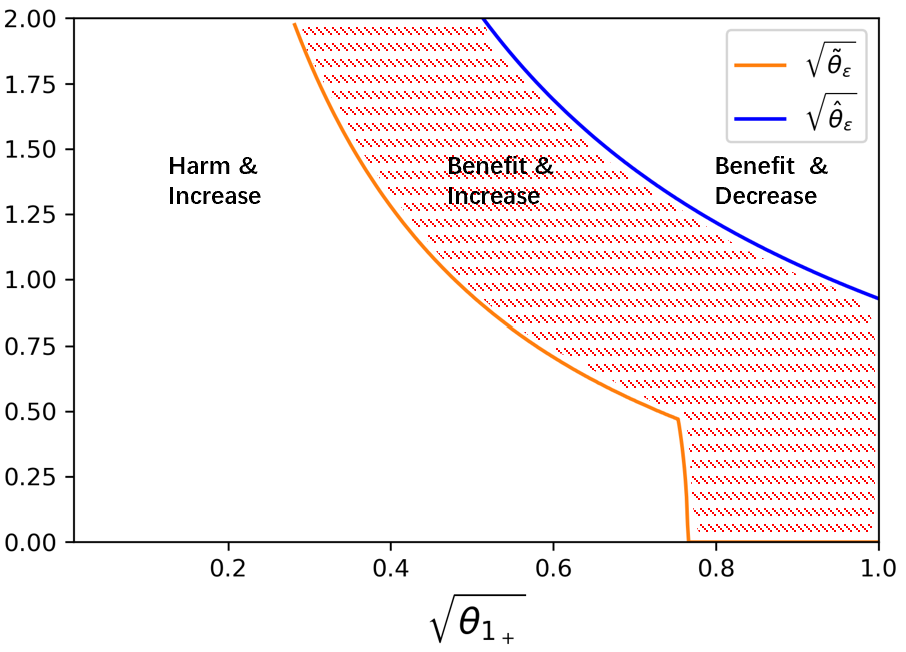}
    }
    \caption{The critical $\Tilde{\theta}_\varepsilon'$ and $\hat{\theta}_\varepsilon'$.}
    \label{figgaminfty-tildehatee}
\end{figure}

We plot the critical $\sqrt{\Tilde{\theta}_\varepsilon}$ and $\sqrt{\hat{\theta}_\varepsilon}$ against $\sqrt{\theta_{1_+}}$ in Figure \ref{figgaminfty-tildehatee}. Only in the dotted region, IT is benefited by Round-Trippers and her profit increases when Round-Trippers receive a less accurate signal.

From Figure \ref{figgaminfty-tildehatee}, if high-speed noise trading is scarce, IT is benefited only when the signal of Round-Tripper is inaccurate to $\sigma_\varepsilon$ is larger than $2\sigma_1$, which is almost impossible. So with few high-speed noise orders, Round-Trippers basically harm IT. Otherwise, with the increase of signal noise, IT will be benefited. If high-speed noise trading is quite active, Round-Trippers benefit IT even if they predict perfectly. What's more, the growth of $\theta_\varepsilon$ may lead Round-Trippers unwilling to provide liquidity, which brings a decrease in IT's profit. \\

\noindent\textbf{HFT's profit.} Round-Tripper's profit always increases with $\theta_{1_+}.$ Given $J\geq1$, there exists a $\hat{\theta}_{1_+}\in(0,1],$ 
\begin{itemize}
    \item if $\theta_{1_+}<\hat{\theta}_{1_+},$ Round-Tripper's profit first increases with $\theta_\varepsilon$ then decreases with it;
    \item if $\theta_{1_+}\geq\hat{\theta}_{1_+},$ Round-Tripper's profit decreases with it.
    \item If $J\geq5,\hat{\theta}_{1_+}=1.$
\end{itemize}

\begin{figure}[!htbp]
    \centering
\subcaptionbox{$J=1$}{
    \includegraphics[width = 0.27\textwidth]{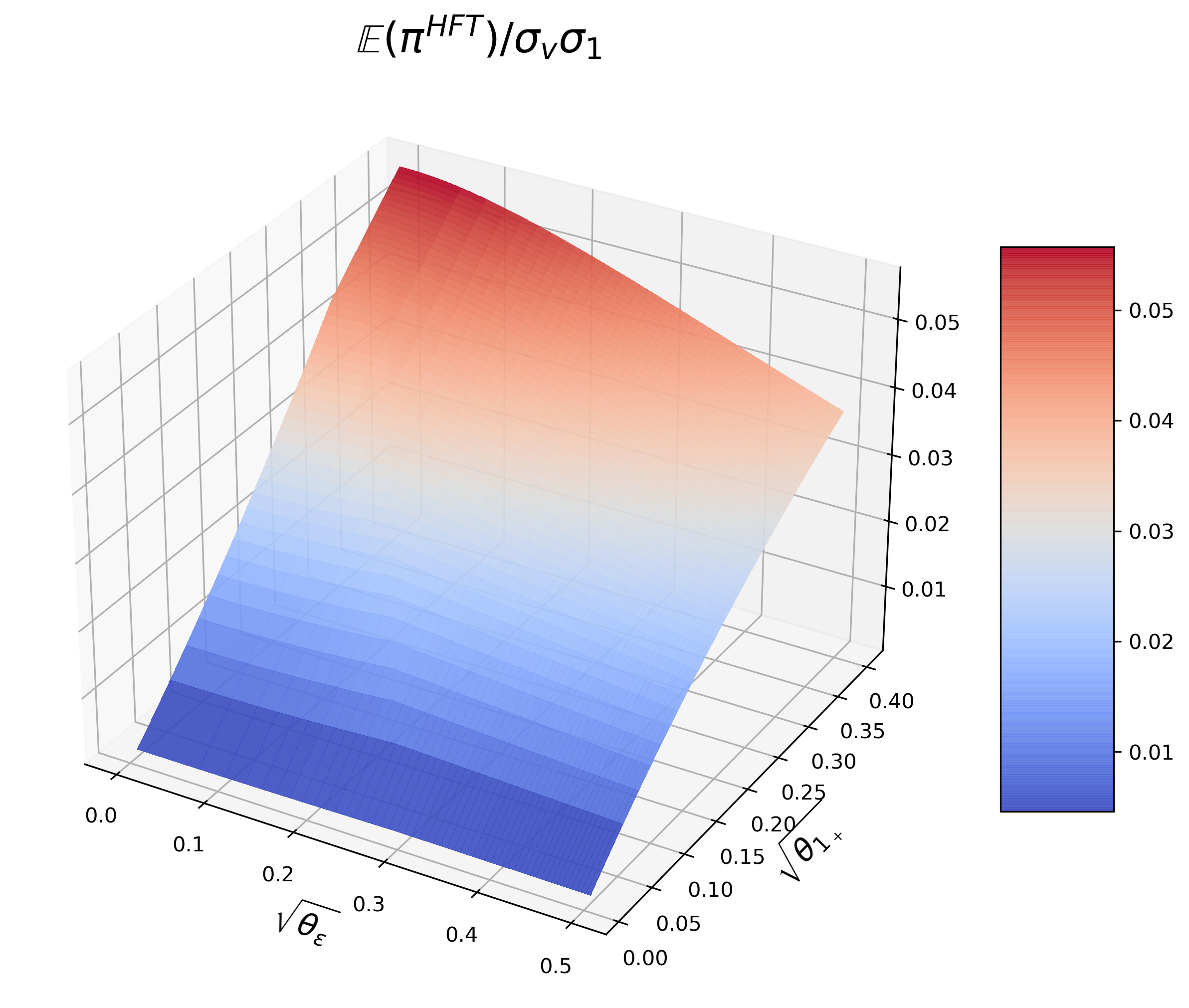}
    }
\subcaptionbox{$J=5$}{
    \includegraphics[width = 0.27\textwidth]{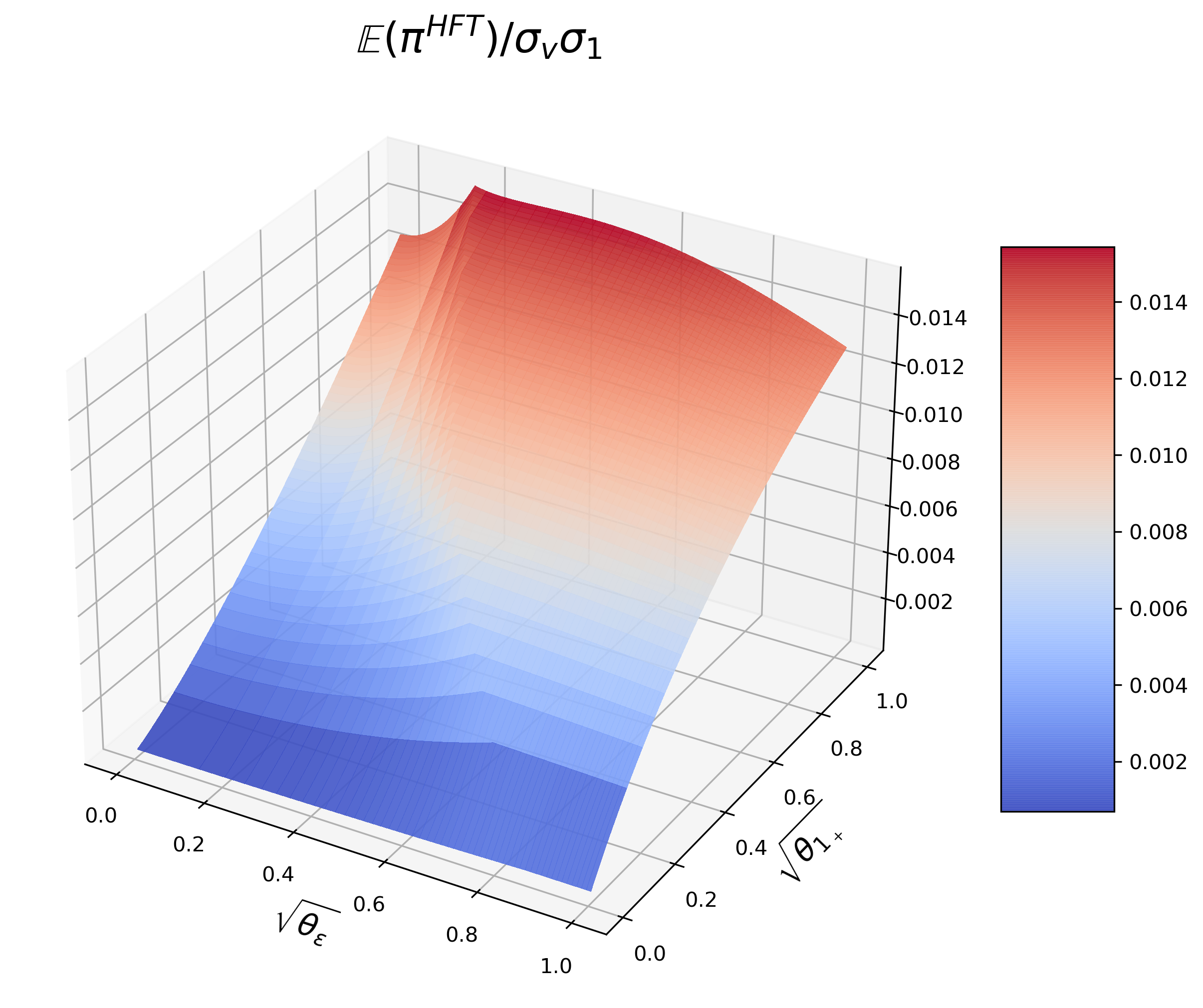}
    }
\subcaptionbox{$J=10$}{
    \includegraphics[width = 0.27\textwidth]{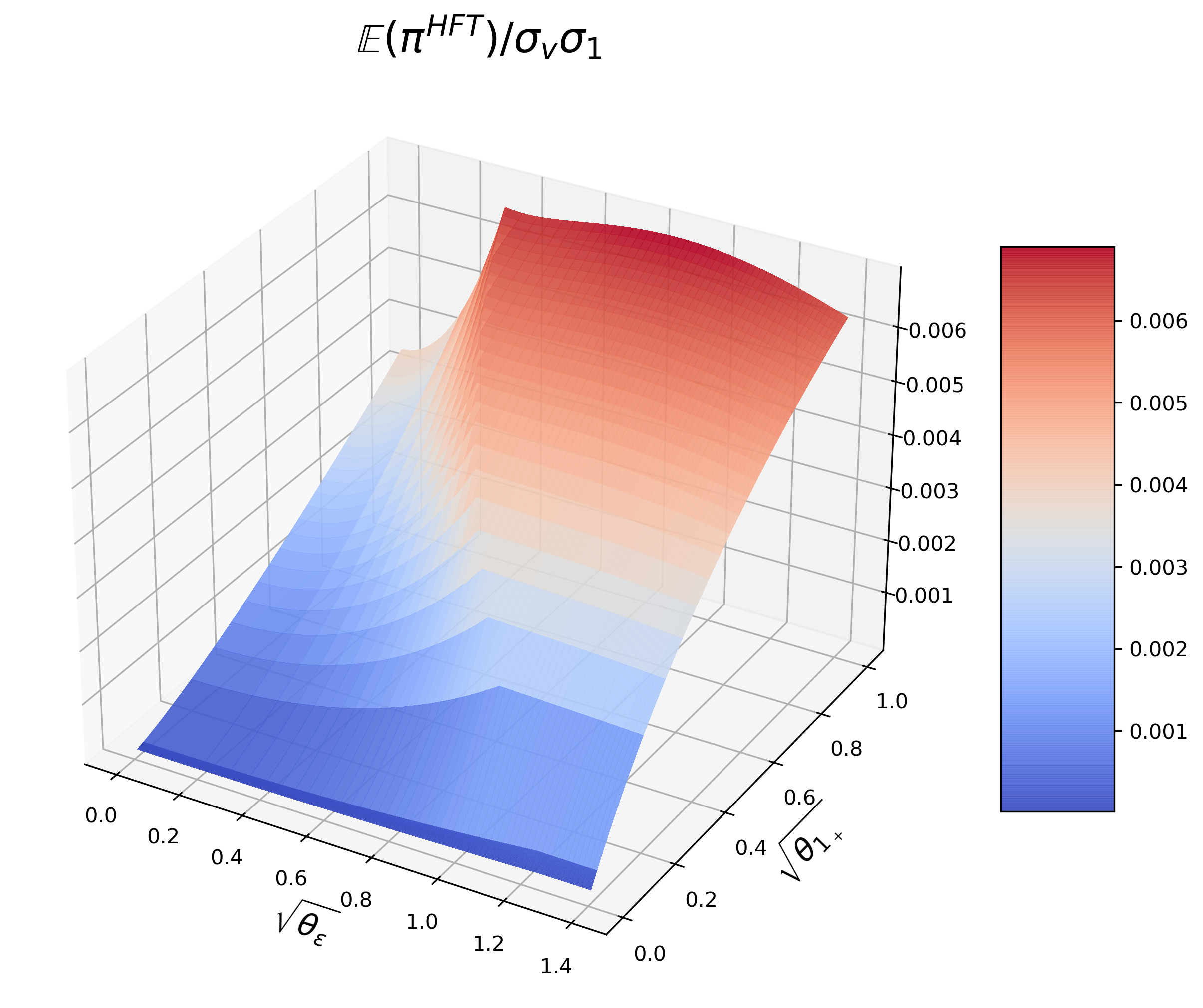}
    }
    \caption{HFT's profit.}
    \label{3DpiHFT}
\end{figure}

When $\theta_\varepsilon$ is relatively small and it grows, IT adds less noise to the order and trades more aggressively, as a result, Round-Tripper receives signals of higher precision and makes more profit from the market impact. Hence, a less accurate signal instead helps Round-Tripper profit more. When there are more than $5$ Round-Trippers in the market, the aforementioned two effects are more obvious and Round-Tripper's profit is bound to increase first, regardless of the market condition. 

\subsection{A market with both types of HFTs}
\label{subsecgam10gam2inf}
In this section, we assume that $J_1\geq1$ and $J_2\geq1,$ there are $J=J_1+J_2$ HFTs in total. In equilibrium, HFTs take following strategies: for $j\leq J_1,$
\begin{equation*}
\begin{aligned}
&x_{1j}=\beta_{11}(\Tilde{i}_{1j}-\mathbb{E}(\Tilde{i}_{1j}|y_1)),\\
&x_{2j}=\beta_{21}(\Tilde{i}_{1j}-\mathbb{E}(\Tilde{i}_{1j}|y_1))+\beta_{22}(\sum_{k\neq j}x_{1k}+u_{1_+})+\beta_{23}x_{1j};\\
\end{aligned}
\end{equation*}
for $j\geq J_1+1,$
\begin{equation*}
\begin{aligned}
&x_{1j}=\beta_{12}(\Tilde{i}_{1j}-\mathbb{E}(\Tilde{i}_{1j}|y_1)),\\
&x_{2j}=-x_{1j}.
\end{aligned}
\end{equation*}
\begin{theorem}[Simplification of the equilibrium with both types of HFTs]
\label{thmgam10gam2inf}
The mixed-strategy equilibrium is characterized by system \eqref{systemmixedgam10gam2inf}, \eqref{systemmixedgam10gam2inf-2} and \eqref{systemmixedgam10gam2inf-3}; the pure-strategy equilibrium is characterized by system \eqref{systempuregam10gam2inf}, \eqref{systempuregam10gam2inf-2} and \eqref{systempuregam10gam2inf-3}.
\end{theorem}

\begin{figure}[!htbp]
    \centering
\subcaptionbox{$J=2$}{
    \includegraphics[width = 0.27\textwidth]{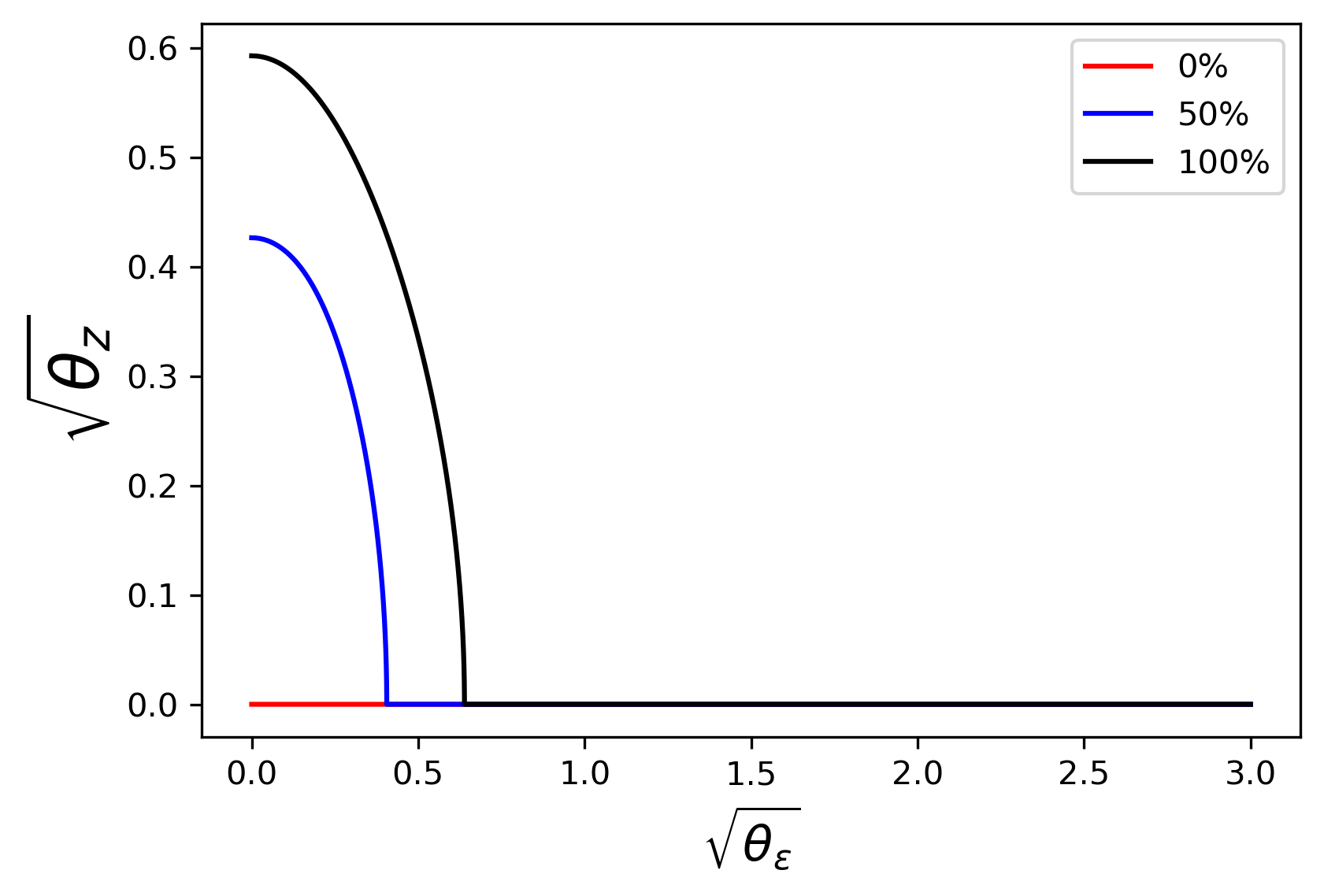}
    }
\subcaptionbox{$J=5$}{
    \includegraphics[width = 0.27\textwidth]{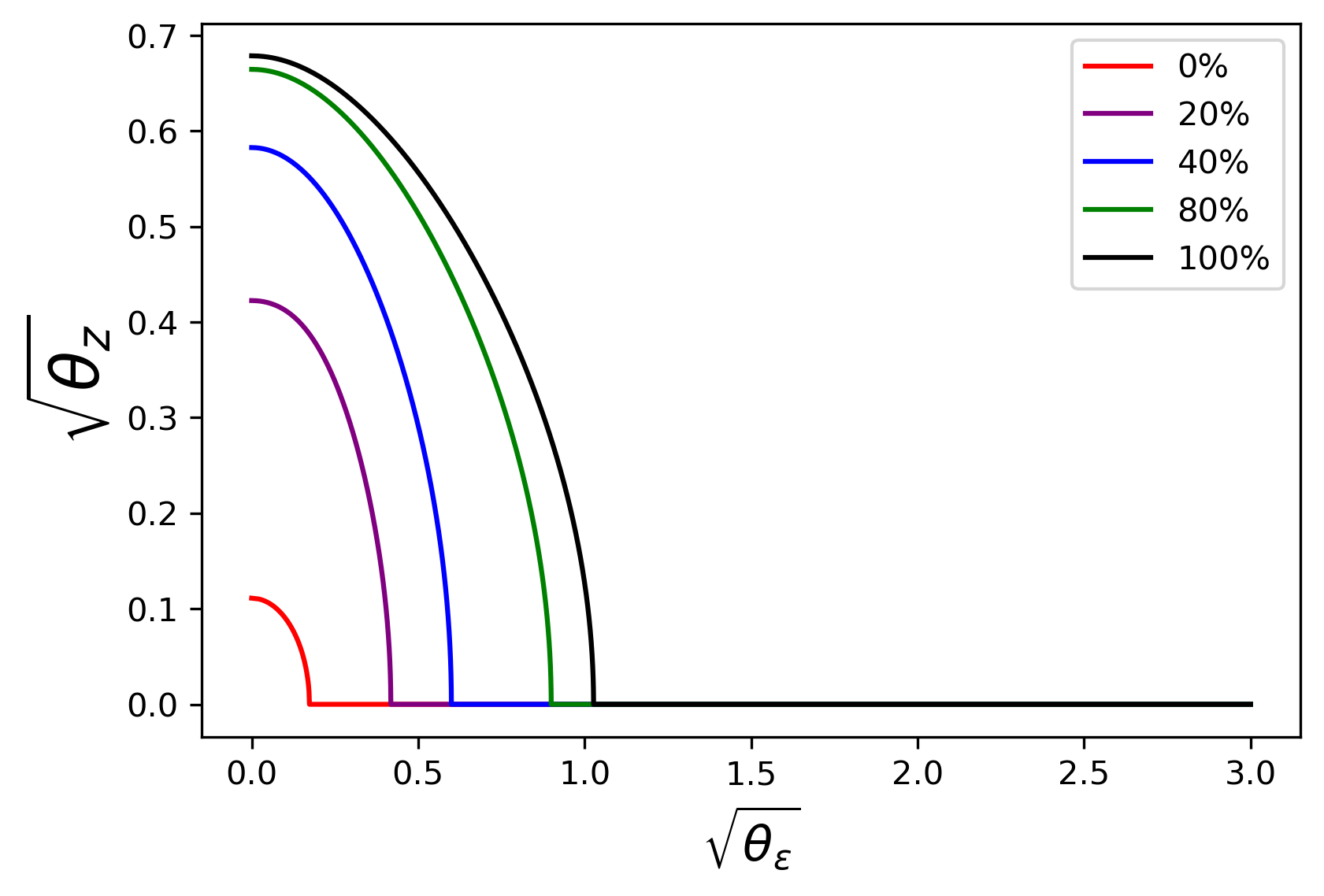}
    }
\subcaptionbox{$J=10$}{
    \includegraphics[width = 0.27\textwidth]{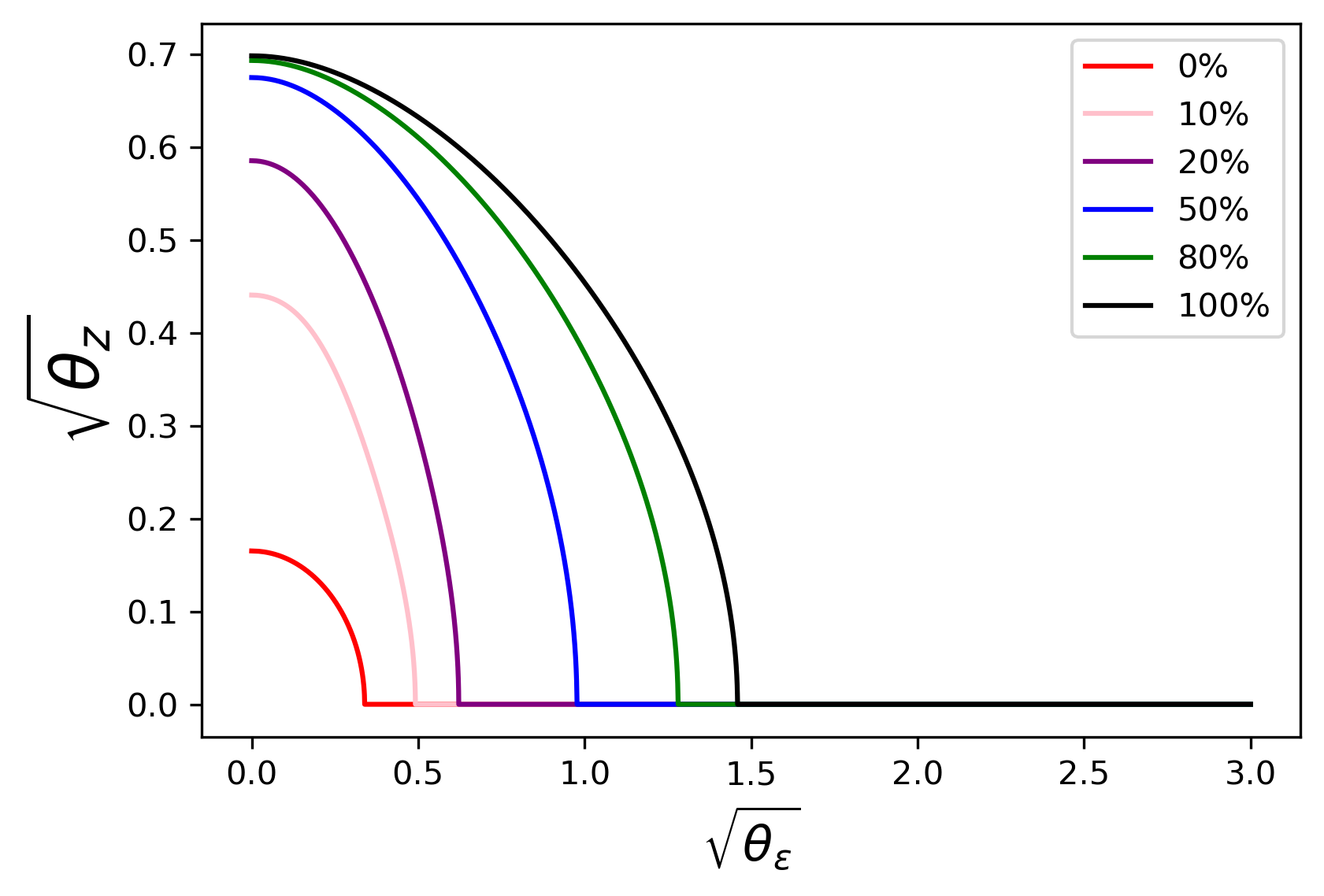}
    }
\caption{IT's mixed strategy with different $J$ and different proportions $\frac{J_1}{J}$ of Small-IT.}
\label{figgam10gam2infz}
\end{figure}

As before, we solve systems in Theorem \ref{thmgam10gam2inf} using numerical methods and obtain some new interesting results. Given $\theta_{1_+}=1$, we show IT's mixed strategy with different numbers of HFTs and different proportions of Small-IT in Figure \ref{figgam10gam2infz}. When $J=2$ and HFTs are all Round-Trippers, IT always take a pure strategy, as mentioned in Section \ref{subsecRT}. However, as long as there are Small-ITs, IT has to adopt the mixed strategy to protect herself. 

IT expands the mixed-strategy area and increases the intensity of endogenous noise, as the proportion of Small-IT increases: if IT needs to randomize when there are $0\%$ Small-ITs, so she does with other proportions.
For the case with $0\%$ Small-ITs, which has been studied in Section \ref{subsecRT}, when $J>3$, it is better for IT to randomize. To sum up, regardless of the specific strategies used by HFTs, IT is suggested to disguise the orders when there are more than $3$ HFTs.

\begin{figure}[!htbp]
    \centering
\subcaptionbox{IT's profit}{
    \includegraphics[width = 0.27\textwidth]{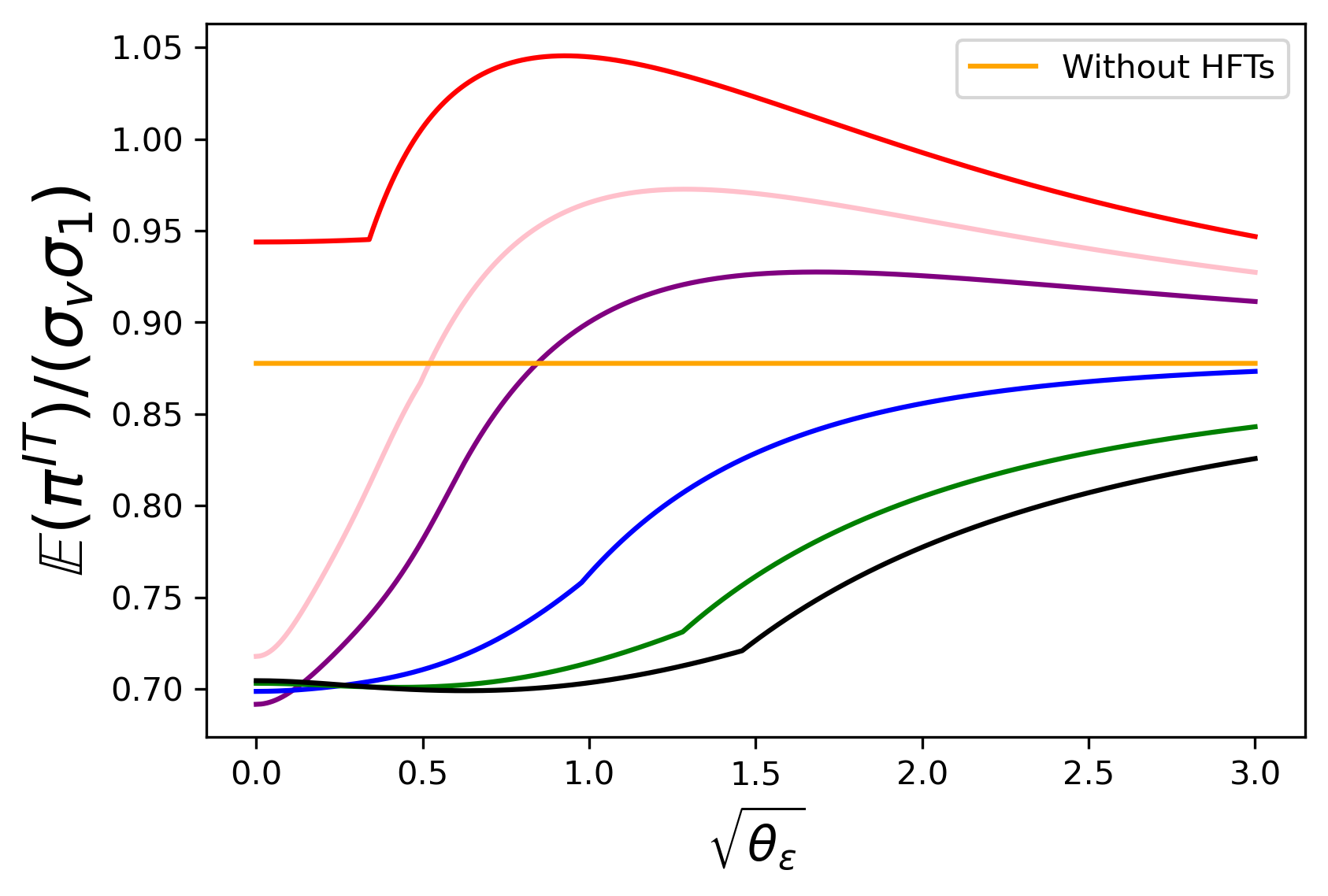}
    }
\subcaptionbox{Small-IT's profit}{
    \includegraphics[width = 0.27\textwidth]{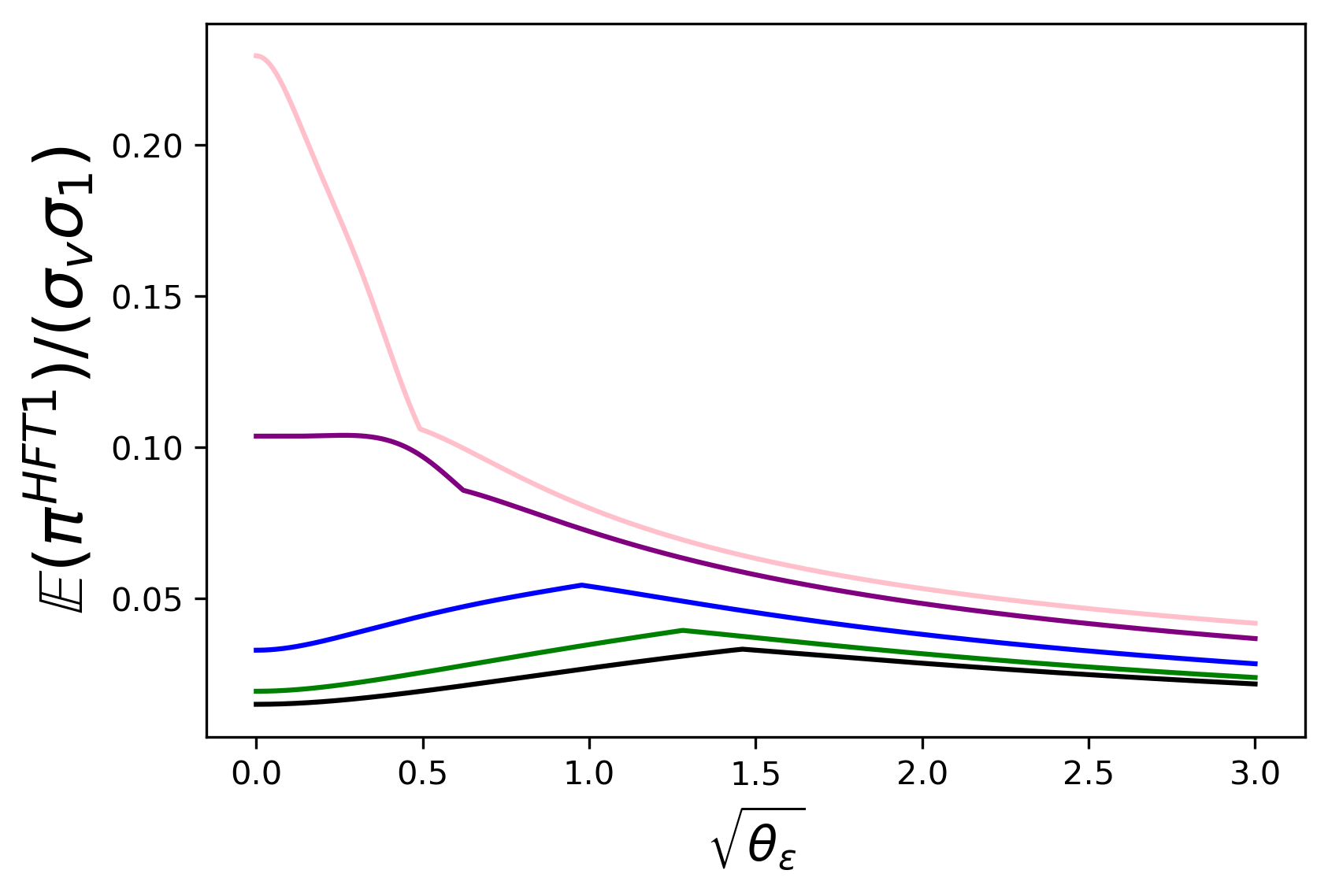}
    }
\subcaptionbox{Round-Tripper's profit}{
    \includegraphics[width = 0.27\textwidth]{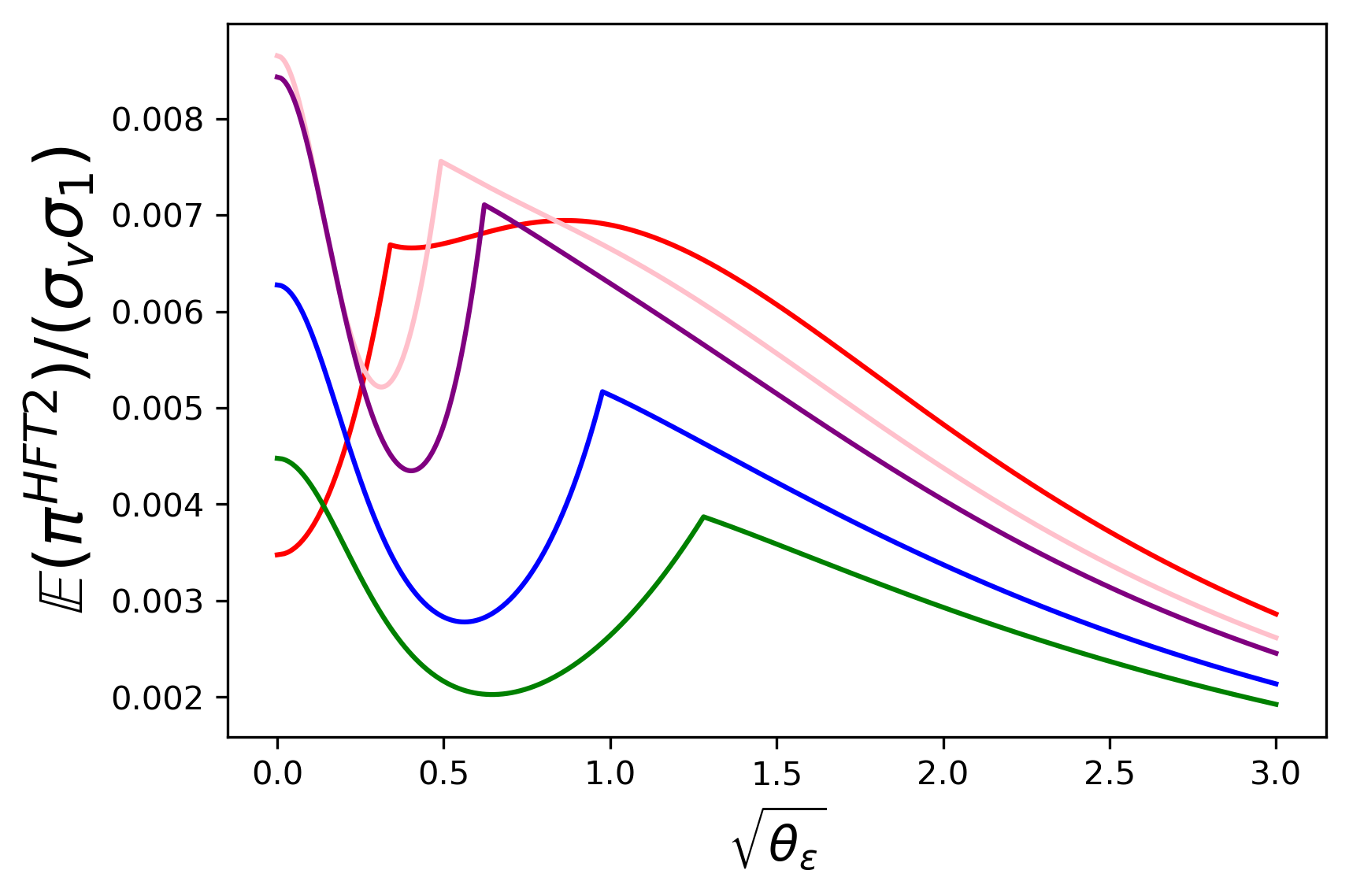}
    }
\caption{$J=10,$ with different proportions $\frac{J_1}{J}$ of Small-IT.}
\label{figgam10gam2inf}
\end{figure}

In the following discussion, $J=10$ is taken as an example. Profits of investors are analyzed in Figure \ref{figgam10gam2inf}.
From (a), in general, IT profits more and is more likely to be benefited with more Round-Trippers, except for the cases where HFTs' signals are of high accuracy. It is because IT adds more endogenous noise to protect herself when there are fewer Round-Trippers, so her profit is higher and it first decreases with $\theta_\varepsilon.$ 

For Small-ITs' profit displayed in (b) of Figure \ref{figgam10gam2inf}, we find that an individual Small-IT profits more with a higher proportion of Round-Trippers. On the one hand, Small-IT receives a more precise signal since IT mixes less. On the other hand, she benefits from the growing liquidity provided by Round-Trippers. 

From (c) of Figure \ref{figgam10gam2inf}, the trend of Round-Tripper's profit with relatively small $\theta_\varepsilon$s is different, when there are Small-ITs or not. Without Small-ITs, as $\theta_\varepsilon$ grows, IT adds less endogenous noise and trades more aggressively, both of which are advantageous for Round-Trippers, so the profit increases. However, with Small-ITs, the advantage of less endogenous noise is shared by them. As a result, Round-Tripper's profit decreases. Round-Tripper's profit increases with $\theta_\varepsilon$ when IT takes pure strategy: the increase of signal noise enables IT to trade more confidently and generates a large price impact.

\begin{figure}[!htbp]
    \centering
\subcaptionbox{Direction of $x_{11}$}{
    \includegraphics[width = 0.27\textwidth]{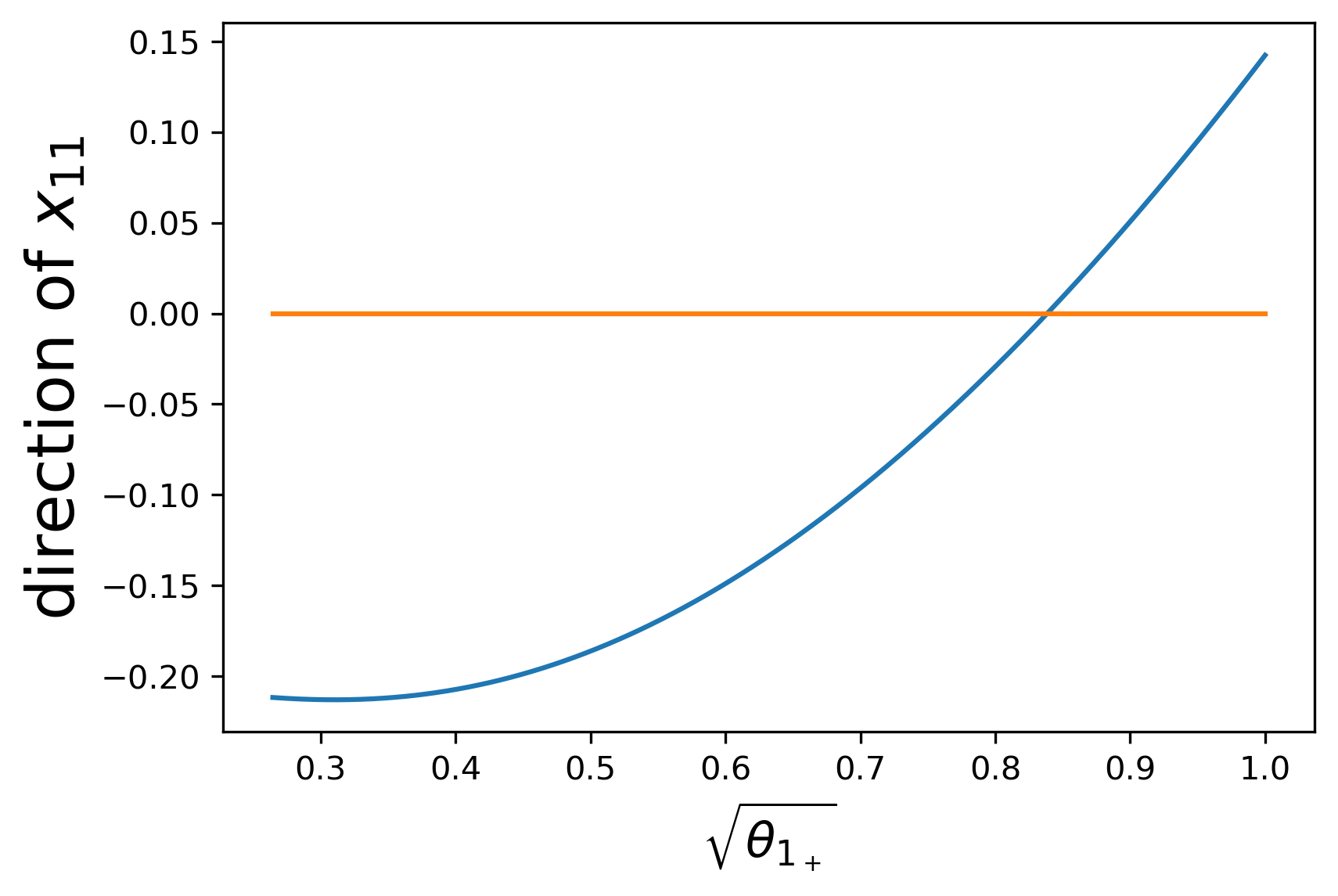}
    }
\subcaptionbox{Direction of $x_{21}$}{
    \includegraphics[width = 0.27\textwidth]{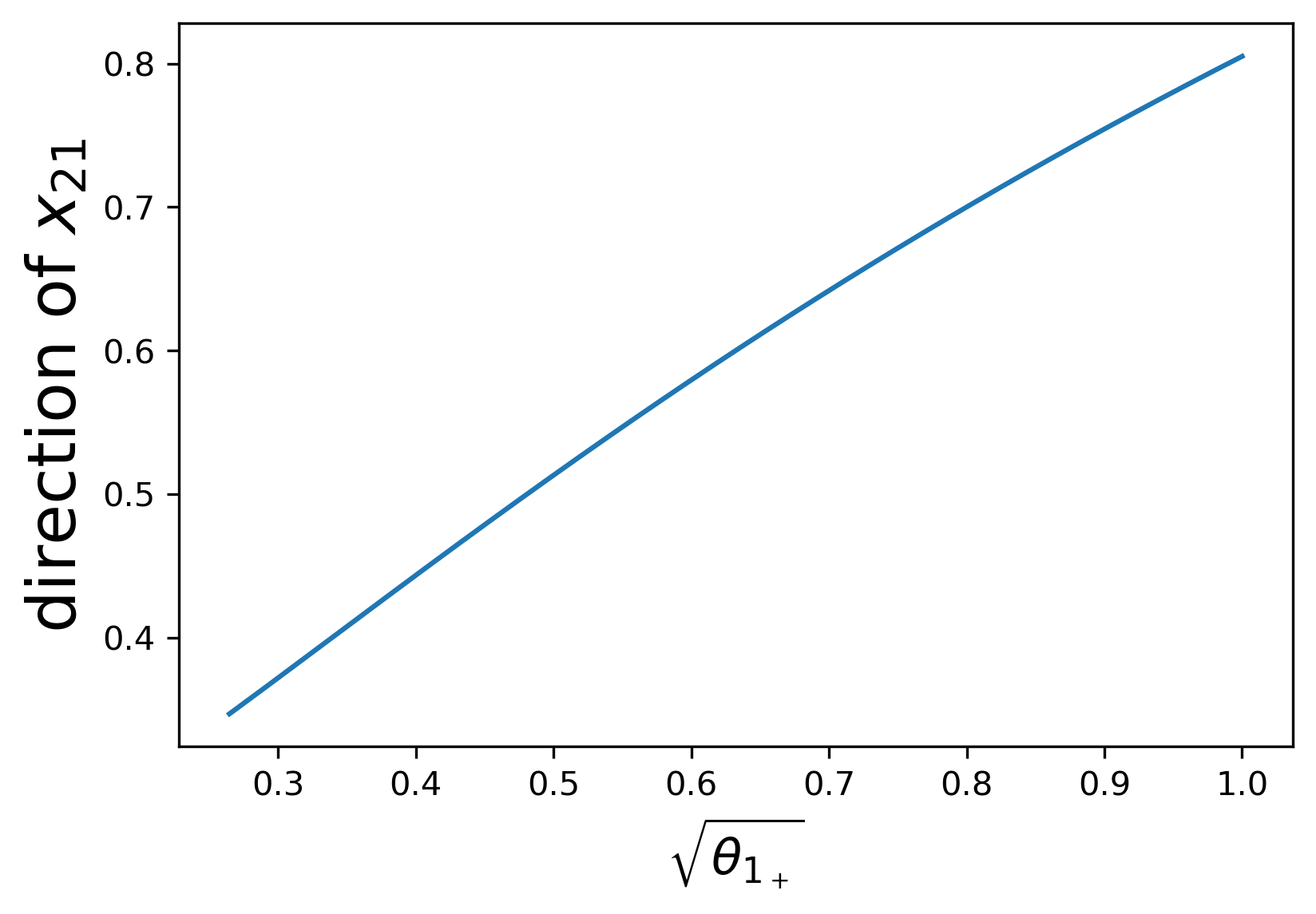}
    }
\subcaptionbox{IT's randomization}{
    \includegraphics[width = 0.27\textwidth]{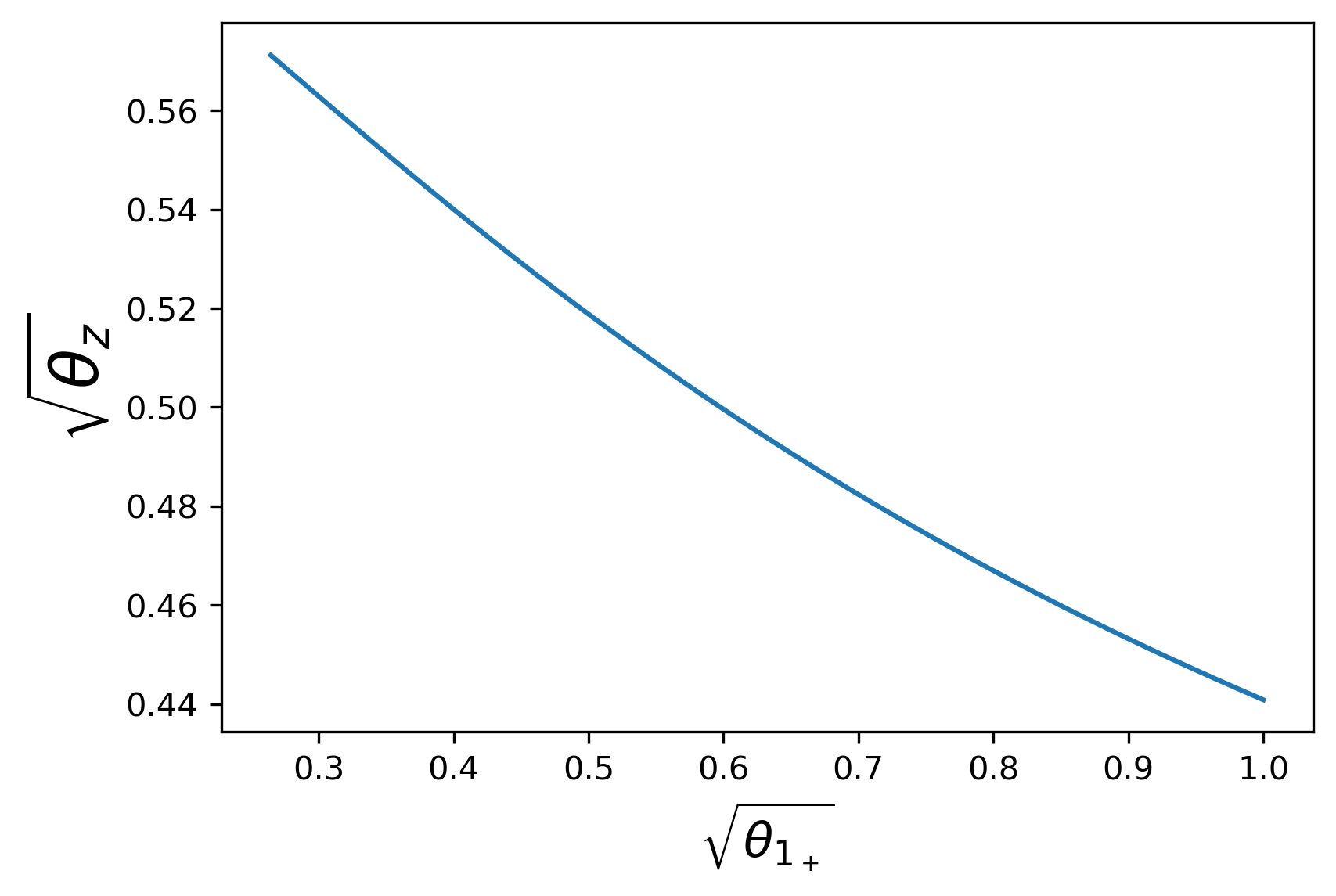}
    }

  \subcaptionbox{IT's profit}{
    \includegraphics[width = 0.27\textwidth]{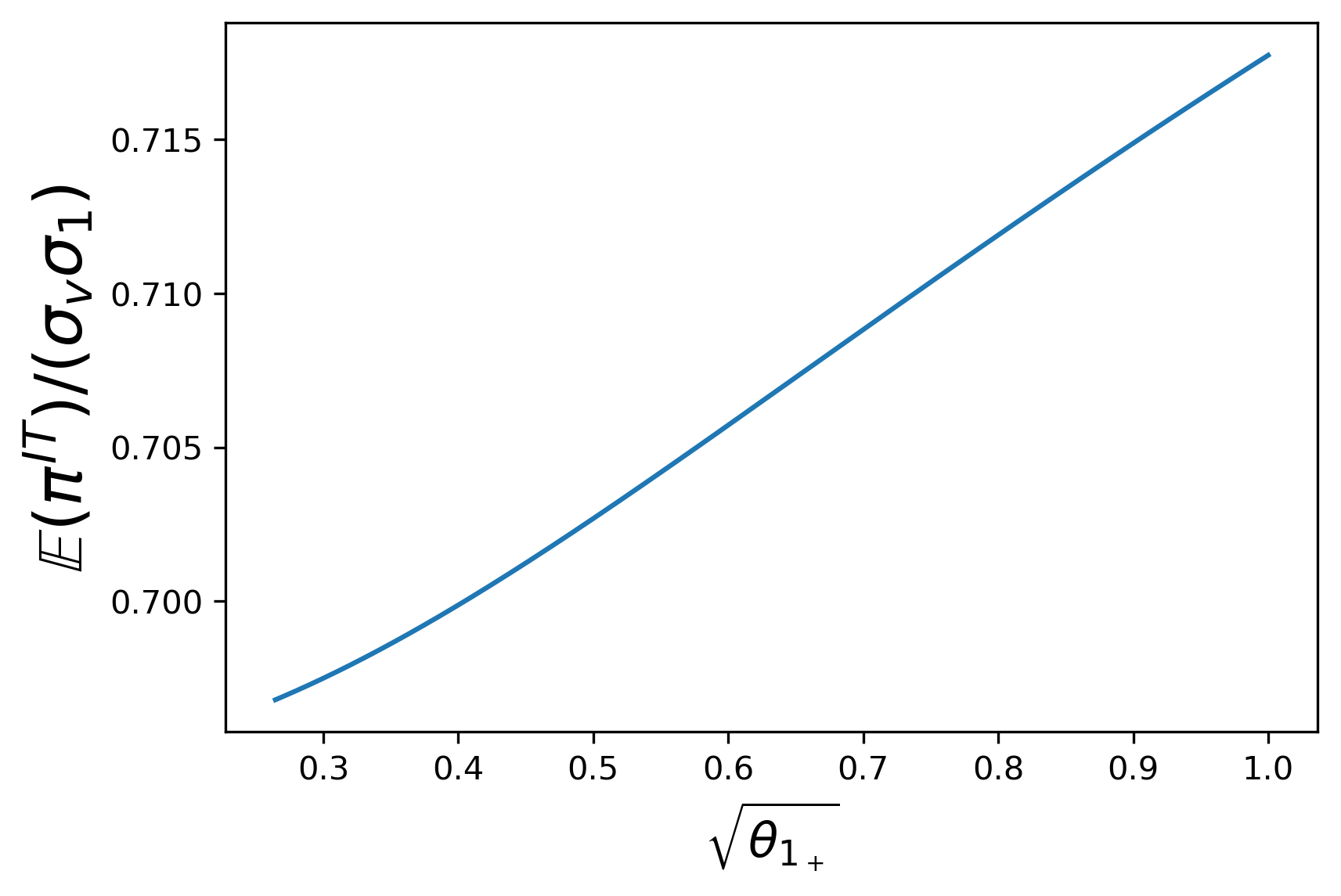}
    }
\subcaptionbox{Profit of type-1 HFT}{
    \includegraphics[width = 0.27\textwidth]{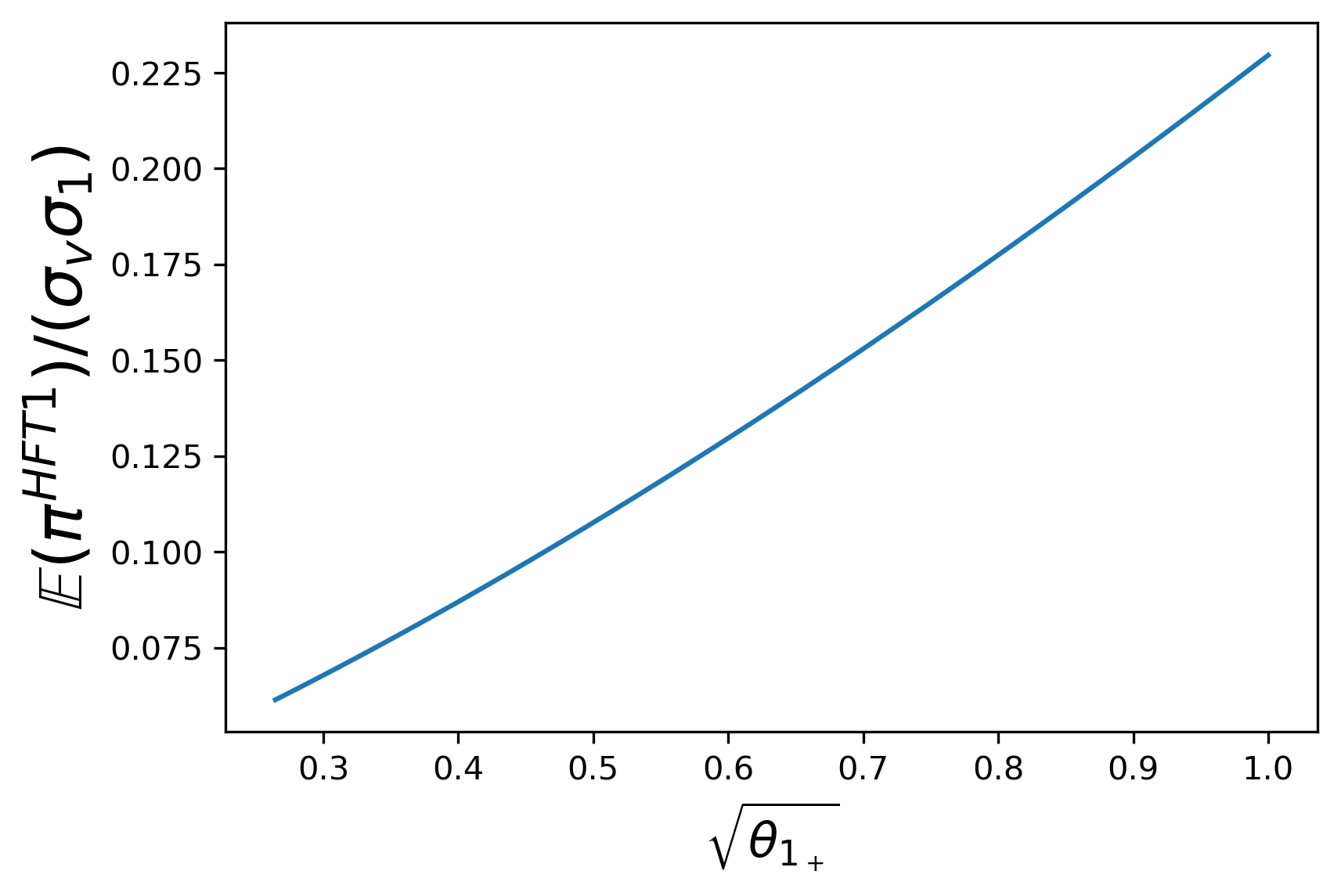}
    }
\subcaptionbox{Round-Tripper's profit}{
    \includegraphics[width = 0.27\textwidth]{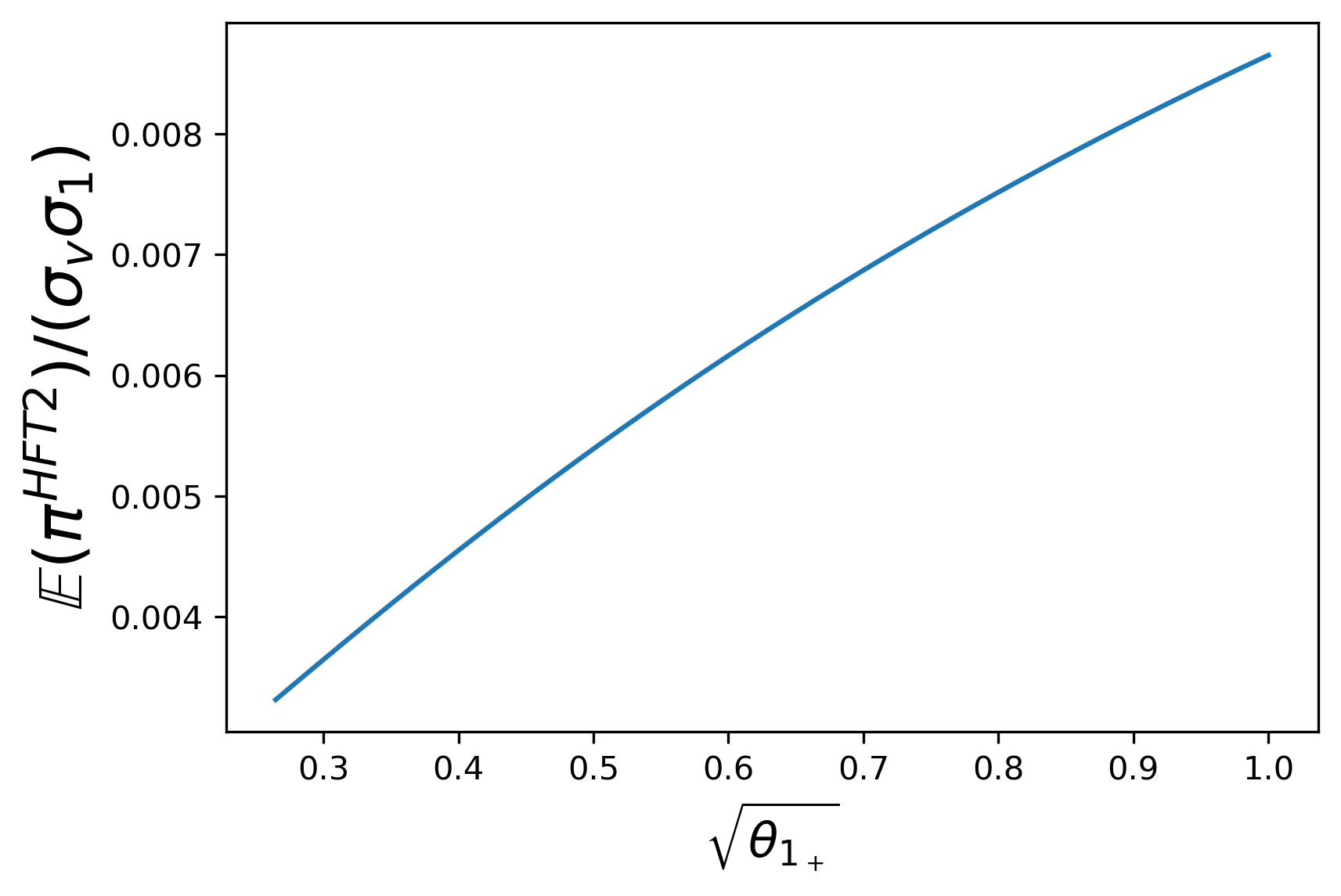}
    }
    \caption{$J_1=1,J_2=9,\theta_\varepsilon=0.$}
\label{figgam10gam2infinversefront}
\end{figure}

Another noteworthy phenomenon is that HFTs with no inventory constraints do not always play the role of Small-IT.
In the presence of both kinds of HFTs, when the size of high-speed noise trading $\theta_{1_+}$ decreases, Small-IT may change her role to a kind of ``inverse Round-Tripper'': trade in the opposite direction as IT at $t=1_{+}$ and same direction at $t=2$. (a) and (b) of figure \ref{figgam10gam2infinversefront} illustrate this. The reason is that HFTs' fast trading brings higher transaction costs with a smaller $\theta_{1_+}$. Consider the case that IT is going to buy at $t=2$, since Round-Tripper must clear all positions at the end, the only profitable strategy is ``buy low \& sell high''. However, other HFTs are free to adjust positions, in order to avoid large price impact, they choose to provide liquidity for Round-Trippers first and then buy back.

\begin{figure}[!htbp]
    \centering
    \includegraphics[width = 0.27\textwidth]{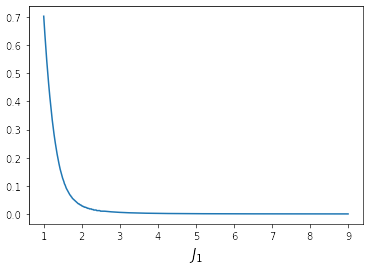}
    \caption{The critical $\theta_{1_+}$ for ``inverse Round-Tripper'', $J=10$ and $\theta_\varepsilon=0$.}
    \label{figgam10gam2infcriticaltheta2IRT}
\end{figure}

 Figure \ref{figgam10gam2infcriticaltheta2IRT} shows the critical value of $\theta_{1_+}$ for ``inverse Round-Tripper''. The value becomes smaller as the number of Round-Trippers grows. It should be noted that when $J_1$ is large, the critical $\theta_{1_+}$ is close to zero but still strictly positive.

\begin{figure}[!htbp]
    \centering
    \includegraphics[width = 0.27\textwidth]{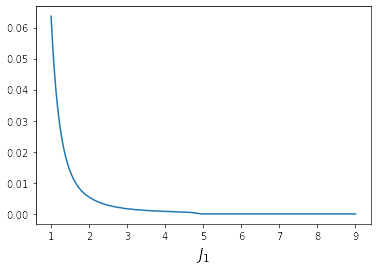}
    \caption{The critical $\theta_{1_+}$ for equilibrium existence, $J=10$ and $\theta_\varepsilon=0$.}
    \label{figgam10gam2infexistence}
\end{figure}

As for the existence of equilibrium, we find that when both kinds of HFTs exist, there is another critical value of $\theta_{1_+}$, when $\theta_{1_+}$ is smaller than this critical value, the equilibrium does not exist. It is illustrated in Figure \ref{figgam10gam2infexistence}.



\subsection{Equilibrium in the case \texorpdfstring{$\theta_{1_+}\rightarrow0$}{}}
\label{sectheta2rightarrow0}
In this section, we investigate the limit equilibrium when $\theta_{1_+}\rightarrow0$, that is, there is little high-speed noise trading. From Figure \ref{figgam10gam2infexistence}, the limit equilibrium does not exist with both-type HFTs, so we only consider the cases with Small-ITs or Round-Trippers, both of which have analytical results.

\begin{proposition}
 \label{proptheta2rightarrow0}
When $J_1=J\geq1$, $J_2=0$ and $\theta_{1_+}\rightarrow0$ (all of the below is almost-sure convergence):
\begin{equation*}
\begin{aligned}
&x_{1j}\rightarrow0,\ x_{2j}\rightarrow\beta_{21}(\Tilde{i}_{1j}-\mathbb{E}(\Tilde{i}_{1j}|y_1));\\
&i_1\rightarrow\alpha_1(v-p_0)+z,\ i_2\rightarrow\alpha_{21}(v-\mathbb{E}(v|y_1))+\alpha_{22}(i_1-\mathbb{E}(i_1|y_1));\\
&p_1\rightarrow p_0+\lambda_1y_1,\ p_{1_+}\rightarrow p_0+\lambda_1y_1,\ p_2 \rightarrow p_1+\lambda_{22}y_2.
\end{aligned}
\end{equation*}
In the mixed-strategy equilibrium,   
\begin{equation*}
\begin{aligned}
& \theta_z=\frac{J-4 \theta_\varepsilon}{J+2+4 \theta_\varepsilon}-\frac{1}{1+\frac{4 \theta_\varepsilon}{J}+\theta_2\left(\frac{J+2+4 \theta_\varepsilon}{J+1}\right)^2}>0, \\
& A_1=\frac{1}{\sqrt{1+\frac{4 \theta_\varepsilon}{J}+\theta_2\left(\frac{J+2+4 \theta_\varepsilon}{J+1}\right)^2}} , \\
& A_{21}=\frac{J+1}{J+2+4 \theta_\varepsilon} \sqrt{1+\frac{4 \theta_\varepsilon}{J}+\theta_2\left(\frac{J+2+4 \theta_\varepsilon}{J+1}\right)^2} , \\
& \alpha_{22}=-\frac{J+1}{J+2+4 \theta_\varepsilon} ,\\
& \beta_{21}=\frac{2(J+1)}{J(J+2+4 \theta_\varepsilon)}, \\
& \Lambda_1=\Lambda_{22}=\frac{J+2+4 \theta_\varepsilon}{2(J+1) \sqrt{1+\frac{4 \theta_\varepsilon}{J}+\theta_2\left(\frac{J+2+4 \theta_\varepsilon}{J+1}\right)^2}} .\\
&
\end{aligned}
\end{equation*}
In the pure-strategy equilibrium, $\theta_z=0,$ $A_1$ solves \eqref{systemgaminf-theta1+righhtarrow0} and
\begin{equation*}
\begin{aligned}
&\Lambda_1=\frac{A_1}{1+A_1^2},\\
&\Lambda_{22}= \sqrt{\frac{A_1^2 J [A_1^2 (\theta_\varepsilon+1)+\theta_\varepsilon]+[2 (A_1^2+1) \theta_\varepsilon+A_1^2]^2}{(A_1^2+1) \theta_2 [A_1^2 (4 \theta_\varepsilon+J+2)+4 \theta_\varepsilon]^2}},\\
&A_{21}=\frac{1}{2\Lambda_{22}},\\
&\beta_{21}=\frac{2 A_1 A_{21}}{A_1^2 (4\theta_\varepsilon+J+2)+4 \theta_\varepsilon},\\
&\alpha_{22}=-\frac{J\beta_{21}}{2}.
\end{aligned}
\end{equation*}

\end{proposition}

\begin{remark}
The mixed-strategy (pure-strategy) equilibrium in Proposition \ref{proptheta2rightarrow0} is just the same as the one with back-runners in Proposition 3 (Proposition 4) in Yang and Zhu (2020) \cite{yang2020back}. That is because, when there is little high-speed noise trading, Small-ITs nearly give up their speed advantage and become back-runners who only trade along with IT in period 2. 
\end{remark}

\begin{proposition}
When $J_2=J\geq1$, $J_1=0$ and $\theta_{1_+}\rightarrow0$, HFTs tend to take no action, therefore, the time-$1_+$ price tends to be the same as the time-$1$ price,
(all of the below is almost-sure convergence):
\begin{equation*}
\begin{aligned}
&x_{1j}\rightarrow0,\ x_{2j}\rightarrow0;\\
&i_1\rightarrow\alpha_1(v-p_0)+z,\ i_2\rightarrow\alpha_{21}(v-\mathbb{E}(v|y_1))+\alpha_{22}(i_1-\mathbb{E}(i_1|y_1));\\
&p_1\rightarrow p_0+\lambda_1y_1,\ p_{1_+}\rightarrow p_0+\lambda_1y_1,\ p_2 \rightarrow p_0+\lambda_1y_1+\lambda_{22}y_2.
\end{aligned}
\end{equation*}
For any $\theta_\varepsilon\geq0,$ there exists $\zeta=\zeta(\theta_\varepsilon,J)$ such that
\begin{equation*}
\lim_{\theta_{1_+}\rightarrow0}\frac{\beta_{12}}{\sqrt{\theta_{1_+}}}=\zeta(\theta_\varepsilon,J).
\end{equation*}
In the mixed-strategy equilibrium, $(A_1,\theta_z)\in\mathbb{R}^+\times\mathbb{R}^+$ solves:
\begin{equation*}
\begin{aligned}
0=&A_1^8 (\theta_\varepsilon J \zeta^2+1)+A_1^6 (3 \theta_z+2) (\theta_\varepsilon J \zeta^2+1)-(\theta_z+1)^2 (\theta_\varepsilon J \theta_z \zeta^2+\theta_\varepsilon J \zeta^2+J^2 \theta_z \zeta^2+\theta_z+1)\\
+&A_1^4 [4 \theta_2 (\theta_\varepsilon J \zeta^2+J^2 \zeta^2+1)+\theta_z (3 \theta_\varepsilon J \theta_z \zeta^2+3 \theta_\varepsilon J \zeta^2-J^2 \zeta^2+3 \theta_z+3)]\\
+&A_1^2\{4 \theta_2 (\theta_\varepsilon J \theta_z \zeta^2+\theta_\varepsilon J \zeta^2+J^2 \theta_z \zeta^2+\theta_z+1)+(\theta_z+1) [-\theta_z (\theta_\varepsilon J \zeta^2+2 J^2 \zeta^2+1)\\
+&\theta_z^2 (\theta_\varepsilon J \zeta^2+1)-2 (\theta_\varepsilon J \zeta^2+1)]\},\\
0=&A_1^8 (\theta_\varepsilon J \zeta^2+1)+3 A_1^6 (\theta_\varepsilon J \theta_z \zeta^2+\theta_z)+A_1^4 [4 \theta_2 (\theta_\varepsilon J \zeta^2+J^2 \zeta^2+1)+(3 \theta_z^2+\theta_z-2) (\theta_\varepsilon J \zeta^2+1)]\\
+&A_1^2 [4 \theta_2 (\theta_\varepsilon J \theta_z \zeta^2+\theta_\varepsilon J \zeta^2+J^2 \theta_z \zeta^2-J^2 \zeta^2+\theta_z+1)+\theta_z (\theta_z+1)^2 (\theta_\varepsilon J \zeta^2+1)]\\
+&(\theta_z+1)^3 (\theta_\varepsilon J \zeta^2+1).
\end{aligned}
\end{equation*}
If it is solved,
\begin{equation*}
\begin{aligned}
&\Lambda_1=\frac{A_1}{A_1^2+\theta_z+1},\\
&\Lambda_{22}=\Lambda_1,\\
&A_{21}=\frac{1}{2\Lambda_{22}},\\
&\alpha_{22}=-\frac{A_1^2+\theta_z+1}{2}.
\end{aligned}
\end{equation*}

In the pure-strategy equilibrium, $\theta_z=0,$ $(\Lambda_{22},A_1,\alpha_{22})$ solves
\begin{equation*}
\begin{aligned}
0=&2 A_1^4 \Lambda_{22}+2 A_1^2 \alpha_{22} \Lambda_{22}-4 A_1 \alpha_{22}^2 \Lambda_{22}^2+A_1-2 (\alpha_{22}+1) \Lambda_{22},\\
0=&4 A_1^2 \Lambda_{22}^2 (\alpha_{22}^2 \theta_\varepsilon J \zeta^2+\alpha_{22}^2+\theta_\varepsilon J \theta_2 \zeta^2+J^2 \theta_2 \zeta^2+\theta_2)+(4 \Lambda_{22}^2 \theta_2-1) (\theta_\varepsilon J \zeta^2+1),\\
0=&\alpha_{22}+\frac{2 A_1 J^2 \Lambda_{22} \theta_2 \zeta^2}{4 A_1^2 \Lambda_{22}^2 (\alpha_{22}^2 \theta_\varepsilon J \zeta^2+\alpha_{22}^2+\theta_\varepsilon J \theta_2 \zeta^2+J^2 \theta_2 \zeta^2+\theta_2)+4 A_1 \alpha_{22} \Lambda_{22} (\theta_\varepsilon J \zeta^2+1)+(4 \Lambda_{22}^2 \theta_2+1) (\theta_\varepsilon J \zeta^2+1)},\\
0<&\Lambda_{22},\\
0<&4 A_1^3 \Lambda_{22}-A_1^2+4 A_1 (\alpha_{22}+1) \Lambda_{22}-4 \alpha_{22}^2 \Lambda_{22}^2.
\end{aligned}
\end{equation*}
If it is solved,
\begin{equation*}
\begin{aligned}
&\Lambda_1=\frac{A_1}{A_1^2+1},\\
&A_{21}=\frac{1}{2\Lambda_{22}}.
\end{aligned}
\end{equation*}

\end{proposition}

\section{Conclusion}
\label{secconclusion}
This paper models the interactions between a normal-speed informed trader and multiple high-frequency anticipatory traders who can predict the former's future order through trading history and trade ahead of her as well. HFTs may play the role of Small-IT or Round-Tripper, according to their inventory aversion. To counteract HFTs' detection, IT may add endogenous noise into her order and take a mixed strategy. By analyzing the market participants' optimization and the equilibria under various conditions, this paper provides
suggestions on IT's randomization: (1) with the promotion of anticipatory traders' speed, IT should widen the range of randomization and increase its intensity; (2) when there are more than $3$ HFTs, IT should better take a mixed strategy, regardless of the specific roles of HFTs; (3) only when high-speed noise trading is quite active and there are no more than $3$ HFTs who are all Round-Trippers, randomization is unnecessary.


    
\newpage
\section*{Appendix}
\label{secappendix}
\appendix

\noindent\textbf{Proof of Theorem \ref{thmj=1}.}
The mixed-strategy equilibrium can be simplified to the following system of $(A_1,\theta_z,\beta_{11}):$
{\scriptsize
\begin{equation}
\label{systemmixedj=1-1}
\begin{aligned}
0<&16 \Gamma+(16 A_{1} \beta_{11})(A_{1}^2 (\beta_{11}^2 (\theta_{\varepsilon}+1)+\theta_{1_+})+\beta_{11}^2 (\theta_{\varepsilon} \theta_{z}+\theta_{\varepsilon}+\theta_{z})+\theta_{1_+} (\theta_{z}+1))-((A_{1}^2+\theta_{z}+1) (4 A_{1}^5 (\theta_{\varepsilon}+1) \Gamma\\
+&A_{1}^4 (\theta_{\varepsilon} (8 \beta_{11} \Gamma^2+4)+8 \beta_{11} \Gamma^2+3)+4 A_{1}^3 \Gamma (2 \beta_{11} (\theta_{\varepsilon}+1)+2 \theta_{\varepsilon} (\theta_{z}+1)+2 \theta_{z}+1)+A_{1}^2 (4 \theta_{\varepsilon} (\theta_{z}+1) (4 \beta_{11} \Gamma^2+1)\\
+&8 \beta_{11} \Gamma^2 (2 \theta_{z}+1)+3 \theta_{z}-1)+4 A_{1} \Gamma (2 \beta_{11}+\theta_{z}+1) (\theta_{\varepsilon} \theta_{z}+\theta_{\varepsilon}+\theta_{z})+8 \beta_{11} \Gamma^2 (\theta_{z}+1) (\theta_{\varepsilon} \theta_{z}+\theta_{\varepsilon}+\theta_{z}))^2)\\
/&(\beta_{11}^2 (A_{1}^2 (\theta_{\varepsilon}+1)+\theta_{\varepsilon} \theta_{z}+\theta_{\varepsilon}+\theta_{z})^2 (A_{1}^2 \Gamma+A_{1}+\Gamma \theta_{z}+\Gamma) (2 A_{1}^2 \Gamma+A_{1}+2 \Gamma (\theta_{z}+1))^2),\\
0<&A_1,\\
0<&\theta_z,\\
0=&\beta_{11}-(A_{1} (8 (A_{1}^2 (\theta_{\varepsilon}+1)+\theta_{\varepsilon} \theta_{z}+\theta_{\varepsilon}+\theta_{z})-((A_{1}^2+\theta_{z}+1)^2 (4 A_{1}^5 (\theta_{\varepsilon}+1) \Gamma+A_{1}^4 (\theta_{\varepsilon} (8 \beta_{11} \Gamma^2+4)+8 \beta_{11} \Gamma^2+3)\\
+&4 A_{1}^3 \Gamma (2 \beta_{11} (\theta_{\varepsilon}+1)+2 \theta_{\varepsilon} (\theta_{z}+1)+2 \theta_{z}+1)+A_{1}^2 (4 \theta_{\varepsilon} (\theta_{z}+1) (4 \beta_{11} \Gamma^2+1)+8 \beta_{11} \Gamma^2 (2 \theta_{z}+1)+3 \theta_{z}-1)\\
+&4 A_{1} \Gamma (2 \beta_{11}+\theta_{z}+1) (\theta_{\varepsilon} \theta_{z}+\theta_{\varepsilon}+\theta_{z})+8 \beta_{11} \Gamma^2 (\theta_{z}+1) (\theta_{\varepsilon} \theta_{z}+\theta_{\varepsilon}+\theta_{z})))/(\beta_{11} (A_{1}^2 \Gamma+A_{1}+\Gamma \theta_{z}+\Gamma) (2 A_{1}^2 \Gamma+A_{1}\\
+&2 \Gamma (\theta_{z}+1)))))/((A_{1}^2 (\theta_{\varepsilon}+1)+\theta_{\varepsilon} \theta_{z}+\theta_{\varepsilon}+\theta_{z})^2 ((16 A_{1} \beta_{11})/(A_{1}^2 (\beta_{11}^2 (\theta_{\varepsilon}+1)+\theta_{1_+})+\beta_{11}^2 (\theta_{\varepsilon} \theta_{z}+\theta_{\varepsilon}+\theta_{z})+\theta_{1_+} (\theta_{z}+1))\\
-&((A_{1}^2+\theta_{z}+1) (4 A_{1}^5 (\theta_{\varepsilon}+1) \Gamma+A_{1}^4 (\theta_{\varepsilon} (8 \beta_{11} \Gamma^2+4)+8 \beta_{11} \Gamma^2+3)+4 A_{1}^3 \Gamma (2 \beta_{11} (\theta_{\varepsilon}+1)+2 \theta_{\varepsilon} (\theta_{z}+1)+2 \theta_{z}+1)\\
+&A_{1}^2 (4 \theta_{\varepsilon} (\theta_{z}+1) (4 \beta_{11} \Gamma^2+1)+8 \beta_{11} \Gamma^2 (2 \theta_{z}+1)+3 \theta_{z}-1)+4 A_{1} \Gamma (2 \beta_{11}+\theta_{z}+1) (\theta_{\varepsilon} \theta_{z}+\theta_{\varepsilon}+\theta_{z})+8 \beta_{11} \Gamma^2 (\theta_{z}+1)\\
&(\theta_{\varepsilon} \theta_{z}+\theta_{\varepsilon}+\theta_{z}))^2)/(\beta_{11}^2 (A_{1}^2 (\theta_{\varepsilon}+1)+\theta_{\varepsilon} \theta_{z}+\theta_{\varepsilon}+\theta_{z})^2 (A_{1}^2 \Gamma+A_{1}+\Gamma \theta_{z}+\Gamma) (2 A_{1}^2 \Gamma+A_{1}+2 \Gamma (\theta_{z}+1))^2)+16 \Gamma)),\\
0=&4 (\theta_{\varepsilon}+1) \Gamma^2 (\theta_{\varepsilon} \beta_{11}^2+\theta_{1_+}) A_{1}^{14}+4 \Gamma (2 \theta_{\varepsilon} (\theta_{\varepsilon}+1) \beta_{11}^2+2 \theta_{\varepsilon} \theta_{1_+}+\theta_{1_+}) A_{1}^{13}+(4 \theta_{\varepsilon} ((6 \theta_{z}+4) \Gamma^2+\theta_{\varepsilon} (6 \theta_{z}+5) \Gamma^2+\theta_{\varepsilon}+1) \beta_{11}^2\\
+&16 (\theta_{\varepsilon}+1) \Gamma^2 \theta_{1_+} \beta_{11}+\theta_{1_+} (8 (3 \theta_{z}+2) \Gamma^2+4 \theta_{\varepsilon} ((6 \theta_{z}+5) \Gamma^2+1)+1)) A_{1}^{12}+4 \Gamma (2 \theta_{\varepsilon} (5 \theta_{z}+\theta_{\varepsilon} (5 \theta_{z}+4)+3) \beta_{11}^2\\
+&2 (\theta_{\varepsilon}+1) \theta_{1_+} \beta_{11}+\theta_{1_+} (5 \theta_{z}+2 \theta_{\varepsilon} (5 \theta_{z}+4)+2)) A_{1}^{11}+4 ((((15 \theta_{z}^2+24 \theta_{z}+4 \theta_{1_+}+4 \theta_2+9) \Gamma^2+4 \theta_{z}+3) \theta_{\varepsilon}^2+((15 \theta_{z}^2+18 \theta_{z}\\
+&8 \theta_{1_+}+8 \theta_2+5) \Gamma^2+4 \theta_{z}+2) \theta_{\varepsilon}+\Gamma^2 (4 \theta_{1_+}+4 \theta_2-\theta_{z})) \beta_{11}^2+4 \Gamma^2 \theta_{1_+} (5 \theta_{z}+\theta_{\varepsilon} (5 \theta_{z}+4)+3) \beta_{11}+\theta_{1_+}\\
&((15 \theta_{z}^2+19 \theta_{z}+4 \theta_2+5) \Gamma^2+\theta_{z}+\theta_{\varepsilon} ((15 \theta_{z}^2+24 \theta_{z}+4 \theta_2+9) \Gamma^2+4 \theta_{z}+3))) A_{1}^{10}+8 \Gamma (((10 \theta_{z}^2+15 \theta_{z}+4 \theta_2+5) \theta_{\varepsilon}^2\\
+&2 (5 \theta_{z}^2+5 \theta_{z}+4 \theta_2+1) \theta_{\varepsilon}+4 \theta_2-\theta_{z}) \beta_{11}^2+\theta_{1_+} (4 \theta_{z}+\theta_{\varepsilon} (4 \theta_{z}+2)+1) \beta_{11}+\theta_{1_+} (5 \theta_{z}^2+3 \theta_{z}+4 \theta_2\\
+&\theta_{\varepsilon} (10 \theta_{z}^2+15 \theta_{z}+4 \theta_2+5)-1)) A_{1}^9+2 (2 (((\theta_{z}+1) (20 \theta_{z}^2+25 \theta_{z}+16 \theta_{1_+}+16 \theta_2+5) \Gamma^2+6 \theta_{z}^2+4 \theta_2+8 \theta_{z}+2) \theta_{\varepsilon}^2\\
+&2 ((4 \theta_{1_+} (4 \theta_{z}+3)+4 \theta_2 (4 \theta_{z}+3)+5 \theta_{z} (2 \theta_{z}^2+3 \theta_{z}+1)) \Gamma^2+3 \theta_{z}^2+4 \theta_2+2 \theta_{z}) \theta_{\varepsilon}+4 \theta_2-\theta_{z}+\Gamma^2 (8 \theta_{1_+} (2 \theta_{z}+1)\\
+&8 \theta_2 (2 \theta_{z}+1)-\theta_{z} (5 \theta_{z}+4))) \beta_{11}^2+8 \Gamma^2 \theta_{1_+} (10 \theta_{z}^2+12 \theta_{z}+2 \theta_{\varepsilon} (5 \theta_{z}^2+8 \theta_{z}+3)+3) \beta_{11}+\theta_{1_+} (40 \Gamma^2 \theta_{z}^3\\
+&70 \Gamma^2 \theta_{z}^2+3 \theta_{z}^2+30 \Gamma^2 \theta_{z}-2 \theta_{z}+8 \theta_2 ((4 \theta_{z}+3) \Gamma^2+1)+2 \theta_{\varepsilon} ((\theta_{z}+1) (20 \theta_{z}^2+25 \theta_{z}+16 \theta_2+5) \Gamma^2\\
+&6 \theta_{z}^2+4 \theta_2+8 \theta_{z}+2)-3)) A_{1}^8+8 \Gamma ((2 (\theta_{z}+1) (6 \theta_2+5 \theta_{z} (\theta_{z}+1)) \theta_{\varepsilon}^2+2 (5 \theta_{z}^3+5 \theta_{z}^2-\theta_{z}+4 \theta_2 (3 \theta_{z}+2)-1) \theta_{\varepsilon}\\
+&4 \theta_2 (3 \theta_{z}+1)-\theta_{z} (4 \theta_{z}+3)) \beta_{11}^2+\theta_{1_+} (6 (\theta_{\varepsilon}+1) \theta_{z}^2+(6 \theta_{\varepsilon}+3) \theta_{z}-1) \beta_{11}+\theta_{1_+} ((\theta_{z}+1) (5 (2 \theta_{\varepsilon}+1) \theta_{z}^2+(10 \theta_{\varepsilon}-3) \theta_{z}-4)\\
+&4 \theta_2 (3 \theta_{z}+3 \theta_{\varepsilon} (\theta_{z}+1)+2))) A_{1}^7+4 (((\theta_{z}+1) ((\theta_{z}+1) (15 \theta_{z}^2+10 \theta_{z}+24 \theta_{1_+}+24 \theta_2-5) \Gamma^2+2 (2 \theta_{z}^2+\theta_{z}+4 \theta_2-1)) \theta_{\varepsilon}^2\\
+&(4 \theta_{z}^3+16 \theta_2 \theta_{z}-6 \theta_{z}+8 \theta_2+\Gamma^2 (\theta_{z}+1) (24 \theta_{1_+} (2 \theta_{z}+1)+24 \theta_2 (2 \theta_{z}+1)+5 (3 \theta_{z}^3+\theta_{z}^2-3 \theta_{z}-1))-2) \theta_{\varepsilon}\\
+&\theta_{z} (8 \theta_2-3 \theta_{z}-2)+2 \Gamma^2 (-\theta_{z} (5 \theta_{z}^2+8 \theta_{z}+3)+2 \theta_{1_+} (6 \theta_{z}^2+6 \theta_{z}+1)+2 \theta_2 (6 \theta_{z}^2+6 \theta_{z}+1))) \beta_{11}^2+4 \Gamma^2 \theta_{1_+} (\theta_{z}+1) \\
&(10 \theta_{z}^2+8 \theta_{z}+2 \theta_{\varepsilon} (5 \theta_{z}^2+7 \theta_{z}+2)+1) \beta_{11}+\theta_{1_+} (\theta_{z}^3-3 \theta_{z}^2+8 \theta_2 \theta_{z}-6 \theta_{z}+4 \theta_2+\Gamma^2 (\theta_{z}+1) (15 \theta_{z}^3+15 \theta_{z}^2+24 \theta_2 \theta_{z}-5 \theta_{z}\\
+&12 \theta_2-5)+\theta_{\varepsilon} (\theta_{z}+1) ((\theta_{z}+1) (15 \theta_{z}^2+10 \theta_{z}+24 \theta_2-5) \Gamma^2+2 (2 \theta_{z}^2+\theta_{z}+4 \theta_2-1))-2)) A_{1}^6+4 \Gamma (2 (\theta_{\varepsilon}^2 (5 \theta_{z}^2+12 \theta_2-5)\\& (\theta_{z}+1)^2
+\theta_{\varepsilon} (5 \theta_{z}^3-5 \theta_{z}^2-13 \theta_{z}+8 \theta_2 (3 \theta_{z}+1)-3) (\theta_{z}+1)+\theta_{z} (-6 \theta_{z}^2+12 \theta_2 \theta_{z}-9 \theta_{z}+8 \theta_2-3)) \beta_{11}^2+2 \theta_{1_+} (\theta_{z}+1) \\
&(4 \theta_{z}^2-\theta_{z}+2 \theta_{\varepsilon} (2 \theta_{z}^2+\theta_{z}-1)-1) \beta_{11}+\theta_{1_+} (\theta_{z}+1) (5 \theta_{z}^3-9 \theta_{z}^2-21 \theta_{z}+8 \theta_2 (3 \theta_{z}+1)+2 \theta_{\varepsilon} (\theta_{z}+1) (5 \theta_{z}^2+12 \theta_2-5)\\
-&7)) A_{1}^5+(4 (\theta_{\varepsilon}^2 ((\theta_{z}+1) (6 \theta_{z}^2-3 \theta_{z}+16 \theta_{1_+}+16 \theta_2-9) \Gamma^2+\theta_{z}^2+4 \theta_2-2 \theta_{z}-3) (\theta_{z}+1)^2+\theta_{\varepsilon} (\theta_{z}^3-5 \theta_{z}^2+(8 \theta_2-7) \theta_{z}\\
+&2 \Gamma^2 (\theta_{z}+1) (3 \theta_{z}^3-6 \theta_{z}^2+16 \theta_2 \theta_{z}-11 \theta_{z}+4 \theta_2+4 \theta_{1_+} (4 \theta_{z}+1)-2)-1) (\theta_{z}+1)+\theta_{z} (2 (\theta_{z}+1) (-5 \theta_{z}^2+8 \theta_2 \theta_{z}-7 \theta_{z}\\
+&4 \theta_2+\theta_{1_+} (8 \theta_{z}+4)-2) \Gamma^2-3 \theta_{z}^2+4 (\theta_2-1) \theta_{z}-1)) \beta_{11}^2+16 \Gamma^2 \theta_{1_+} (\theta_{z}+1)^2 (\theta_{z} (5 \theta_{z}+2)+\theta_{\varepsilon} (5 \theta_{z}^2+6 \theta_{z}+1)) \beta_{11}\\
+&\theta_{1_+} (\theta_{z}+1) (\theta_{z}^3-13 \theta_{z}^2+16 \theta_2 \theta_{z}-17 \theta_{z}+8 \Gamma^2 (\theta_{z}+1) (3 \theta_{z}^3-\theta_{z}^2-6 \theta_{z}+\theta_2 (8 \theta_{z}+2)-2)+4 \theta_{\varepsilon} (\theta_{z}+1) ((\theta_{z}+1) \\
&(6 \theta_{z}^2-3 \theta_{z}+16 \theta_2-9) \Gamma^2+\theta_{z}^2+4 \theta_2-2 \theta_{z}-3)-3)) A_{1}^4+4 \Gamma (\theta_{z}+1) (2 (\theta_{z} \theta_{\varepsilon}+\theta_{\varepsilon}+\theta_{z}) (-4 \theta_{z}^2+(4 \theta_2-5) \theta_{z}\\
+&\theta_{\varepsilon} (\theta_{z}+1) (\theta_{z}^2-3 \theta_{z}+4 \theta_2-4)-1) \beta_{11}^2+2 \theta_{1_+} (\theta_{z} \theta_{\varepsilon}+\theta_{\varepsilon}+\theta_{z}) (\theta_{z}^2-1) \beta_{11}+\theta_{1_+} (\theta_{z}+1) (\theta_{z}^3-8 \theta_{z}^2+(8 \theta_2-11) \theta_{z}\\
+&2 \theta_{\varepsilon} (\theta_{z}+1) (\theta_{z}^2-3 \theta_{z}+4 \theta_2-4)-2)) A_{1}^3+4 (\theta_{z}+1)^2 ((\theta_{z} \theta_{\varepsilon}+\theta_{\varepsilon}+\theta_{z}) ((-5 \theta_{z}^2+(4 \theta_{1_+}+4 \theta_2-6) \theta_{z}-1) \Gamma^2-\theta_{z}\\
+&\theta_{\varepsilon} (\theta_{z}+1) (\Gamma^2 (\theta_{z}^2-4 \theta_{z}+4 \theta_{1_+}+4 \theta_2-5)-1)) \beta_{11}^2+4 \Gamma^2 \theta_{1_+} \theta_{z} (\theta_{z}+1) (\theta_{z} \theta_{\varepsilon}+\theta_{\varepsilon}+\theta_{z}) \beta_{11}+\theta_{1_+} (\theta_{z}+1)\\
&((\theta_{z}^3-4 \theta_{z}^2+4 \theta_2 \theta_{z}-6 \theta_{z}-1) \Gamma^2-\theta_{z}+\theta_{\varepsilon} (\theta_{z}+1) (\Gamma^2 (\theta_{z}^2-4 \theta_{z}+4 \theta_2-5)-1))) A_{1}^2-8 \Gamma (\theta_{z}+1)^3 (\theta_{z} \theta_{\varepsilon}+\theta_{\varepsilon}+\theta_{z})\\ &((\theta_{z} \theta_{\varepsilon}+\theta_{\varepsilon}+\theta_{z}) \beta_{11}^2+\theta_{1_+} (\theta_{z}+1)) A_{1}-4 \Gamma^2 (\theta_{z}+1)^4 (\theta_{z} \theta_{\varepsilon}+\theta_{\varepsilon}+\theta_{z}) ((\theta_{z} \theta_{\varepsilon}+\theta_{\varepsilon}+\theta_{z}) \beta_{11}^2+\theta_{1_+} (\theta_{z}+1)),\\
\end{aligned}
\end{equation}

\begin{equation}
\label{systemmixedj=1-2}
\begin{aligned}
0=&-2 (\theta_{1_+} (A_{1}^2+\theta_{z}+1)^2 (4 (\theta_{\varepsilon}+1) \Gamma A_{1}^4+(4 \theta_{\varepsilon}+3) A_{1}^3+4 \Gamma (\theta_{\varepsilon} \beta_{11}+\beta_{11}+2 \theta_{\varepsilon}+2 \theta_{\varepsilon} \theta_{z}+2 \theta_{z}+1) A_{1}^2\\
+&(3 \theta_{z}+4 \theta_{\varepsilon} (\theta_{z}+1)-1) A_{1}+4 \Gamma (\beta_{11}+\theta_{z}+1) (\theta_{z} \theta_{\varepsilon}+\theta_{\varepsilon}+\theta_{z}))^2 A_{1}^4+16 \beta_{11}^2 \theta_2 ((\theta_{\varepsilon}+1) A_{1}^2+\theta_{\varepsilon}+\theta_{\varepsilon} \theta_{z}+\theta_{z})^2\\
&(\Gamma A_{1}^2+A_{1}+\Gamma+\Gamma \theta_{z})^2 (2 \Gamma A_{1}^2+A_{1}+2 \Gamma (\theta_{z}+1))^2 A_{1}^2+4 \beta_{11}^2 \theta_{\varepsilon} (A_{1}^2+\theta_{z}+1)^2 (\Gamma A_{1}^2+A_{1}+\Gamma+\Gamma \theta_{z})^2 ((2 \theta_{\varepsilon}+1) A_{1}^3\\
+&4 \beta_{11} (\theta_{\varepsilon}+1) \Gamma A_{1}^2+(\theta_{z}+2 \theta_{\varepsilon} (\theta_{z}+1)-1) A_{1}+4 \beta_{11} \Gamma (\theta_{z} \theta_{\varepsilon}+\theta_{\varepsilon}+\theta_{z}))^2 A_{1}^2+4 \beta_{11}^2 \theta_{z} (\Gamma A_{1}^2+A_{1}+\Gamma+\Gamma \theta_{z})^2 \\
&(4 (\theta_{\varepsilon}+1) \Gamma A_{1}^4
+(4 \theta_{\varepsilon}+3) A_{1}^3+4 \Gamma (\theta_{\varepsilon} \beta_{11}+\beta_{11}+2 \theta_{\varepsilon}+2 \theta_{\varepsilon} \theta_{z}+2 \theta_{z}+1) A_{1}^2+(3 \theta_{z}+4 \theta_{\varepsilon} (\theta_{z}+1)-1) A_{1}\\
+&4 \Gamma (\beta_{11}+\theta_{z}+1) (\theta_{\varepsilon}(\theta_{z}+1)+\theta_{z}))^2 A_{1}^2+4 \beta_{11}^2 (\Gamma A_{1}^2+A_{1}+\Gamma+\Gamma \theta_{z})^2 (2 (\theta_{\varepsilon}+1) \Gamma A_{1}^6+(3 \theta_{\varepsilon}+2) A_{1}^5+2 \Gamma (3 \theta_{z} \theta_{\varepsilon}+\theta_{\varepsilon}\\
+&2 \beta_{11} (\theta_{\varepsilon}+1)+3 \theta_{z}) A_{1}^4+2 (3 \theta_{z} \theta_{\varepsilon}+\theta_{\varepsilon}+2 \theta_{z}-1) A_{1}^3+2 \Gamma (3 \theta_{z}^2+4 \beta_{11} \theta_{z}+\theta_{\varepsilon} (3 \theta_{z}^2+4 \beta_{11} \theta_{z}+2 \theta_{z}+2 \beta_{11}-1)-1) A_{1}^2\\
+&(2 (\theta_{z}-1) \theta_{z}+\theta_{\varepsilon} (3 \theta_{z}^2+2 \theta_{z}-1)) A_{1}+2 \Gamma (\theta_{z} \theta_{\varepsilon}+\theta_{\varepsilon}+\theta_{z}) (\theta_{z}^2+2 \beta_{11} \theta_{z}-1))^2 A_{1}^2+4 \beta_{11}^2 (\Gamma A_{1}^2+A_{1}+\Gamma+\Gamma \theta_{z})^2 \\
&(2 (\theta_{\varepsilon}+1) \Gamma A_{1}^6+(3 \theta_{\varepsilon}+2) A_{1}^5+2 \Gamma (\theta_{z} \theta_{\varepsilon}+\theta_{\varepsilon}+2 \beta_{11} (\theta_{\varepsilon}+1)+\theta_{z}) A_{1}^4+(\theta_{z}+2 \theta_{\varepsilon} (\theta_{z}+1)-2) A_{1}^3\\
+&2 \Gamma (-\theta_{z}^2+2 (\beta_{11}-1) \theta_{z}+\theta_{\varepsilon} (2 \beta_{11}-\theta_{z}-1) (\theta_{z}+1)-1) A_{1}^2-(\theta_{z}+1) (\theta_{z} \theta_{\varepsilon}+\theta_{\varepsilon}+\theta_{z}) A_{1}\\
-&2 \Gamma (\theta_{z}+1)^2 (\theta_{z} \theta_{\varepsilon}+\theta_{\varepsilon}+\theta_{z}))^2)+4 (A_{1}^2+\theta_{z}+1) (\Gamma A_{1}^2+A_{1}+\Gamma+\Gamma \theta_{z}) (2 (\theta_{\varepsilon}+1) \Gamma A_{1}^6+(3 \theta_{\varepsilon}+2) A_{1}^5+2 \Gamma (\theta_{z} \theta_{\varepsilon}+\theta_{\varepsilon}\\
+&2 \beta_{11} (\theta_{\varepsilon}+1)+\theta_{z}) A_{1}^4+(\theta_{z}+2 \theta_{\varepsilon} (\theta_{z}+1)-2) A_{1}^3+2 \Gamma (-\theta_{z}^2+2 (\beta_{11}-1) \theta_{z}+\theta_{\varepsilon} (2 \beta_{11}-\theta_{z}-1) (\theta_{z}+1)-1) A_{1}^2\\
-&(\theta_{z}+1) (\theta_{z} \theta_{\varepsilon}+\theta_{\varepsilon}+\theta_{z}) A_{1}-2 \Gamma (\theta_{z}+1)^2 (\theta_{z} \theta_{\varepsilon}+\theta_{\varepsilon}+\theta_{z})) (4 \beta_{11}^2 (\theta_{\varepsilon}+1) \Gamma^2 A_{1}^8+4 (\theta_{\varepsilon}+1) \Gamma ((\theta_{\varepsilon}+2) \beta_{11}^2+\theta_{1_+}) A_{1}^7\\
+&(8 (\theta_{\varepsilon}+1)^2 \Gamma^2 \beta_{11}^3+4 (\theta_{\varepsilon}^2+2 (2 \theta_{z} \Gamma^2+\Gamma^2+1) \theta_{\varepsilon}+\Gamma^2+4 \Gamma^2 \theta_{z}+1) \beta_{11}^2+(4 \theta_{\varepsilon}+3) \theta_{1_+}) A_{1}^6+4 \Gamma (2 (\theta_{\varepsilon}+1)^2 \beta_{11}^3\\
+&(3 (\theta_{z}+1) \theta_{\varepsilon}^2+(9 \theta_{z}+4) \theta_{\varepsilon}+6 \theta_{z}) \beta_{11}^2+(\theta_{\varepsilon}+1) \theta_{1_+} \beta_{11}+\theta_{1_+} (3 \theta_{z}+3 \theta_{\varepsilon} (\theta_{z}+1)+2)) A_{1}^5\\
+&2 (4 (\theta_{\varepsilon}+1) \Gamma^2 (3 \theta_{z}+3 \theta_{\varepsilon} (\theta_{z}+1)+1) \beta_{11}^3+2 (2 (\theta_{z}+1) \theta_{\varepsilon}^2+6 \Gamma^2 \theta_{z}^2 \theta_{\varepsilon}+(6 \Gamma^2+4) \theta_{z} \theta_{\varepsilon}+\theta_{\varepsilon}+2 \theta_{z}\\
+&\Gamma^2 (6 \theta_{z}^2+3 \theta_{z}-1)-1) \beta_{11}^2+\theta_{1_+} (3 \theta_{z}+4 \theta_{\varepsilon} (\theta_{z}+1)+1)) A_{1}^4+4 \Gamma (4 (\theta_{\varepsilon}+1) (\theta_{z} \theta_{\varepsilon}+\theta_{\varepsilon}+\theta_{z}) \beta_{11}^3+(6 \theta_{z}^2+3 \theta_{\varepsilon}^2 (\theta_{z}+1)^2\\
+&\theta_{\varepsilon} (9 \theta_{z}^2+8 \theta_{z}-1)-2) \beta_{11}^2+\theta_{1_+} (2 \theta_{z}+2 \theta_{\varepsilon} (\theta_{z}+1)+1) \beta_{11}+\theta_{1_+} (\theta_{z}+1) (3 \theta_{z}+3 \theta_{\varepsilon} (\theta_{z}+1)+1)) A_{1}^3+(8 \Gamma^2 (3 \theta_{\varepsilon}^2 (\theta_{z}+1)^2\\
+&\theta_{z} (3 \theta_{z}+2)+\theta_{\varepsilon} (6 \theta_{z}^2+8 \theta_{z}+2)) \beta_{11}^3+4 ((4 \theta_{z}^3+3 \theta_{z}^2-2 \theta_{z}-1) \Gamma^2+\theta_{\varepsilon}^2 (\theta_{z}+1)^2+(\theta_{z}-1) \theta_{z}+\theta_{\varepsilon} (2 \theta_{z}^2+\theta_{z}-1)\\
&(2 (\theta_{z}+1) \Gamma^2+1)) \beta_{11}^2+\theta_{1_+} (\theta_{z}+1) (3 \theta_{z}+4 \theta_{\varepsilon} (\theta_{z}+1)-1)) A_{1}^2+4 \Gamma (\theta_{z} \theta_{\varepsilon}+\theta_{\varepsilon}+\theta_{z}) (2 (\theta_{z} \theta_{\varepsilon}+\theta_{\varepsilon}+\theta_{z}) \beta_{11}^3+(\theta_{z}+1) \\
&(\theta_{z} \theta_{\varepsilon}+\theta_{\varepsilon}+2 \theta_{z}-2) \beta_{11}^2+\theta_{1_+} (\theta_{z}+1) \beta_{11}+\theta_{1_+} (\theta_{z}+1)^2) A_{1}+4 \beta_{11}^2 \Gamma^2 (\theta_{z}+1) (\theta_{z} \theta_{\varepsilon}+\theta_{\varepsilon}+\theta_{z}) (\theta_{z}^2\\
+&2 \beta_{11} (\theta_{z} \theta_{\varepsilon}+\theta_{\varepsilon}+\theta_{z})-1))+(2 \Gamma A_{1}^2+A_{1}+2 \Gamma (\theta_{z}+1)) (4 (\theta_{\varepsilon}+1) \Gamma A_{1}^4+(4 \theta_{\varepsilon}+3) A_{1}^3\\
+&4 \Gamma (\theta_{\varepsilon} \beta_{11}+\beta_{11}+2 \theta_{\varepsilon}+2 \theta_{\varepsilon} \theta_{z}+2 \theta_{z}+1) A_{1}^2+(3 \theta_{z}+4 \theta_{\varepsilon} (\theta_{z}+1)-1) A_{1}+4 \Gamma (\beta_{11}+\theta_{z}+1) (\theta_{z} \theta_{\varepsilon}+\theta_{\varepsilon}+\theta_{z})) (4 (\theta_{\varepsilon}+1) \\
&\Gamma^2 (\theta_{\varepsilon} \beta_{11}^2+\theta_{1_+}) A_{1}^{14}+4 \Gamma (2 \theta_{\varepsilon} (\theta_{\varepsilon}+1) \beta_{11}^2+2 \theta_{\varepsilon} \theta_{1_+}+\theta_{1_+}) A_{1}^{13}+(4 \theta_{\varepsilon} ((6 \theta_{z}+2) \Gamma^2+\theta_{\varepsilon} (6 \theta_{z}+3) \Gamma^2+\theta_{\varepsilon}+1) \beta_{11}^2\\
+&16 (\theta_{\varepsilon}+1) \Gamma^2 \theta_{1_+} \beta_{11}+\theta_{1_+} (8 (3 \theta_{z}+1) \Gamma^2+4 \theta_{\varepsilon} ((6 \theta_{z}+3) \Gamma^2+1)+1)) A_{1}^{12}+4 \Gamma (2 \theta_{\varepsilon} (5 \theta_{z}+\theta_{\varepsilon} (5 \theta_{z}+2)+1) \beta_{11}^2\\
+&2 (\theta_{\varepsilon}+1) \theta_{1_+} \beta_{11}+\theta_{1_+} (5 \theta_{z}+2 \theta_{\varepsilon} (5 \theta_{z}+2)-1)) A_{1}^{11}+4 ((((15 \theta_{z}^2+16 \theta_{z}+4 \theta_{1_+}+4 \theta_2+1) \Gamma^2+4 \theta_{z}+1) \theta_{\varepsilon}^2\\
+&((15 \theta_{z}^2+12 \theta_{z}+8 \theta_{1_+}+8 \theta_2-1) \Gamma^2+4 \theta_{z}) \theta_{\varepsilon}+\Gamma^2 (4 \theta_{1_+}+4 \theta_2+\theta_{z})) \beta_{11}^2+4 \Gamma^2 \theta_{1_+} (5 \theta_{z}+\theta_{\varepsilon} (5 \theta_{z}+3)+2) \beta_{11}\\
+&\theta_{1_+} ((15 \theta_{z}^2+11 \theta_{z}+4 \theta_2-1) \Gamma^2+\theta_{\varepsilon} (15 \theta_{z}^2+16 \theta_{z}+4 \theta_2+1) \Gamma^2+\theta_{\varepsilon}+4 \theta_{\varepsilon} \theta_{z}+\theta_{z}-1)) A_{1}^{10}\\
+&8 \Gamma (((10 \theta_{z}^2+9 \theta_{z}+4 \theta_2-1) \theta_{\varepsilon}^2+2 (5 \theta_{z}^2+3 \theta_{z}+4 \theta_2-1) \theta_{\varepsilon}+4 \theta_2+\theta_{z}) \beta_{11}^2+\theta_{1_+} (4 (\theta_{\varepsilon}+1) \theta_{z}-1) \beta_{11}+\theta_{1_+} (5 \theta_{z}^2-\theta_{z}+4 \theta_2\\
+&\theta_{\varepsilon} (10 \theta_{z}^2+9 \theta_{z}+4 \theta_2-1)-3)) A_{1}^9+2 (2 (((\theta_{z}+1) (20 \theta_{z}^2+15 \theta_{z}+16 \theta_{1_+}+16 \theta_2-5) \Gamma^2+6 \theta_{z}^2+4 \theta_2+4 \theta_{z}-2) \theta_{\varepsilon}^2\\
+&2 ((10 \theta_{z}^3+15 \theta_{z}^2+3 \theta_{z}+4 \theta_{1_+} (4 \theta_{z}+3)+4 \theta_2 (4 \theta_{z}+3)-2) \Gamma^2+3 \theta_{z}^2+4 \theta_2+\theta_{z}-1) \theta_{\varepsilon}+4 \theta_2+\theta_{z}+\Gamma^2 (8 \theta_{1_+} (2 \theta_{z}+1)\\
+&8 \theta_2 (2 \theta_{z}+1)+\theta_{z} (5 \theta_{z}+4))) \beta_{11}^2+16 \Gamma^2 \theta_{1_+} (\theta_{z} (5 \theta_{z}+4)+\theta_{\varepsilon} (5 \theta_{z}^2+6 \theta_{z}+1)) \beta_{11}+\theta_{1_+} (2 (20 \theta_{z}^3+25 \theta_{z}^2+\theta_{z}\\
+&4 \theta_2 (4 \theta_{z}+3)-4) \Gamma^2+3 \theta_{z}^2+8 \theta_2-4 \theta_{z}+2 \theta_{\varepsilon} ((\theta_{z}+1) (20 \theta_{z}^2+15 \theta_{z}+16 \theta_2-5) \Gamma^2+6 \theta_{z}^2+4 \theta_2+4 \theta_{z}-2)-1)) A_{1}^8\\
+&8 \Gamma ((2 (\theta_{z}+1) (5 \theta_{z}^2+3 \theta_{z}+6 \theta_2-2) \theta_{\varepsilon}^2+2 (5 \theta_{z}^3+7 \theta_{z}^2+\theta_{z}+4 \theta_2 (3 \theta_{z}+2)-1) \theta_{\varepsilon}+4 \theta_2 (3 \theta_{z}+1)+\theta_{z} (4 \theta_{z}+3)) \beta_{11}^2\\
+&\theta_{1_+} (6 \theta_{z}^2-3 \theta_{z}+6 \theta_{\varepsilon} (\theta_{z}^2-1)-5) \beta_{11}+\theta_{1_+} (5 \theta_{z}^3+\theta_{z}^2-5 \theta_{z}+4 \theta_2 (3 \theta_{z}+2)+2 \theta_{\varepsilon} (\theta_{z}+1) (5 \theta_{z}^2+3 \theta_{z}+6 \theta_2-2)-1)) A_{1}^7\\
+&4 (((\theta_{z}+1) ((\theta_{z}+1) (15 \theta_{z}^2+10 \theta_{z}+24 \theta_{1_+}+24 \theta_2-5) \Gamma^2+2 (2 \theta_{z}^2+\theta_{z}+4 \theta_2-1)) \theta_{\varepsilon}^2+((\theta_{z}+1) (15 \theta_{z}^3+25 \theta_{z}^2+48 \theta_2 \theta_{z}\\
+&9 \theta_{z}+24 \theta_2+24 \theta_{1_+} (2 \theta_{z}+1)-1) \Gamma^2+2 (2 \theta_{z}+1) (\theta_{z}^2+\theta_{z}+4 \theta_2)) \theta_{\varepsilon}+\theta_{z} (8 \theta_2+3 \theta_{z}+2)+2 \Gamma^2 (\theta_{z} (5 \theta_{z}^2+8 \theta_{z}+3)\\
+&2 \theta_{1_+} (6 \theta_{z}^2+6 \theta_{z}+1)+2 \theta_2 (6 \theta_{z}^2+6 \theta_{z}+1))) \beta_{11}^2+8 \Gamma^2 \theta_{1_+} (\theta_{z}+1) (5 \theta_{z}^2+\theta_{z}+\theta_{\varepsilon} (5 \theta_{z}^2+4 \theta_{z}-1)-1) \beta_{11}\\
+&\theta_{1_+} (\theta_{z}^3+8 \theta_2 \theta_{z}+2 \theta_{z}+4 \theta_2+\Gamma^2 (\theta_{z}+1) (15 \theta_{z}^3+15 \theta_{z}^2+24 \theta_2 \theta_{z}-\theta_{z}+12 \theta_2-1)+\theta_{\varepsilon} (\theta_{z}+1) ((\theta_{z}+1) \\
&(15 \theta_{z}^2+10 \theta_{z}+24 \theta_2-5) \Gamma^2+2 (2 \theta_{z}^2+\theta_{z}+4 \theta_2-1))+3)) A_{1}^6+4 \Gamma (2 (\theta_{\varepsilon}^2 (5 \theta_{z}^2+4 \theta_{z}+12 \theta_2-1) (\theta_{z}+1)^2\\
+&\theta_{\varepsilon} ((5 \theta_{z}+1) (\theta_{z}+1)^2+8 \theta_2 (3 \theta_{z}+1)) (\theta_{z}+1)+\theta_{z} (6 \theta_{z}^2+9 \theta_{z}+4 \theta_2 (3 \theta_{z}+2)+3)) \beta_{11}^2+2 \theta_{1_+} (\theta_{z}+1) (4 \theta_{z}^2-7 \theta_{z}\\
+&4 \theta_{\varepsilon} (\theta_{z}^2-\theta_{z}-2)-3) \beta_{11}+\theta_{1_+} (\theta_{z}+1) (5 \theta_{z}^3+3 \theta_{z}^2+3 \theta_{z}+8 \theta_2 (3 \theta_{z}+1)+2 \theta_{\varepsilon} (\theta_{z}+1) (5 \theta_{z}^2+4 \theta_{z}+12 \theta_2-1)+5)) A_{1}^5\\
+&(4 (\theta_{\varepsilon}^2 ((\theta_{z}+1) (6 \theta_{z}^2+7 \theta_{z}+16 \theta_{1_+}+16 \theta_2+1) \Gamma^2+(\theta_{z}+1)^2+4 \theta_2) (\theta_{z}+1)^2+\theta_{\varepsilon} (\theta_{z}^3+5 \theta_{z}^2+(8 \theta_2+5) \theta_{z}\\
+&2 \Gamma^2 (\theta_{z}+1) (3 \theta_{z}^3+9 \theta_{z}^2+16 \theta_2 \theta_{z}+7 \theta_{z}+4 \theta_2+4 \theta_{1_+} (4 \theta_{z}+1)+1)+1) (\theta_{z}+1)+\theta_{z} (2 (\theta_{z}+1) (5 \theta_{z}^2+7 \theta_{z}+\theta_{1_+} (8 \theta_{z}+4)\\
+&\theta_2 (8 \theta_{z}+4)+2) \Gamma^2+3 \theta_{z}^2+4 (\theta_2+1) \theta_{z}+1)) \beta_{11}^2+16 \Gamma^2 \theta_{1_+} (\theta_{z}+1)^2 (5 \theta_{z}^2-2 \theta_{z}+\theta_{\varepsilon} (5 \theta_{z}^2+2 \theta_{z}-3)-1) \beta_{11}\\
\end{aligned}
\end{equation}

\begin{equation}
\begin{aligned}
\label{systemmixedj=1-3}
+&\theta_{1_+} (\theta_{z}+1) (\theta_{z}^3+7 \theta_{z}^2+16 \theta_2 \theta_{z}+15 \theta_{z}+8 \Gamma^2 (\theta_{z}+1) (3 \theta_{z}^3+4 \theta_{z}^2+2 \theta_{z}+\theta_2 (8 \theta_{z}+2)+1)+4 \theta_{\varepsilon} (\theta_{z}+1) ((\theta_{z}+1)\\
&(6 \theta_{z}^2+7 \theta_{z}+16 \theta_2+1) \Gamma^2+(\theta_{z}+1)^2+4 \theta_2)+9)) A_{1}^4+4 \Gamma (\theta_{z}+1) (2 (\theta_{z} \theta_{\varepsilon}+\theta_{\varepsilon}+\theta_{z}) (4 \theta_{z}^2+(4 \theta_2+5) \theta_{z}+\theta_{\varepsilon} (\theta_{z}+1)\\&
(\theta_{z}^2+3 \theta_{z}+4 \theta_2+2)+1) \beta_{11}^2+2 \theta_{1_+} (\theta_{z} \theta_{\varepsilon}+\theta_{\varepsilon}+\theta_{z}) (\theta_{z}^2-2 \theta_{z}-3) \beta_{11}+\theta_{1_+} (\theta_{z}+1) (\theta_{z}^3+5 \theta_{z}^2+(8 \theta_2+7) \theta_{z}\\
+&2 \theta_{\varepsilon} (\theta_{z}+1) (\theta_{z}^2+3 \theta_{z}+4 \theta_2+2)+3)) A_{1}^3+4 (\theta_{z}+1)^2 ((\theta_{z} \theta_{\varepsilon}+\theta_{\varepsilon}+\theta_{z}) ((5 \theta_{z}^2+(4 \theta_{1_+}+4 \theta_2+6) \theta_{z}+1) \Gamma^2+\theta_{z}\\
+&\theta_{\varepsilon} (\theta_{z}+1) ((\theta_{z}^2+4 \theta_{z}+4 \theta_{1_+}+4 \theta_2+3) \Gamma^2+1)) \beta_{11}^2+4 \Gamma^2 \theta_{1_+} (\theta_{z} \theta_{\varepsilon}+\theta_{\varepsilon}+\theta_{z}) (\theta_{z}^2-1) \beta_{11}+\theta_{1_+} (\theta_{z}+1) ((\theta_{z}^3+4 \theta_{z}^2\\
+&4 (\theta_2+1) \theta_{z}+1) \Gamma^2+\theta_{z}+\theta_{\varepsilon} (\theta_{z}+1) ((\theta_{z}^2+4 \theta_{z}+4 \theta_2+3) \Gamma^2+1))) A_{1}^2+8 \Gamma (\theta_{z}+1)^3 (\theta_{z} \theta_{\varepsilon}+\theta_{\varepsilon}+\theta_{z}) \\&((\theta_{z} \theta_{\varepsilon}+\theta_{\varepsilon}+\theta_{z}) \beta_{11}^2+\theta_{1_+} (\theta_{z}+1)) A_{1}+4 \Gamma^2 (\theta_{z}+1)^4 (\theta_{z} \theta_{\varepsilon}+\theta_{\varepsilon}+\theta_{z}) ((\theta_{z} \theta_{\varepsilon}+\theta_{\varepsilon}+\theta_{z}) \beta_{11}^2+\theta_{1_+} (\theta_{z}+1))).
\end{aligned}
\end{equation}
}
If it is solved,
\begin{equation}
\label{systemmixedj=1-4}
\begin{aligned}
\Lambda_1&=\frac{A_1}{1+A_1^2+\theta_z},\\
\Lambda_{1_+}&=\frac{A_{1} \beta_{11}}{A_{1}^2 [\beta_{11}^2 (\theta_{\varepsilon}+1)+\theta_{1_+}]+\beta_{11}^2 (\theta_{\varepsilon} \theta_{z}+\theta_{\varepsilon}+\theta_{z})+\theta_{1_+} (\theta_{z}+1)},\\
\Lambda_{21}&=A_{1} \{4 A_{1}^4 (\theta_{\varepsilon}+1) \Gamma+A_{1}^3 (4 \theta_{\varepsilon}+3)+4 A_{1}^2 \Gamma (\beta_{11} \theta_{\varepsilon}+\beta_{11}+2 \theta_{\varepsilon} \theta_{z}+2 \theta_{\varepsilon}+2 \theta_{z}+1)\\
&+A_{1} [4 \theta_{\varepsilon} (\theta_{z}+1)+3 \theta_{z}-1]+4 \Gamma (\beta_{11}+\theta_{z}+1) (\theta_{\varepsilon} \theta_{z}+\theta_{\varepsilon}+\theta_{z})\}\\
&\quad/\{2 \beta_{11} [A_{1}^2 (\theta_{\varepsilon}+1)+\theta_{\varepsilon} \theta_{z}+\theta_{\varepsilon}+\theta_{z}] [2 A_{1}^2 \Gamma+A_{1}+2 \Gamma (\theta_{z}+1)]\},\\
\Lambda_{22}&=\Lambda_1,\\
A_{21}&=\frac{1}{2\Lambda_{22}},\\
\alpha_{22}&=-\frac{1+A_1^2+\theta_z}{2},\\
\beta_{21}&=\frac{A_{1} (A_{1}^2+\theta_{z}+1)^2}{4 [A_{1}^2 (\theta_{\varepsilon}+1)+\theta_{\varepsilon} \theta_{z}+\theta_{\varepsilon}+\theta_{z}] (A_{1}^2 \Gamma+A_{1}+\Gamma \theta_{z}+\Gamma)},\\
\beta_{22}&=-\frac{\Lambda_{21}}{2(\Gamma+\Lambda_{22})},\\
\beta_{23}&=-\frac{2\Gamma+\Lambda_{21}}{2(\Gamma+\Lambda_{22})}.
\end{aligned}
\end{equation}

The pure-strategy equilibrium can be simplified to the following system of $(\Lambda_{21},\Lambda_{22},A_1,\beta_{11})$:
{\scriptsize
\begin{equation}
\label{systempurej=1}
\begin{aligned}
0=&2 (A_{1}^2+1) (\Gamma+\Lambda_{22}) (\beta_{11} ((3 \theta_{\varepsilon}+2) \Lambda_{21} \Lambda_{22}+2 (\theta_{\varepsilon}+1) \Gamma (\Lambda_{21}+\Lambda_{22})) A_{1}^3-2 (\theta_{\varepsilon}+1) (\Gamma+\Lambda_{22}) A_{1}^2+\beta_{11} \theta_{\varepsilon} (3 \Lambda_{21} \Lambda_{22}\\
+&2 \Gamma (\Lambda_{21}+\Lambda_{22})) A_{1}-2 \theta_{\varepsilon} (\Gamma+\Lambda_{22})) (-2 \beta_{11} (\Gamma+\Lambda_{22}) (\beta_{11} ((3 \theta_{\varepsilon}+2) \Lambda_{21} \Lambda_{22}+2 (\theta_{\varepsilon}+1) \Gamma (\Lambda_{21}+\Lambda_{22})) A_{1}^3\\
-&2 (\theta_{\varepsilon}+1) (\Gamma+\Lambda_{22}) A_{1}^2+\beta_{11} \theta_{\varepsilon} (3 \Lambda_{21} \Lambda_{22}+2 \Gamma (\Lambda_{21}+\Lambda_{22})) A_{1}-2 \theta_{\varepsilon} (\Gamma+\Lambda_{22})) A_{1}^3-2 \beta_{11} (\Gamma+\Lambda_{22}) (\beta_{11} ((3 \theta_{\varepsilon}+2) \Lambda_{21} \Lambda_{22}\\
+&2 (\theta_{\varepsilon}+1) \Gamma (\Lambda_{21}+\Lambda_{22})) A_{1}^3-2 (\theta_{\varepsilon}+1) (\Gamma+\Lambda_{22}) A_{1}^2+\beta_{11} \theta_{\varepsilon} (3 \Lambda_{21} \Lambda_{22}+2 \Gamma (\Lambda_{21}+\Lambda_{22})) A_{1}-2 \theta_{\varepsilon} (\Gamma+\Lambda_{22})) A_{1}\\
-&2 (A_{1}^2+1)^2 \beta_{11} \theta_{\varepsilon} (\beta_{11} (4 (\theta_{\varepsilon}+1) \Gamma+2 \theta_{\varepsilon} \Lambda_{21}+\Lambda_{21}) A_{1}^2-A_{1}+2 \beta_{11} \theta_{\varepsilon} (2 \Gamma+\Lambda_{21})) \Lambda_{22} (\Gamma+\Lambda_{22})\\
-&(A_{1}^2+1)^2 \Lambda_{21} \Lambda_{22} ((4 (\theta_{\varepsilon}+1) \Gamma+(4 \theta_{\varepsilon}+3) \Lambda_{22}) A_{1}^2+4 \theta_{\varepsilon} (\Gamma+\Lambda_{22})) \theta_{1_+})+A_{1} (A_{1}^2+1) \beta_{11} (4 (A_{1}^2+1)^2 \theta_{\varepsilon} (-\beta_{11} (4 (\theta_{\varepsilon}+1) \Gamma\\
+&2 \theta_{\varepsilon} \Lambda_{21}+\Lambda_{21}) A_{1}^2+A_{1}-2 \beta_{11} \theta_{\varepsilon} (2 \Gamma+\Lambda_{21}))^2 \Lambda_{22}^2 (\Gamma+\Lambda_{22})^2+4 A_{1}^2 (\beta_{11} ((3 \theta_{\varepsilon}+2) \Lambda_{21} \Lambda_{22}+2 (\theta_{\varepsilon}+1) \Gamma (\Lambda_{21}+\Lambda_{22})) A_{1}^3\\
-&2 (\theta_{\varepsilon}+1) (\Gamma+\Lambda_{22}) A_{1}^2+\beta_{11} \theta_{\varepsilon} (3 \Lambda_{21} \Lambda_{22}+2 \Gamma (\Lambda_{21}+\Lambda_{22})) A_{1}-2 \theta_{\varepsilon} (\Gamma+\Lambda_{22}))^2 (\Gamma+\Lambda_{22})^2+4 (\beta_{11} ((3 \theta_{\varepsilon}+2) \Lambda_{21} \Lambda_{22}\\
+&2 (\theta_{\varepsilon}+1) \Gamma (\Lambda_{21}+\Lambda_{22})) A_{1}^3-2 (\theta_{\varepsilon}+1) (\Gamma+\Lambda_{22}) A_{1}^2+\beta_{11} \theta_{\varepsilon} (3 \Lambda_{21} \Lambda_{22}+2 \Gamma (\Lambda_{21}+\Lambda_{22})) A_{1}-2 \theta_{\varepsilon} (\Gamma+\Lambda_{22}))^2 (\Gamma+\Lambda_{22})^2\\
+&4 (A_{1}^2+1)^2 \Lambda_{22}^2 ((4 (\theta_{\varepsilon}+1) \Gamma+(4 \theta_{\varepsilon}+3) \Lambda_{22}) A_{1}^2+4 \theta_{\varepsilon} (\Gamma+\Lambda_{22}))^2 \theta_2 (\Gamma+\Lambda_{22})^2+(A_{1}^2+1)^2 \Lambda_{21}^2 \Lambda_{22}^2 ((4 (\theta_{\varepsilon}+1) \Gamma\\
+&(4 \theta_{\varepsilon}+3) \Lambda_{22}) A_{1}^2+4 \theta_{\varepsilon} (\Gamma+\Lambda_{22}))^2 \theta_{1_+})+\Lambda_{21} ((2 \beta_{11} (\Gamma+\Lambda_{22}) (\beta_{11} ((3 \theta_{\varepsilon}+2) \Lambda_{21} \Lambda_{22}+2 (\theta_{\varepsilon}+1) \Gamma (\Lambda_{21}+\Lambda_{22})) A_{1}^3\\
-&2 (\theta_{\varepsilon}+1) (\Gamma+\Lambda_{22}) A_{1}^2+\beta_{11} \theta_{\varepsilon} (3 \Lambda_{21} \Lambda_{22}+2 \Gamma (\Lambda_{21}+\Lambda_{22})) A_{1}-2 \theta_{\varepsilon} (\Gamma+\Lambda_{22})) A_{1}^3+2 \beta_{11} (\Gamma+\Lambda_{22}) (\beta_{11} ((3 \theta_{\varepsilon}+2) \Lambda_{21} \Lambda_{22}\\
+&2 (\theta_{\varepsilon}+1) \Gamma (\Lambda_{21}+\Lambda_{22})) A_{1}^3-2 (\theta_{\varepsilon}+1) (\Gamma+\Lambda_{22}) A_{1}^2+\beta_{11} \theta_{\varepsilon} (3 \Lambda_{21} \Lambda_{22}+2 \Gamma (\Lambda_{21}+\Lambda_{22})) A_{1}-2 \theta_{\varepsilon} (\Gamma+\Lambda_{22})) A_{1}\\
+&2 (A_{1}^2+1)^2 \beta_{11} \theta_{\varepsilon} (\beta_{11} (4 (\theta_{\varepsilon}+1) \Gamma+2 \theta_{\varepsilon} \Lambda_{21}+\Lambda_{21}) A_{1}^2-A_{1}+2 \beta_{11} \theta_{\varepsilon} (2 \Gamma+\Lambda_{21})) \Lambda_{22} (\Gamma+\Lambda_{22})+(A_{1}^2+1)^2 \Lambda_{21} \Lambda_{22}\\&
((4 (\theta_{\varepsilon}+1) \Gamma+(4 \theta_{\varepsilon}+3) \Lambda_{22}) A_{1}^2+4 \theta_{\varepsilon} (\Gamma+\Lambda_{22})) \theta_{1_+})^2-(A_{1}^2+1) (((\theta_{\varepsilon}+1) \beta_{11}^2+\theta_{1_+}) A_{1}^2+\beta_{11}^2 \theta_{\varepsilon}+\theta_{1_+}) (4 (A_{1}^2+1)^2 \theta_{\varepsilon}\\
&(-\beta_{11} (4 (\theta_{\varepsilon}+1) \Gamma+2 \theta_{\varepsilon} \Lambda_{21}+\Lambda_{21}) A_{1}^2+A_{1}-2 \beta_{11} \theta_{\varepsilon} (2 \Gamma+\Lambda_{21}))^2 \Lambda_{22}^2 (\Gamma+\Lambda_{22})^2+4 A_{1}^2 (\beta_{11} ((3 \theta_{\varepsilon}+2) \Lambda_{21} \Lambda_{22}\\
+&2 (\theta_{\varepsilon}+1) \Gamma (\Lambda_{21}+\Lambda_{22})) A_{1}^3-2 (\theta_{\varepsilon}+1) (\Gamma+\Lambda_{22}) A_{1}^2+\beta_{11} \theta_{\varepsilon} (3 \Lambda_{21} \Lambda_{22}+2 \Gamma (\Lambda_{21}+\Lambda_{22})) A_{1}-2 \theta_{\varepsilon} (\Gamma+\Lambda_{22}))^2 (\Gamma+\Lambda_{22})^2\\
+&4 (\beta_{11} ((3 \theta_{\varepsilon}+2) \Lambda_{21} \Lambda_{22}+2 (\theta_{\varepsilon}+1) \Gamma (\Lambda_{21}+\Lambda_{22})) A_{1}^3-2 (\theta_{\varepsilon}+1) (\Gamma+\Lambda_{22}) A_{1}^2+\beta_{11} \theta_{\varepsilon} (3 \Lambda_{21} \Lambda_{22}+2 \Gamma (\Lambda_{21}+\Lambda_{22})) A_{1}\\
-&2 \theta_{\varepsilon} (\Gamma+\Lambda_{22}))^2 (\Gamma+\Lambda_{22})^2+4 (A_{1}^2+1)^2 \Lambda_{22}^2 ((4 (\theta_{\varepsilon}+1) \Gamma+(4 \theta_{\varepsilon}+3) \Lambda_{22}) A_{1}^2+4 \theta_{\varepsilon} (\Gamma+\Lambda_{22}))^2 \theta_2 (\Gamma+\Lambda_{22})^2\\
+&(A_{1}^2+1)^2 \Lambda_{21}^2 \Lambda_{22}^2 ((4 (\theta_{\varepsilon}+1) \Gamma+(4 \theta_{\varepsilon}+3) \Lambda_{22}) A_{1}^2+4 \theta_{\varepsilon} (\Gamma+\Lambda_{22}))^2 \theta_{1_+})),\\
0=&-4 (A_{1}^2+1)^2 ((4 (\theta_{\varepsilon}+1) \Gamma+(4 \theta_{\varepsilon}+3) \Lambda_{22}) A_{1}^2+4 \theta_{\varepsilon} (\Gamma+\Lambda_{22})) (\beta_{11} ((3 \theta_{\varepsilon}+2) \Lambda_{21} \Lambda_{22}+2 (\theta_{\varepsilon}+1) \Gamma (\Lambda_{21}+\Lambda_{22})) A_{1}^3\\
-&2 (\theta_{\varepsilon}+1) (\Gamma+\Lambda_{22}) A_{1}^2+\beta_{11} \theta_{\varepsilon} (3 \Lambda_{21} \Lambda_{22}+2 \Gamma (\Lambda_{21}+\Lambda_{22})) A_{1}-2 \theta_{\varepsilon} (\Gamma+\Lambda_{22})) (((\theta_{\varepsilon}+1) \beta_{11}^2+\theta_{1_+}) A_{1}^2+\beta_{11}^2 \theta_{\varepsilon}+\theta_{1_+})\\ &(\Gamma+\Lambda_{22})^2-2 A_{1} (A_{1}^2+1) \beta_{11} ((4 (\theta_{\varepsilon}+1) \Gamma+(4 \theta_{\varepsilon}+3) \Lambda_{22}) A_{1}^2+4 \theta_{\varepsilon} (\Gamma+\Lambda_{22})) (-2 \beta_{11} (\Gamma+\Lambda_{22}) (\beta_{11} ((3 \theta_{\varepsilon}+2) \Lambda_{21} \Lambda_{22}\\
+&2 (\theta_{\varepsilon}+1) \Gamma (\Lambda_{21}+\Lambda_{22})) A_{1}^3-2 (\theta_{\varepsilon}+1) (\Gamma+\Lambda_{22}) A_{1}^2+\beta_{11} \theta_{\varepsilon} (3 \Lambda_{21} \Lambda_{22}+2 \Gamma (\Lambda_{21}+\Lambda_{22})) A_{1}-2 \theta_{\varepsilon} (\Gamma+\Lambda_{22})) A_{1}^3\\
-&2 \beta_{11} (\Gamma+\Lambda_{22}) (\beta_{11} ((3 \theta_{\varepsilon}+2) \Lambda_{21} \Lambda_{22}+2 (\theta_{\varepsilon}+1) \Gamma (\Lambda_{21}+\Lambda_{22})) A_{1}^3-2 (\theta_{\varepsilon}+1) (\Gamma+\Lambda_{22}) A_{1}^2+\beta_{11} \theta_{\varepsilon} (3 \Lambda_{21} \Lambda_{22}\\
+&2 \Gamma (\Lambda_{21}+\Lambda_{22})) A_{1}-2 \theta_{\varepsilon} (\Gamma+\Lambda_{22})) A_{1}-2 (A_{1}^2+1)^2 \beta_{11} \theta_{\varepsilon} (\beta_{11} (4 (\theta_{\varepsilon}+1) \Gamma+2 \theta_{\varepsilon} \Lambda_{21}+\Lambda_{21}) A_{1}^2-A_{1}\\
+&2 \beta_{11} \theta_{\varepsilon} (2 \Gamma+\Lambda_{21})) \Lambda_{22} (\Gamma+\Lambda_{22})-(A_{1}^2+1)^2 \Lambda_{21} \Lambda_{22} ((4 (\theta_{\varepsilon}+1) \Gamma+(4 \theta_{\varepsilon}+3) \Lambda_{22}) A_{1}^2+4 \theta_{\varepsilon} (\Gamma+\Lambda_{22})) \theta_{1_+}) (\Gamma+\Lambda_{22})\\
+&(2 \beta_{11} (\Gamma+\Lambda_{22}) (\beta_{11} ((3 \theta_{\varepsilon}+2) \Lambda_{21} \Lambda_{22}+2 (\theta_{\varepsilon}+1) \Gamma (\Lambda_{21}+\Lambda_{22})) A_{1}^3-2 (\theta_{\varepsilon}+1) (\Gamma+\Lambda_{22}) A_{1}^2+\beta_{11} \theta_{\varepsilon} (3 \Lambda_{21} \Lambda_{22}\\
+&2 \Gamma (\Lambda_{21}+\Lambda_{22})) A_{1}-2 \theta_{\varepsilon} (\Gamma+\Lambda_{22})) A_{1}^3+2 \beta_{11} (\Gamma+\Lambda_{22}) (\beta_{11} ((3 \theta_{\varepsilon}+2) \Lambda_{21} \Lambda_{22}+2 (\theta_{\varepsilon}+1) \Gamma (\Lambda_{21}+\Lambda_{22})) A_{1}^3\\
-&2 (\theta_{\varepsilon}+1) (\Gamma+\Lambda_{22}) A_{1}^2+\beta_{11} \theta_{\varepsilon} (3 \Lambda_{21} \Lambda_{22}+2 \Gamma (\Lambda_{21}+\Lambda_{22})) A_{1}-2 \theta_{\varepsilon} (\Gamma+\Lambda_{22})) A_{1}+2 (A_{1}^2+1)^2 \beta_{11} \theta_{\varepsilon} \\
&(\beta_{11} (4 (\theta_{\varepsilon}+1) \Gamma+2 \theta_{\varepsilon} \Lambda_{21}+\Lambda_{21}) A_{1}^2-A_{1}+2 \beta_{11} \theta_{\varepsilon} (2 \Gamma+\Lambda_{21})) \Lambda_{22} (\Gamma+\Lambda_{22})+(A_{1}^2+1)^2 \Lambda_{21} \Lambda_{22} ((4 (\theta_{\varepsilon}+1) \Gamma\\
+&(4 \theta_{\varepsilon}+3) \Lambda_{22}) A_{1}^2+4 \theta_{\varepsilon} (\Gamma+\Lambda_{22})) \theta_{1_+})^2-(A_{1}^2+1) (((\theta_{\varepsilon}+1) \beta_{11}^2+\theta_{1_+}) A_{1}^2+\beta_{11}^2 \theta_{\varepsilon}+\theta_{1_+}) (4 (A_{1}^2+1)^2 \theta_{\varepsilon} (-\beta_{11} (4 (\theta_{\varepsilon}+1) \Gamma\\
+&2 \theta_{\varepsilon} \Lambda_{21}+\Lambda_{21}) A_{1}^2+A_{1}-2 \beta_{11} \theta_{\varepsilon} (2 \Gamma+\Lambda_{21}))^2 \Lambda_{22}^2 (\Gamma+\Lambda_{22})^2+4 A_{1}^2 (\beta_{11} ((3 \theta_{\varepsilon}+2) \Lambda_{21} \Lambda_{22}+2 (\theta_{\varepsilon}+1) \Gamma (\Lambda_{21}+\Lambda_{22})) A_{1}^3\\
-&2 (\theta_{\varepsilon}+1) (\Gamma+\Lambda_{22}) A_{1}^2+\beta_{11} \theta_{\varepsilon} (3 \Lambda_{21} \Lambda_{22}+2 \Gamma (\Lambda_{21}+\Lambda_{22})) A_{1}-2 \theta_{\varepsilon} (\Gamma+\Lambda_{22}))^2 (\Gamma+\Lambda_{22})^2+4 (\beta_{11} ((3 \theta_{\varepsilon}+2) \Lambda_{21} \Lambda_{22}\\
+&2 (\theta_{\varepsilon}+1) \Gamma (\Lambda_{21}+\Lambda_{22})) A_{1}^3-2 (\theta_{\varepsilon}+1) (\Gamma+\Lambda_{22}) A_{1}^2+\beta_{11} \theta_{\varepsilon} (3 \Lambda_{21} \Lambda_{22}+2 \Gamma (\Lambda_{21}+\Lambda_{22})) A_{1}-2 \theta_{\varepsilon} (\Gamma+\Lambda_{22}))^2 (\Gamma+\Lambda_{22})^2\\
+&4 (A_{1}^2+1)^2 \Lambda_{22}^2 ((4 (\theta_{\varepsilon}+1) \Gamma+(4 \theta_{\varepsilon}+3) \Lambda_{22}) A_{1}^2+4 \theta_{\varepsilon} (\Gamma+\Lambda_{22}))^2 \theta_2 (\Gamma+\Lambda_{22})^2+(A_{1}^2+1)^2 \Lambda_{21}^2 \Lambda_{22}^2 ((4 (\theta_{\varepsilon}+1) \Gamma\\
+&(4 \theta_{\varepsilon}+3) \Lambda_{22}) A_{1}^2+4 \theta_{\varepsilon} (\Gamma+\Lambda_{22}))^2 \theta_{1_+}),\\
0=&A_1-(-\frac{2 A_1^2 \beta_{11} (\theta_{\varepsilon}+1) (2 \Gamma (\Lambda_{21}-\Lambda_{22})+\Lambda_{21} \Lambda_{22})+A_1 \Lambda_{22}+2 \beta_{11} \theta_{\varepsilon} (2 \Gamma (\Lambda_{21}-\Lambda_{22})+\Lambda_{21} \Lambda_{22})}{(A_1^2+1) \Lambda_{22} (A_1^2 (4 (\theta_{\varepsilon}+1) \Gamma+(4 \theta_{\varepsilon}+3) \Lambda_{22})+4 \theta_{\varepsilon} (\Gamma+\Lambda_{22}))}-\frac{A_1}{A_1^2 \Lambda_{22}+\Lambda_{22}}+2)\\
/&(4 (\frac{A_1}{A_1^2+1}-((A_1^3 (4 (\theta_{\varepsilon}+1) \Gamma+(4 \theta_{\varepsilon}+3) \Lambda_{22})+2 A_1^2 \beta_{11} (\theta_{\varepsilon}+1) (2 \Gamma (\Lambda_{21}-\Lambda_{22})+\Lambda_{21} \Lambda_{22})+A_1 (4 \theta_{\varepsilon} (\Gamma+\Lambda_{22})+\Lambda_{22})\\
+&2 \beta_{11} \theta_{\varepsilon} (2 \Gamma (\Lambda_{21}-\Lambda_{22})+\Lambda_{21} \Lambda_{22}))^2)/(4 (A_1^2+1)^2 \Lambda_{22} (A_1^2 (4 (\theta_{\varepsilon}+1) \Gamma+(4 \theta_{\varepsilon}+3) \Lambda_{22})+4 \theta_{\varepsilon} (\Gamma+\Lambda_{22}))^2))),\\
0=&(A_{1} (A_{1}^2 (\beta_{11}^2 (\theta_{\varepsilon}+1)+\theta_{1_+})+\beta_{11}^2 \theta_{\varepsilon}+\theta_{1_+}) (A_{1}^3 \beta_{11} (\theta_{\varepsilon}+1) (2 \Gamma+\Lambda_{21}) (2 \Gamma (\Lambda_{21}-\Lambda_{22})+\Lambda_{21} \Lambda_{22})\\
-&2 A_{1}^2 (\Gamma+\Lambda_{22}) (2 (\theta_{\varepsilon}+1) \Gamma-(\theta_{\varepsilon}+1) \Lambda_{21}+(4 \theta_{\varepsilon}+3) \Lambda_{22})+A_{1} \beta_{11} \theta_{\varepsilon} (2 \Gamma+\Lambda_{21}) (2 \Gamma (\Lambda_{21}-\Lambda_{22})+\Lambda_{21} \Lambda_{22})\\
-&2 \theta_{\varepsilon} (\Gamma+\Lambda_{22}) (2 \Gamma-\Lambda_{21}+4 \Lambda_{22})))/((A_{1}^2 (\theta_{\varepsilon}+1)+\theta_{\varepsilon}) (A_{1}^2 (4 (\theta_{\varepsilon}+1) \Gamma+(4 \theta_{\varepsilon}+3) \Lambda_{22})+4 \theta_{\varepsilon} (\Gamma+\Lambda_{22})) \\&(A_{1}^2 (\beta_{11}^2 (\theta_{\varepsilon}+1)+\theta_{1_+}) (-(4 \Gamma (\Lambda_{21}-\Lambda_{22})+\Lambda_{21}^2))+4 A_{1} \beta_{11} (\Gamma+\Lambda_{22})-(\beta_{11}^2 \theta_{\varepsilon}+\theta_{1_+}) (4 \Gamma (\Lambda_{21}-\Lambda_{22})+\Lambda_{21}^2)))+\beta_{11},\\
\end{aligned}
\end{equation}

\begin{equation}
\label{systempurej=1-2}
\begin{aligned}
0<&\Lambda_{22},\\
0<&A_{1}/(A_{1}^2+1)-((A_{1}^3 (4 (\theta_{\varepsilon}+1) \Gamma+(4 \theta_{\varepsilon}+3) \Lambda_{22})+2 A_{1}^2 \beta_{11} (\theta_{\varepsilon}+1) (2 \Gamma (\Lambda_{21}-\Lambda_{22})+\Lambda_{21} \Lambda_{22})+A_{1} (4 \theta_{\varepsilon} (\Gamma+\Lambda_{22})+\Lambda_{22})\\
+&2 \beta_{11} \theta_{\varepsilon} (2 \Gamma (\Lambda_{21}-\Lambda_{22})+\Lambda_{21} \Lambda_{22}))^2)/(4 (A_{1}^2+1)^2 \Lambda_{22} (A_{1}^2 (4 (\theta_{\varepsilon}+1) \Gamma+(4 \theta_{\varepsilon}+3) \Lambda_{22})+4 \theta_{\varepsilon} (\Gamma+\Lambda_{22}))^2),\\
0<&\frac{4 A_{1} \beta_{11}}{A_{1}^2 (\beta_{11}^2 (\theta_{\varepsilon}+1)+\theta_{1_+})+\beta_{11}^2 \theta_{\varepsilon}+\theta_{1_+}}-\frac{(2 \Gamma+\Lambda_{21})^2}{\Gamma+\Lambda_{22}}+4 \Gamma.
\end{aligned}
\end{equation}
}
If it is solved,
\begin{equation}
\label{systempurej=1-3}
\begin{aligned}
\Lambda_1&=\frac{A_1}{1+A_1^2},\\
\Lambda_{1_+}&=\frac{A_{1} \beta_{11} }{A_{1}^2 [\beta_{11}^2 (\theta_{\varepsilon}+1)+\theta_{1_+}]+\beta_{11}^2 \theta_{\varepsilon}+\theta_{1_+}},\\
A_{21}&=\frac{1}{2\Lambda_{22}},\\
\alpha_{22}&=-\frac{2 A_{1}^2 \beta_{11} (\theta_{\varepsilon}+1) [2 \Gamma (\Lambda_{21}-\Lambda_{22})+\Lambda_{21} \Lambda_{22}]+A_{1} \Lambda_{22}+2 \beta_{11} \theta_{\varepsilon} [2 \Gamma (\Lambda_{21}-\Lambda_{22})+\Lambda_{21} \Lambda_{22}]}{2 \Lambda_{22} \{A_{1}^2 [4 (\theta_{\varepsilon}+1) \Gamma+(4 \theta_{\varepsilon}+3) \Lambda_{22}]+4 \theta_{\varepsilon} (\Gamma+\Lambda_{22})\}},\\
\beta_{21}&=\frac{\frac{A_{1}^2 \beta_{11} [2 \Gamma (\Lambda_{21}-\Lambda_{22})+\Lambda_{21} \Lambda_{22}]}{2 (\Gamma+\Lambda_{22})}+A_{1}}{A_{1}^2 [4 (\theta_{\varepsilon}+1) \Gamma+(4 \theta_{\varepsilon}+3) \Lambda_{22}]+4 \theta_{\varepsilon} (\Gamma+\Lambda_{22})},\\
\beta_{22}&=-\frac{\Lambda_{21}}{2(\Gamma+\Lambda_{22})},\\
\beta_{23}&=-\frac{2\Gamma+\Lambda_{21}}{2(\Gamma+\Lambda_{22})}.
\end{aligned}
\end{equation}

\noindent\textbf{Proof of Proposition \ref{propj=1gam=inf}.} $\beta_{21},\beta_{22},\beta_{23}$ follows
\begin{equation}
\label{j=1beta212223}
\begin{aligned}
&\beta_{21} = \frac{(1-\lambda_{22}\alpha_{21})\eta_{21}-\lambda_{22}\alpha_{22}\mu_{21}}{2(\lambda_{22}+\gamma)},\\
&\beta_{22} = \frac{(1-\lambda_{22}\alpha_{21})\eta_{22}-\lambda_{22}\alpha_{22}\mu_{22}-\lambda_{21}}{2(\lambda_{22}+\gamma)},\\
&\beta_{23}=-\frac{\lambda_{21}+2\gamma}{2(\lambda_{22}+\gamma)},
\end{aligned}
\end{equation}
where $(\eta_{21},\eta_{22}),(\mu_{21},\mu_{22})$ are calculated through \eqref{etamu} for $J_1=1,J_2=0.$ Substitute \eqref{j=1beta212223} into
\begin{equation*}
\beta_{11}=\frac{\eta+2(\lambda_{22}+\gamma)\beta_{23}\beta_{21}}{2(\lambda_{1_+}+\gamma-(\lambda_{22}+\gamma)\beta_{23}^2)},
\end{equation*}
if $\gamma=\infty$, we have 
\begin{equation*}
  \beta_{11}=\frac{\lambda_{22}(\alpha_{21}\eta+\alpha_{22}\mu)}{2(\lambda_{1_+}+\lambda_{22}-\lambda_{21})},  
\end{equation*}
which is exactly the equation for $\beta_{11}$ where $x_1^*=\beta_{11}(\Tilde{i}_1-\mathbb{E}(\Tilde{i}_1|y_1))$ satisfies \eqref{j=1HFTobjgam=inf}. What's more, by \eqref{j=1beta212223}, we have
\begin{equation*}
    \beta_{21}=\beta_{22}=0,\ \beta_{23}=-1.
\end{equation*}

\newpage
\noindent\textbf{Proof of Theorem \ref{thmgam0}.}
The mixed-strategy equilibrium can be simplified to the following system of $(A_1,\theta_z,\beta_{11},\beta_{21},\beta_{22},\beta_{23}):$
{\scriptsize
\begin{equation}
\label{systemmixedgam0}
\begin{aligned}
0=&J (\theta_{\varepsilon} (A_1^2+\theta_z+1) (A_1^8+(3 \theta_z-\beta_{21} J) A_1^6+ A_1^4(4 \theta_{1_+} \beta_{22}^2-8 \beta_{23} \theta_{1_+} \beta_{22}+3 \theta_z^2+4 \beta_{23}^2 \theta_{1_+}+4 \theta_2+\theta_z\\
-&\beta_{21} (2 \theta_z J+J)-2)
+(\beta_{21} J (4 (2 J-3) \theta_{1_+} \beta_{22}^2-8 \beta_{23} (J-3) \theta_{1_+} \beta_{22}-\theta_z^2-12 \beta_{23}^2 \theta_{1_+}-4 \theta_2+1)\\
+&(\theta_z+1) (4 \theta_{1_+} \beta_{22}^2-8 \beta_{23} \theta_{1_+} \beta_{22}+\theta_z^2+4 \beta_{23}^2 \theta_{1_+}+4 \theta_2+\theta_z)) A_1^2+(\theta_z+1)^2 (\beta_{21} J+\theta_z+1))\\
+&J (4 (-(J-2) \theta_{1_+} \beta_{22}^2+\beta_{23} (J-4) \theta_{1_+} \beta_{22}+2 \beta_{23}^2 \theta_{1_+}+\theta_2) A_1^6+(4 \theta_{1_+} (2 \beta_{21} J^2-(3 \beta_{21}+2 \theta_z+1) J+4 \theta_z+1) \beta_{22}^2\\
-&4 \beta_{23} \theta_{1_+} (2 \beta_{21} J^2-(6 \beta_{21}+2 \theta_z+1) J+8 \theta_z+2) \beta_{22}-4 \beta_{21} J \theta_2+\beta_{21} J \theta_z+8 \theta_2 \theta_z\\
+&4 \beta_{23}^2 A_1^4\theta_{1_+} (-3 \beta_{21} J+4 \theta_z+1))
+2 (-2 \theta_{1_+} ((J-2) \theta_z^2+(-2 \beta_{21} J^2+3 \beta_{21} J+J-1) \theta_z+1) \beta_{22}^2\\
+&2 \beta_{22} \beta_{23} \theta_{1_+}((J-4) \theta_z^2+(-2 \beta_{21} J^2+6 \beta_{21} J+J-2) \theta_z+2) 
+\beta_{21} J \theta_z^2+2 \theta_2 \theta_z^2-2 \theta_2+\beta_{21} J \theta_z-2 \beta_{21} J \theta_2 \theta_z\\
+&2 \beta_{23}^2 \theta_{1_+} (2 \theta_z^2-3 \beta_{21} J \theta_z+\theta_z-1)) A_1^2+\beta_{21} J \theta_z (\theta_z+1)^2)) \beta_{11}^2\\
-&J \theta_{1_+} (\beta_{23} (5 A_1^8+(-8 \beta_{21} \theta_{\varepsilon}+2 J)+15 \theta_z+4) A_1^6+(12 J \theta_{\varepsilon}+J) \beta_{21}^2-8 (4 \theta_z J+J+2  (\theta_z+1)) \beta_{21}\\
+&15 \theta_z^2+4 \theta_2+9 \theta_z-6) A_1^4+(12 J (\theta_z \theta_{\varepsilon}+\theta_{\varepsilon}+J \theta_z) \beta_{21}^2-8 (\theta_z+1) (\theta_z \theta_{\varepsilon}+\theta_{\varepsilon}-J+2 J \theta_z) \beta_{21}\\
+&(\theta_z+1) (5 \theta_z^2+\theta_z+4 \theta_2-4)) A_1^2+(\theta_z+1)^3)+\beta_{22} ((J-5) A_1^8+(8 \beta_{21} \theta_{\varepsilon}-4 \beta_{21} (J-4) J-15 \theta_z+J (3 \theta_z+2)-4) A_1^6\\
+&(4 (J-3) J \theta_{\varepsilon}+J) \beta_{21}^2+4 (4 \theta_{\varepsilon} (\theta_z+1)-J (2 \theta_z J+J-8 \theta_z-2)) \beta_{21}+4 (J-1) \theta_2+3 (\theta_z+1) ((J-5) \theta_z+2)) A_1^4\\
+&(4 (J-3) J (\theta_z \theta_{\varepsilon}+\theta_{\varepsilon}+J \theta_z) \beta_{21}^2+4 (\theta_z+1) (2 \theta_{\varepsilon} (\theta_z+1)-J ((J-4) \theta_z+2)) \beta_{21}\\
+&(\theta_z+1) (-5 \theta_z^2-\theta_z-4 \theta_2+J (\theta_z^2-\theta_z+4 \theta_2-2)+4)) A_1^2-(J+1) (\theta_z+1)^3)) \beta_{11}-\beta_{11}^3 (\beta_{23}+\beta_{22} (J-1))\\
&J^2 (J ((4 \theta_{1_+} \beta_{22}^2-8 \beta_{23} \theta_{1_+} \beta_{22}+4 \beta_{23}^2 \theta_{1_+}+4 \theta_2-\theta_z) A_1^4-2 \theta_z (-2 \theta_{1_+} \beta_{22}^2+4 \beta_{23} \theta_{1_+} \beta_{22}-2 \beta_{23}^2 \theta_{1_+}-2 \theta_2\\
+&\theta_z+1) A_1^2-\theta_z (\theta_z+1)^2)+\theta_{\varepsilon} (A_1^2+\theta_z+1) (A_1^6+(2 \theta_z+1) A_1^4+(4 \theta_{1_+} \beta_{22}^2-8 \beta_{23} \theta_{1_+} \beta_{22}+\theta_z^2+4 \beta_{23}^2 \theta_{1_+}+4 \theta_2-1) A_1^2\\
-&(\theta_z+1)^2))+\theta_{1_+} (A_1^2-\beta_{21} J+\theta_z+1) (A_1^8+(3 \theta_z-4 \beta_{21} J) A_1^6+(4 J \theta_{\varepsilon}+J) \beta_{21}^2-8 J \theta_z \beta_{21}+3 \theta_z^2+4 \theta_2+\theta_z-2) A_1^4\\
+&(4 J (\theta_z \theta_{\varepsilon}+\theta_{\varepsilon}+J \theta_z) \beta_{21}^2-4 J (\theta_z^2-1) \beta_{21}+(\theta_z+1) (\theta_z^2+\theta_z+4 \theta_2)) A_1^2+(\theta_z+1)^3),\\
0=&4 A_1^2 \beta_{11} J \theta_{1_+} (\beta_{22}-\beta_{23}) (A_1^4+A_1^2 (-2 \beta_{21} \theta_{\varepsilon}+J)+2 \theta_z+1)-2 \beta_{21} \theta_{\varepsilon} \theta_z+\theta_{\varepsilon}+J \theta_z)+\theta_z (\theta_z+1))\\
+&\beta_{11}^2 J (J (A_1^4 (4 \beta_{22}^2 \theta_{1_+}-8 \beta_{22} \beta_{23} \theta_{1_+}+4 \beta_{23}^2 \theta_{1_+}+4 \theta_2-\theta_z)-2 A_1^2 \theta_z (-2 \beta_{22}^2 \theta_{1_+}+4 \beta_{22} \beta_{23} \theta_{1_+}-2 \beta_{23}^2 \theta_{1_+}-2 \theta_2+\theta_z+1)\\
-&\theta_z (\theta_z+1)^2)+\theta_{\varepsilon} (A_1^2+\theta_z+1) (A_1^6+A_1^4 (2 \theta_z+1)+A_1^2 (4 \beta_{22}^2 \theta_{1_+}-8 \beta_{22} \beta_{23} \theta_{1_+}+4 \beta_{23}^2 \theta_{1_+}+4 \theta_2+\theta_z^2-1)-(\theta_z+1)^2))\\
+&\theta_{1_+} (A_1^8+A_1^6 (-4 \beta_{21} J+3 \theta_z+2)+A_1^4 (4 \beta_{21}^2 J \theta_{\varepsilon}+J)-4 \beta_{21} (2 J \theta_z+J)+4 \theta_2+3 \theta_z (\theta_z+1))\\
+&A_1^2 (4 \beta_{21}^2 J \theta_{\varepsilon} \theta_z+\theta_{\varepsilon}+J \theta_z)
-4 \beta_{21} J \theta_z (\theta_z+1)+(\theta_z+1) (4 \theta_2+\theta_z^2-\theta_z-2))-(\theta_z+1)^3),\\
0=&2 A_1 \beta_{11} \beta_{22} (J-1) \theta_{1_+} (A_1^2+\theta_z) (J-2)^+-A_1^5 (\beta_{11}^2 \theta_{\varepsilon} (J-1)+\theta_{1_+})+\theta_{\varepsilon}+1) \theta_{1_+})\\
+&2 A_1^3 (\beta_{11}^2 \theta_{\varepsilon} (J-1) (2 \beta_{21} \theta_{\varepsilon}+J)-\theta_z-1)+\beta_{11} (J-1) \theta_{1_+} (\beta_{22} (-J)+\beta_{22}+\beta_{23})+\theta_{1_+} (\beta_{21} (2 \theta_{\varepsilon}+J+1)-\theta_z-1))\\
+&A_1 (\beta_{11}^2 \theta_{\varepsilon} (J-1) (4 \beta_{21} \theta_{\varepsilon} \theta_z+\theta_{\varepsilon}+J \theta_z)-(\theta_z+1)^2)+2 \beta_{11} (J-1) \theta_{1_+} \theta_z (\beta_{22} (-J)+\beta_{22}+\beta_{23})\\
+&\theta_{1_+} (4 \beta_{21} \theta_{\varepsilon} (\theta_z+1)+2 \beta_{21} (J+1) \theta_z-(\theta_z+1)^2))\\
+&\theta_{1_+} \theta_{\varepsilon} \theta_z+\theta_{\varepsilon}+\theta_z)),\\
0=&2 A_1 \beta_{11}^3 \beta_{22} \theta_{\varepsilon} (J-1) J (J-2)^+ (\theta_{\varepsilon} (A_1^2+\theta_z+1)+J (A_1^2+\theta_z))+A_1^5 (\beta_{11}^2 \theta_{\varepsilon} (J-1) (2 \theta_{\varepsilon}+J)+2 \theta_{\varepsilon}+1) \theta_{1_+})\\
+&\theta_{1_+})-2 A_1^3 (\beta_{11}^3 \beta_{22} \theta_{\varepsilon} (J-3) (J-1) J \theta_{\varepsilon}+J)-\beta_{11}^2 \theta_{\varepsilon} (J-1) (2 \theta_{\varepsilon} (\theta_z+1)+J \theta_z)-\beta_{11} \theta_{\varepsilon}+1) J \theta_{1_+} (2 \beta_{22}-\beta_{23})\\
+&\theta_{1_+} (\beta_{21} \theta_{\varepsilon} J+\beta_{21} J-2 \theta_{\varepsilon}+1) \theta_z-2 \theta_{\varepsilon}-1))+2 \theta_{1_+} \theta_{\varepsilon} \theta_z+\theta_{\varepsilon}+\theta_z)+\theta_{1_+})\\
+&A_1 (-2 \beta_{11}^3 \beta_{22} \theta_{\varepsilon} (J-3) (J-1) J (\theta_z \theta_{\varepsilon}+J)+\theta_{\varepsilon})+\beta_{11}^2 \theta_{\varepsilon} (J-1) (\theta_z+1) (2 \theta_{\varepsilon} (\theta_z+1)+J (\theta_z-1))\\
+&2 \beta_{11} J \theta_{1_+} (2 \beta_{22}-\beta_{23}) \theta_{\varepsilon} \theta_z+\theta_{\varepsilon}+\theta_z)-2 \theta_{1_+} \theta_{\varepsilon} \theta_z+\theta_{\varepsilon}+\theta_z) (\beta_{21} J-\theta_z-1))+\theta_{1_+} \theta_{\varepsilon} \theta_z+\theta_{\varepsilon}+\theta_z)),\\
0=&A_1^3+A_1 (\beta_{11} \beta_{23} J-\beta_{21} J+\theta_z+1),\\
0=&\beta_{11}-\frac{-\frac{A_1 \beta_{11}^2 (J-1) J (A_1^2+\theta_z)}{\beta_{11}^2 J (\theta_{\varepsilon} (A_1^2+\theta_z+1)+J (A_1^2+\theta_z))+\theta_{1_+} (A_1^2+\theta_z+1)}+\frac{\beta_{11} \beta_{22} (J-1) \theta_z+\beta_{21} \theta_{\varepsilon} \theta_z+\theta_{\varepsilon}+\theta_z))}{A_1^2+\theta_z+1}+A_1}{2 (A_1^2 \theta_{\varepsilon}+1)+\theta_{\varepsilon} \theta_z+\theta_{\varepsilon}+\theta_z) (\frac{A_1 \beta_{11} J}{\beta_{11}^2 J (\theta_{\varepsilon} (A_1^2+\theta_z+1)+J (A_1^2+\theta_z))+\theta_{1_+} (A_1^2+\theta_z+1)}-\frac{A_1 \beta_{23}^2}{A_1^2+\theta_z+1})},\\
0<&\frac{A_1 \beta_{11} J}{\beta_{11}^2 J (\theta_{\varepsilon} (A_1^2+\theta_z+1)+J (A_1^2+\theta_z))+\theta_{1_+} (A_1^2+\theta_z+1)}-\frac{A_1 \beta_{23}^2}{A_1^2+\theta_z+1},\\
0<&A_1,\\
0<&\theta_z.
\end{aligned}
\end{equation}}
If it is solved,
\begin{equation}
\label{systemmixedgam0-2}
\begin{aligned}
&\Lambda_1=\frac{A_1 }{A_1^2 +\theta_z+1},\\
&\Lambda_{1_+}=\frac{A_1 \beta_{11}J}{\beta_{11}^2 J [\theta_{\varepsilon} (A_1^2+\theta_z+1)+J (A_1^2+\theta_z)]+\theta_{1_+} (A_1^2+\theta_z+1)},\\
&\Lambda_{21}=\frac{A_1-\frac{A_1 J  [\beta_{11}(\beta_{22} (J-1)+\beta_{23})+\beta_{21}]}{A_1^2 + \theta_z+1}}{\beta_{11}J},\\
&\Lambda_{22}=\Lambda_1,\\
&A_{21}=\frac{1}{2\Lambda_{22}},\\
&\alpha_{22}=-\frac{A_1^2+\theta_z+1}{2}.
\end{aligned}
\end{equation}

\newpage
The pure-strategy equilibrium can be simplified to the following system of $(\Lambda_{21},\Lambda_{22},A_1,\beta_{11},\beta_{21},\beta_{22}):$
{\scriptsize
\begin{equation}
\label{systempuregam0}
\begin{aligned}
0=&-A_1^2 \beta_{11}^4 \theta_{\varepsilon} J^3 \Lambda_{21} (\Lambda_{22} (\beta_{22} (J-1) \Lambda_{22}+\Lambda_{21}))^2-\beta_{11}^2 J (4 \theta_{\varepsilon} (\Lambda_{22}^2 (\beta_{22} \Lambda_{22} (2 (A_1^2+1) (J+1) \Lambda_{21}^2 \theta_{1_+}\\
-&2 A_1 \beta_{21} (J-1) J \Lambda_{22}+J-1)
+\Lambda_{21} (4 (A_1^2+1) \Lambda_{22}^2 \theta_2+A_1 (\beta_{21} J \Lambda_{22} (A_1 \beta_{21} J \Lambda_{22}-2)+A_1 \Lambda_{21}^2 \theta_{1_+})+\Lambda_{21}^2 \theta_{1_+})\\
+&(A_1^2+1) \beta_{22}^2 (J+1)^2 \Lambda_{21} \Lambda_{22}^2 \theta_{1_+}))
+A_1^2 J \Lambda_{21} (\theta_{1_+} (\Lambda_{22} (3 \beta_{22} J \Lambda_{22}+\beta_{22} \Lambda_{22}+3 \Lambda_{21}))^2+16 \Lambda_{22}^4 \theta_2 )\\
+&4 \beta_{11} J \Lambda_{22}(\beta_{21} \Lambda_{22} (4 (A_1^2+1) \theta_{\varepsilon} \Lambda_{21} \Lambda_{22} \theta_{1_+} (\beta_{22} (J+1) \Lambda_{22}+\Lambda_{21})\\
+&A_1^2 J \Lambda_{21} \theta_{1_+} (\Lambda_{22} (3 \beta_{22} J \Lambda_{22}+\beta_{22} \Lambda_{22}+3 \Lambda_{21}))-2 \theta_{\varepsilon} \Lambda_{22})+2 A_1 \beta_{21}^2 \theta_{\varepsilon} J \Lambda_{22}^2\Lambda_{22}\\
+&A_1 \theta_{1_+} (\beta_{22} J \Lambda_{22}+\Lambda_{21}) (\Lambda_{22} (3 \beta_{22} J \Lambda_{22}+\beta_{22} \Lambda_{22}+3 \Lambda_{21}))+4 A_1 \Lambda_{22}^2 \theta_2 \Lambda_{22})\\
-&4 \theta_{1_+} \Lambda_{22}^2 (\Lambda_{21} \Lambda_{22} (4 (A_1^2+1) \Lambda_{22} \theta_2+\beta_{21} J (A_1 (A_1 \beta_{21} \Lambda_{22} (4 \theta_{\varepsilon}+J)+2)+4 \beta_{21} \theta_{\varepsilon} \Lambda_{22}))\\
+&2 \beta_{22} J \Lambda_{22} (A_1 \beta_{21} J \Lambda_{22}+1)+\Lambda_{21})\\
-&2 A_1 \beta_{11}^3 \theta_{\varepsilon} J^2 (\Lambda_{22} (\beta_{22} (J-1) \Lambda_{22}+\Lambda_{21})) (\Lambda_{22}^2 (2 A_1 \beta_{21} J \Lambda_{21}+\beta_{22} (-J)+\beta_{22})-\Lambda_{21} \Lambda_{22}),\\
0=&-A_1^2 \beta_{11}^4 \theta_{\varepsilon} J^3 (\Lambda_{22} (\beta_{22} (J-1) \Lambda_{22}+\Lambda_{21}))^2-4 A_1^2 \beta_{11}^3 \beta_{21} \theta_{\varepsilon} J^3 \Lambda_{22} \Lambda_{22} (\Lambda_{22} (\beta_{22} (J-1) \Lambda_{22}+\Lambda_{21}))\\
-&\beta_{11}^2 J (4 \theta_{\varepsilon} (
+\Lambda_{22}^2 (A_1^2 (\beta_{21}^2 J^2 \Lambda_{22}^2+\theta_{1_+} (\beta_{22} (J+1) \Lambda_{22}+\Lambda_{21})^2+4 \Lambda_{22}^2 \theta_2)+\theta_{1_+} (\beta_{22} (J+1) \Lambda_{22}+\Lambda_{21})^2+4 \Lambda_{22}^2 \theta_2-1))\\
+&A_1^2 J (\theta_{1_+} (\Lambda_{22} (3 \beta_{22} J \Lambda_{22}+\beta_{22} \Lambda_{22}+3 \Lambda_{21}))^2+16 \Lambda_{22}^2 \theta_2 \Lambda_{22}^2))+4 \beta_{11} \beta_{21} J \Lambda_{22} \theta_{1_+}\Lambda_{22}(\Lambda_{22} (4 (A_1^2+1) \theta_{\varepsilon} \Lambda_{21}+3A_1^2 J \Lambda_{21})\\
+&\beta_{22} \Lambda_{22}^2 (4 (A_1^2+1) \theta_{\varepsilon} (J+1)+A_1^2 J (3 J+1)))-4 \theta_{1_+} \Lambda_{22}^2 (\Lambda_{22}^2 (\beta_{21}^2 J (4 (A_1^2+1) \theta_{\varepsilon}+A_1^2 J)+4 (A_1^2+1) \theta_2)-1),\\
0=&(-4 A_1^2 \Lambda_{22}^2+2 A_1 \Lambda_{22}+\beta_{11} J (\Lambda_{22} (\beta_{22} (J-1) \Lambda_{22}+\Lambda_{21}))+2 \Lambda_{22} (\beta_{21} J-2) \Lambda_{22})\\
/&(8 (A_1^2+1) \Lambda_{22}^2(\frac{A_1}{A_1^2+1}-\frac{(2 A_1 \Lambda_{22}+J (\beta_{11} \Lambda_{22} (\beta_{22} (J-1) \Lambda_{22}+\Lambda_{21})+2 \beta_{21} \Lambda_{22}^2))^2}{16 (A_1^2+1)^2 \Lambda_{22}^3}))+A_1,\\
0=&4 A_1^2 \beta_{11} \beta_{22} (J-1) \Lambda_{22} \theta_{1_+}\Lambda_{22} (J-2)^++\Lambda_{22} (\beta_{22} (J-1) \Lambda_{22}+\Lambda_{21}))+2 \beta_{11}^2 \beta_{21} \theta_{\varepsilon} (J-1) \Lambda_{22} (4 \theta_{\varepsilon} \Lambda_{22}+J 3 \Lambda_{22})\\
-&\beta_{11} \theta_{1_+} (\Lambda_{22} (\beta_{22} (7 J^2-13 J+6) \Lambda_{22}+(3 J-2) \Lambda_{21}))+2 \beta_{21} \theta_{1_+} \Lambda_{22}(\Lambda_{22} (4 \theta_{\varepsilon}+J+2)))-2 A_1 \Lambda_{22} (\beta_{11}^2 \theta_{\varepsilon} (J-1)+\theta_{1_+})\\
+&8 \beta_{21} \theta_{\varepsilon} \Lambda_{22}^2 (\beta_{11}^2 \theta_{\varepsilon} (J-1)+\theta_{1_+}),\\
0=&4 \beta_{11}^2 \beta_{22} \theta_{\varepsilon} (J-1) \Lambda_{22}^2(J-2)^+ (A_1^2 \theta_{\varepsilon}+A_1^2 J+\theta_{\varepsilon})+A_1^2 (\beta_{11}^2 \theta_{\varepsilon} (J-1) (\beta_{22} (\Lambda_{22}^2 ((11-3 J) J-2 \theta_{\varepsilon} (J-5)))\\
+&\Lambda_{21} \Lambda_{22}(2 \theta_{\varepsilon}+J))+2 \beta_{11} \beta_{21} \theta_{\varepsilon} (J-1) \Lambda_{22} (2 \theta_{\varepsilon}+J) \Lambda_{22}+4 (\theta_{\varepsilon}+1) \theta_{1_+} \Lambda_{22} (\beta_{22} (J \Lambda_{22}+\Lambda_{22})+\Lambda_{21}))\\
-&2 A_1 \beta_{11} \theta_{\varepsilon} (J-1) \Lambda_{22}+2 \theta_{\varepsilon} (\beta_{11}^2 \theta_{\varepsilon} (J-1) (\Lambda_{22} \Lambda_{21}-\beta_{22} (J-5) \Lambda_{22}^2)\\
+&2 \beta_{11} \beta_{21} \theta_{\varepsilon} (J-1) \Lambda_{22}^2+2 \theta_{1_+}\Lambda_{22} (\beta_{22} (J+1) \Lambda_{22}+\Lambda_{21})),\\
0=&\beta_{11}-\frac{2 (-(\beta_{22} (J-1) \Lambda_{22}+\Lambda_{21}) (A_1^2 (\beta_{11} \beta_{22} (J-1)+\beta_{21} (\theta_{\varepsilon}+1))+\beta_{21} \theta_{\varepsilon})-\frac{A_1^3 \beta_{11}^2 (J-1) J}{\beta_{11}^2 J (A_1^2 \theta_{\varepsilon}+A_1^2 J+\theta_{\varepsilon})+(A_1^2+1) \theta_{1_+}}+A_1)}{(A_1^2 (\theta_{\varepsilon}+1)+\theta_{\varepsilon}) (\frac{4 A_1 \beta_{11} J}{\beta_{11}^2 J (A_1^2 \theta_{\varepsilon}+A_1^2 J+\theta_{\varepsilon})+(A_1^2+1) \theta_{1_+}}-\frac{(\beta_{22} (J-1) \Lambda_{22}+\Lambda_{21})^2}{\Lambda_{22}})},\\
0<&\Lambda_{22},\\
0<&\frac{A_1}{A_1^2+1}-\frac{(2 A_1 \Lambda_{22}+J (\beta_{11} \Lambda_{22} (\beta_{22} (J-1) \Lambda_{22}+\Lambda_{21})+2 \beta_{21} \Lambda_{22}^2))^2}{16 (A_1^2+1)^2 \Lambda_{22}^3},\\
0<&\frac{4 A_1 \beta_{11} J}{\beta_{11}^2 J (A_1^2 \theta_{\varepsilon}+A_1^2 J+\theta_{\varepsilon})+(A_1^2+1) \theta_{1_+}}-\frac{(\beta_{22} (J-1) \Lambda_{22}+\Lambda_{21})^2}{\Lambda_{22}}.
\end{aligned}
\end{equation}}
If it is solved,
\begin{equation}
\label{systempuregam0-2}
\begin{aligned}
&\Lambda_1=\frac{A_1 }{A_1^2 +1},\\
&\Lambda_{1_+}=\frac{A_1 \beta_{11}J}{\beta_{11}^2 J [\theta_{\varepsilon} (A_1^2+1)+JA_1^2]+\theta_{1_+} (A_1^2+1)},\\
&A_{21}=\frac{1}{2\Lambda_{22}},\\
&\alpha_{22}=-\frac{\beta_{11}J\Lambda_{21}+J\Lambda_{22}[\beta_{21}+\beta_{23}\beta_{11}+(J-1)\beta_{22}\beta_{11}]}{2 \Lambda_{22}},\\
&\beta_{23}=-\frac{\beta_{22} (J-1) \Lambda_{22}+\Lambda_{21}}{2\Lambda_{22}}.
\end{aligned}
\end{equation}

\newpage
\noindent\textbf{Proof of Theorem \ref{thmgaminf}.} 
The mixed-strategy equilibrium can be simplified to the following system of $(A_1,\theta_z,\beta_{12})\in(\mathbb{R}^+,\mathbb{R}^+,\mathbb{R}^+):$
\begin{equation}
\label{systemmixedgaminf}
\begin{aligned}
 0&=\beta_{12}^3 J^2 (J (A_1^4 (4 \theta_{1_+}+4 \theta_2-\theta_z)-2 A_1^2 \theta_z (-2 \theta_{1_+}-2 \theta_2+\theta_z+1)-\theta_z (\theta_z+1)^2)\\
 &+\theta_\varepsilon (A_1^2+\theta_z+1) (A_1^6+A_1^4 (2 \theta_z+1)+A_1^2 (4 \theta_{1_+}+4 \theta_2+\theta_z^2-1)-(\theta_z+1)^2))\\
 &+\beta_{12}^2 J (A_1^2+\theta_z+1) (4 A_1^2 J (\theta_{1_+} (2 A_1^2+2 \theta_z-1)+\theta_2 (A_1^2+\theta_z-1))\\
 &+\theta_\varepsilon (A_1^2+\theta_z+1) (A_1^6+A_1^4 (2 \theta_z-1)+A_1^2 (4 \theta_{1_+}+4 \theta_2+\theta_z^2-1)+(\theta_z+1)^2))\\
 &+\beta_{12} J \theta_{1_+} (A_1^2+\theta_z+1) (5 A_1^6+A_1^4 (10 \theta_z-1)+A_1^2 (4 \theta_2+5 \theta_z^2-5)+(\theta_z+1)^2)\\
 &+\theta_{1_+} (A_1^2+\theta_z+1)^2 (A_1^6+A_1^4 (2 \theta_z-1)+A_1^2 (4 \theta_2+\theta_z^2-1)+(\theta_z+1)^2),\\
0&=4 A_1^2 \beta_{12} J \theta_{1_+} (A_1^4+2 A_1^2 \theta_z+A_1^2+\theta_z^2+\theta_z)+\beta_{12}^2 J (J (A_1^4 (4 \theta_{1_+}+4 \theta_2-\theta_z)\\
-&2 A_1^2 \theta_z (-2 \theta_{1_+}-2 \theta_2+\theta_z+1)-\theta_z (\theta_z+1)^2)+\theta_\varepsilon (A_1^2+\theta_z+1) (A_1^6+A_1^4 (2 \theta_z+1)\\
+&A_1^2 (4 \theta_{1_+}+4 \theta_2+\theta_z^2-1)-(\theta_z+1)^2))\\
&+\theta_{1_+} (A_1^2+\theta_z+1) (A_1^6+A_1^4 (2 \theta_z+1)+A_1^2 (4 \theta_2+\theta_z^2-1)-(\theta_z+1)^2),\\
0&=A_1^4 \theta_{1_+} (4 \theta_\varepsilon+J+2)+2 A_1^2 \theta_{1_+} (4 \theta_\varepsilon (\theta_z+1)+J \theta_z+J+2 \theta_z+1)+\beta_{12}^2 (A_1^4 J (\theta_\varepsilon+J) (4 \theta_\varepsilon+J+2)\\
&+A_1^2 J (8 \theta_\varepsilon^2 (\theta_z+1)+2 \theta_\varepsilon (5 J \theta_z+J+2 \theta_z+1)+J (J+2) (2 \theta_z-1))\\
&+J (4 \theta_\varepsilon^2 (\theta_z+1)^2+\theta_\varepsilon (\theta_z+1) (J (5 \theta_z-3)+2 \theta_z)+J (J+2) (\theta_z-1) \theta_z))\\
&+\theta_{1_+} (\theta_z+1) (4 \theta_\varepsilon (\theta_z+1)+J \theta_z+J+2 \theta_z),\\
0&<A_1,\\
0&<-\frac{A_1(\beta_{12}^2 J (\theta_\varepsilon (A_1^2+\theta_z+1)+J (A_1^2+\theta_z-1))+\theta_{1_+} (A_1^2+\theta_z+1))}{\beta_{12} J (\beta_{12}^2 J (\theta_\varepsilon (A_1^2+\theta_z+1)+J (A_1^2+\theta_z))+\theta_{1_+} (A_1^2+\theta_z+1))},\\
0&<\theta_z.   
\end{aligned}
\end{equation}
If it is solved:
\begin{equation}
\label{systemmixedgaminf-2}
\begin{aligned}
&\Lambda_1=\frac{A_1}{A_1^2+\theta_z+1},\\
&\Lambda_{1_+}=\frac{A_1 \beta_{12} J}{\beta_{12}^2 J [\theta_\varepsilon (A_1^2+\theta_z+1)+J (A_1^2+\theta_z)]+\theta_{1_+} (A_1^2+\theta_z+1)},\\
&\Lambda_{22}=\Lambda_1,\\
&\Lambda_{21}=\Lambda_{22}+\frac{A_1}{\beta_{12} J},\\
&A_{21}=\frac{1}{2\Lambda_{22}},\\
&\alpha_{22}=-\frac{A^2+\theta_z+1}{2}.\\
\end{aligned}
\end{equation}

Now we prove that $\beta_{12}>0.$ By the third equality in \eqref{systemmixedgaminf}, we have
\begin{equation*}
\begin{aligned}
f(\theta_z)&=A_1^4 J (\theta_\varepsilon+J) (4 \theta_\varepsilon+J+2)+\theta_z (8 A_1^2 \theta_\varepsilon^2 J+10 A_1^2 \theta_\varepsilon J^2+4 A_1^2 \theta_\varepsilon J+2 A_1^2 J^2 (J+2)\\
&+8 \theta_\varepsilon^2 J+2 \theta_\varepsilon J^2+2 \theta_\varepsilon J-J^2 (J+2))+8 A_1^2 \theta_\varepsilon^2 J+2 A_1^2 \theta_\varepsilon J^2+2 A_1^2 \theta_\varepsilon J-A_1^2 J^2 (J+2)\\
&+\theta_z^2 (4 \theta_\varepsilon^2 J+5 \theta_\varepsilon J^2+2 \theta_\varepsilon J+J^2 (J+2))+4 \theta_\varepsilon^2 J-3 \theta_\varepsilon J^2<0,    
\end{aligned}
\end{equation*}
Denote it as $f(\theta_z)=a\theta_z^2+b\theta_z+c,$ we have
\begin{equation*}
\begin{aligned}
&-\frac{b}{2a}-(1-A_1^2)=-\frac{16 \theta_\varepsilon^2+6 \theta_\varepsilon (2 J+1)+J (J+2)}{2 (4 \theta_\varepsilon^2+\theta_\varepsilon (5 J+2)+J (J+2))}<0,\\ &f(1-A_1^2)=4 \theta_\varepsilon J (1 + 4 \theta_\varepsilon + J)>0.    
\end{aligned}
\end{equation*}
Thus, we must have the equilibrium value $\theta_z<1-A_1^2,$ i.e., $A_1^2+\theta_z<1.$ 

Also by the third equality, we can  express $\beta_{12}^2$ through $A_1,\theta_z$. Substitute into the second inequality, 
\begin{equation*}
\begin{aligned}
&\beta_{12}^2 J (\theta_\varepsilon (A_1^2+\theta_z+1)+J (A_1^2+\theta_z-1))+\theta_{1_+} (A_1^2+\theta_z+1)\\
=&\frac{J^2\theta_{1_+}((A_1^2+\theta_z)^2-1)}{-f(\theta_z)/J}<0,
\end{aligned}
\end{equation*}
so this inequality holds if and only if $\beta_{12}>0.$

The pure-strategy equilibrium conditions can be simplified to the following system of $(\Lambda_{21},\Lambda_{22},A_1,\beta_{12})\in(\mathbb{R}^+,\mathbb{R}^+,\mathbb{R}^+,\mathbb{R}^+):$
{\footnotesize
\begin{equation}
\label{systempuregaminf}
\begin{aligned}
0&=2 A_1^4 \Lambda_{22}+\beta_{12} (A_1^2 J (\Lambda_{22}-\Lambda_{21})+J (\Lambda_{21}-\Lambda_{22}))-A_1 \beta_{12}^2 J^2 (\Lambda_{21}-\Lambda_{22})^2+A_1-2 \Lambda_{22},\\
0&=\beta_{12}^2 (A_1^3 (-J) (3 \theta_\varepsilon+J+2)-3 A_1 \theta_\varepsilon J)+A_1^3 \theta_{1_+}+\beta_{12}^3 (A_1^4 J (\theta_\varepsilon+J) (4 \theta_\varepsilon+J+2) (\Lambda_{21}-\Lambda_{22})\\
&+A_1^2 \theta_\varepsilon J (8 \theta_\varepsilon+5 J+2) (\Lambda_{21}-\Lambda_{22})+4 \theta_\varepsilon^2 J (\Lambda_{21}-\Lambda_{22}))+\beta_{12} (A_1^4 \theta_{1_+} (4 \theta_\varepsilon+J+2) (\Lambda_{21}-\Lambda_{22})\\
&+A_1^2 \theta_{1_+} (8 \theta_\varepsilon+J+2) (\Lambda_{21}-\Lambda_{22})+4 \theta_\varepsilon \theta_{1_+} (\Lambda_{21}-\Lambda_{22}))+A_1 \theta_{1_+},\\
0&=\beta_{12}^2 (A_1^3 (-J) (3 \theta_\varepsilon+J+2)-3 A_1 \theta_\varepsilon J)+A_1^3 \theta_{1_+}+\beta_{12}^3 (A_1^4 J (\theta_\varepsilon+J) (4 \theta_\varepsilon+J+2) (\Lambda_{21}-\Lambda_{22})\\
&+A_1^2 \theta_\varepsilon J (8 \theta_\varepsilon+5 J+2) (\Lambda_{21}-\Lambda_{22})+4 \theta_\varepsilon^2 J (\Lambda_{21}-\Lambda_{22}))+\beta_{12} (A_1^4 \theta_{1_+} (4 \theta_\varepsilon+J+2) (\Lambda_{21}-\Lambda_{22})\\
&+A_1^2 \theta_{1_+} (8 \theta_\varepsilon+J+2) (\Lambda_{21}-\Lambda_{22})+4 \theta_\varepsilon \theta_{1_+} (\Lambda_{21}-\Lambda_{22}))+A_1 \theta_{1_+},\\
0&=A_1^2 \beta_{12}^4 \theta_\varepsilon J^3 (\Lambda_{21}-\Lambda_{22})^2+\beta_{12}^2 J (\theta_\varepsilon (4 (A_1^2+1) \Lambda_{22}^2 (\theta_{1_+}+\theta_2)-1)\\
&+A_1^2 J (\Lambda_{21}^2 \theta_{1_+}+2 \Lambda_{21} \Lambda_{22} \theta_{1_+}+\Lambda_{22}^2 (\theta_{1_+}+4 \theta_2)))+\theta_{1_+} (4 (A_1^2+1) \Lambda_{22}^2 \theta_2-1),\\
0&<\Lambda_{22},\\
0&<\beta_{12}^2 (\theta_\varepsilon J (\Lambda_{22}-\Lambda_{21})-A_1^2 J (\theta_\varepsilon+J) (\Lambda_{21}-\Lambda_{22}))+A_1^2 \theta_{1_+} (-(\Lambda_{21}-\Lambda_{22}))+A_1 \beta_{12} J+\theta_{1_+} (\Lambda_{22}-\Lambda_{21}),\\
0&<4 A_1^3 \Lambda_{22}+A_1^2+2 A_1 \beta_{12} J (\Lambda_{22}-\Lambda_{21})+4 A_1 \Lambda_{22}-\beta_{12}^2 J^2 (\Lambda_{21}-\Lambda_{22})^2.
\end{aligned}
\end{equation}}
If it is solved:
\begin{equation}
\label{systempuregaminf-2}
\begin{aligned}
&\Lambda_1=\frac{A_1}{A_1^2+1},\\
&\Lambda_{1_+}=\frac{A_1 \beta_{12} J }{\beta_{12}^2 J (A_1^2 \theta_\varepsilon+A_1^2 J+\theta_\varepsilon)+\theta_{1_+}(A_1^2+1) },\\
&A_{21}=\frac{1}{2\Lambda_{22}},\\
&\alpha_{22}=-\frac{\beta_{12} J (\Lambda_{21}-\Lambda_{22})}{2 \Lambda_{22}}.
\end{aligned}
\end{equation}

By SOC \eqref{SOC4}, we have $\Lambda_1>0,$ so $A_1>0.$ By the second equality in \eqref{systempuregaminf}, we have
\begin{equation*}
\begin{aligned}
 \Lambda_{22}=&-(-A_1^4 \beta_{12} \Lambda_{21} (4 \theta_\varepsilon+J+2) (\beta_{12}^2 J (\theta_\varepsilon+J)+\theta_{1_+})-A_1^3 (\theta_{1_+}-\beta_{12}^2 J (3 \theta_\varepsilon+J+2))\\
 &-A_1^2 \beta_{12} \Lambda_{21} (\beta_{12}^2 \theta_\varepsilon J (8 \theta_\varepsilon+5 J+2)+\theta_{1_+} (8 \theta_\varepsilon+J+2))-A_1 (\theta_{1_+}-3 \beta_{12}^2 \theta_\varepsilon J)-4 \beta_{12} \theta_\varepsilon \Lambda_{21} (\beta_{12}^2 \theta_\varepsilon J+\theta_{1_+}))\\
 &/(-A_1^4\beta_{12} (4 \theta_\varepsilon+J+2) (\beta_{12}^2 J (\theta_\varepsilon+J)+\theta_{1_+})-A_1^2 \beta_{12} (\beta_{12}^2 \theta_\varepsilon J (8 \theta_\varepsilon+5 J+2)+\theta_{1_+} (8 \theta_\varepsilon+J+2))\\
 &-4 \beta_{12} \theta_\varepsilon (\beta_{12}^2 \theta_\varepsilon J+\theta_{1_+})).   
\end{aligned}
\end{equation*}

Substitute it into the second inequality, we have
\begin{equation*}
    \frac{A_1 (A_1^2+1) (\beta_{12}^2 \theta_\varepsilon J+\theta_{1_+})}{\beta_{12} (A_1^2 (4 \theta_\varepsilon+J+2)+4 \theta_\varepsilon)}>0,
\end{equation*}
consequently, $\beta_{12}>0.$
By \eqref{lam32}, we know that $\Lambda_{21}$ has the same sign with 
\begin{equation*}
\beta_{12} J (A_1 (\alpha_{22} \beta_{12} \theta_\varepsilon+\theta_2)+A_{21} \beta_{12} \theta_\varepsilon),
\end{equation*}
If $\alpha_{22}>0,$ we have $\Lambda_{21}>0.$ If not, by \eqref{alpha22},
\begin{equation*}
\alpha_{22}=-\frac{\beta_{12} J (\Lambda_{21}-\Lambda_{22})}{2 \Lambda_{22}}.
\end{equation*}
then $\alpha_{22}<0$ implies $\Lambda_{21}>\Lambda_{22}>0,$ so we have $\Lambda_{21}>0.$

\newpage
\noindent\textbf{Proof of Theorem \ref{thmgam10gam2inf}.}
The mixed-strategy equilibrium can be simplified to the following system of \\$(A_1,\theta_z,\beta_{21},\beta_{11},\beta_{22},\beta_{23},\beta_{12}):$
{\scriptsize
\begin{equation}
\label{systemmixedgam10gam2inf}
\begin{aligned}
0=&(A_{1} (A_{1}^2-\beta_{11} \beta_{22} J_{1}^2-\beta_{21} J_{1}+\beta_{11} \beta_{22} J_{1}-\beta_{11} \beta_{23} J_{1}+\beta_{12} J_{2}-\beta_{12} \beta_{22} J_{1} J_{2}+\theta_z+1))/((\beta_{11} J_{1}+\beta_{12} J_{2}) (A_{1}^2+\theta_z+1))\\
-&(A_{1} (-2 (\beta_{23}+\beta_{22} (J_{1}-1)) J_{1} (J_{1} \theta_z (A_{1}^2+\theta_z+1)+\theta_{\varepsilon} (-A_{1}^4+2 \beta_{12} (\beta_{22}-\beta_{23}-1) J_{2} A_{1}^2+(\theta_z+1)^2)) \beta_{11}^2\\
+&J_{1} (\theta_z A_{1}^4+2 (2 (\beta_{22}-\beta_{23}-1) \theta_{\varepsilon} (\beta_{22} J_{1}-1) J_{2} \beta_{12}^2+(-2 J_{1} \beta_{22}+\beta_{22}-\beta_{23}+1) J_{2} \theta_z \beta_{12}+\theta_z^2+2 \beta_{22}^2 J_{1} \theta_{1_+}\\
-&2 \beta_{22} \beta_{23} J_{1} \theta_{1_+}+2 \theta_2+\theta_z) A_{1}^2-(2 \beta_{12} (\beta_{23}+\beta_{22} (2 J_{1}-1)-1) J_{2}-\theta_z-1) \theta_z (\theta_z+1)+2 \beta_{21} (\theta_{\varepsilon} (A_{1}^4\\
+&2 \beta_{12} (2 \beta_{23}+\beta_{22} (J_{1}-2)+1) J_{2} A_{1}^2-(\theta_z+1)^2)-J_{1} \theta_z (A_{1}^2+\theta_z+1))) \beta_{11}-2 \beta_{22} J_{1} \theta_{1_+} (-A_{1}^4+2 \beta_{21} J_{1} A_{1}^2+(\theta_z+1)^2)\\
+&\beta_{12} J_{2} (\theta_z A_{1}^4+2 (2 \theta_{\varepsilon} J_{1} \beta_{21}^2-J_{1} \theta_z \beta_{21}+\theta_z^2+2 \beta_{22} J_{1} \theta_{1_+}+2 \theta_2+\theta_z) A_{1}^2-(2 \beta_{21} J_{1}-\theta_z-1) \theta_z (\theta_z+1))-2 \beta_{12}^2 (\beta_{22} J_{1}-1) J_{2} \\
&(J_{2} \theta_z (A_{1}^2+\theta_z+1)+\theta_{\varepsilon} (-A_{1}^4+2 \beta_{21} J_{1} A_{1}^2+(\theta_z+1)^2))))/(4 \beta_{12} J_{2} \theta_{1_+} (A_{1}^4+(2 \theta_z-2 \beta_{21} J_{1}) A_{1}^2+\theta_z^2-2 \beta_{21} J_{1} \theta_z-1) A_{1}^2\\
+&\theta_{1_+} (A_{1}^8+(3 \theta_z-4 \beta_{21} J_{1}) A_{1}^6+(4 J_{1} (\theta_{\varepsilon}+J_{1}) \beta_{21}^2-8 J_{1} \theta_z \beta_{21}+3 \theta_z^2+4 \theta_2+\theta_z-2) A_{1}^4+(4 J_{1} (\theta_z \theta_{\varepsilon}+\theta_{\varepsilon}+J_{1} \theta_z) \beta_{21}^2\\
-&4 J_{1} (\theta_z^2-1) \beta_{21}+(\theta_z+1) (\theta_z^2+\theta_z+4 \theta_2)) A_{1}^2+(\theta_z+1)^3)+\beta_{12}^2 J_{2} (4 A_{1}^2 \beta_{21}^2 J_{1} (A_{1}^2+\theta_z+1) \theta_{\varepsilon}^2+(A_{1}^8+(3 \theta_z-4 \beta_{21} J_{1}) A_{1}^6\\
+&(4 J_{1} (J_{1}+J_{2}) \beta_{21}^2-8 J_{1} \theta_z \beta_{21}+3 \theta_z^2+4 \theta_{1_+}+4 \theta_2+\theta_z-2) A_{1}^4
+(4 J_{1} (J_{1}+J_{2}) \theta_z \beta_{21}^2-4 J_{1} (\theta_z^2-1) \beta_{21}+(\theta_z+1) (\theta_z^2+\theta_z\\
+&4 \theta_{1_+}
+4 \theta_2)) A_{1}^2+(\theta_z+1)^3) \theta_{\varepsilon}+J_{2} ((4 \theta_{1_+}+4 \theta_2+\theta_z) A_{1}^4+2 \theta_z (2 \theta_{1_+}+2 \theta_2+\theta_z+1) A_{1}^2+\theta_z (\theta_z+1)^2))+\beta_{11}^2 J_{1} (4 A_{1}^2 \beta_{12}^2 \\
&(-\beta_{22}+\beta_{23}+1)^2 
J_{2} (A_{1}^2+\theta_z+1) \theta_{\varepsilon}^2+(A_{1}^8+(4 \beta_{12} (-\beta_{22}+\beta_{23}+1) J_{2}+3 \theta_z) A_{1}^6+(4 \theta_{1_+} \beta_{22}^2-8 \beta_{23} \theta_{1_+} \beta_{22}+3 \theta_z^2\\
+&4 \beta_{12}^2 (-\beta_{22}+\beta_{23}+1)^2 J_{2} (J_{1}+J_{2})
+4 \beta_{23}^2 \theta_{1_+}+4 \theta_2+8 \beta_{12} (-\beta_{22}+\beta_{23}+1) J_{2} \theta_z+\theta_z-2) A_{1}^4+(4 \beta_{12}^2 J_{2} (J_{1}+J_{2})\theta_z\\
& (-\beta_{22}+\beta_{23}+1)^2-4 \beta_{12} (\beta_{22}-\beta_{23}-1) J_{2} (\theta_z^2-1)
+(\theta_z+1) (4 \theta_{1_+} \beta_{22}^2-8 \beta_{23} \theta_{1_+} \beta_{22}+\theta_z^2+4 \beta_{23}^2 \theta_{1_+}+4 \theta_2+\theta_z)) A_{1}^2\\
+&(\theta_z+1)^3) 
\theta_{\varepsilon}+J_{1} ((4 \theta_{1_+} \beta_{22}^2-8 \beta_{23} \theta_{1_+} \beta_{22}+4 \beta_{23}^2 \theta_{1_+}+4 \theta_2+\theta_z) A_{1}^4
+2 \theta_z (2 \theta_{1_+} \beta_{22}^2-4 \beta_{23} \theta_{1_+} \beta_{22}+2 \beta_{23}^2 \theta_{1_+}+2 \theta_2+\theta_z+1) A_{1}^2\\
+&\theta_z (\theta_z+1)^2))
+2 \beta_{11} J_{1} (2 ((\beta_{22}-\beta_{23}-1) \theta_{\varepsilon} J_{2} \beta_{12}^2+\beta_{21} \theta_{\varepsilon} J_{2} \beta_{12}
+(\beta_{22}-\beta_{23}) \theta_{1_+}) A_{1}^6+(-4 (\beta_{22}-\beta_{23}-1) \theta_{\varepsilon} \\
&J_{2} (\beta_{21} (\theta_{\varepsilon}+J_{1}+J_{2})-\theta_z) \beta_{12}^2
+J_{2} (4 \beta_{22} \theta_{1_+}-4 \beta_{23} \theta_{1_+}+4 \theta_2+4 \beta_{21} \theta_{\varepsilon} \theta_z+\theta_z) \beta_{12}
-4 (\beta_{22}-\beta_{23}) \theta_{1_+} (\beta_{21} (\theta_{\varepsilon}+J_{1})-\theta_z)) A_{1}^4\\
+&2 (-(\beta_{22}-\beta_{23}-1) \theta_{\varepsilon} J_{2} (-\theta_z^2
+2 \beta_{21} ((J_{1}+J_{2}) \theta_z+\theta_{\varepsilon} (\theta_z+1))+1) \beta_{12}^2+J_{2} (\theta_z (2 \beta_{22} \theta_{1_+}-2 \beta_{23} \theta_{1_+}+2 \theta_2+\theta_z+1)\\
+&\beta_{21} \theta_{\varepsilon} (\theta_z^2-1)) \beta_{12}
-(\beta_{22}-\beta_{23}) \theta_{1_+} (-\theta_z^2+2 \beta_{21} (\theta_z \theta_{\varepsilon}+\theta_{\varepsilon}+J_{1} \theta_z)+1)) A_{1}^2+\beta_{12} J_{2} \theta_z (\theta_z+1)^2)),\\
0=&4 \beta_{12} J_{2} \theta_{1_+} (A_{1}^2+\theta_z) (A_{1}^2-2 \beta_{21} J_{1}+\theta_z+1) A_{1}^2+\theta_{1_+} (A_{1}^8+(-4 \beta_{21} J_{1}+3 \theta_z+2) A_{1}^6+(4 J_{1} (\theta_{\varepsilon}+J_{1}) \beta_{21}^2-4 (2 \theta_z J_{1}+J_{1}) \beta_{21}\\
+&4 \theta_2+3 \theta_z (\theta_z+1)) A_{1}^4+(4 J_{1} (\theta_z \theta_{\varepsilon}+\theta_{\varepsilon}+J_{1} \theta_z) \beta_{21}^2-4 J_{1} \theta_z (\theta_z+1) \beta_{21}+(\theta_z+1) (\theta_z^2-\theta_z+4 \theta_2-2)) A_{1}^2-(\theta_z+1)^3)\\
+&\beta_{12}^2 J_{2} (4 A_{1}^2 \beta_{21}^2 J_{1} (A_{1}^2+\theta_z+1) \theta_{\varepsilon}^2+(A_{1}^8+(-4 \beta_{21} J_{1}+3 \theta_z+2) A_{1}^6+(4 J_{1} (J_{1}+J_{2}) \beta_{21}^2-4 (2 \theta_z J_{1}+J_{1}) \beta_{21}+3 \theta_z^2+4 \theta_{1_+}\\
+&4 \theta_2+3 \theta_z) A_{1}^4+(\theta_z^3+4\beta_{21}J_1\theta_z(\beta_{21}J_1-1-\theta_z)+4 \beta_{21}^2 J_{1} J_{2} \theta_z-3 \theta_z+4 \theta_{1_+} (\theta_z+1)+4 \theta_2 (\theta_z+1)-2) A_{1}^2-(\theta_z+1)^3) \theta_{\varepsilon}\\
+&J_{2} ((4 \theta_{1_+}+4 \theta_2-\theta_z) A_{1}^4-2 \theta_z (-2 \theta_{1_+}-2 \theta_2+\theta_z+1) A_{1}^2-\theta_z (\theta_z+1)^2))+2 \beta_{11} J_{1} (2 ((\beta_{22}-\beta_{23}-1) \theta_{\varepsilon} J_{2} \beta_{12}^2+\beta_{21} \theta_{\varepsilon} J_{2} \beta_{12}\\
+&(\beta_{22}-\beta_{23}) \theta_{1_+}) A_{1}^6+(-2 (\beta_{22}-\beta_{23}-1) \theta_{\varepsilon} J_{2} (2 \beta_{21} (\theta_{\varepsilon}+J_{1}+J_{2})-2 \theta_z-1) \beta_{12}^2+J_{2} (4 \beta_{22} \theta_{1_+}-4 \beta_{23} \theta_{1_+}+4 \theta_2-\theta_z+2 \beta_{21} \\
&(2 \theta_z \theta_{\varepsilon}+\theta_{\varepsilon})) \beta_{12}-2 (\beta_{22}-\beta_{23}) \theta_{1_+} (2 \beta_{21} (\theta_{\varepsilon}+J_{1})-2 \theta_z-1)) A_{1}^4-2 ((\beta_{22}-\beta_{23}-1) \theta_{\varepsilon} J_{2} (2 \beta_{21} ((J_{1}+J_{2}) \theta_z+\theta_{\varepsilon} (\theta_z+1))\\
-&\theta_z (\theta_z+1)) \beta_{12}^2+J_{2} \theta_z (-2 \beta_{22} \theta_{1_+}+2 \beta_{23} \theta_{1_+}-2 \theta_2+\theta_z-\beta_{21} \theta_{\varepsilon} (\theta_z+1)+1) \beta_{12}+(\beta_{22}-\beta_{23}) \theta_{1_+} (2 \beta_{21} ((\theta_z+1) \theta_{\varepsilon}\\
+&J_{1} \theta_z)-\theta_z (\theta_z+1))) A_{1}^2-\beta_{12} J_{2} \theta_z (\theta_z+1)^2)+\beta_{11}^2 J_{1} (4 A_{1}^2 \beta_{12}^2 (-\beta_{22}+\beta_{23}+1)^2 J_{2} (A_{1}^2+\theta_z+1) \theta_{\varepsilon}^2+(A_{1}^8\\
+&(4 \beta_{12} (-\beta_{22}+\beta_{23}+1) J_{2}+3 \theta_z+2) A_{1}^6+(4 \theta_{1_+} \beta_{22}^2-8 \beta_{23} \theta_{1_+} \beta_{22}+3 \theta_z^2+4 \beta_{12}^2 (-\beta_{22}+\beta_{23}+1)^2 J_{2} (J_{1}+J_{2})+4 \beta_{23}^2 \theta_{1_+}+4 \theta_2\\
+&3 \theta_z-4 \beta_{12} (\beta_{22}-\beta_{23}-1) J_{2} (2 \theta_z+1)) A_{1}^4+(\theta_z^3+4 \beta_{12} J_{2} \theta_z^2+4 \beta_{12}^2 J_{2}^2 \theta_z+4 \beta_{12} J_{2} \theta_z+4 \beta_{12}^2 J_{1} J_{2} \theta_z+4 \theta_2 \theta_z\\
+&4 \beta_{12} \beta_{23} J_{2} (2 \beta_{12} (J_{1}+J_{2})+\theta_z+1) \theta_z-3 \theta_z+4 \theta_2+4 \beta_{22}^2 (J_{2} (J_{1}+J_{2}) \theta_z \beta_{12}^2+\theta_{1_+} (\theta_z+1))+4 \beta_{23}^2 (J_{2} (J_{1}+J_{2}) \theta_z \beta_{12}^2\\
+&\theta_{1_+} (\theta_z+1))-4 \beta_{22} (\beta_{12} J_{2} \theta_z (2 \beta_{12} (J_{1}+J_{2})+\theta_z+1)+2 \beta_{23} (J_{2} (J_{1}+J_{2}) \theta_z \beta_{12}^2+\theta_{1_+} (\theta_z+1)))-2) A_{1}^2-(\theta_z+1)^3) \theta_{\varepsilon}\\
+&J_{1} ((4 \theta_{1_+} \beta_{22}^2-8 \beta_{23} \theta_{1_+} \beta_{22}+4 \beta_{23}^2 \theta_{1_+}+4 \theta_2-\theta_z) A_{1}^4-2 \theta_z (-2 \theta_{1_+} \beta_{22}^2+4 \beta_{23} \theta_{1_+} \beta_{22}-2 \beta_{23}^2 \theta_{1_+}-2 \theta_2+\theta_z+1) A_{1}^2-\theta_z (\theta_z+1)^2)),\\
0=&\beta_{11}-(-((A_{1}^2+\theta_z) (\beta_{11} (J_{1}-1)+\beta_{12} J_{2}) (\beta_{11} J_{1}+\beta_{12} J_{2}))/(\beta_{11}^2 J_{1} (\theta_{\varepsilon} (A_{1}^2+\theta_z+1)+J_{1} (A_{1}^2+\theta_z))\\
+&2 \beta_{11} \beta_{12} J_{1} J_{2} (A_{1}^2+\theta_z)+\beta_{12}^2 J_{2} (\theta_{\varepsilon} (A_{1}^2+\theta_z+1)+J_{2} (A_{1}^2+\theta_z))+\theta_{1_+} (A_{1}^2+\theta_z+1))\\
+&\frac{2 \beta_{23} (A_{1}^2 (\beta_{11} \beta_{22} (J_{1}-1)+\beta_{12} \beta_{22} J_{2}+\beta_{21} (\theta_{\varepsilon}+1))+\beta_{22} \theta_z (\beta_{11} (J_{1}-1)+\beta_{12} J_{2})+\beta_{21} (\theta_{\varepsilon} \theta_z+\theta_{\varepsilon}+\theta_z))}{A_{1}^2+\theta_z+1}+1)\\
/&(2 (A_{1}^2 (\theta_{\varepsilon}+1)+\theta_{\varepsilon} \theta_z+\theta_{\varepsilon}+\theta_z) ((\beta_{11} J_{1}+\beta_{12} J_{2})/(\beta_{11}^2 J_{1} (\theta_{\varepsilon} (A_{1}^2+\theta_z+1)+J_{1} (A_{1}^2+\theta_z))+2 \beta_{11} \beta_{12} J_{1} J_{2} (A_{1}^2+\theta_z)\\
+&\beta_{12}^2 J_{2} (\theta_{\varepsilon} (A_{1}^2+\theta_z+1)+J_{2} (A_{1}^2+\theta_z))+\theta_{1_+} (A_{1}^2+\theta_z+1))-(\beta_{23}^2)/(A_{1}^2+\theta_z+1))),\\
\end{aligned}
\end{equation}
\begin{equation}
\label{systemmixedgam10gam2inf-2}
\begin{aligned}
0=&2 \beta_{11} \beta_{22} (J_{1}-1) (A_{1}^2+\theta_z) (J_1-2)^+ (\beta_{11} \beta_{12} \theta_{\varepsilon} J_{2}+\beta_{12}^2 (-\theta_{\varepsilon}) J_{2}-\theta_{1_+})+A_{1}^4 \beta_{12}^2 \theta_{\varepsilon} J_{2}+A_{1}^4 \theta_{1_+}+2 \beta_{11} (J_{1}-1) (A_{1}^2+\theta_z)\\
&(-\beta_{12}^2-\theta_{\varepsilon} J_{2} (\beta_{22} (-J_{1})+\beta_{22}+\beta_{23}+1)-3 \beta_{12} \beta_{21} \theta_{\varepsilon} J_{2}-\theta_{1_+} (\beta_{22} (-J_{1})+\beta_{22}+\beta_{23}))-4 A_{1}^2 \beta_{12}^2 \beta_{21} \theta_{\varepsilon}^2 J_{2}-2 A_{1}^2 \beta_{12}^2 \beta_{21} \theta_{\varepsilon} J_{1} J_{2}\\
-&4 A_{1}^2 \beta_{12}^2 \beta_{21} \theta_{\varepsilon} J_{2}^2-2 A_{1}^2 \beta_{12}^2 \beta_{21} \theta_{\varepsilon} J_{2}+2 A_{1}^2 \beta_{12}^2 \theta_{\varepsilon} J_{2} \theta_z+2 A_{1}^2 \beta_{12}^2 \theta_{\varepsilon} J_{2}+2 A_{1}^2 \beta_{12} J_{2} \theta_{1_+}-4 A_{1}^2 \beta_{21} \theta_{\varepsilon} \theta_{1_+}-2 A_{1}^2 \beta_{21} J_{1} \theta_{1_+}-2 A_{1}^2 \beta_{21} \theta_{1_+}\\
+&2 A_{1}^2 \theta_{1_+} \theta_z
+2 A_{1}^2 \theta_{1_+}+\beta_{11}^2 \theta_{\varepsilon} (J_{1}-1) (A_{1}^4+2 A_{1}^2 (\beta_{12} J_{2} (\beta_{22} (-J_{1})+\beta_{22}+\beta_{23}+1)-2 \beta_{21} (\theta_{\varepsilon}+J_{1})+\theta_z+1)-2 \beta_{12} \beta_{22} J_{1} J_{2} \theta_z\\
+&2 \beta_{12} \beta_{22} J_{2} \theta_z+2 \beta_{12} \beta_{23} J_{2} \theta_z+2 \beta_{12} J_{2} \theta_z-4 \beta_{21} \theta_{\varepsilon} \theta_z-4 \beta_{21} \theta_{\varepsilon}-4 \beta_{21} J_{1} \theta_z+\theta_z^2+2 \theta_z+1)-4 \beta_{12}^2 \beta_{21} \theta_{\varepsilon}^2 J_{2} \theta_z-4 \beta_{12}^2 \beta_{21} \theta_{\varepsilon}^2 J_{2}\\
-&2 \beta_{12}^2 \beta_{21} \theta_{\varepsilon} J_{1} J_{2} \theta_z-4 \beta_{12}^2 \beta_{21} \theta_{\varepsilon} J_{2}^2 \theta_z-2 \beta_{12}^2 \beta_{21} \theta_{\varepsilon} J_{2} \theta_z+\beta_{12}^2 \theta_{\varepsilon} J_{2} \theta_z^2+2 \beta_{12}^2 \theta_{\varepsilon} J_{2} \theta_z+\beta_{12}^2 \theta_{\varepsilon} J_{2}+2 \beta_{12} J_{2} \theta_{1_+} \theta_z\\
-&4 \beta_{21} \theta_{\varepsilon} \theta_{1_+} \theta_z-4 \beta_{21} \theta_{\varepsilon} \theta_{1_+}-2 \beta_{21} J_{1} \theta_{1_+} \theta_z-2 \beta_{21} \theta_{1_+} \theta_z+\theta_{1_+} \theta_z^2+2 \theta_{1_+} \theta_z+\theta_{1_+},\\
0=&\beta_{12}^2 \theta_{\varepsilon} J_{2}^2 A_{1}^4+2 \beta_{12}^2 \theta_{\varepsilon}^2 J_{2} A_{1}^4+2 \beta_{12}^2 \theta_{\varepsilon} J_{2} A_{1}^4+2 \theta_{\varepsilon} \theta_{1_+} A_{1}^4+2 \theta_{1_+} A_{1}^4+2 \beta_{12}^3 \beta_{22} \theta_{\varepsilon} J_{2}^3 A_{1}^2+2 \beta_{12}^3 \beta_{22} \theta_{\varepsilon}^2 J_{2}^2 A_{1}^2-2 \beta_{12}^2 \beta_{21} \theta_{\varepsilon} J_{2}^2 A_{1}^2\\
+&2 \beta_{12}^3 \beta_{22} \theta_{\varepsilon} J_{2}^2 A_{1}^2
+4 \beta_{12}^2 \theta_{\varepsilon}^2 J_{2} A_{1}^2+2 \beta_{12}^2 \theta_{\varepsilon} J_{2} A_{1}^2-2 \beta_{12}^2 \beta_{21} \theta_{\varepsilon}^2 J_{1} J_{2} A_{1}^2-2 \beta_{12}^2 \beta_{21} \theta_{\varepsilon} J_{1} J_{2} A_{1}^2+4 \theta_{\varepsilon} \theta_{1_+} A_{1}^2-2 \beta_{21} J_{1} \theta_{1_+} A_{1}^2\\
-&2 \beta_{21} \theta_{\varepsilon} J_{1} \theta_{1_+} A_{1}^2+2 \beta_{12} J_{2} \theta_{1_+} A_{1}^2+2 \beta_{12} \beta_{22} J_{2} \theta_{1_+} A_{1}^2+2 \beta_{12} \theta_{\varepsilon} J_{2} \theta_{1_+} A_{1}^2+2 \beta_{12} \beta_{22} \theta_{\varepsilon} J_{2} \theta_{1_+} A_{1}^2+2 \theta_{1_+} A_{1}^2+2 \beta_{12}^2 \theta_{\varepsilon} J_{2}^2 \theta_z A_{1}^2+4 \beta_{12}^2 \theta_{\varepsilon}^2 J_{2} \theta_z A_{1}^2\\
+&4 \beta_{12}^2 \theta_{\varepsilon} J_{2} \theta_z A_{1}^2+4 \theta_{\varepsilon} \theta_{1_+} \theta_z A_{1}^2+4 \theta_{1_+} \theta_z A_{1}^2+2 \beta_{12}^3 \beta_{22} \theta_{\varepsilon}^2 J_{2}^2-\beta_{12}^2 \theta_{\varepsilon} J_{2}^2+\beta_{12}^2 \theta_{\varepsilon} J_{2}^2 \theta_z^2+2 \beta_{12}^2 \theta_{\varepsilon}^2 J_{2} \theta_z^2+2 \beta_{12}^2 \theta_{\varepsilon} J_{2} \theta_z^2\\
+&2 \theta_{\varepsilon} \theta_{1_+} \theta_z^2+2 \theta_{1_+} \theta_z^2+2 \beta_{12}^2 \theta_{\varepsilon}^2 J_{2}-2 \beta_{12}^2 \beta_{21} \theta_{\varepsilon}^2 J_{1} J_{2}+2 \theta_{\varepsilon} \theta_{1_+}-2 \beta_{21} \theta_{\varepsilon} J_{1} \theta_{1_+}+2 \beta_{12} \theta_{\varepsilon} J_{2} \theta_{1_+}+2 \beta_{12} \beta_{22} \theta_{\varepsilon} J_{2} \theta_{1_+}+2 \beta_{12}^3 \beta_{22} \theta_{\varepsilon} J_{2}^3 \theta_z\\
+&2 \beta_{12}^3 \beta_{22} \theta_{\varepsilon}^2 J_{2}^2 \theta_z-2 \beta_{12}^2 \beta_{21} \theta_{\varepsilon} J_{2}^2 \theta_z+2 \beta_{12}^3 \beta_{22} \theta_{\varepsilon} J_{2}^2 \theta_z+4 \beta_{12}^2 \theta_{\varepsilon}^2 J_{2} \theta_z+2 \beta_{12}^2 \theta_{\varepsilon} J_{2} \theta_z-2 \beta_{12}^2 \beta_{21} \theta_{\varepsilon}^2 J_{1} J_{2} \theta_z-2 \beta_{12}^2 \beta_{21} \theta_{\varepsilon} J_{1} J_{2} \theta_z\\
+&4 \theta_{\varepsilon} \theta_{1_+} \theta_z-2 \beta_{21} J_{1} \theta_{1_+} \theta_z-2 \beta_{21} \theta_{\varepsilon} J_{1} \theta_{1_+} \theta_z+2 \beta_{12} J_{2} \theta_{1_+} \theta_z+2 \beta_{12} \beta_{22} J_{2} \theta_{1_+} \theta_z+2 \beta_{12} \theta_{\varepsilon} J_{2} \theta_{1_+} \theta_z+2 \beta_{12} \beta_{22} \theta_{\varepsilon} J_{2} \theta_{1_+} \theta_z+2 \theta_{1_+} \theta_z\\
-&2 \beta_{11}^3 \beta_{22} \theta_{\varepsilon} J_{1} (J_{1}^2-4 J_{1}+3) (J_{1} (A_{1}^2+\theta_z)+\theta_{\varepsilon} (A_{1}^2+\theta_z+1))+\beta_{11} (2 \theta_{\varepsilon} J_{2} (-A_{1}^2 (\beta_{22} J_{2} J_{1}^2+(\beta_{23}+1) (\theta_{\varepsilon}+1) J_{1}\\
-&2 \beta_{22} (\theta_{\varepsilon}+3 J_{2}+1) J_{1}+(3 \beta_{22}+\beta_{23}+1) J_{2})-(\beta_{22} J_{2} J_{1}^2+(-6 J_{2} \beta_{22}-2 \beta_{22}+\beta_{23}+1) J_{1}+(3 \beta_{22}+\beta_{23}+1) J_{2}) \theta_z\\
+&(2 \beta_{22}-\beta_{23}-1) \theta_{\varepsilon} J_{1} (\theta_z+1)) \beta_{12}^2+\theta_{\varepsilon} J_{2} (A_{1}^2+\theta_z+1) ((2 J_{1}-3) A_{1}^2+2 \beta_{21} \theta_{\varepsilon} (J_{1}-1)-2 J_{1}+2 J_{1} \theta_z-3 \theta_z+1) \beta_{12}\\
+&2 (2 \beta_{22}-\beta_{23}) J_{1} \theta_{1_+} ((\theta_{\varepsilon}+1) A_{1}^2+\theta_{\varepsilon}+\theta_{\varepsilon} \theta_z+\theta_z))+\beta_{11}^2 \theta_{\varepsilon} (J_{1}-1) (2 \theta_{\varepsilon} (A_{1}^2+\theta_z+1) (A_{1}^2+\beta_{12} (-J_{1} \beta_{22}+2 \beta_{22}+\beta_{23}+1) J_{2}\\
+&\theta_z+1)+J_{1} (A_{1}^4+2 (\beta_{12} \beta_{22} (7-2 J_{1}) J_{2}+\theta_z) A_{1}^2+\theta_z^2+2 \beta_{12} \beta_{22} (7-2 J_{1}) J_{2} \theta_z-1))+2 \beta_{11} \beta_{22} \theta_{\varepsilon} (J_{1}-1)\\
&(\beta_{11} J_{1}+\beta_{12} J_{2}) (\beta_{11} J_{1} (A_{1}^2+\theta_z)+\beta_{12} J_{2} (A_{1}^2+\theta_z)+\beta_{11} \theta_{\varepsilon} (A_{1}^2+\theta_z+1)) (J_1-2)^+,\\
0=&\frac{A_{1}^2+\beta_{11} \beta_{23} J_{1}-\beta_{12} \beta_{22} J_{2}+2 \beta_{12} \beta_{23} J_{2}+\beta_{12} J_{2}-\beta_{21} J_{1}+\theta_z+1}{2 \beta_{11} J_{1}+2 \beta_{12} J_{2}},\\
0=&\beta_{12}-(A_{1} (((A_{1}^2+\theta_z) (-(2 (\beta_{11} J_{1}+\beta_{12} (J_{2}-1)) ((A_{1}^2+\theta_z+1)^2 (\beta_{11} J_{1}+\beta_{12} J_{2})^2-(\theta_{\varepsilon} (A_{1}^2+\theta_z+1)^2 (\beta_{11}^2 J_{1}+\beta_{12}^2 J_{2})\\
+&(A_{1}^2+\theta_z)^2 (\beta_{11} J_{1}+\beta_{12} J_{2})^2+(A_{1}^2+\theta_z) (\beta_{11} J_{1}+\beta_{12} J_{2})^2+\theta_{1_+} (A_{1}^2+\theta_z+1)^2) (A_{1}^2-\beta_{11} \beta_{22} J_{1}^2+\beta_{11} \beta_{22} J_{1}-\beta_{11} \beta_{23} J_{1}\\
-&\beta_{12} \beta_{22} J_{1} J_{2}+\beta_{12} J_{2}-\beta_{21} J_{1}+\theta_z+1)))/(\theta_{\varepsilon} (A_{1}^2+\theta_z+1)^2 (\beta_{11}^2 J_{1}+\beta_{12}^2 J_{2})+(A_{1}^2+\theta_z)^2 (\beta_{11} J_{1}+\beta_{12} J_{2})^2\\
+&(A_{1}^2+\theta_z) (\beta_{11} J_{1}+\beta_{12} J_{2})^2+\theta_{1_+} (A_{1}^2+\theta_z+1)^2)-\beta_{12} J_{2} (A_{1}^2+\beta_{11} J_{1} (\beta_{22}-\beta_{23}-1)-\beta_{21} J_{1}+\theta_z+1)\\
-&\beta_{11} J_{1} (A_{1}^2-\beta_{12} \beta_{22} J_{2}+\beta_{12} \beta_{23} J_{2}+\beta_{12} J_{2}-\beta_{21} J_{1}+\theta_z+1)+2 J_{1} (\beta_{11} J_{1}+\beta_{12} J_{2}) (\beta_{11} (\beta_{22} (J_{1}-1)+\beta_{23})\\
+&\beta_{12} \beta_{22} (J_{2}-1)+\beta_{21})-\beta_{21} J_{1} (\beta_{11} J_{1}+\beta_{12} J_{2})-2 \beta_{12} (J_{2}-1) (\beta_{11} J_{1}+\beta_{12} J_{2})))\\
/&((A_{1}^2+\theta_z+1) (\beta_{11} J_{1}+\beta_{12} J_{2}))+1))/(4 (A_{1}^2 (\theta_{\varepsilon}+1)+\theta_{\varepsilon} \theta_z+\theta_{\varepsilon}+\theta_z) \\
&(\frac{A_{1} (\beta_{11} J_{1}+\beta_{12} J_{2})}{\beta_{11}^2 J_{1} (\theta_{\varepsilon} (A_{1}^2+\theta_z+1)+J_{1} (A_{1}^2+\theta_z))+2 \beta_{11} \beta_{12} J_{1} J_{2} (A_{1}^2+\theta_z)+\beta_{12}^2 J_{2} (\theta_{\varepsilon} (A_{1}^2+\theta_z+1)+J_{2} (A_{1}^2+\theta_z))+\theta_{1_+} (A_{1}^2+\theta_z+1)}\\
-&\frac{A_{1} (A_{1}^2-\beta_{11} \beta_{22} J_{1}^2+\beta_{11} \beta_{22} J_{1}-\beta_{11} \beta_{23} J_{1}-\beta_{12} \beta_{22} J_{1} J_{2}+\beta_{12} J_{2}-\beta_{21} J_{1}+\theta_z+1)}{(A_{1}^2+\theta_z+1) (\beta_{11} J_{1}+\beta_{12} J_{2})}+\frac{A_{1}-A_{1} \beta_{22} J_{1}}{A_{1}^2+\theta_z+1})),\\
0<&(\beta_{11} J_{1}+\beta_{12} J_{2})/(\beta_{11}^2 J_{1} (\theta_{\varepsilon} (A_{1}^2+\theta_z+1)+J_{1} (A_{1}^2+\theta_z))+2 \beta_{11} \beta_{12} J_{1} J_{2} (A_{1}^2+\theta_z)+\beta_{12}^2 J_{2} (\theta_{\varepsilon} (A_{1}^2+\theta_z+1)+J_{2} (A_{1}^2+\theta_z))\\
+&\theta_{1_+} (A_{1}^2+\theta_z+1))-\frac{\beta_{23}^2}{A_{1}^2+\theta_z+1},\\
0<&\frac{A_{1} (\beta_{11} J_{1}+\beta_{12} J_{2})}{\beta_{11}^2 J_{1} (\theta_{\varepsilon} (A_{1}^2+\theta_z+1)+J_{1} (A_{1}^2+\theta_z))+2 \beta_{11} \beta_{12} J_{1} J_{2} (A_{1}^2+\theta_z)+\beta_{12}^2 J_{2} (\theta_{\varepsilon} (A_{1}^2+\theta_z+1)+J_{2} (A_{1}^2+\theta_z))+\theta_{1_+} (A_{1}^2+\theta_z+1)}\\
-&\frac{A_{1} (A_{1}^2-\beta_{11} \beta_{22} J_{1}^2+\beta_{11} \beta_{22} J_{1}-\beta_{11} \beta_{23} J_{1}-\beta_{12} \beta_{22} J_{1} J_{2}+\beta_{12} J_{2}-\beta_{21} J_{1}+\theta_z+1)}{(A_{1}^2+\theta_z+1) (\beta_{11} J_{1}+\beta_{12} J_{2})}+\frac{A_{1}-A_{1} \beta_{22} J_{1}}{A_{1}^2+\theta_z+1},\\
0<&A_1,\\
0<&\theta_z.
\end{aligned}
\end{equation}}
If it is solved,
\begin{equation}
\label{systemmixedgam10gam2inf-3}
\begin{aligned}
\Lambda_1=&\frac{A_1 }{A_1^2 +\theta_z+1},\\
\Lambda_{1_+}=&A_{1} (\beta_{11} J_{1}+\beta_{12} J_{2})/\{\beta_{11}^2 J_{1} [\theta_{\varepsilon} (A_{1}^2+\theta_z+1)+J_{1} (A_{1}^2+\theta_z)]\\
+&2 \beta_{11} \beta_{12} J_{1} J_{2} (A_{1}^2+\theta_z)+\beta_{12}^2 J_{2} [\theta_{\varepsilon} (A_{1}^2+\theta_z+1)+J_{2} (A_{1}^2+\theta_z)]+\theta_{1_+} (A_{1}^2+\theta_z+1)\},\\
\Lambda_{21}=&\frac{A_{1} (A_{1}^2-\beta_{11} \beta_{22} J_{1}^2+\beta_{11} \beta_{22} J_{1}-\beta_{11} \beta_{23} J_{1}-\beta_{12} \beta_{22} J_{1} J_{2}+\beta_{12} J_{2}-\beta_{21} J_{1}+\theta_z+1)}{(A_{1}^2+\theta_z+1) (\beta_{11} J_{1}+\beta_{12} J_{2})},\\
\Lambda_{22}=&\Lambda_1,\\
A_{21}=&\frac{1}{2\Lambda_{22}},\\
\alpha_{22}=&-\frac{A_1^2+\theta_z+1}{2}.
\end{aligned}
\end{equation}

The pure-strategy equilibrium can be simplified to the following system of $(\Lambda_{21},\Lambda_{22},A_1,\beta_{11},\beta_{21},\beta_{22},\beta_{12}):$
{\scriptsize
\begin{equation}
\label{systempuregam10gam2inf}
\begin{aligned}
0=&4 A_{1}^2 \theta_{\varepsilon} J_{2}^3 \Lambda_{21} (\Lambda_{21}+(\beta_{22} J_{1}-1) \Lambda_{22})^2 \beta_{12}^4+4 A_{1} \theta_{\varepsilon} J_{2}^2 (\Lambda_{21}+(\beta_{22} J_{1}-1) \Lambda_{22}) (\Lambda_{21} (3 A_{1} \beta_{11} J_{1} \Lambda_{21}-2)\\
+&(-2 \beta_{22} J_{1}+A_{1} (\beta_{11} (3 J_{1} \beta_{22}+\beta_{22}-4)-2 \beta_{21}) \Lambda_{21} J_{1}+2) \Lambda_{22}) \beta_{12}^3+J_{2} (4 J_{2} \Lambda_{21} (4 \theta_2 \Lambda_{22}^2\\
+&(\Lambda_{21}+\beta_{22} J_{1} \Lambda_{22}+\Lambda_{22})^2 \theta_{1_+}) A_{1}^2
+4 (A_{1}^2+1) J_{1} \Lambda_{21} (\beta_{11} \theta_{\varepsilon} (\Lambda_{21}+(J_{1} \beta_{22}+\beta_{22}-2) \Lambda_{22})-2 \beta_{21} \theta_{\varepsilon} \Lambda_{22})^2\\
+&\theta_{\varepsilon} (9 A_{1}^2 \beta_{11}^2 J_{1}^2 \Lambda_{21}^3
+6 A_{1} \beta_{11} J_{1} (A_{1} J_{1} (\beta_{11} (3 J_{1} \beta_{22}+\beta_{22}-4)-2 \beta_{21}) \Lambda_{22}-2) \Lambda_{21}^2
+(\Lambda_{22} (\Lambda_{22} (J_{1} (4 J_{2} (\beta_{11}(1-\beta_{22})+2 \beta_{21})^2\\
+&J_{1} (\beta_{11} (3 J_{1} \beta_{22}+\beta_{22}-4)-2 \beta_{21})^2)+16 (\theta_{1_+}+\theta_2)) A_{1}^2-4 J_{1} (\beta_{11} (6 J_{1} \beta_{22}+\beta_{22}-7)-2 \beta_{21}) A_{1}\\
+&16 \Lambda_{22} (\theta_{1_+}+\theta_2))+4)
\Lambda_{21}+4 (\beta_{22} J_{1}-1) \Lambda_{22} (A_{1} J_{1} (2 \beta_{21}-\beta_{11} (3 J_{1} \beta_{22}+\beta_{22}-4)) \Lambda_{22}+2))) \beta_{12}^2\\
-&4 A_{1} J_{2} (A_{1} (\beta_{22}-1) \theta_{\varepsilon} J_{1}^2 \Lambda_{21} \Lambda_{22} (\Lambda_{21}+\beta_{22} (J_{1}-1) \Lambda_{22}) \beta_{11}^3+\theta_{\varepsilon} J_{1} \Lambda_{22} (-\Lambda_{21} (\beta_{22}+2 A_{1} \beta_{21} J_{1} \Lambda_{21}-1)\\
-&(\beta_{22}-1) \beta_{22} (J_{1}-1) \Lambda_{22}-2 A_{1} \beta_{21} (\beta_{22} (J_{1}-2)+1) J_{1} \Lambda_{21} \Lambda_{22}) \beta_{11}^2
-J_{1} (4 A_{1} \beta_{21}^2 \theta_{\varepsilon} J_{1} \Lambda_{21} \Lambda_{22}^2\\
-&2 \beta_{21} \theta_{\varepsilon} (\Lambda_{21}+\beta_{22} (J_{1}-2) \Lambda_{22}+\Lambda_{22}) \Lambda_{22}
+A_{1} \Lambda_{21} (8 \theta_2 \Lambda_{22}^2+(\Lambda_{21}+\beta_{22} J_{1} \Lambda_{22}+\Lambda_{22}) (3 \Lambda_{21}+\beta_{22} \Lambda_{22}+3 \beta_{22} J_{1} \Lambda_{22}) \theta_{1_+})) \beta_{11}\\
+&2 (2 \beta_{21}^2 \theta_{\varepsilon} J_{1} \Lambda_{22}^2+2 \theta_2 \Lambda_{22}^2+A_{1} \beta_{21} J_{1} \Lambda_{21} (\Lambda_{21}+\beta_{22} J_{1} \Lambda_{22}+\Lambda_{22}) \theta_{1_+} \Lambda_{22}+(\Lambda_{21}+\beta_{22} J_{1} \Lambda_{22}) (\Lambda_{21}+\beta_{22} J_{1} \Lambda_{22}+\Lambda_{22}) \theta_{1_+})) \beta_{12}\\
-&2 A_{1} J_{1} (\theta_{\varepsilon} J_{1} (\Lambda_{21}+\beta_{22} (J_{1}-1) \Lambda_{22})^2 \beta_{11}^3+4 \beta_{21} \theta_{\varepsilon} J_{1} \Lambda_{22} (\Lambda_{21}+\beta_{22} (J_{1}-1) \Lambda_{22}) \beta_{11}^2+2 (2 \beta_{21}^2 \theta_{\varepsilon} J_{1} \Lambda_{22}^2+4 \theta_2 \Lambda_{22}^2\\
+&(\Lambda_{21}+\beta_{22} J_{1} \Lambda_{22}) (3 \Lambda_{21}+\beta_{22} \Lambda_{22}+3 \beta_{22} J_{1} \Lambda_{22}) \theta_{1_+}) \beta_{11}-4 \beta_{21} \Lambda_{22} (\Lambda_{21}+\beta_{22} J_{1} \Lambda_{22}) \theta_{1_+})\\
+&A_{1}^2 \Lambda_{21} (\theta_{\varepsilon} J_{1}^3 (\Lambda_{21}+\beta_{22} (J_{1}-1) \Lambda_{22})^2 \beta_{11}^4+4 \beta_{21} \theta_{\varepsilon} J_{1}^3 \Lambda_{22} (\Lambda_{21}+\beta_{22} (J_{1}-1) \Lambda_{22}) \beta_{11}^3+J_{1} (4 \beta_{21}^2 \theta_{\varepsilon} J_{1}^2 \Lambda_{22}^2+16 (\theta_{\varepsilon}+J_{1}) \theta_2 \Lambda_{22}^2\\
+&(J_{1} (3 \Lambda_{21}+\beta_{22} \Lambda_{22}+3 \beta_{22} J_{1} \Lambda_{22})^2+4 \theta_{\varepsilon} (\Lambda_{21}+\beta_{22} (J_{1}+1) \Lambda_{22})^2) \theta_{1_+}) \beta_{11}^2-4 \beta_{21} J_{1} \Lambda_{22} (J_{1} (3 \Lambda_{21}+\beta_{22} \Lambda_{22}+3 \beta_{22} J_{1} \Lambda_{22})\\
+&4 \theta_{\varepsilon} (\Lambda_{21}+\beta_{22} (J_{1}+1) \Lambda_{22})) \theta_{1_+} \beta_{11}+4 \Lambda_{22}^2 \theta_{1_+} (J_{1} (4 \theta_{\varepsilon}+J_{1}) \beta_{21}^2+4 \theta_2))+4 (\theta_{\varepsilon} J_{1} (\theta_{1_+} \Lambda_{21}^3\\
+&\Lambda_{22} (\beta_{22} (J_{1}+(J_{1}+1) \Lambda_{21} (2 \Lambda_{21}+\beta_{22} (J_{1}+1) \Lambda_{22}) \theta_{1_+}-1)+4 \Lambda_{21} \Lambda_{22} \theta_2)) \beta_{11}^2\\
+&2 \beta_{21} \theta_{\varepsilon} J_{1} \Lambda_{22} (1-2 \Lambda_{21} (\Lambda_{21}+\beta_{22} (J_{1}+1) \Lambda_{22}) \theta_{1_+}) \beta_{11}+\theta_{1_+} (4 \Lambda_{21} (\theta_{\varepsilon} J_{1} \beta_{21}^2+\theta_2) \Lambda_{22}^2+2 \beta_{22} J_{1} \Lambda_{22}+\Lambda_{21})),\\
0=&A_{1}^2 \beta_{11}^4 \theta_{\varepsilon} J_{1}^3 (\beta_{22} (J_{1}-1) \Lambda_{22}+\Lambda_{21})^2+4 A_{1}^2 \beta_{11}^3 \theta_{\varepsilon} J_{1}^2 \Lambda_{22} (\beta_{22} (J_{1}-1) \Lambda_{22}+\Lambda_{21}) (\beta_{21} J_{1}-\beta_{12} (\beta_{22}-1) J_{2})\\
+&\beta_{11}^2 J_{1} (\theta_{\varepsilon} (A_{1}^2 (\beta_{12}^2 J_{2} (\Lambda_{22}^2 (J_{1} (3 \beta_{22} J_{1}+\beta_{22}-4)^2+4 (\beta_{22}-1)^2 J_{2})+6 J_{1} \Lambda_{21} \Lambda_{22} (3 \beta_{22} J_{1}+\beta_{22}-4)+9 J_{1} \Lambda_{21}^2)\\
+&8 \beta_{12} \beta_{21} J_{1} J_{2} \Lambda_{22} (\beta_{22} (J_{1}-2) \Lambda_{22}+\Lambda_{21}+\Lambda_{22})+4 (\beta_{21}^2 J_{1}^2 \Lambda_{22}^2+\theta_{1_+} (\beta_{22} (J_{1}+1) \Lambda_{22}+\Lambda_{21})^2+4 \Lambda_{22}^2 \theta_2))\\
+&4 \theta_{1_+} (\beta_{22} (J_{1}+1) \Lambda_{22}+\Lambda_{21})^2+16 \Lambda_{22}^2 \theta_2-4)+4 (A_{1}^2+1) \beta_{12}^2 \theta_{\varepsilon}^2 J_{2} (\Lambda_{22} (\beta_{22} J_{1}+\beta_{22}-2)+\Lambda_{21})^2\\
+&A_{1}^2 J_{1} (\theta_{1_+} (\beta_{22} (3 J_{1}+1) \Lambda_{22}+3 \Lambda_{21})^2+16 \Lambda_{22}^2 \theta_2))+4 \beta_{11} J_{1} (A_{1}^2 (\beta_{12}^3 \theta_{\varepsilon} J_{2}^2 (\Lambda_{22} (\beta_{22} J_{1}-1)+\Lambda_{21})\\
&(\Lambda_{22} (3 \beta_{22} J_{1}+\beta_{22}-4)+3 \Lambda_{21})-\beta_{12}^2 \beta_{21} \theta_{\varepsilon} J_{2} \Lambda_{22} (4 \theta_{\varepsilon} (\Lambda_{22} (\beta_{22} J_{1}+\beta_{22}-2)+\Lambda_{21})+J_{1} \Lambda_{22} (3 \beta_{22} J_{1}+\beta_{22}-4)\\
+&4 (\beta_{22}-1) J_{2} \Lambda_{22}+3 J_{1} \Lambda_{21})+\beta_{12} J_{2} (4 \beta_{21}^2 \theta_{\varepsilon} J_{1} \Lambda_{22}^2+\theta_{1_+} (\beta_{22} J_{1} \Lambda_{22}+\Lambda_{21}+\Lambda_{22}) (3 \beta_{22} J_{1} \Lambda_{22}+\beta_{22} \Lambda_{22}+3 \Lambda_{21})+8 \Lambda_{22}^2 \theta_2)\\
-&\beta_{21} \Lambda_{22} \theta_{1_+} (4 \theta_{\varepsilon} (\beta_{22} (J_{1}+1) \Lambda_{22}+\Lambda_{21})+J_{1} (3 \beta_{22} J_{1} \Lambda_{22}+\beta_{22} \Lambda_{22}+3 \Lambda_{21})))\\
+&4 \beta_{21} \theta_{\varepsilon} \Lambda_{22} (-\beta_{12}^2 \theta_{\varepsilon} J_{2} (\Lambda_{22} (\beta_{22} J_{1}+\beta_{22}-2)+\Lambda_{21})
-\theta_{1_+} (\beta_{22} (J_{1}+1) \Lambda_{22}+\Lambda_{21})))+4 (A_{1}^2 \beta_{12}^4 \theta_{\varepsilon} J_{2}^3 (\Lambda_{22} (\beta_{22} J_{1}-1)+\Lambda_{21})^2\\
-&2 A_{1}^2 \beta_{12}^3 \beta_{21} \theta_{\varepsilon} J_{1} J_{2}^2 \Lambda_{22} (\Lambda_{22} (\beta_{22} J_{1}-1)+\Lambda_{21})+\beta_{12}^2 J_{2} (4 (A_{1}^2+1) \beta_{21}^2 \theta_{\varepsilon}^2 J_{1} \Lambda_{22}^2\\
+&\theta_{\varepsilon} (\Lambda_{22}^2 (A_{1}^2 (\beta_{21}^2 J_{1} (J_{1}+4 J_{2})+4 (\theta_{1_+}+\theta_2))+4 (\theta_{1_+}+\theta_2))-1)+A_{1}^2 J_{2} (\theta_{1_+} (\beta_{22} J_{1} \Lambda_{22}+\Lambda_{21}+\Lambda_{22})^2+4 \Lambda_{22}^2 \theta_2))\\
-&2 A_{1}^2 \beta_{12} \beta_{21} J_{1} J_{2} \Lambda_{22} \theta_{1_+} (\beta_{22} J_{1} \Lambda_{22}+\Lambda_{21}+\Lambda_{22})+\theta_{1_+} (\Lambda_{22}^2 (\beta_{21}^2 J_{1} (4 (A_{1}^2+1) \theta_{\varepsilon}+A_{1}^2 J_{1})+4 (A_{1}^2+1) \theta_2)-1)),\\
0=&\frac{(A_{1}^2+1)  (-2 (A_{1}^2+1) \Lambda_{22} +A_{1} +\frac{1}{2} \beta_{11} J_{1}  (\beta_{22} (J_{1}-1) \Lambda_{22}+\Lambda_{21})+ (\beta_{12} J_{2} (\Lambda_{22} (\beta_{22} J_{1}-1)+\Lambda_{21})+\beta_{21} J_{1} \Lambda_{22}))}{4 A_{1} (A_{1}^2+1) \Lambda_{22} -\frac{1}{4}  (2 A_{1}+\beta_{11} J_{1} (\beta_{22} (J_{1}-1) \Lambda_{22}+\Lambda_{21})+2 (\beta_{12} J_{2} (\beta_{22} J_{1} \Lambda_{22}+\Lambda_{21}-\Lambda_{22})+\beta_{21} J_{1} \Lambda_{22}))^2}+A_{1},\\
0=&-4 A_{1}^2 \beta_{11} \beta_{22} (J_{1}-1) \Lambda_{22} (J_1-2)^+ (\beta_{11} \beta_{12} \theta_{\varepsilon} J_{2}+\beta_{12}^2 (-\theta_{\varepsilon}) J_{2}-\theta_{1_+})+A_{1}^2 (\beta_{11}^3 (-\theta_{\varepsilon}) (J_{1}-1) J_{1} (\beta_{22} (J_{1}-1) \Lambda_{22}+\Lambda_{21})\\
+&2 \beta_{11}^2 \theta_{\varepsilon} (J_{1}-1) \Lambda_{22} (\beta_{12} J_{2} (\beta_{22} (2 J_{1}-3)-1)+\beta_{21} (4 \theta_{\varepsilon}+3 J_{1}))-\beta_{11} (\beta_{12}^2 \theta_{\varepsilon} J_{2} ((J_{1}-1) \Lambda_{22} (7 \beta_{22} J_{1}-6 \beta_{22}-4)\\
+&(3 J_{1}-2) \Lambda_{21})-12 \beta_{12} \beta_{21} \theta_{\varepsilon} (J_{1}-1) J_{2} \Lambda_{22}+\theta_{1_+} (\beta_{22} (7 J_{1}^2-13 J_{1}+6) \Lambda_{22}+(3 J_{1}-2) \Lambda_{21}))\\
+&2 (\beta_{12}^3 (-\theta_{\varepsilon}) J_{2}^2 (\Lambda_{22} (\beta_{22} J_{1}-1)+\Lambda_{21})+\beta_{12}^2 \beta_{21} \theta_{\varepsilon} J_{2} \Lambda_{22} (4 \theta_{\varepsilon}+J_{1}+4 J_{2}+2)-\beta_{12} J_{2} \theta_{1_+} (\beta_{22} J_{1} \Lambda_{22}+\Lambda_{21}+\Lambda_{22})\\
+&\beta_{21} \Lambda_{22} \theta_{1_+} (4 \theta_{\varepsilon}+J_{1}+2)))-2 A_{1} (\beta_{11}^2 \theta_{\varepsilon} (J_{1}-1)+\beta_{12}^2 \theta_{\varepsilon} J_{2}+\theta_{1_+})+8 \beta_{21} \theta_{\varepsilon} \Lambda_{22} (\beta_{11}^2 \theta_{\varepsilon} (J_{1}-1)+\beta_{12}^2 \theta_{\varepsilon} J_{2}+\theta_{1_+}),\\
0=&4 \beta_{11} \beta_{22} \theta_{\varepsilon} (J_{1}-1) \Lambda_{22} (J_1-2)^+ (\beta_{11} (A_{1}^2 \theta_{\varepsilon}+A_{1}^2 J_{1}+\theta_{\varepsilon})+A_{1}^2 \beta_{12} J_{2})+A_{1}^2 (\beta_{11}^2 \theta_{\varepsilon} (J_{1}-1) (2 \theta_{\varepsilon} (\Lambda_{21}-\beta_{22} (J_{1}-5) \Lambda_{22})\\
+&J_{1} (\beta_{22} (11-3 J_{1}) \Lambda_{22}+\Lambda_{21}))+\beta_{11} \theta_{\varepsilon} (\beta_{12} J_{2} ((3 J_{1}-4) \Lambda_{21}-(J_{1}-1) \Lambda_{22} (\beta_{22} (J_{1}-14)+2))+2 \beta_{21} (J_{1}-1) \Lambda_{22} (2 \theta_{\varepsilon}+J_{1}))\\
+&2 (\beta_{12}^2 \theta_{\varepsilon} J_{2} (\Lambda_{22} (\beta_{22} (2 \theta_{\varepsilon} (J_{1}+1)+J_{1} J_{2}+2 J_{1}+2 J_{2}+2)-2 \theta_{\varepsilon}-J_{2}-2)+\Lambda_{21} (2 \theta_{\varepsilon}+J_{2}+2))+\beta_{12} \beta_{21} \theta_{\varepsilon} (J_{1}-2) J_{2} \Lambda_{22}\\
+&2 (\theta_{\varepsilon}+1) \theta_{1_+} (\beta_{22} (J_{1}+1) \Lambda_{22}+\Lambda_{21})))+2 A_{1} \theta_{\varepsilon} (\beta_{11} (J_{1}-1)+\beta_{12} J_{2})+2 \theta_{\varepsilon} (\beta_{11}^2 \theta_{\varepsilon} (J_{1}-1) (\Lambda_{21}-\beta_{22} (J_{1}-5) \Lambda_{22})\\
+&2 \beta_{11} \beta_{21} \theta_{\varepsilon} (J_{1}-1) \Lambda_{22}+2 (\beta_{12}^2 \theta_{\varepsilon} J_{2} (\Lambda_{22} (\beta_{22} J_{1}+\beta_{22}-1)+\Lambda_{21})+\theta_{1_+} (\beta_{22} (J_{1}+1) \Lambda_{22}+\Lambda_{21}))),\\
\end{aligned}
\end{equation}

\begin{equation}
\label{systempuregam10gam2inf-2}
\begin{aligned}
0=&\beta_{11}-(-(\beta_{22} (J_{1}-1) \Lambda_{22}+\Lambda_{21}) (A_{1}^2 (\beta_{11} \beta_{22} (J_{1}-1)+\beta_{12} \beta_{22} J_{2}+\beta_{21} (\theta_{\varepsilon}+1))+\beta_{21} \theta_{\varepsilon})\\
-&\frac{A_{1}^3 (\beta_{11} (J_{1}-1)+\beta_{12} J_{2}) (\beta_{11} J_{1}+\beta_{12} J_{2})}{\beta_{11}^2 J_{1} (A_{1}^2 \theta_{\varepsilon}+A_{1}^2 J_{1}+\theta_{\varepsilon})+2 A_{1}^2 \beta_{11} \beta_{12} J_{1} J_{2}+\beta_{12}^2 J_{2} (A_{1}^2 \theta_{\varepsilon}+A_{1}^2 J_{2}+\theta_{\varepsilon})+(A_{1}^2+1) \theta_{1_+}}+A_{1})\\
/&(2 (A_{1}^2 (\theta_{\varepsilon}+1)+\theta_{\varepsilon}) (\frac{A_{1} (\beta_{11} J_{1}+\beta_{12} J_{2})}{\beta_{11}^2 J_{1} (A_{1}^2 \theta_{\varepsilon}+A_{1}^2 J_{1}+\theta_{\varepsilon})+2 A_{1}^2 \beta_{11} \beta_{12} J_{1} J_{2}+\beta_{12}^2 J_{2} (A_{1}^2 \theta_{\varepsilon}+A_{1}^2 J_{2}+\theta_{\varepsilon})+(A_{1}^2+1) \theta_{1_+}}\\
-&\frac{(\beta_{22} (J_{1}-1) \Lambda_{22}+\Lambda_{21})^2}{4 \Lambda_{22}})),\\
0=&\beta_{12}-(A_{1} (A_{1} (4 (\beta_{11} J_{1}+\beta_{12} (J_{2}-1)) (\Lambda_{21}\\
-&\frac{A_{1} (\beta_{11} J_{1}+\beta_{12} J_{2})}{\beta_{11}^2 J_{1} (A_{1}^2 \theta_{\varepsilon}+A_{1}^2 J_{1}+\theta_{\varepsilon})+2 A_{1}^2 \beta_{11} \beta_{12} J_{1} J_{2}+\beta_{12}^2 J_{2} (A_{1}^2 \theta_{\varepsilon}+A_{1}^2 J_{2}+\theta_{\varepsilon})+(A_{1}^2+1) \theta_{1_+}})+\beta_{11} J_{1} (\beta_{22} (J_{1}-1) \Lambda_{22}-3 \Lambda_{21})\\
-&2 \beta_{12} (J_{2} (-\beta_{22} J_{1} \Lambda_{22}+\Lambda_{21}+\Lambda_{22})+2 \Lambda_{22} (\beta_{22} J_{1}-1))+2 \beta_{21} J_{1} \Lambda_{22})+2))/(8 (A_{1}^2 (\theta_{\varepsilon}+1)+\theta_{\varepsilon})\\
&(\frac{A_{1} (\beta_{11} J_{1}+\beta_{12} J_{2})}{\beta_{11}^2 J_{1} (A_{1}^2 \theta_{\varepsilon}+A_{1}^2 J_{1}+\theta_{\varepsilon})+2 A_{1}^2 \beta_{11} \beta_{12} J_{1} J_{2}+\beta_{12}^2 J_{2} (A_{1}^2 \theta_{\varepsilon}+A_{1}^2 J_{2}+\theta_{\varepsilon})+(A_{1}^2+1) \theta_{1_+}}-\beta_{22} J_{1} \Lambda_{22}-\Lambda_{21}+\Lambda_{22})),\\
0<&\frac{A_{1}}{A_{1}^2+1}-\frac{(2 A_{1}+\beta_{11} J_{1} (\beta_{22} (J_{1}-1) \Lambda_{22}+\Lambda_{21})+2 (\beta_{12} J_{2} (\beta_{22} J_{1} \Lambda_{22}+\Lambda_{21}-\Lambda_{22})+\beta_{21} J_{1} \Lambda_{22}))^2}{16 (A_{1}^2+1)^2 \Lambda_{22}},\\
0<&\frac{A_{1} (\beta_{11} J_{1}+\beta_{12} J_{2})}{\beta_{11}^2 J_{1} (A_{1}^2 \theta_{\varepsilon}+A_{1}^2 J_{1}+\theta_{\varepsilon})+2 A_{1}^2 \beta_{11} \beta_{12} J_{1} J_{2}+\beta_{12}^2 J_{2} (A_{1}^2 \theta_{\varepsilon}+A_{1}^2 J_{2}+\theta_{\varepsilon})+(A_{1}^2+1) \theta_{1_+}}-\frac{(\beta_{22} (J_{1}-1) \Lambda_{22}+\Lambda_{21})^2}{4 \Lambda_{22}},\\
0<&\frac{A_{1} (\beta_{11} J_{1}+\beta_{12} J_{2})}{\beta_{11}^2 J_{1} (A_{1}^2 \theta_{\varepsilon}+A_{1}^2 J_{1}+\theta_{\varepsilon})+2 A_{1}^2 \beta_{11} \beta_{12} J_{1} J_{2}+\beta_{12}^2 J_{2} (A_{1}^2 \theta_{\varepsilon}+A_{1}^2 J_{2}+\theta_{\varepsilon})+(A_{1}^2+1) \theta_{1_+}}-\beta_{22} J_{1} \Lambda_{22}-\Lambda_{21}+\Lambda_{22},\\
0<&\Lambda_{22}.
\end{aligned}
\end{equation}}
If it is solved,
\begin{equation}
\label{systempuregam10gam2inf-3}
\begin{aligned}
&\Lambda_1=\frac{A_1 }{A_1^2 +1},\\
&\Lambda_{1_+}=\frac{A_{1} (\beta_{11} J_{1}+\beta_{12} J_{2})}{\beta_{11}^2 J_{1} (A_{1}^2 \theta_{\varepsilon}+A_{1}^2 J_{1}+\theta_{\varepsilon})+2 A_{1}^2 \beta_{11} \beta_{12} J_{1} J_{2}+\beta_{12}^2 J_{2} (A_{1}^2 \theta_{\varepsilon}+A_{1}^2 J_{2}+\theta_{\varepsilon})+(A_{1}^2+1) \theta_{1_+}},\\
&A_{21}=\frac{1}{2\Lambda_{22}},\\
&\alpha_{22}=-\frac{\beta_{11} J_{1} [\beta_{22} (J_{1}-1) \Lambda_{22}+\Lambda_{21}]+2 [\beta_{12} J_{2} (\beta_{22} J_{1} \Lambda_{22}+\Lambda_{21}-\Lambda_{22})+\beta_{21} J_{1} \Lambda_{22}]}{4 \Lambda_{22}},\\
&\beta_{23}=-\frac{\beta_{22} (J_1-1) \Lambda_{22}+\Lambda_{21}}{2\Lambda_{22}}.
\end{aligned}
\end{equation}

\noindent\textbf{Proof of Proposition \ref{proptheta2rightarrow0}.} Through some numerical experiments, we find that $\beta_1\rightarrow0,$ and the convergence rate 
\begin{equation*}
\beta_{11}\sim O(\theta_{1_+}^a),\ 0<a\leq1.
\end{equation*}
Then we derive the exact value of $a$. Substitute \eqref{beta23j} into \eqref{SOC2}, it becomes
\begin{equation}
    \label{SOC2-beta23j}
    4\lambda_{1_+}\lambda_{22}>\lambda_{21}^2+4\lambda_{21}\lambda_{22}\beta_{22}(J-1)+\lambda_{22}^2\beta_{22}^2(J-1)^2.
\end{equation}
If $\frac{1}{2}<a<1,$
\begin{equation*}
\lambda_{1_+}\sim O(\frac{\beta_{11}}{\theta_{1_+}}),\ \lambda_{21}\geq O(\frac{\beta_{11}}{\theta_{1_+}}),\ \lambda_{22}\sim O(1),
\end{equation*}
so the SOC \eqref{SOC2-beta23j} fails if $\theta_{1_+}$ is small enough.

If $a=\frac{1}{2},$
\begin{equation*}
\lambda_{1_+}\sim O(\frac{1}{\sqrt{\theta_{1_+}}}),\ \lambda_{21}\geq O(\frac{1}{\sqrt{\theta_{1_+}}}),\ \lambda_{22}\sim O(1),
\end{equation*}
so the SOC \eqref{SOC2-beta23j} can fail either.

If $0<a<\frac{1}{2},$
\begin{equation*}
\lambda_{1_+}\sim O(\frac{1}{\sqrt{\beta_{11}}}),\ \lambda_{21}\geq O(\frac{1}{\sqrt{\beta_{11}}}),\ \lambda_{22}\sim O(1),
\end{equation*}
so the SOC \eqref{SOC2-beta23j} can fail either.

Consequently, we must have $\beta_{11}\sim O(\theta_{1_+}).$ Actually, \eqref{SOC2-beta23j} requires that $\lambda_{21}$ converges to a finite value, which implies
\begin{equation*}
\beta_{22}\beta_{11}\rightarrow0,\beta_{23}\beta_{11}\rightarrow0.
\end{equation*}
The limit equilibrium conditions are given by
\begin{equation}
\label{theta2rightarrow0lams}
\begin{aligned}
&\lambda_1=\frac{\alpha_1\sigma_v^2}{\alpha_1^2\sigma_v^2+\sigma_z^2+\sigma_1^2},\\
&\lambda_{22}=\frac{\sigma_v^2x_1}{\sigma_v^2x_1^2+\sigma_z^2\kappa_2^2+\sigma_1^2x_3^2+\sigma_\varepsilon^2J\beta_{21}^2+\sigma_2^2},\\
\end{aligned}
\end{equation}
where 
\begin{equation*}
\begin{aligned}
&x_1=\alpha_{21}+\frac{\alpha_1}{\alpha_1^2\sigma_v^2+\sigma_z^2+\sigma_1^2}(-\alpha_{21}\alpha_1\sigma_v^2+(\alpha_{22}+J\beta_{21})\sigma_1^2),\\
&x_2=\frac{x_1-\alpha_{21}}{\alpha_1},\\
&x_3=\frac{\alpha_{21}\alpha_1\sigma_v^2+(\alpha_{22}+J\beta_{21})(\alpha_1^2\sigma_v^2+\sigma_z^2)}{\alpha_1^2\sigma_v^2+\sigma_z^2+\sigma_1^2}.
\end{aligned}
\end{equation*}
And
\begin{equation}
\label{theta2rightarrow0gamj=gambeta21}
\beta_{21}=\frac{(1-\lambda_{22}\alpha_{21})\eta-\lambda_{22}(\alpha_{22}+(J-1)\beta_{21})\mu}{2(\lambda_{22}+\gamma)}.
\end{equation}

\begin{equation}
\label{theta2rightarrow0gamj=gamalpha21}
\alpha_{21}=\frac{1}{2\lambda_{22}}
\end{equation}
\begin{equation}
\label{theta2rightarrow0gamj=gamalpha22}
\alpha_{22}=-\frac{J\beta_{21}}{2}.
\end{equation}
In period 1, the mixed-strategy equilibrium conditions are still \eqref{z1} and \eqref{z2}. The pure-strategy equilibrium conditions are still \eqref{alpha1} and \eqref{SOC4}.

In mixed-strategy equilibrium, substitute \eqref{theta2rightarrow0lams} and \eqref{theta2rightarrow0gamj=gamalpha22} into \eqref{theta2rightarrow0gamj=gamalpha21}, we have
\begin{equation*}
\theta_z=\frac{4 A_1^2 \beta_{21}^2 \theta_\varepsilon J+A_1^2 \beta_{21}^2 J^2+4 A_1^2 \theta_2-4 A_{21}^2+4 \beta_{21}^2 \theta_\varepsilon J+4 \theta_2}{4 A_{21}^2-4 \beta_{21}^2 \theta_\varepsilon J-\beta_{21}^2 J^2-4 \theta_2}.
\end{equation*}
Then \eqref{z2} becomes
\begin{equation*}
A_{1}=\frac{4 A_{21}^2-4 \beta_{21}^2 \theta_\varepsilon J-\beta_{21}^2 J^2+2 \beta_{21} J-4 \theta_2}{4 A_{21}}.
\end{equation*}
\eqref{z1} becomes
\begin{equation*}
A_{21}=\frac{1}{2} \sqrt{\beta_{21}^2 J (4 \theta_\varepsilon+J)+4 \theta_2}.
\end{equation*}
We have represented all variables in $\beta_{21}$, \eqref{theta2rightarrow0gamj=gambeta21} becomes:
\begin{equation*}
\beta_{21}J(J+2+4\theta_\varepsilon)-2(J+1)=0.
\end{equation*}
Then other variables can be calculated directely.

In pure-strategy equilibrium, $\sigma_z=0.$ Substitute \eqref{theta2rightarrow0lams} and \eqref{theta2rightarrow0gamj=gamalpha22} into \eqref{theta2rightarrow0gamj=gambeta21}, we have
\begin{equation*}
\beta_{21}=\frac{2 A_1 A_{21}}{A_1^2 (4\theta_\varepsilon+J+2)+4 \theta_\varepsilon }.
\end{equation*}
Then from \eqref{theta2rightarrow0gamj=gamalpha22}, 
\begin{equation*}
A_{21}=\frac{1}{2} \sqrt{\frac{(A_1^2+1) \theta_2 (A_1^2 (4 \theta_\varepsilon+J+2)+4 \theta_\varepsilon)^2}{A_1^2 J (A_1^2 (\theta_\varepsilon+1)+\theta_\varepsilon)+(2 (A_1^2+1) \theta_\varepsilon+A_1^2)^2}}.
\end{equation*}
We have represented all variables in $A_{1},$ substitute them into \eqref{alpha1}, the system for $A_1$ is as below:
\begin{equation}
\label{systemgaminf-theta1+righhtarrow0}
\begin{aligned}
0=&A_1^{14}(J+4 \theta_\varepsilon+J \theta_\varepsilon+4 \theta_\varepsilon^2+1)(J+4 \theta_\varepsilon+2)^2\\
+&A_1^{12}(J+4 \theta_\varepsilon+2)(-3 J+4 \theta_\varepsilon+4 J \theta_\varepsilon+40 \theta_\varepsilon^2+48 \theta_\varepsilon^3+12 J \theta_\varepsilon^2-J^2-2)\\
+&A_1^{10}(64 \theta_\varepsilon^4-64 \theta_\varepsilon-4 \theta_{2}-24 J^2 \theta_\varepsilon^2-88 J \theta_\varepsilon-4 J \theta_{2}-192 \theta_\varepsilon^2-128 \theta_\varepsilon^3-8 J-32 \theta_\varepsilon \theta_{2}-160 J \theta_\varepsilon^2\\
-&32 J^2 \theta_\varepsilon
-48 J \theta_\varepsilon^3-2 J^3 \theta_\varepsilon-J^2 \theta_{2}-96 \theta_\varepsilon^2 \theta_{2}-128 \theta_\varepsilon^3 \theta_{2} -64 \theta_\varepsilon^4 \theta_{2}-5 J^2-J^3-4 J^2 \theta_\varepsilon^2 \theta_{2}-24 J \theta_\varepsilon \theta_{2}\\
-&48 J \theta_\varepsilon^2 \theta_{2}-4 J^2 \theta_\varepsilon \theta_{2}-32 J \theta_\varepsilon^3 \theta_{2}-4)\\
+&A_1^{8}(8 J-24 J^2 \theta_\varepsilon^2-4 J \theta_{2}-192 \theta_\varepsilon^2-512 \theta_\varepsilon^3-320 \theta_\varepsilon^4-32 \theta_\varepsilon \theta_{2} -160 J \theta_\varepsilon^2-192 J \theta_\varepsilon^3-2 J^2 \theta_{2}-192 \theta_\varepsilon^2 \theta_{2} \\
-&384 \theta_\varepsilon^3 \theta_{2}-256 \theta_\varepsilon^4 \theta_{2}+5 J^2+J^3-16 J^2 \theta_\varepsilon^2 \theta_{2}-48 J \theta_\varepsilon \theta_{2} 144 J \theta_\varepsilon^2 \theta_{2}-12 J^2 \theta_\varepsilon \theta_{2}-128 J \theta_\varepsilon^3 \theta_{2}+4)\\
+&A_1^6(32 \theta_\varepsilon+12 J^2 \theta_\varepsilon^2+44 J \theta_\varepsilon+96 \theta_\varepsilon^2-128 \theta_\varepsilon^3-320 \theta_\varepsilon^4  +80 J \theta_\varepsilon^2+16 J^2 \theta_\varepsilon-48 J \theta_\varepsilon^3+J^3 \theta_\varepsilon-J^2 \theta_{2}\\
-&96 \theta_\varepsilon^2 \theta_{2}-384 \theta_\varepsilon^3 \theta_{2}-384 \theta_\varepsilon^4 \theta_{2}-24 J^2 \theta_\varepsilon^2 \theta_{2}-24 J \theta_\varepsilon \theta_{2} -144 J \theta_\varepsilon^2 \theta_{2}-12 J^2 \theta_\varepsilon \theta_{2}-192 J \theta_\varepsilon^3 \theta_{2})\\
+&A_1^4(-4 \theta_\varepsilon(-24 \theta_\varepsilon-20 J \theta_\varepsilon-64 \theta_\varepsilon^2-16 \theta_\varepsilon^3-24 J \theta_\varepsilon^2-3 J^2 \theta_\varepsilon+J^2 \theta_{2}\\
+&32 \theta_\varepsilon^2 \theta_{2}+64 \theta_\varepsilon^3 \theta_{2}+12 J \theta_\varepsilon \theta_{2}+32 J \theta_\varepsilon^2 \theta_{2}+4 J^2 \theta_\varepsilon \theta_{2}))\\
+&A_1^2(-4 \theta_\varepsilon^2(-32 \theta_\varepsilon-12 J \theta_\varepsilon-48 \theta_\varepsilon^2+J^2 \theta_{2}+16 \theta_\varepsilon^2 \theta_{2}+8 J \theta_\varepsilon \theta_{2}))+64\theta_\varepsilon^4,\\
0<&\Lambda_1-\frac{1}{2A_{21}}(\frac{A_1A_{21}-\alpha_{22}}{A_1^2+1})^2.
\end{aligned}
\end{equation}


\newpage
\bibliographystyle{unsrt}
\bibliography{bibfile}
\addcontentsline{toc}{section}{Reference} 

\end{document}